\documentclass[fleqn,usenatbib]{mnras}

\RequirePackage{rotating}

\usepackage{newtxtext,newtxmath}

\usepackage[T1]{fontenc}
\usepackage{ae,aecompl}

\usepackage{graphicx}	

\usepackage{savesym}
\savesymbol{Bbbk}
\usepackage{amsmath}	
\usepackage{amssymb}	
\restoresymbol{Bbbk1}{Bbbk}

\usepackage[dvipsnames]{xcolor}
\usepackage{pdflscape}
\usepackage{multirow}
\usepackage{upgreek}

\usepackage{soul}

\newcommand{\um}{$\upmu$m}

\newcommand{\dsec}{$\rlap{.}^s$}

\title[The JCMT GBS Core Catalogue]{The JCMT Gould Belt Survey Complete Core Catalogue: Core mass function variations between nearby molecular clouds}

\author[K. Pattle et al.]{
Kate Pattle,$^{1,2,3}$\thanks{E-mail: k.pattle@ucl.ac.uk}
James Di Francesco,$^{4,5}$
Jenny Hatchell,$^{6}$
Helen Kirk,$^{4,5}$
Sarah Sadavoy,$^{7}$
\newauthor 
Derek Ward-Thompson,$^{8}$
Doug Johnstone,$^{4,5}$
Sammohith Nittala,$^{9,4}$
Ronan Kerr,$^{10,4}$
Jared Keown,$^{4,5}$
\newauthor
Harold Butner,$^{11}$
Simon Coud\'e,$^{12,13}$,
Malcolm Currie,$^{14,15}$
Rachel Friesen,$^{10}$
Tim Jenness,$^{17,15}$
\newauthor
Lewis Knee,$^{4}$
and Glenn White$^{18,14}$
\\
$^{1}$Department of Physics and Astronomy, University College London, Gower Street, London WC1E 6BT, UK\\
$^{2}$Institute of Astronomy, Department of Physics, National Tsing Hua University, Guangfu Road, Hsinchu 30013, Taiwan\\
$^{3}$Centre for Astronomy, School of Natural Sciences, University of Galway, University Road, Galway H91 TK33, Ireland\\
$^{4}$NRC Herzberg Astronomy and Astrophysics Research Centre, 5071 West Saanich Road, Victoria, BC V9E 2E7, Canada\\
$^{5}$Department of Physics and Astronomy, University of Victoria, 3800 Finnerty Road, Victoria, BC, V8P 5C2, Canada\\
$^{6}$Physics and Astronomy, University of Exeter, Stocker Road, Exeter EX4 4QL, UK\\
$^{7}$Department of Physics, Engineering Physics and Astronomy, Queens University, Kingston, ON, K7L 3N6, Canada \\
$^{8}$Jeremiah Horrocks Institute, University of Lancashire, Preston PR1 2HE, UK\\
$^{9}$Center for the Physics of Materials, Department of Physics, McGill University, 3600 Rue University, Montreal, QC, H3A 2T8, Canada\\
$^{10}$Dunlap Institute for Astronomy \& Astrophysics, University of Toronto, Toronto, ON M5S 3H4, Canada\\
$^{11}$Department of Physics and Astronomy, James Madison University, MSC 4502, 901 Carrier Drive, Harrisonburg, Virginia 22807, USA\\
$^{12}$Department of Earth, Environment, and Physics, Worcester State University, Worcester, MA 01602, USA\\
$^{13}$Center for Astrophysics | Harvard \& Smithsonian, 60 Garden Street, Cambridge, MA 02138, USA \\
$^{14}$RAL Space, Rutherford Appleton Laboratory, Harwell Campus, Didcot, Oxfordshire, OX11 0QX, UK\\
$^{15}$East Asian Observatory, 660 N. A`oh\={o}k\={u} Place, University Park, Hilo, HI 96720, USA\\
$^{16}$Vera C. Rubin Observatory, 950 North Cherry Avenue, Tucson, AZ 85719, USA\\
$^{17}$Department of Physical Sciences, The Open University, Milton Keynes MK7 6AA, UK
}

\date{Accepted XXX. Received YYY; in original form ZZZ}

\pubyear{2025}

\begin{document}
\label{firstpage}
\pagerange{\pageref{firstpage}--\pageref{lastpage}}
\maketitle

\begin{abstract}
We present a catalogue of dense cores identified in James Clerk Maxwell Telescope (JCMT) Gould Belt Survey SCUBA-2 observations of nearby star-forming clouds.  We identified 2257 dense cores using the \textit{getsources} algorithm, of which 59\% are starless, and 41\% are potentially protostellar.  71\% of the starless cores are prestellar core candidates, suggesting a prestellar core lifetime similar to that of Class 0/I YSOs.  Higher-mass clouds have a higher fraction of prestellar cores compared to protostars, suggesting a longer average prestellar core lifetime.  We assessed completeness by inserting critically-stable Bonnor--Ebert spheres into a blank SCUBA-2 field: completeness scales as distance squared, with an average mass recovery fraction of $73\pm6$\% for recovered sources.  We calculated core masses and radii, and assessed their gravitational stability using the Bonnor--Ebert criterion.   Maximum starless core mass scales with cloud complex mass with an index $0.58\pm 0.13$, consistent with the behaviour of maximum stellar masses in embedded clusters.  We performed least-squares and Monte Carlo modelling of the core mass functions (CMFs) of our starless and prestellar core samples.  The CMFs can be characterised using log-normal distributions: we do not sample the full range of core masses needed to create the stellar Initial Mass Function (IMF).  The CMFs of the clouds are not consistent with being drawn from a single underlying distribution.  The peak mass of the starless core CMF increases with cloud mass; the prestellar CMF of the more distant clouds has a peak mass $\sim 3\times$ the log-normal peak for the system IMF, implying a $\sim 33$\% prestellar core-to-star efficiency.  
\end{abstract}




\section{Introduction}
\label{sec:intro}

\begin{figure*}
    \includegraphics[width=\textwidth]{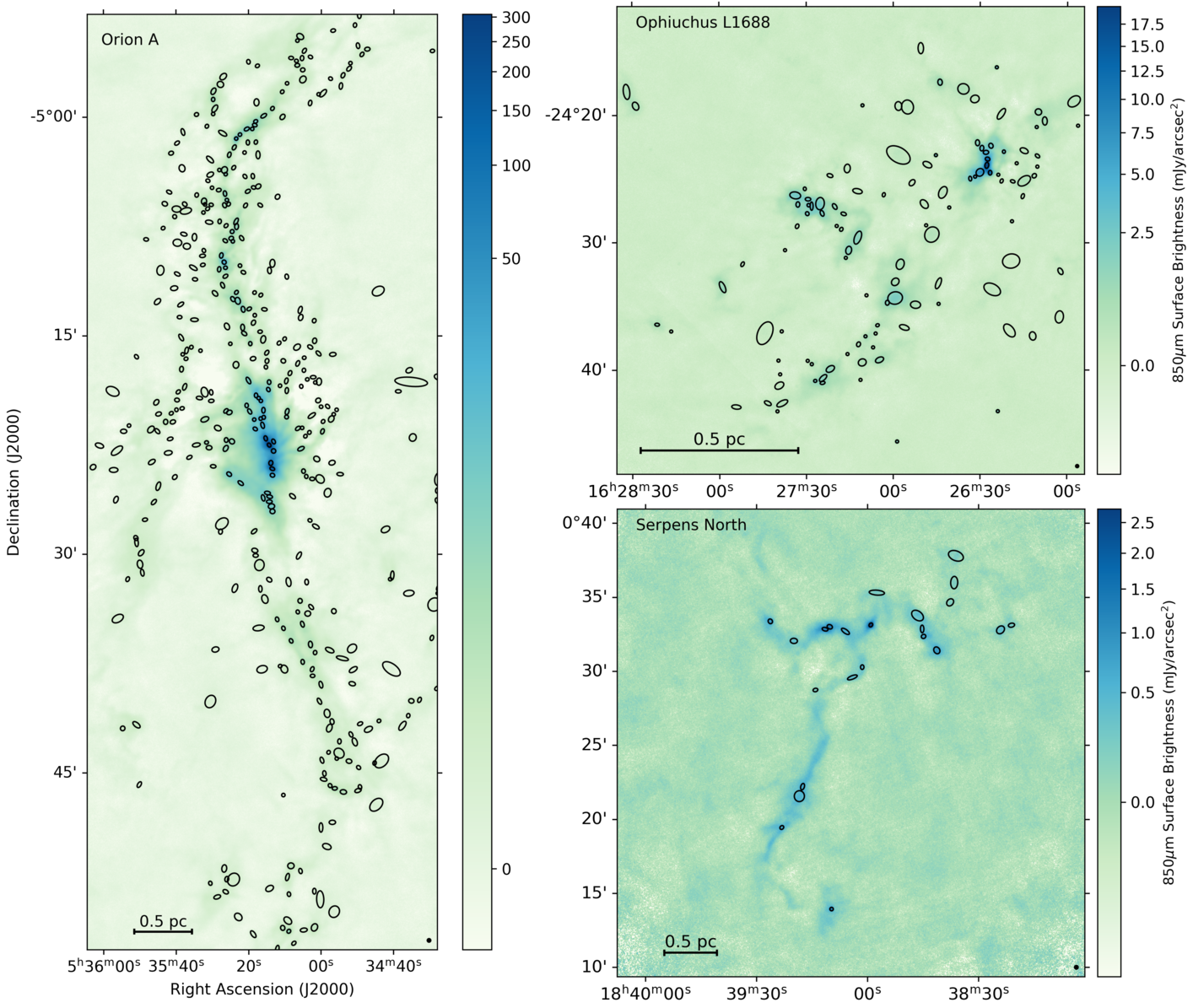}
    \caption{Extracts from three of the regions which we observed with SCUBA-2 as part of the JCMT Gould Belt Survey: the Integral Filament in Orion A (left), the L1688 region of the Ophiuchus Molecular Cloud (top right), and the Serpens North region (bottom right).  Each panel shows SCUBA-2 850\,\um\ emission, fourth-root scaled in the case of Orion A, and cube-root scaled for the other regions.  Cores identified in our catalogue are marked with black ellipses, with ellipse diameter marking the FWHM size of the core.  A 0.5\,pc scale bar is shown in the lower left-hand corner of each panel; the JCMT 850\,\um\ beam size is shown as a filled black circle in the lower right-hand corner.  The full set of regions observed are shown in Appendix~\ref{sec:appendix_images}.}
    \label{fig:fig1}
\end{figure*}

This paper presents a catalogue of cores identified from the Submillimetre Common-User Bolometer Array 2 (SCUBA-2; \citealt{holland2013}) 850\,$\upmu$m and 450\,$\upmu$m data taken by the James Clerk Maxwell Telescope (JCMT) Gould Belt Survey (GBS), a JCMT Legacy Program designed to map across regions of high extinction in 13 star-forming molecular clouds within 500 pc distance that are visible from Maunakea, Hawai`i. 
The JCMT GBS was first presented to the community in an overview paper by \citet{wardthompson2007}.  Since then, several papers have presented results from the Heterodyne Array Receiver Program (HARP; \citealt{buckle2009}) and SCUBA-2 components of the GBS (see Table~\ref{tab:fields} for references).  The GBS originally had a planned polarimetric mapping component, which due to the delayed commissioning of SCUBA-2's POL-2 polarimeter \citep{friberg2016} instead became the basis for the JCMT B-Fields in Star-forming Region Observations (BISTRO; \citealt{wardthompson2017}) Survey.  In this current paper, we examine the GBS SCUBA-2 data for all of the target clouds, to identify the cores within them in a self-consistent way.

Cores are the interface between larger molecular clouds and the star formation within them.  Physically, they are compact locations of high gas and dust density in clouds.  Functionally, they have been identified from observations of the optically thin millimetre/submillimetre continuum emission from dust (e.g. \citealt{ladd1991}) or specific, largely optically thin, line emission of molecules excited at moderate densities (e.g., 10$^4$ cm$^{-3}$ or higher) that are still abundant in cold ($\sim\,10\,$K), dense environments (e.g. \citealt{benson1989}).   In these investigations, locally bright emission is considered to indicate the presence of a locally dense configuration of gas and dust.  Cores can therefore teach us about the star-formation process, as gas and dust can be accumulated within them to a point where they collapse from their own gravity to form young stellar objects (YSOs), which themselves eventually become stars.  Those cores without YSOs are named ``starless cores" \citep{benson1989} and the subset of these that appear to be gravitationally bound are named ``prestellar cores" \citep{wardthompson1994}.  Meanwhile, those with YSOs are named ``protostellar cores", with the detected mass comprising the extended envelopes of the embedded Class 0 or I YSOs \citep[see][for reviews]{difrancesco2007,wardthompson2007,andre2014}.

A robust catalogue of cores within a star-forming cloud can be useful because it provides the locations of objects with which further, more detailed investigations of ongoing star formation can be performed, e.g., examinations of the internal physical or chemical structures of specific cores.  Such catalogues are also useful for identifying the populations of cores, i.e., snapshots of the current star-forming activity in various clouds.  Moreover, comparing core populations between clouds of different character (e.g., mass, temperature, dense gas fraction, metallicity) can in principle provide us with insights into any similarities and differences of star formation over different environments.  A key goal of the JCMT GBS was to address this latter point, by acquiring extensive and highly sensitive submillimetre continuum maps of nearby star-forming clouds to identify their core populations.

Many previous investigations of molecular cloud core populations have focused on the number distributions of core masses, i.e., their Core Mass Functions (CMFs).  The earliest studies of core populations in nearby clouds from their millimetre or submillimetre emission \citep{motte1998,testi1998,johnstone2000,johnstone2001,stanke2006} found remarkable similarities between the shape of the CMFs and that of the Initial Mass Function (IMF; \citealt{salpeter1955}; \citealt{kroupa2001}; \citealt{chabrier2003}), with power-law slopes at the high-mass ends of the CMFs being statistically similar to that of the IMF, i.e., $-1.35$ in log-log space \citep{salpeter1955}.  Attention was given to the higher-mass ends of the CMFs due to the relatively limited sensitivities of the instruments involved, and little distinction was made between starless and protostellar cores in these studies.  However, an early multi-cloud comparison of the CMFs of five nearby clouds by \citet{sadavoy2010} drew on archival SCUBA data and distinguished between their starless and protostellar cores, finding consistency within errors between slopes of the starless CMFs of the five clouds and the Salpeter IMF slope, although accurately distinguishing between starless and protostellar cores remains difficult.

More recently, more sensitive instruments such as SCUBA-2 on the JCMT, SPIRE and PACS on the \emph{Herschel} Space Observatory, and the Atacama Large Millimetre Array (ALMA) have enabled more thorough investigations of the CMF and its relationship to the IMF (e.g. \citealt{offner2014}).  These studies have allowed the the high-mass end of the CMF to be well-constrained, and characterizations of them to go beyond the high-mass power-law slope to probe the lower-mass regime where the majority of cores must lie.  Notably, Herschel data of the Aquila Rift showed that its prestellar CMF has a log-normal shape consistent with that of the Chabrier system IMF \citep{chabrier2003}, but one shifted to higher overall masses by a factor of ~2--4 \citep{andre2010,konyves2015}.  Similar CMF results were found with Herschel data of other clouds, e.g., Orion B \citep{konyves2020} and the Cepheus Flare clouds \citep{difrancesco2020}.  In contrast, ALMA studies of more distant Giant Molecular Clouds (GMCs) have suggested the presence of ``top-heavy" CMFs, e.g., in W43-MM1 \citep{motte2018}.  A recent study by the ALMA-IMF Large Project \citep{motte2022,ginsburg2022}, which identified core populations in the protocluster regions of 15 GMCs at 2.0--5.5 kpc distance, has revealed other ``top-heavy" CMFs, including all core types, but curiously the high-mass-end slopes of these are more Salpeter-like for arguably more-evolved protocluster regions \citep{pouteau2023}.  In addition, a recent ALMA-IMF study of the CMFs specifically of W43, i.e., its three subregions MM1, MM2 and MM3, shows that the slope of W43's prestellar CMF is consistent with that of the high-mass end of the Salpeter IMF \citep{nony2023}. 

Recent investigations of the stellar IMF based on \textit{Gaia} samples conclude that it varies little between nearby star-forming regions -- the low-mass Taurus molecular cloud has the same IMF as the massive Orion Nebula cluster \citep{luhman:2018}.  However, if a sub-sample is drawn from the IMF, then the maximum stellar mass increases with the size of the sample \citep{weidnerkroupa:2006, WeidnerKroupaBonnell:2010, Elmegreen:2006}.  Only the largest clouds (with masses above around $10^4$\,M$_{\odot}$) sample the full mass distribution and include the highest mass stars.  In a hierarchical star-formation model, clusters build from star formation in smaller components, such as the clouds and filaments that are evident in the SCUBA-2 and Herschel data \citep{Parker:2017}.  Each of these components contributes a fraction of the mass of the final cluster.  These smaller samples of stars do not fully represent the IMF, and we might expect that smaller samples of cores might likewise under-represent the full mass distribution of the CMF.  This possibility is something we can test with the GBS dataset, which covers a range of cloud masses from 200 to 20\,000 ${\rm M}_{\odot}$.

In this work, we describe the core populations of 12 nearby clouds drawn from the JCMT GBS SCUBA-2 data, in order to provide further insight into their nature and their CMFs.  In Section~\ref{sec:dr}, we summarize the observations and data reduction steps taken to produce the maps used in the subsequent analysis.  In Section~\ref{sec:source_extraction}, we describe how the cores were extracted from the dataset, and how catalogue completeness was determined.  In Section~\ref{sec:cat}, we present the JCMT GBS core catalogue, describing its contents, core classification, and determination of individual core properties such as temperature, mass and density. In Section~\ref{sec:properties}, we describe the overall characteristics of the core populations, including mass vs. size, relative numbers of starless and protostellar sources, and core stability.  In Section~\ref{sec:cmfs} we construct CMFs for the clouds that we survey, and in Section~\ref{sec:cmf_implications} we discuss the implications of these CMFs.  In Section~\ref{sec:conclusions}, we summarize the conclusions of this paper.

\section{Observations and Data Reduction}
\label{sec:dr}

The JCMT Gould Belt Survey carried out SCUBA-2 observations of nearby molecular clouds between October 2011 and January 2015.  The observing strategy is described in more detail by \citet{kirk2018}.  In brief, each field was observed between four and six times depending on the weather conditions to obtain comparable sensitivities at 850~$\upmu$m.  Weather has a larger impact on the 450~$\upmu$m data, and hence there is a larger variation in the sensitivities of those maps.  Several observing modes were tested during the science verification phase, but the main survey used the PONG1800 \citep{Kackley2010} mapping mode.  We consider only the PONG1800 observations here.
The PONG1800 mode produces a circular field of 0.5$^{\circ}$ diameter with near-uniform sensitivity \citep{holland2013}.  For large GBS fields, several PONG1800 maps were stitched together using the \textsc{ccd-pack} \textit{makemos} tool in the \textit{Starlink} software package.  A summary of the regions mapped, their central coordinates, coverage area, and associated publications are given in Table~\ref{tab:fields}.  The data used in this paper were taken under project codes MJLSG31 (Orion A), MJLSG32 (Ophiuchus), MJLSG33 (Aquila and Serpens), MJLSG34 (Lupus), MJLSG35 (Corona Australis), MJLSG36 (IC 5146), MJLSG37 (Auriga and Taurus), MJLSG38 (Perseus), MJLSG39 (Pipe Nebula), MJLSG40 (Cepheus) and MJLSG41 (Orion B).  

All ground-based submillimetre telescopes are unable to map large-scale emission structures, due to the bright and varying emission from the Earth's atmosphere.  Newer instrumentation and observing techniques have resulted in better recovery of such larger structures than was previously possible, and the JCMT GBS team spent significant effort testing different data reduction schemes to maximize the recovery of large-scale emission, as summarized by \cite{kirk2018}.  Since these reduction techniques were developed alongside the science analysis of the survey data, some published papers from the survey listed in Table~\ref{tab:fields} used earlier reduction methods.  Maps of all regions using the best reduction method are available for public download through \citet{kirk2018}, or directly at the DOI\footnote{{\tt \url{https://doi.org/10.11570/18.0005}}}.  In this work, we extract sources for the JCMT GBS core catalogue from this specific set of maps.

\begin{figure}
    \centering
    \includegraphics[width=\columnwidth]{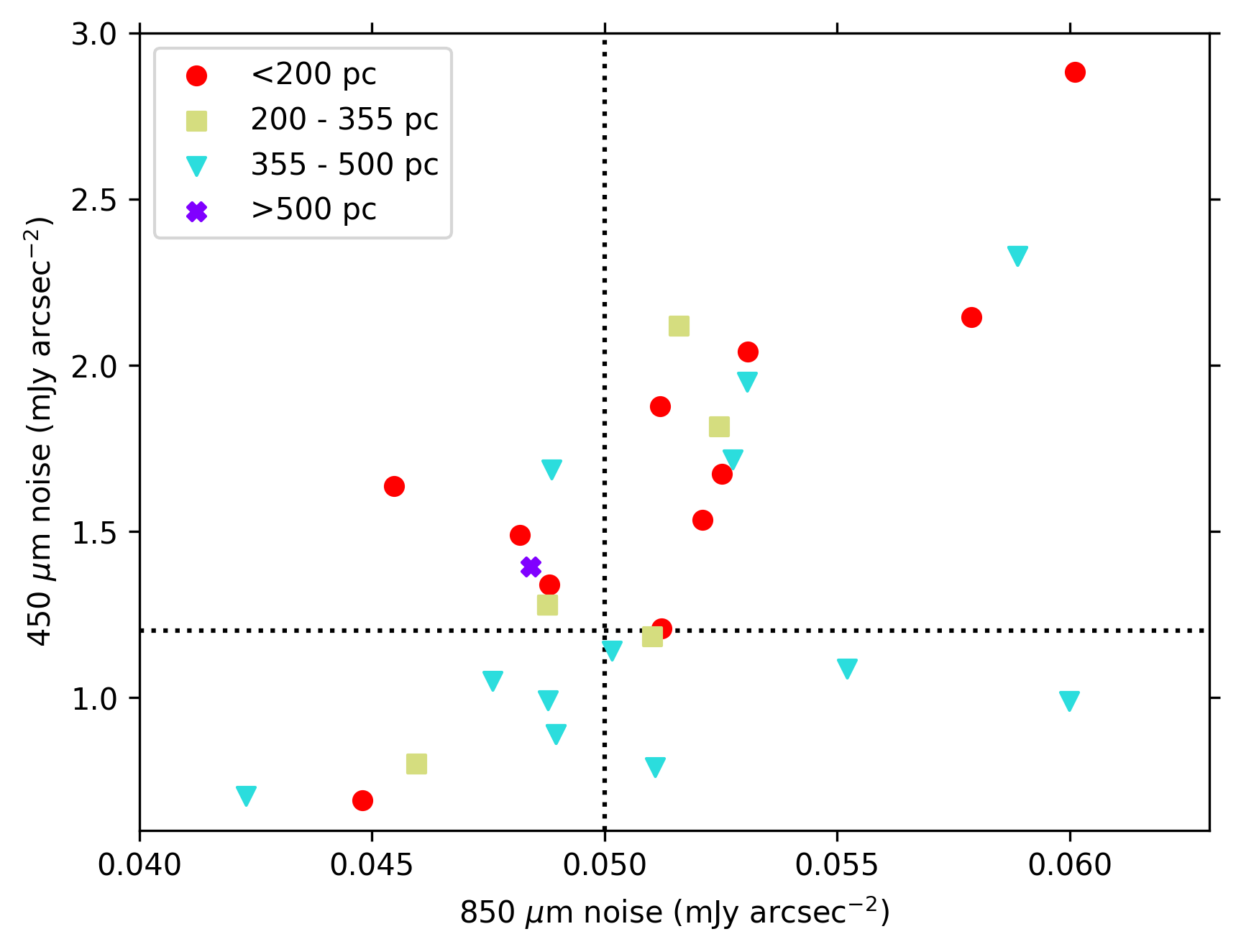}
    \caption{Representative rms values per mosaic at 850~$\upmu$m and 450~$\upmu$m.  The dashed lines show the median rms per observing field (see text for more details).  The points are colour-coded to the approximate distance to each cloud, as used in later analysis.}
    \label{fig:mosaic_rms}
\end{figure}

Figure~\ref{fig:mosaic_rms} shows representative 1-$\sigma$ rms values for each mosaic at 850~$\upmu$m and 450~$\upmu$m.  We calculated these values by taking the median of the rms values per observing area (PONG1800 area) included in each mosaic.  The dashed lines show the median rms of 0.050 mJy arcsec$^{-2}$ and 1.2 mJy arcsec$^{-2}$ at 850~$\upmu$m and 450~$\upmu$m respectively as measured from all individual PONG1800 observing areas.  Note that the larger mosaics, such as in Orion and Perseus, had more high-priority fields observed in better weather conditions with fewer integrations needed to reach the same 850~$\upmu$m noise level.  With their greater sensitivity to the sky conditions, the 450~$\upmu$m observations tend therefore to have lower noise levels in the larger mosaics. Since Figure~\ref{fig:mosaic_rms} plots the noise over a per-mosaic area rather than per-PONG area, and there are a greater number of small-area mosaics compared to the large-area mosaics, the median of all 450~$\upmu$m noise levels is at a noticeably lower value than the median of the per-region values shown in the figure.

\subsection{Mapping Completeness}
 
\begin{figure}
    \centering
    \includegraphics[width=\columnwidth]{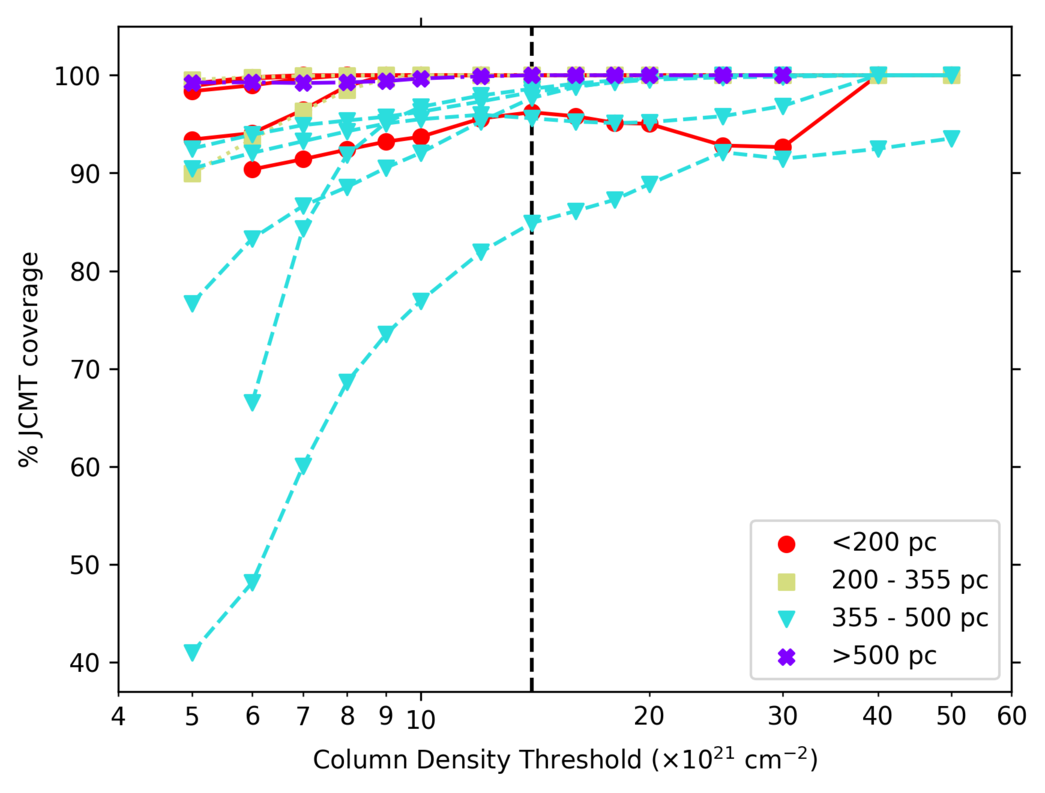}
    \caption{The fraction of material at a given \textit{Herschel} Gould Belt Survey column density or higher mapped by the JCMT GBS.  The vertical dashed line shows a column density of $N({\rm H}_{2}) = 1.4\times10^{22}$~cm$^{-2}$, which corresponds approximately to the JCMT GBS mapping goal of $A_{V} \gtrsim 7$~mag.  Each line represents a different cloud complex that we consider.}
    \label{fig:Av_completeness}
\end{figure}

One important aspect that all surveys need to consider is how complete or how representative their mapped areas are.  For the JCMT
GBS, our goal was to map as much as possible of the highest column density material in the Gould Belt (nearby) molecular clouds visible from the JCMT.  At the time, evidence suggested that most dense cores were found in clouds with $A_V \gtrsim 5-7$~mag \citep[e.g.,][]{Johnstone04,Hatchell05,Kirk06,Enoch06,Froebrich10,Belloche11}, so the GBS team targeted these areas, as were best known at the time.  Here, we examine how complete our final mapped areas are, in the context of column densities of the Gould Belt clouds as derived by the Herschel Gould Belt Survey \citep{andre2010,Palmeirim2013,schneider2013,polychroni2013,rygl2013,konyves2015,bresnahan2018,Arzoumanian2019,difrancesco2020,ladjelate2020,Pezzuto2021,fiorellino2021}, which generally mapped all of the clouds in the JCMT GBS over a larger area.  For further details see the notes in Table~\ref{tab:map_completeness}.
In general, the JCMT GBS maps cover all or nearly all of the higher column density material seen in the corresponding Herschel field: many of the JCMT maps lie near the 100\% line running across the top of Figure~\ref{fig:Av_completeness}, and are difficult to distinguish.  
To convert between column density and $A_V$, the Herschel GBS assumes that $N({\rm H}_2)$ (cm$^{-2}$) = $0.94 \times 10^{21}$ $A_V$ (mag) \citep{Bohlin78}. This produces good agreement with 2MASS-based extinction maps in some regions \citep[e.g.][]{konyves2015}, but elsewhere can produce discrepancies of up to a factor $\sim 2$ \citep[e.g.][]{konyves2020,difrancesco2020,Pezzuto2021}. Figure~\ref{fig:Av_completeness} therefore shows a line at $N({\rm H}_{2}) = 1.4\times 10^{22}$ cm$^{-2}$, or a Herschel GBS $A_V$ of 14~mag, broadly corresponding to a 2MASS-derived $A_V$ of about 7~mag.  As can be seen in Figure~\ref{fig:Av_completeness}, our mapping completeness above this level is $>90$\% everywhere other than in the extremely dispersed Auriga molecular cloud (see Figure~\ref{fig:ir3_auriga} in Appendix~\ref{sec:appendix_images}), where it is $>80$\%, indicating that the JCMT GBS met its mapping completeness goals.
Table~\ref{tab:map_completeness} in Appendix~\ref{sec:appendix_images} provides the full set of mapping completeness values shown in Figure~\ref{fig:Av_completeness}.

\begin{table*} 
\centering 
\caption{Regions observed by the Gould Belt Survey.} 
\begin{tabular}{c cc c c cc} 
\hline 
Region & R.A. & Dec. & Coverage & JCMT GBS Publication & Distance & Dist. Ref. \\ \cline{2-3} 
 & \multicolumn{2}{c}{(J2000)} & (deg$^{2}$) & & (pc) & \\ 
\hline 
Auriga-California & 04$^{h}$20$^{m}$40\dsec29 & $+$37$^{\circ}$44$^{\prime}$53\farcs7 & 1.68 &  \citet{broekhovenfiene2018} & 470 & \citet{zucker2019} \\
\rule{0pt}{2ex} & & & & & & \\
Cepheus L1228 &  20$^{h}$57$^{m}$51\dsec78 & $+$77$^{\circ}$38$^{\prime}$18\farcs4 & 3.34 & \citet{pattle2017} & 352 & \citet{zucker2019} \\ 
Cepheus L1251 & 22$^{h}$33$^{m}$43\dsec74 & $+$75$^{\circ}$14$^{\prime}$52\farcs0 & 0.56 & \citet{pattle2017} & 352 & \citet{zucker2019} \\ 
Cepheus South & 20$^{h}$50$^{m}$03\dsec94 & $+$67$^{\circ}$57$^{\prime}$44\farcs1 & 2.01 & \citet{pattle2017}  & 341 & \citet{zucker2020} \\ 
\rule{0pt}{2ex} & & & & & & \\
Corona Australis & 19$^{h}$05$^{m}$57\dsec95 & $-$37$^{\circ}$04$^{\prime}$34\farcs0 & 1.51 &  \citet{pattle2025}  & 151 & \citet{zucker2019} \\ 
\rule{0pt}{2ex} & & & & & & \\
IC 5146 & 21$^{h}$49$^{m}$41\dsec31 & $+$47$^{\circ}$26$^{\prime}$24\farcs9 & 1.56 & \citet{johnstone2017}  & 741 & \citet{zucker2020} \\ 
\rule{0pt}{2ex} & & & & & & \\
Lupus & 15$^{h}$42$^{m}$27\dsec00 & $-$34$^{\circ}$22$^{\prime}$55\farcs7 & 1.59 & \citet{mowat2017} & 151 & \citet{zucker2019} \\ 
\rule{0pt}{2ex} & & & & & & \\
Ophiuchus L1688 & \multirow{2}{*}{16$^{h}$32$^{m}$05\dsec01} & \multirow{2}{*}{$-$24$^{\circ}$27$^{\prime}$14\farcs8} & \multirow{2}{*}{3.95} & \multirow{2}{*}{\citet{pattle2015}} & 139 & \citet{zucker2019} \\
Ophiuchus L1689 &  & &  &  & 147 & \citet{ortizleon2017} \\ 
Oph/Sco N2 & 16$^{h}$47$^{m}$38\dsec77 & $-$12$^{\circ}$02$^{\prime}$58\farcs5 & 0.58 & -- & 134 & \citet{zucker2020} \\
Oph/Sco N3 & 16$^{h}$50$^{m}$51\dsec73 &  $-$15$^{\circ}$21$^{\prime}$40\farcs0 & 0.57 & -- & 151 & \citet{zucker2020} \\
Oph/Sco N6 & 16$^{h}$21$^{m}$13\dsec76 & $-$20$^{\circ}$08$^{\prime}$20\farcs9 & 0.57 & -- & -- & -- \\
\rule{0pt}{2ex} & & & & & & \\
Orion A & 05$^{h}$37$^{m}$58\dsec03 & $-$06$^{\circ}$55$^{\prime}$21\farcs4 &  6.18 &  \citet{salji2015}  & 432 & \citet{zucker2020} \\
 & & & & \citet{salji2015a} & & \citet{kounkel2017} \\
 & & & & \citet{coude2016} & & \\
 & & & & \citet{mairs2016} & & \\
 & & & & \citet{lane2016} & & \\
Orion B L1622 & 05$^{h}$54$^{m}$32\dsec80 & $+$01$^{\circ}$49$^{\prime}$31\farcs7 & 0.57 & \citet{kirk2016} & 423 & \citet{zucker2019} \\
 & & & & \citet{kirk2016a} & & \\
Orion B N2023 & 05$^{h}$42$^{m}$05\dsec02 & $-$01$^{\circ}$44$^{\prime}$04\farcs8 & 2.10 & \citet{kirk2016} & 423 & \citet{zucker2019} \\
 & & & & & & \citet{kounkel2017} \\
 & & & & \citet{kirk2016a} & & \\
Orion B N2068 & 05$^{h}$46$^{m}$51\dsec30 & $+$00$^{\circ}$19$^{\prime}$16\farcs5 & 1.71 & \citet{kirk2016} & 423 & \citet{zucker2019} \\
 & & & & \citet{kirk2016a} & & \\
 \rule{0pt}{2ex} & & & & & & \\
Perseus IC348 & 03$^{h}$43$^{m}$42\dsec75 &  $+$32$^{\circ}$22$^{\prime}$04\farcs2 & 1.99 & \citet{chen2016} & 321 & \citet{ortizleon2018} \\ 
 & & & & & & \citet{zucker2020} \\
Perseus West & 03$^{h}$30$^{m}$53\dsec54 & $+$30$^{\circ}$45$^{\prime}$55\farcs6 & 3.98 & \citet{hatchell2013} (NGC1333) & 294 & \citet{zucker2019} \\ 
 & & & & \citet{sadavoy2013} (B1) & & \\
 & & & & \citet{dodds2015} (NGC1333) & & \\
 & & & & \citet{chen2016} & & \\
 \rule{0pt}{2ex} & & & & & & \\
Pipe B59 & 17$^{h}$11$^{m}$31\dsec94 & $-$27$^{\circ}$26$^{\prime}$27\farcs7 & 0.57 & -- & 180 & \citet{zucker2020} \\ 
Pipe E1 & 17$^{h}$34$^{m}$05\dsec46 & $-$25$^{\circ}$39$^{\prime}$19\farcs9 & 0.56 & -- & 180 & \citet{zucker2020} \\ 
\rule{0pt}{2ex} & & & & & & \\
Serpens Aquila & 18$^{h}$30$^{m}$52\dsec27 & $-$02$^{\circ}$05$^{\prime}$50\farcs3 & 1.68 & \citet{rumble2016} (W40) & 484 & \citet{zucker2019} \\
Serpens East & 18$^{h}$37$^{m}$29\dsec09 & $-$01$^{\circ}$27$^{\prime}$05\farcs4 & 1.32 &  -- & 484 & \citet{zucker2019} \\ 
Serpens Main & 18$^{h}$29$^{m}$36\dsec20 & $+$00$^{\circ}$52$^{\prime}$05\farcs4 & 1.12 & --  & 436 & \citet{ortizleon2017a} \\ 
Serpens MWC297 & 18$^{h}$28$^{m}$13\dsec80 & $-$03$^{\circ}$43$^{\prime}$55\farcs3 & 0.59 & \citet{rumble2015} & 383 & \citet{herczeg2019} \\ 
Serpens North & 18$^{h}$39$^{m}$05\dsec20 & $+$00$^{\circ}$27$^{\prime}$56\farcs6 & 0.57 & -- & 484 & \citet{zucker2019} \\ 
\rule{0pt}{2ex} & & & & & & \\
Taurus South & 04$^{h}$17$^{m}$30\dsec23 & $+$27$^{\circ}$50$^{\prime}$08\farcs0 & 2.86 & -- & 141 & \citet{zucker2019} \\ 
Taurus L1495 & 04$^{h}$29$^{m}$20\dsec91 & $+$24$^{\circ}$35$^{\prime}$42\farcs8 & 2.72 & \citet{buckle2015} & 141 & \citet{zucker2019} \\
 & & & & \citet{wardthompson2016} & & \\
Taurus TMC1 & 04$^{h}$40$^{m}$01\dsec45 & $+$26$^{\circ}$00$^{\prime}$42\farcs2 & 1.67 & -- & 141 & \citet{zucker2019} \\ 
\hline
\end{tabular}
\label{tab:fields}
\end{table*}

\section{Source Extraction}
\label{sec:source_extraction}

\subsection{\textit{getsources}}
\label{sec:getsources}

We identified sources in the final JCMT GBS data release data products at 450\,$\upmu$m and 850\,$\upmu$m from \citet{kirk2018} using \textit{getsources} \citep{menshchikov2012}, an algorithm developed to identify and characterise sources in multi-wavelength submillimetre data sets.  We use version 1.140127 of \textit{getsources}, the same version used by the Herschel Gould Belt Survey for their source extractions, for consistency with and to aid with future comparisons with their catalogues \citep[e.g.,][]{konyves2015,konyves2020,difrancesco2020}.  We included SCUBA-2 maps at both 850\,$\upmu$m and 450\,$\upmu$m from the final JCMT GBS data release \citep{kirk2018} as input to \textit{getsources}.

The \textit{getsources} algorithm consists of two distinct stages. In the first ``detection" stage, \textit{getsources} smooths the input maps to successively lower resolution, subtracts maps at adjacent resolutions, and identifies positions of significant residual emission in the difference maps. These latter maps allow sources to be assembled and evaluated over ranges of scale and at each wavelength.  At the end of the monochromatic evaluation, \emph{getsources} combines the output from each wavelength to build an initial catalogue.  In the second ``measurement" stage, \textit{getsources} determines the fluxes and sizes of detected sources using the original input images at each wavelength at their native resolutions.  It further uses information from data at higher resolution to assist in deblending sources that overlap at lower resolution.   Background levels determined via linear interpolation under source footprints are subtracted to determine the measured fluxes at each wavelength. Unlike for the Herschel Gould Belt Survey catalogues, aperture corrections were not applied to the measured fluxes, because SCUBA-2 aperture corrections were not available in \textit{getsources}. Possible effects of this are discussed in Section~\ref{sec:mass_recovery_frac}, below.

\subsection{Source selection criteria}
\label{sec:source_selection}

The \textit{getsources} algorithm initially identified 4546 sources across the GBS regions.  Applying the source selection criteria described below, determined through extensive visual inspection of the initial \textit{getsources} output, resulted in a final core catalogue of 2257 sources considered reliable.

\textit{getsources} determines the significance of a detection at a given wavelength using a metric known as ``monochromatic significance", an analogue to peak signal-to-noise ratio (SNR) determined by measuring SNRs over the multiple size scales on which \textit{getsources} makes measurements \citep{menshchikov2012}. 
Similarly, the overall significance of a detection is quantified by its ``global significance", which is the quadrature sum of the monochromatic significances. 
\textit{Getsources} also provides a ``global goodness parameter'' combining the global significance and global SNR parameters.
We removed sources for which any of the following criteria were met:
\begin{enumerate}
    \item Global goodness $<$ 1 
    \item 850\,\um\ monochromatic significance $\leq$ 7
    \item Global significance $\leq$ 10
\end{enumerate}
These criteria were chosen as \textit{getsources} documentation advises that only sources with a global goodness $>1$ and monochromatic significances $>7$ be retained.  The global significance criterion results from the quadrature sum of the 850\,\um\ and 450\,\um\ monochromatic significances; a global significance of 10 approximately equates to monochromatic significances of 7 in both wavelengths.
For blended sources, the \textit{getsources} monochromatic significance should be a better measure of source reliability than a conventional SNR determined from the final \textit{getsources} catalogue values \citep{menshchikov2012}.  As a backstop for the more stringent criteria (i)--(iii), we also excluded sources where:
\begin{enumerate}
    \setcounter{enumi}{3}
    \item 850\,\um\ peak flux density SNR $\leq$ 2
    \item 850\,\um\ total (integrated) flux density SNR $\leq$ 1
\end{enumerate}
We further excluded sources where \textit{getsources} had judged either the peak or total flux density (or the uncertainties on either of these parameters) to be too small to measure at 850 \um: 
\begin{enumerate}
    \setcounter{enumi}{5}
    \item 850\,\um\ peak flux density OR 850\,\um\ peak flux density uncertainty OR 850\,\um\ total flux density OR 850\,\um\ total flux density uncertainty = 9.999E-31
\end{enumerate}
The output from \textit{getsources} includes a ``monochromatic flag" (FM) parameter, which provides information on the reliability of a source based on its size, SNR, sub-structure, significance, etc., at a given wavelength.  We excluded sources where
\begin{enumerate}
    \setcounter{enumi}{6}
    \item 850\,\um\ FM = 1 AND (450\,\um\ FM = 31 OR 450\,\um\ FM > 200)
\end{enumerate}
FM = 1 indicates that a source is larger than the characteristic size scale, while FM = 31 indicates both FM = 1 holds for a source and that its total flux and peak flux density both have SNR < 1 (FM =  30).  FM > 200 indicates a monochromatic significance < 3.5.  This final criterion thus excludes large diffuse 850\,\um\ sources if there is no hint of a detection at 450\,\um.

\subsection{Known CO artefacts}

As the SCUBA-2 850 \um\ band includes the rest frequency of the CO $J=3-2$ line, observed 850\,$\upmu$m fluxes may be artificially increased at locations of bright CO line emission \citep{drabek2012,sadavoy2013}.  Maps of CO $J=3-2$ emission obtained with the JCMT HARP instrument by the JCMT GBS can used to remove CO emission in some locations -- see \citet{kirk2018} for more information.  However, the JCMT GBS HARP coverage was not as extensive as the SCUBA-2 coverage.  We thus do not use CO-subtracted 850\,\um\ maps in this work, both so that all of the regions that we consider are treated consistently, and in order to avoid edge effects at the interfaces between regions with CO data and without.  We expect the contribution of CO emission to the 850\,\um\ fluxes in our catalogue to be generally $< 20$\% \citep[e.g.,][]{pattle2015}.  However, CO contamination is most significant toward bright outflows \citep{johnstone2003,coude2016}, and a small number of knots of bright compact CO emission have been identified by \textit{getsources} in our maps.  Knowing these locations from experience (near the bright protostellar sources L1147-mm and IRAS 16293-2422; \citealt{pattle2015,pattle2017}), five associated sources were flagged and removed from the final catalogue by hand.

\subsection{Catalogue completeness}
\label{sec:completeness}

\begin{figure}
    \centering
    \includegraphics[width=\columnwidth]{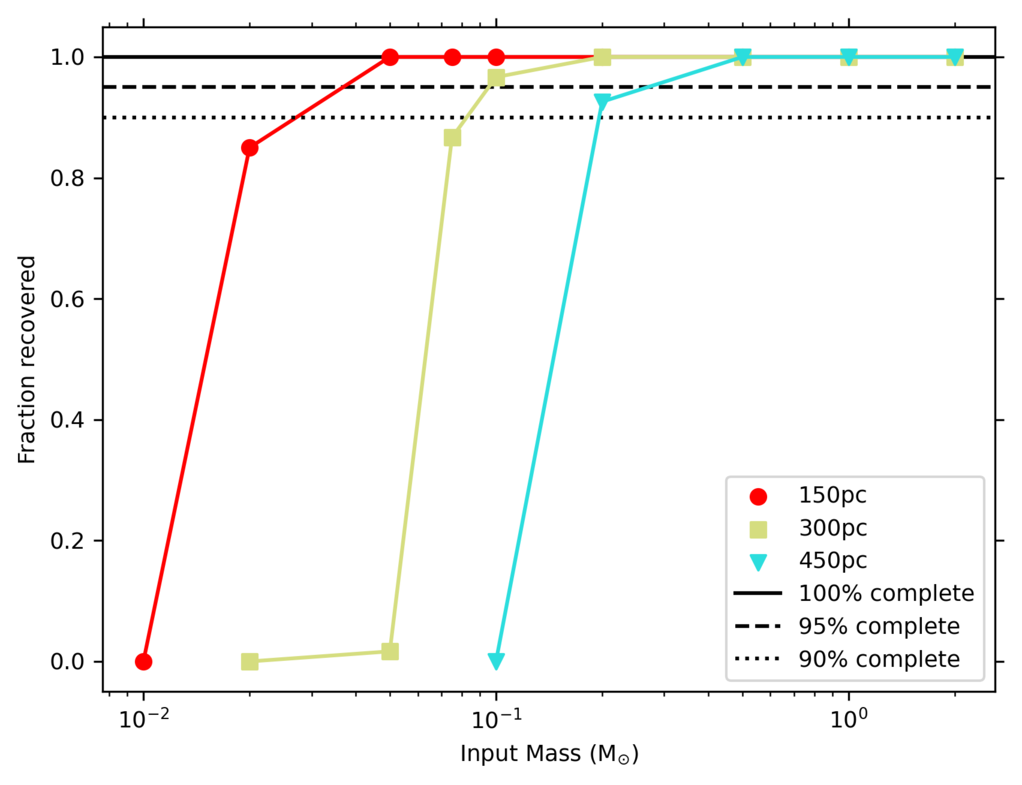}
    \caption{Fraction of input BEC spheres returned by \textit{getsources} as a function of input source mass, for sources at distances of 150\,pc (red circles), 300\,pc (green squares) and 450\,pc (blue triangles).  Solid black line marks 100\% completeness; dashed line marks 95\% completeness; dotted line marks 90\% completeness.}
    \label{fig:getsources_return_frac}
\end{figure}

\begin{figure}
    \centering
    \includegraphics[width=\columnwidth]{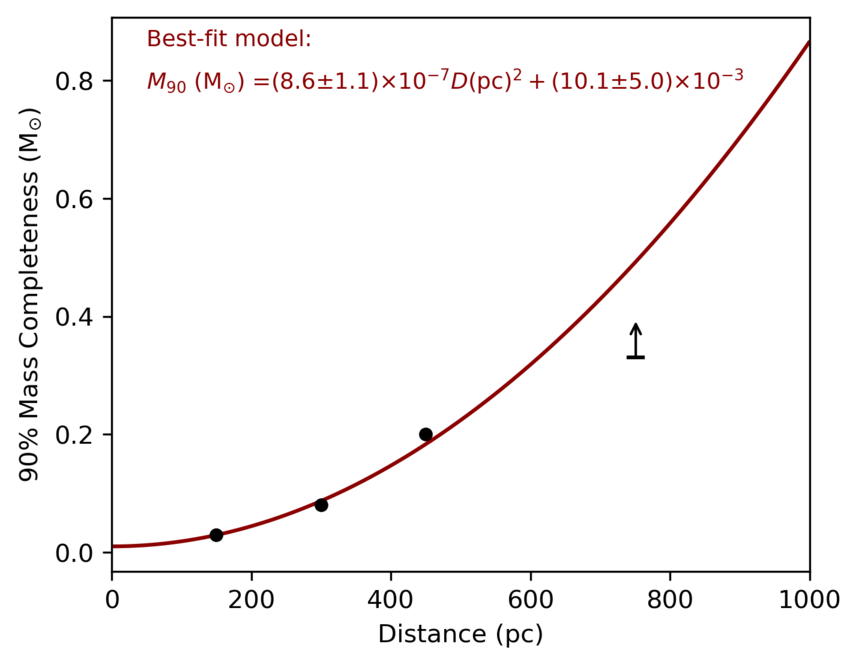}
    \caption{Our estimated 90\% mass completenesses as a function of distance.  A quadratic fit to the data is shown.  The lowest-mass source detected in IC 5146 is shown as a lower limit for mass completeness at this distance.}
    \label{fig:completeness_extrapolation}
\end{figure}

\begin{figure}
    \centering
    \includegraphics[width=\columnwidth]{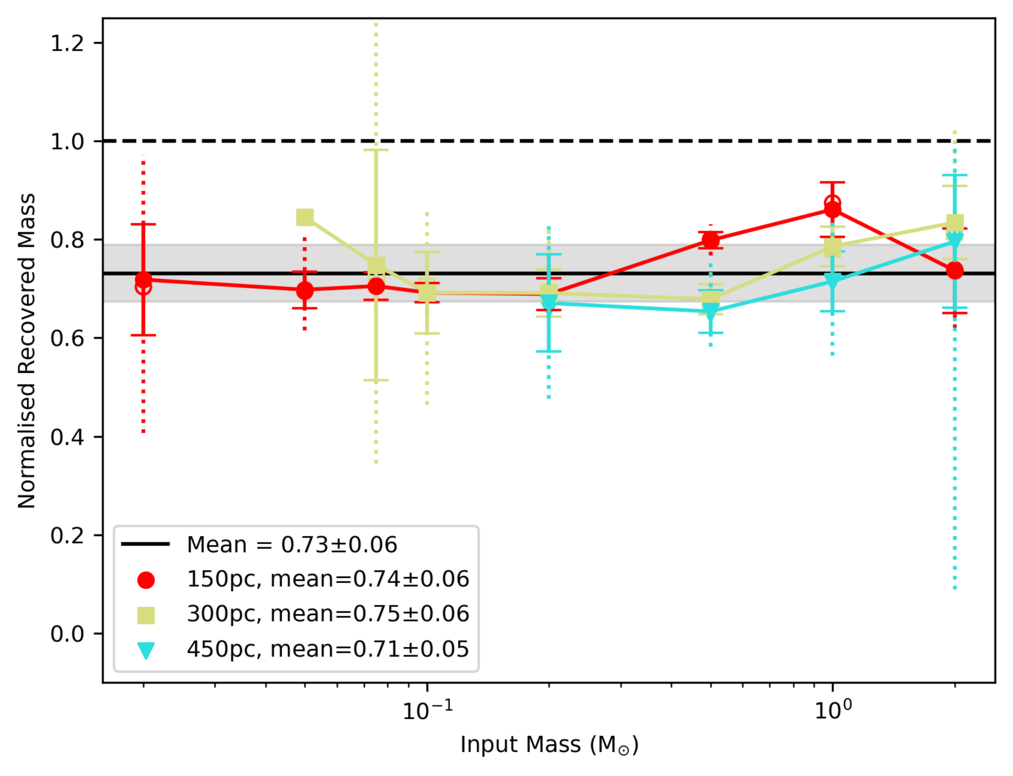}
    \caption{Fraction of input source mass recovered by \textit{getsources}, as a function of input source mass.  Colour and shape of data points are as in Figure~\ref{fig:getsources_return_frac}.  Symbols mark mean recovered mass fraction; solid error bars mark 1-$\sigma$ variation around the mean, while dotted error bars mark the full range of values.  The solid black line marks the `mean of means': the average mass recovery fraction; grey shading marks the 1-$\sigma$ variation on this value.  Dashed line marks a recovered mass fraction of 1.0.}
    \label{fig:getsources_normed_mass}
\end{figure}

In order to interpret the distribution of core properties which we measure, it is vital to understand the mass completeness of our catalogue: both the fraction of sources that are recovered as a function of core mass (the source recovery fraction), and the fraction of true source mass that is recovered in our catalogue (the mass recovery fraction).

We expect mass completeness to be a strong function of distance.  For example, lower-mass sources should be fainter and smaller, and (for a given core temperature) will therefore be significantly easier to detect in nearer clouds.  Throughout this work, we thus divide the sample clouds into four distance ranges:
\begin{enumerate}
    \item \textbf{Near ($<$200\,pc), representative distance 150\,pc:} Corona Australis, Lupus, Ophiuchus, Oph/Sco North, Pipe and Taurus.
    \item \textbf{Mid-distance clouds (200--355\,pc), representative distance 300\,pc:} Cepheus and Perseus.
    \item \textbf{Far clouds (355--500\,pc), representative distance 450\,pc:} Auriga-California, Orion A and B, and Serpens.
    \item \textbf{Very far clouds ($>$500\,pc):} IC 5146.
\end{enumerate}
IC 5146 the only cloud at a distance $> 500$\,pc, placed at 751\,pc by \citet{zucker2020} using \textit{Gaia} measurements.  The distance to this cloud has been revised upward significantly since the GBS was designed, from a previous value of 460\,pc \citep{lada1999}, and the cloud is typically no longer considered to be a member of the Gould Belt \citep[e.g.][]{dzib2018}.  Given the lack of clouds at comparable distance for comparison, and the relatively small number of sources (71) detected in the region, we include this cloud and its cores in our catalogue, but largely exclude it from our discussion of typical core properties.

\subsubsection{Method for determining mass completeness}

We determined the completeness of our core catalogue by running a series of tests in which we inserted artificial critically-stable Bonnor--Ebert (BEC) spheres \citep{ebert1955,bonnor1956} into the Oph/Sco N6 field.  This field contains no detectable sources, and has previously been used for GBS completeness testing \citep{kirk2018}.  We tested masses in the range 0.01--2\,M$_{\odot}$, placing our input BEC spheres at the representative distances for the near-, mid- and far-distance clouds: 150\,pc, 300\,pc, and 450\,pc.  We converted the mass distributions of our BEC spheres into flux densities using the \citet{hildebrand1983} relationship, taking dust opacity index $\beta = 1.8$, consistent with Planck observations \citep{juvela2015} and joint \textit{Herschel}/SCUBA-2 fits to GBS data \citep{chen2016,sadavoy2013}.  The process for creating these BEC sphere models is described in detail in Appendix~\ref{app_compl}, and the python code used to do so is publicly available\footnote{BEC code is available at: \url{https://github.com/KatePattle/bonnor-ebert-sphere}}.

We repeated the data reduction process for Oph/Sco N6, using the \textit{fakemap} parameter\footnote{The \textit{fakemap} parameter allows the user to provide an image of the sky that will produce corresponding additional astronomical signal in the SCUBA-2 bolometer time series; see {\tt \url{http://starlink.eao.hawaii.edu/docs/sun258.pdf}}} in \textit{makemap} \citep{chapin2013} to insert each of our fields of model BEC spheres.  We then used the \textit{getsources} algorithm, with the same parameters as described in Section~\ref{sec:getsources} above, to search for the cores which we had inserted.  We then applied the same selection criteria to the \textit{getsources} catalogues as described in Section~\ref{sec:source_selection}, above.  Our completeness testing procedure is described in detail in Appendix~\ref{app_compl}.  Here, we highlight the key results.

\subsubsection{Source recovery completeness fraction}

Figure~\ref{fig:getsources_return_frac} shows the fraction of input sources that are recovered as a function of input mass, for each of our three representative distances.  As expected, more massive sources are more easily recovered, and mass completeness is better at nearer distances.  We find $> 90$\% completeness limits for the three distance ranges of 0.03\,M$_{\odot}$, 0.075\,M$_{\odot}$ and 0.2\,M$_{\odot}$ for 150\,pc, 300\,pc and 450\,pc, respectively.
These limits are shown as a function of distance in Figure~\ref{fig:completeness_extrapolation}, and are quite well-fitted by a quadratic function, $M_{90}\,({\rm M}_{\odot})\approx (8.6\pm1.1)\times 10^{-7}\,D({\rm pc})^{2}$.  We choose to fit a quadratic function as the flux density of a source of given mass and temperature scales as $D^{-2}$ \citep{hildebrand1983}, where $D$ is the distance of the source.

The mass completeness limit in IC 5146 is $\geq 0.33$\,M$_{\odot}$, as this is the lowest-mass source that we detect (see Section~\ref{sec:mass}, below).  Extrapolation of our best-fit quadratic model suggests a $>$ 90\% mass completeness limit in IC 5146 of $\approx 0.5\,$M$_{\odot}$, consistent with our observations.

There are a number of reasons to expect that these completeness limits are somewhat conservative.  In our completeness testing, we place our model sources directly onto the noisy background of the Oph/Sco N6 field.  In reality, many sources are embedded within filaments or other structures, as can be seen in Figure~\ref{fig:fig1}, and so may be sufficiently boosted above the background to be detected.  We further note that we have performed our completeness testing for critically-stable Bonnor--Ebert spheres, whereas gravitationally bound and collapsing cores should be more centrally condensed, and so more easily detectable.  Moreover, the Oph/Sco N6 field was observed in Band 2 weather; regions observed in Band 1 weather will have somewhat better SNR, particularly at 450\,$\upmu$m, and so fainter sources will be more easily detected in these regions. 
Particularly in the far-distance fields, a non-negligible number of sources are detected below the nominal completeness limit (cf. Section~\ref{sec:cmfs}, below).

\subsubsection{Mass recovery fraction}
\label{sec:mass_recovery_frac}

We calculated the recovered mass of each of our recovered sources from the 850\,\um\ flux densities returned by \textit{getsources} and the assumed temperature and dust properties described above (cf. eqs.~\ref{eq:mass} and \ref{eq:opacity}, below).  Figure~\ref{fig:getsources_normed_mass} shows the fraction of input mass recovered for a given source as a function of the true input mass, for each of our three representative distances. 

Our global mass recovery fraction -- the typical fraction of input mass recovered above the 90\% source recovery completeness limit -- is $0.73\pm 0.06$.   The mean recovered mass fractions for each input mass and distance are shown in Figure~\ref{fig:getsources_normed_mass}.  We took the mean of these mean values, for input masses above the 90\% completeness level in each field, in order to calculate our global mass recovery fraction.  We therefore excluded the single 0.05\,M$_{\odot}$ source recovered at 300\,pc as the 90\% completeness limit at 300\,pc is 0.075\,M$_{\odot}$.

There are several effects that are likely to be resulting in loss of input flux, and so the decrease in recovered mass. (1) A BEC sphere is characterised by a relatively flat central plateau at small radii, with a power-law drop-off in density beyond a critical radius \citep{ebert1955,bonnor1956}.  We further convolve the BEC sphere surface brightness profiles which we generate with the JCMT beam \citep{dempsey2013}.  At low masses, we simply expect the fainter material associated with the source to be undetectable above the noise in the field.  (2) In more extended higher-mass sources, we expect that some fraction of the flux loss is likely due to the effects of performing submillimetre continuum measurements below the atmosphere.  The SCUBA-2 iterative map-making process \textit{makemap} \citep{chapin2013} is unable to distinguish between atmospheric and astrophysical signal on sizes comparable to the SCUBA-2 array ($\sim 600^{\prime\prime}$; \citealt{holland2013}), and we expect this behaviour to result in loss of extended structure in both our real and our synthetic observations. This effect has been discussed extensively in previous SCUBA-2 papers \citep{sadavoy2013,pattle2015,mairs2015,kirk2018}.   We further note that as we have chosen to model BEC spheres, as we go to higher masses, the sources both get larger and have lower peak brightnesses: these effects may conspire to make the sources harder to recover.
A direct comparison with the Gaussian mass (total flux) recovery results from \citet{kirk2018} suggests that for all but the few highest mass BEC models, filtering has a $< 10$\% effect when the core's peak flux is at or above 5 times the local noise.
(3) The lack of appropriate SCUBA-2 aperture corrections available in \textit{getsources} may result in some loss of extended emission.

Figure~\ref{fig:getsources_normed_mass} shows a peak in mass recovery fraction at 1\,M$_{\odot}$ at 150\,pc, followed by a slight drop-off for 2\,M$_{\odot}$, likely due to a combination of these effects.  The 300\,pc and 450\,pc sources appear to show a similar trend, displaced to higher masses; both have their highest mass recovery fraction at the highest mass tested, 2\,M$_{\odot}$, with the 2\,M$_{\odot}$ mass recovery fraction being slightly higher at 300\,pc than at 450\,pc, although the two agree within their respective error bars.  However, despite this variation, the fraction of mass recovered is quite constant across the range of masses which we consider.  Thus, we adopt our mean mass recovery fraction, $0.73\pm 0.06$, wherever a correction is required in the following analysis.  Masses are presented as measured by \textit{getsources} (without any correction for flux loss) unless otherwise stated.

\subsubsection{Comparison to \textit{Herschel} Gould Belt Survey completeness}
\label{sec:herschel_completeness}

The dense core catalogues produced by the \textit{Herschel} GBS \citep{andre2010} provide a natural point of comparison to our JCMT GBS catalogue.  However, direct comparison between individual sources in the JCMT and \textit{Herschel} catalogues is non-trivial, due to the many differences between observations made by the two instruments, particularly the differing instrumental responses to large-scale structure between the \textit{Herschel} photometers and SCUBA-2 \citep{sadavoy2013,wardthompson2016}, and the differences in wavelength between the two instruments (note that the emission peak within a single source may vary with wavelength; e.g., \citealt{encalada2024}).  Nonetheless, we find that the mass completeness limits of our catalogue and the \textit{Herschel} GBS catalogues are comparable.

\textit{Herschel} Gould Belt Survey observations are cirrus confusion limited \citep{andre2010}, meaning that their source completeness is dependent on environment \citep{konyves2015}.  Most \textit{Herschel} GBS papers model their completeness for prestellar cores on a high-column-density background, with cores modelled as having critically-stable Bonnor-Ebert-like density profiles, with a significant drop in temperature towards the centre \citep{konyves2015}.

The masses at which the \textit{Herschel} GBS achieves 80--90\% completeness for recovery of prestellar cores in dense environments are comparable to but somewhat higher than our own values at comparable distances due to their need to disentangle dense cores from the extended structure to which SCUBA-2 is not sensitive.  The 80--90\% mass completeness limit for prestellar cores is consistently found to occur at 0.1\,M$_{\odot}$ in nearby (130--200\,pc) clouds \citep{marsh2016,benedettini2018,bresnahan2018,ladjelate2020,kirk2024}, 0.3--0.4 at 300--400\,pc \citep{difrancesco2020,konyves2020,Pezzuto2021}, and 0.8\,M$_{\odot}$ at the maximum distance the HGBS considers, 484\,pc \citep{fiorellino2021}.  We note that these mass completeness limits are typically corrected for the estimated 20--30\% underestimation of source mass that arises from fitting a single-temperature modified blackbody model to the spectral energy distribution of a non-isothermal core \citep{konyves2015}.

The \textit{Herschel} GBS typically recovers significantly larger numbers of low-mass cores in low-column-density regions than are detectable in SCUBA-2 observations \citep{wardthompson2016,konyves2020}, due to SCUBA-2's lack of sensitivity to extended emission.   \citet{marsh2016} found a mass completeness limit in Taurus of $>85$\% at 0.015\,$M_{\odot}$ for unbound starless cores on a low-column-density background at a distance of 140\,pc, two orders of magnitude better than their completeness for deeply embedded prestellar cores at the same distance, and comparable to but notably better than our 90\% mass completeness at 150\,pc of 0.03\,M$_{\odot}$.

The complexity of interpreting the differences in source identification and completeness means that a detailed core-by-core comparison of the JCMT and \textit{Herschel} GBS catalogues is beyond the scope of this work.  However, we have chosen to use the same source extraction algorithm as the \textit{Herschel} GBS in order to perform this work in the future; we note that the similar completeness levels between the two surveys suggests that such a comparison would be meaningful.

\begin{table*} 
\setlength\tabcolsep{3pt}\centering 
\caption{Measured properties of the first ten JCMT GBS sources in our catalogue.  The abbreviation ``mJy/sqa'' is used for the unit mJy\,arcsec$^{-2}$.  The full catalogue is available in the online material associated with this paper.} 
\begin{tabular}{c c c r@{\,$\pm$\,}lr@{\,$\pm$\,}l r@{\,$\times$\,}lc @{\extracolsep{4pt}} r@{\,$\pm$\,}lr@{\,$\pm$\,}l r@{\,$\times$\,}lc} 
\hline 
Running & & Source ID (JCMTLSG...) & \multicolumn{7}{c}{850$\upmu$m} & \multicolumn{7}{c}{450$\upmu$m} \\ \cline{4-10} \cline{11-17}
No. & Region & R.A., Dec. (J2000) & \multicolumn{2}{c}{Peak F.D.} & \multicolumn{2}{c}{Total F.D.} & \multicolumn{2}{c}{FWHM} & {P.A.} & \multicolumn{2}{c}{Peak F.D.} & \multicolumn{2}{c}{Total F.D.} & \multicolumn{2}{c}{FWHM} & {P.A.} \\ 
 & & HHMMSS.S+DDMMSS & \multicolumn{2}{c}{(mJy/sqa)} & \multicolumn{2}{c}{(mJy)} & \multicolumn{2}{c}{(arcsec)} & (deg) & \multicolumn{2}{c}{(mJy/sqa)} & \multicolumn{2}{c}{(mJy)} & \multicolumn{2}{c}{(arcsec)} & (deg) \\ 
\hline
1 & Aquila &  183004.0-020306 & 3.70 & 0.08 & 5.8 & 0.1 & 16.3 & 14.1 & 29 & 13.2 & 0.3 & 61.1 & 0.9 & 18.0 & 14.2 & 22 \\ 
2 & Aquila &  183002.4-020249 & 1.44 & 0.08 & 3.1 & 0.1 & 23.2 & 14.1 & 164 & 5.8 & 0.3 & 17.6 & 0.6 & 16.0 & 9.7 & 178 \\ 
3 & Aquila &  182937.6-015101 & 0.65 & 0.05 & 0.79 & 0.06 & 14.1 & 14.1 & -- & 3.0 & 0.1 & 4.6 & 0.2 & 9.6 & 9.6 & -- \\ 
4 & Aquila &  182908.1-013049 & 0.82 & 0.05 & 0.93 & 0.05 & 14.1 & 14.1 & -- & 4.5 & 0.1 & 5.5 & 0.2 & 9.6 & 9.6 & -- \\ 
5 & Aquila &  183109.5-020624 & 0.62 & 0.04 & 0.76 & 0.04 & 14.1 & 14.1 & -- & 3.0 & 0.2 & 5.6 & 0.2 & 11.2 & 9.6 & 92 \\ 
6 & Aquila &  183001.4-021027 & 0.69 & 0.03 & 1.05 & 0.03 & 15.5 & 14.1 & 167 & 3.0 & 0.1 & 5.8 & 0.2 & 11.4 & 9.6 & 171 \\ 
7 & Aquila &  183121.3-020658 & 0.53 & 0.08 & 0.62 & 0.08 & 14.1 & 14.1 & -- & 2.3 & 0.3 & 4.0 & 0.4 & 11.6 & 9.6 & 16 \\ 
8 & Aquila &  183110.4-020350 & 0.51 & 0.06 & 0.67 & 0.06 & 20.3 & 14.1 & 3 & 2.3 & 0.2 & 4.0 & 0.2 & 15.3 & 9.6 & 2 \\ 
9 & Aquila &  182903.6-013907 & 0.56 & 0.03 & 0.60 & 0.03 & 14.1 & 14.1 & -- & 2.4 & 0.1 & 2.5 & 0.1 & 9.6 & 9.6 & -- \\ 
10 & Aquila &  183121.0-020623 & 0.76 & 0.08 & 1.5 & 0.1 & 20.7 & 14.1 & 12 & 3.0 & 0.2 & 11.1 & 0.3 & 18.5 & 11.8 & 15 \\ 
\hline
\end{tabular}
\label{tab:measured_sample}
\end{table*}

\begin{table*} 
\centering 
\caption{Derived properties of the first ten JCMT GBS sources in our catalogue.  The full catalogue is available in the online material associated with this paper.} 
\begin{tabular}{c c c c r@{\,$\pm$\,}l r@{\,$\pm$\,}l r@{\,$\pm$\,}l r@{\,$\pm$\,}l r@{\,$\pm$\,}l r@{\,$\pm$\,}l r@{\,$\pm$\,}l} 
\hline 
Running & Type & $R_{\rm deconv}$ & $T$ & \multicolumn{2}{c}{$M$} & \multicolumn{2}{c}{$N({\rm H}_{2})$} & \multicolumn{2}{c}{$n({\rm H}_{2})$} & \multicolumn{2}{c}{$M_{\textsc{be}}$} & \multicolumn{2}{c}{$\alpha_{BE}$} & \multicolumn{2}{c}{$M_{\textsc{be}}(10\,{\rm K})$} & \multicolumn{2}{c}{$\alpha_{BE}(10\,{\rm K})$} \\ 
No.  & & (pc) & (K) & \multicolumn{2}{c}{(M$_{\odot}$)} & \multicolumn{2}{c}{($\times 10^{22}\,$cm$^{-2}$)} &  \multicolumn{2}{c}{($\times 10^{4}\,$cm$^{-3}$)} & \multicolumn{2}{c}{(M$_{\odot}$)} & \multicolumn{2}{c}{(--)} & \multicolumn{2}{c}{(M$_{\odot}$)} & \multicolumn{2}{c}{(--)} \\ 
\hline 
1 & P & 0.013 & 17.5 & 10.7 & 0.2 & 87 & 1 & 1610 & 30 & 0.59 & 0.07 & 0.055 & 0.006 & 0.34 & 0.04 & 0.012 & 0.001 \\ 
2 & P & 0.027 & 17.4 & 5.8 & 0.2 & 11.5 & 0.5 & 105 & 4 & 0.95 & 0.09 & 0.16 & 0.02 & 0.55 & 0.05 & 0.035 & 0.004 \\ 
3 & P & -- & 16.2 & 1.7 & 0.1 & \multicolumn{2}{c}{--} & \multicolumn{2}{c}{--}  & 0.21 & 0.05 & 0.13 & 0.03 & 0.13 & 0.03 & 0.033 & 0.007 \\ 
4 & H & -- & 15.5 & 2.1 & 0.1 & \multicolumn{2}{c}{--} & \multicolumn{2}{c}{--}  & 0.21 & 0.04 & 0.10 & 0.02 & 0.13 & 0.03 & 0.028 & 0.006 \\ 
5 & H & -- & 31.9 & 0.60 & 0.03 & \multicolumn{2}{c}{--} & \multicolumn{2}{c}{--}  & 0.42 & 0.09 & 0.7 & 0.2 & 0.13 & 0.03 & 0.035 & 0.008 \\ 
6 & C & 0.010 & 17.3 & 1.96 & 0.07 & 25.1 & 0.8 & 590 & 20 & 0.51 & 0.06 & 0.26 & 0.03 & 0.30 & 0.04 & 0.056 & 0.007 \\ 
7 & C & -- & 57.8 & 0.24 & 0.03 & \multicolumn{2}{c}{--} & \multicolumn{2}{c}{--}  & 0.8 & 0.2 & 3.2 & 0.8 & 0.13 & 0.03 & 0.04 & 0.01 \\ 
8 & C & 0.022 & 28.2 & 0.62 & 0.05 & 1.8 & 0.2 & 20 & 2 & 1.3 & 0.1 & 2.2 & 0.3 & 0.48 & 0.05 & 0.14 & 0.02 \\ 
9 & C & -- & 15.6 & 1.34 & 0.08 & \multicolumn{2}{c}{--} & \multicolumn{2}{c}{--}  & 0.21 & 0.04 & 0.15 & 0.03 & 0.13 & 0.03 & 0.04 & 0.01 \\ 
10 & C & 0.023 & 58.5 & 0.58 & 0.04 & 1.6 & 0.1 & 17 & 1 & 2.8 & 0.3 & 4.9 & 0.6 & 0.49 & 0.05 & 0.063 & 0.008 \\ 
\hline
\end{tabular}
\label{tab:derived_sample}
\end{table*}

\section{The JCMT GBS Catalogue}
\label{sec:cat}

The JCMT Gould Belt Survey core catalogue contains 2257 sources.  A sample of the measured source properties is given in Table~\ref{tab:measured_sample}.  The key derived source properties for the same source sample is given in Table~\ref{tab:derived_sample}.  The data available in the catalogue is summarized in Table~\ref{tab:columns}.  The full catalogue is available online, along with the output of the \textit{getsources} algorithm without our selection criteria applied.

\begin{table*}
\centering
\caption{A list of the quantities and information given in the JCMT Gould Belt Survey core catalogue, available as online material associated with this paper.  See Section~\ref{sec:be} for a detailed description of columns 31--39, and Section~\ref{sec:classification} for a detailed description of columns 40--44.}
\begin{tabular}{ccc}
\hline
Column number & Units & Description \\
\hline
(1) & -- & Running number  \\
(2) & -- & Region \\
(3) & --  & Source ID\hyperlink{fna}{\textsuperscript{a}} \\
(4, 5) & mJy\,arcsec$^{-2}$ & Peak 850 \um\ flux density and associated uncertainty \\
(6, 7) & mJy & Total 850 \um\ flux density and associated uncertainty \\
(8, 9) & arcsec & 850 \um\ major and minor FWHMs \\
(10) & deg & 850 \um\ position angle\hyperlink{fnb}{\textsuperscript{b}} \\
(11, 12) & mJy\,arcsec$^{-2}$ & Peak 450 \um\ flux density and associated uncertainty \\
(13, 14) & mJy & Total 450 \um\ flux density and associated uncertainty \\
(15, 16) & arcsec & 450 \um\ major and minor FWHMs \\
(17) & deg & 450 \um\ position angle\hyperlink{fnb}{\textsuperscript{b}} \\
(18) & -- & Source classification\hyperlink{fnc}{\textsuperscript{c}} \\
(19) &  -- & Resolved/unresolved flag\hyperlink{fnd}{\textsuperscript{d}} \\
(20) & pc & Deconvolved radius \\
(21, 22) & K & Assumed temperature and associated uncertainty \\
(23, 24) &  M$_{\odot}$ & Derived mass and associated uncertainty \\
(25, 26) & cm$^{-2}$ & Column density of H$_{2}$ and associated uncertainty \\
(29, 30) & cm$^{-3}$ & Volume density of H$_{2}$ and associated uncertainty \\
(31, 32) & M$_{\odot}$ & Critical Bonnor--Ebert (BEC) mass and associated uncertainty \\
(33, 34) & -- & BEC stability ratio and associated uncertainty \\
(35, 36) &  M$_{\odot}$ & BEC mass at 10\,K and associated uncertainty \\
(37, 38) & -- & BEC stability ratio at 10\,K and associated uncertainty \\
(39) & -- & Boundedness flag\hyperlink{fne}{\textsuperscript{e}} \\
(40) & -- & Best match to source in \textit{Spitzer} 24\,$\upmu$m catalogue \\
(41) & -- & Best match to source in \textit{Spitzer} protostellar catalogues \\
(42) & -- & Best match to source in WISE all-sky YSO catalogue \citep{marton2016} \\
(43) & -- & Best match to source in NED database \\
(44) & -- & For sources with a NED match, classification of source in SIMBAD database \\ 
\hline
\end{tabular}
\begin{flushleft}
{\footnotesize{\hypertarget{fna}{\textsuperscript{a}} Format ``JCMTLSG HHMMSS.S+DDMMSS"; where the source name comprises the J2000 coordinates of the source.}}\\
{\footnotesize{\hypertarget{fnb}{\textsuperscript{b}} Position angles are measured East of North.}}\\
{\footnotesize{\hypertarget{fnc}{\textsuperscript{c}} See Section~\ref{sec:classification}.}}\\
{\footnotesize{\hypertarget{fnd}{\textsuperscript{d}} 1 = resolved at 850\,$\upmu$m, and 0 = unresolved at 850\,$\upmu$m.}}\\
{\footnotesize{\hypertarget{fne}{\textsuperscript{e}} $-1$ = not a starless core; 0 = unbound, 1 = candidate starless core, bound at 10\,K, 2 = robust prestellar core, bound at $\geq15$\,K.}}
\end{flushleft}
\label{tab:columns}
\end{table*}

\subsection{Source categorization}
\label{sec:classification}

We categorize our sources as starless cores (`C'); protostellar (`P'), i.e. those sources cross-matched with a known YSO or YSO candidate; heated (`H'), i.e. those sources cross-matched with a \textit{Spitzer} 24\,$\upmu$m detection and so potentially protostellar; or as a potential contaminant extragalactic source (`G').  A source is categorized as a starless core by elimination, i.e. if it is not categorized as protostellar, heated or potentially extragalactic.

\subsubsection{Potential extragalactic contaminants}

To identify any potential extragalactic contaminants in our catalogue, we queried the NED database\footnote{\url{https://ned.ipac.caltech.edu/}} within a circular area around each peak source position of radius equal to the geometric mean of the major and minor FWHM axes of the source.  If the query returned any objects classified as galaxies, galaxy groups or galaxy clusters (i.e., having a `G', `GClstr', `GGroup', `GPair', `GTrpl' or `G\_Lens' designation in NED), we initially categorized the source as a potential extragalactic contaminant.

Of the 2257 sources in our catalogue, 65 had NED matches.  We note that all of these sources are associated with nearby molecular clouds, and so their true status as extragalactic sources is doubtful.  When comparing against protostellar catalogues (as described in Section~\ref{sec:classify_protostellar}, below), 45 of the sources with NED matches were also identified with a protostellar source.  We classified these 45 sources as protostellar, as the more probable identification.  We investigated the remaining 20 NED matches individually, comparing the name given in the NED database with their equivalent entries (if any) in the SIMBAD\footnote{\url{http://simbad.u-strasbg.fr/simbad/}} database \citep{wenger2000}.  One source, JCMTLSG183004.0-020306, is known to be a small protostellar cluster \citep{kern2016}, and was clearly misidentified in NED.  This source was also associated with 24\,$\upmu$m emission (cf. Section~\ref{sec:classify_24um}).  We thus classified this source as protostellar.  Of the remaining 19 sources, two were identified in SIMBAD as radio sources, two as low-mass stars, and one as a Herbig--Haro object.  The remainder had no match in SIMBAD, and appear to have been classified as galaxies in the 2MASS eXtended \citep{skrutskie2006} or allWISE \citep{cutri2014} catalogues.  We excluded all 19 of these sources from further consideration in the interest of ensuring an uncontaminated catalogue of dense cores.  However, all sources appear in the final catalogue, with their NED and, where relevant, SIMBAD identifiers noted. 

\subsubsection{Protostellar sources}
\label{sec:classify_protostellar}

A source is categorized as protostellar if it contains at least one protostar or YSO within its area.  We searched the \textit{Spitzer} and WISE protostellar catalogues for matches within a circular area around each peak source position with a radius equal to the geometric mean of the major and minor FWHM axes of the source.

The revised \textit{Spitzer} c2d catalogue \citep{dunham2015} lists protostars and YSOs detected by \textit{Spitzer} in Lupus, Ophiuchus, Perseus, Serpens (except Serpens East), and Chamaeleon (not covered by our observations).  Protostars in Orion A and B were surveyed by \citet{megeath2012}, and in Taurus by \citep{rebull2010}.  The \textit{Spitzer} Gould Belt Survey lists protostars in Auriga-California \citep{broekhoven-fiene2014}, Cepheus \citep{kirk2009}, Corona Australis \citep{peterson2011}, IC 5146 \citep{harvey2008}, and Ophiuchus North \citep{hatchell2012}.  The only region observed by the GBS not covered by a \textit{Spitzer} protostellar catalogue is Serpens East. 

We further performed cross-matching with the Wide-Field Infrared Survey Explorer (WISE) All-Sky Survey YSO catalogue \citep{marton2016}.  We cross-matched our sources with their list of Class I and II sources only.  We did not consider their Class III sources, as these will no longer be embedded.

We did not cross-match with \textit{Herschel} protostellar catalogues \citep[e.g.][]{konyves2015,bresnahan2018,konyves2020} because these catalogues are not available for all of the regions that we consider, and so we could not do so self-consistently.

\begin{figure}
    \centering
    \includegraphics[width=\columnwidth]{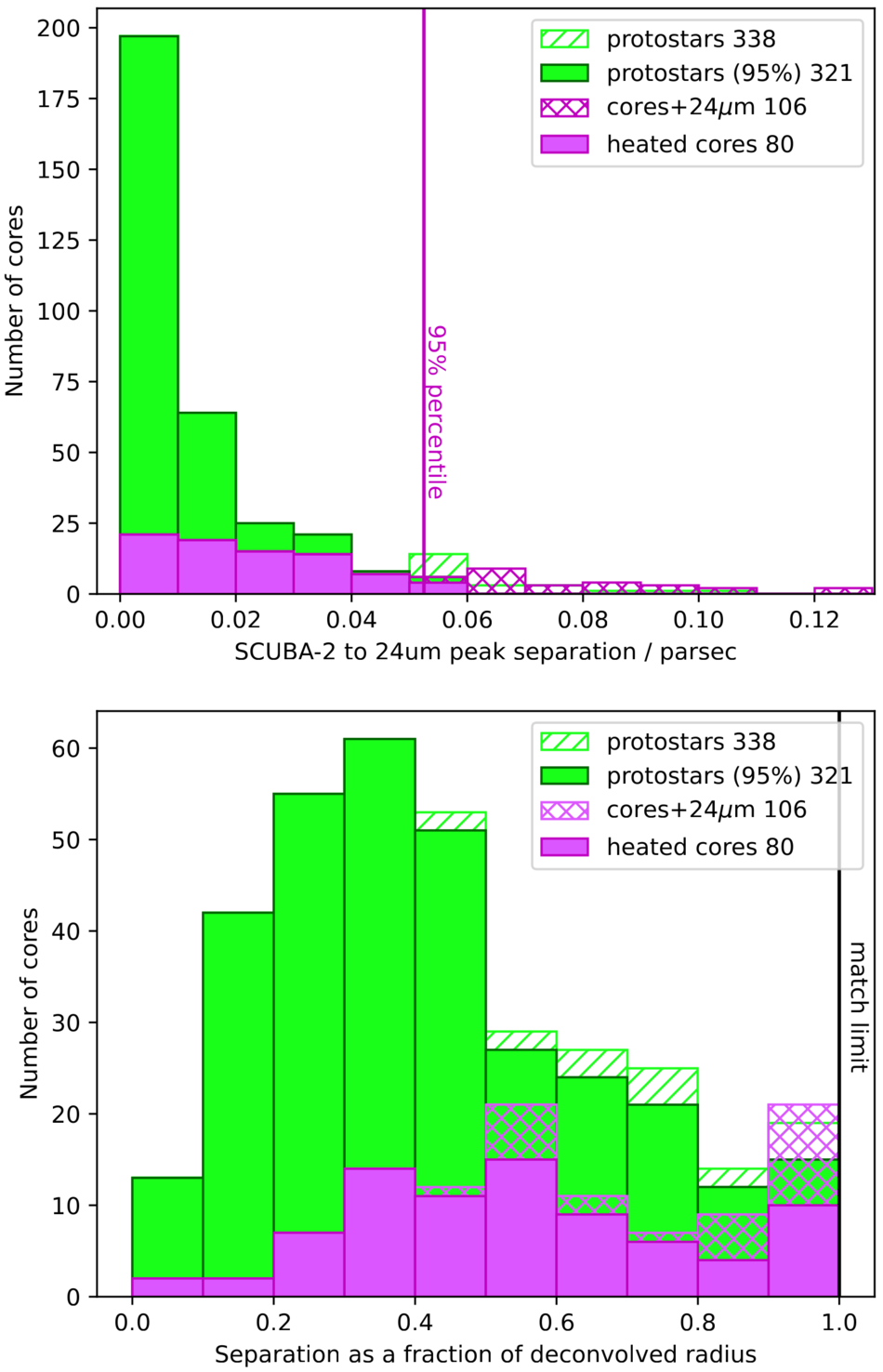}
    \caption{The distribution of separations between the SCUBA-2 core peak and the nearest 24 \micron\ \textit{Spitzer} source as a function of distance (top panel) and as a fraction of core effective diameter (bottom panel).  Cores already classified as protostellar are shown in green and cores reclassified from starless to `heated' (i.e. potentially protostellar) due to a 24 \micron\ counterpart are shown in magenta, before (hashed) and after (solid) applying a 95\% percentile separation cut.  Only those starless cores which remain associated with 24\,$\upmu$m emission after the 95\% separation cut are reclassified as heated in the final catalogue.}
    \label{fig:24um}
\end{figure}

\begin{figure}
    \centering
    \includegraphics[width=0.47\textwidth]{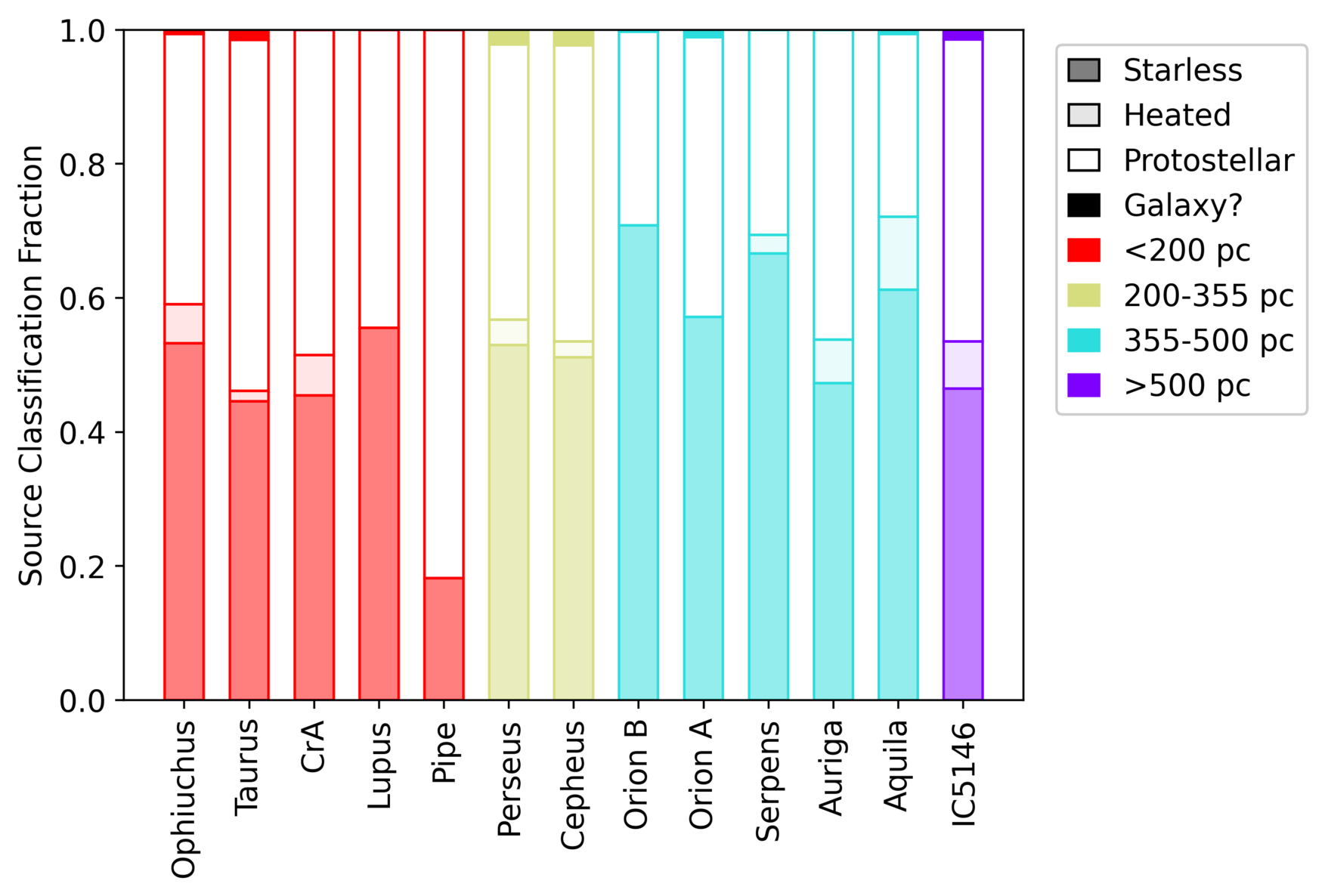}
    \caption{Bar chart summarizing source classification statistics for each cloud complex.  The complexes are plotted in order of average distance from the Sun.  Bars are colour-coded by distance, with opacity denoting source classification.} 
    \label{fig:class_bar}
\end{figure}

\subsubsection{Cross-matching with \textit{Spitzer} 24 \micron\ detections}
\label{sec:classify_24um}

Cores containing low-luminosity embedded protostars sometimes escape classification as YSOs but can still be identified by their \textit{Spitzer} 24~\micron\ emission, as can bright galaxies and AGB stars.  To remove these potential contaminants from the starless core catalogue, the \textit{Spitzer} Cores to Disks and \textit{Spitzer} Gould Belt full high reliability catalogues were searched for 24~\micron\ matches with offsets from the SCUBA-2 peak within the core deconvolved radius 
or half the JCMT 850 \micron\ beam FWHM, whichever was the greater.  To be considered as a match, the 24 \micron\ source had to have an SNR of at least 3 (quality flag \verb+MP1_Q_det_c = `A'+ or \verb+`B'+; \citealp{evans2007}).  

The separation distributions of 24 \micron\ matches for protostellar cores and for starless cores both peak at small separations of $\sim 0.01$~pc as shown in Fig.~\ref{fig:24um} (top panel), but whereas for protostellar cores the peak is at small fractions (0.3--0.4) of the deconvolved radius, for starless cores the peak is further out (0.5--0.6) and rises again at separations close to the deconvolved radius (Fig.~\ref{fig:24um} bottom panel).   This second peak suggests that some of the 24 \micron\ matches are bright cloud rims or background sources.  To reduce such false positives due to background confusion, we calculated the 95\% percentile of the separation distribution for 24~\micron\ matches with protostellar cores (bona fide 24~\micron\ matches) and applied this as an additional separation cutoff to the starless core candidates.  This criterion was applied to physical (projected) distance, rather than angular distance, so that the distribution reflects the real physical separations between dust peaks and 24~\micron\ emission in protostars e.g. due to outflow cavities.  As a result, we only consider separations below 0.0525~pc as genuine associations.  Using this revised criterion reduced the number of reclassified cores by roughly a quarter.  In total, 80 cores (just under 5\%) were reclassified from starless to `heated' (i.e. potentially protostellar) due to 24 \micron\ associations, with a separation distribution shown in Fig.~\ref{fig:24um}.  10 of these cores were later excluded by the selection criteria described in Section~\ref{sec:source_selection}, leaving 70 heated cores in the final catalogue.

We note that the \textit{Spitzer} MIPS 24\,$\upmu$m channel is saturated in the brightest parts of the Orion Molecular Cloud.  We manually flagged the Orion BN/KL region in Orion A (JCMTLSG 053514.3-052231) and the centre of Orion B NGC 2074 (JCMTLSG 054144.6-015540) as protostellar.

\subsubsection{Final classification}

\begin{table*} 
\centering 
\caption{A summary of the classification of the sources in our catalogue by region.}
\begin{tabular}{c c c c c c c c} 
\hline 
Region & Total Sources & Starless & Protostellar & 24$\mu$m-bright (`Heated') & Galaxy & Resolved & Unresolved \\ 
\hline 
Aquila & 312 & 191 & 85 & 34 & 2 & 284 & 28 \\ 
Auriga & 93 & 44 & 43 & 6 & 0 & 79 & 14 \\ 
Cepheus L1228 & 8 & 5 & 2 & 1 & 0 & 7 & 1 \\ 
Cepheus L1251 & 14 & 8 & 5 & 0 & 1 & 14 & 0 \\ 
Cepheus South & 21 & 9 & 12 & 0 & 0 & 16 & 5 \\ 
CrA & 33 & 15 & 16 & 2 & 0 & 25 & 8 \\ 
IC5146 & 71 & 33 & 32 & 5 & 1 & 67 & 4 \\ 
Lupus & 9 & 5 & 4 & 0 & 0 & 9 & 0 \\ 
Ophiuchus L1688 & 119 & 60 & 52 & 6 & 1 & 91 & 28 \\ 
Oph L1689/1709/1712 & 34 & 22 & 10 & 2 & 0 & 27 & 7 \\ 
Oph/Scorpius N2 & 1 & 0 & 0 & 1 & 0 & 1 & 0 \\ 
Orion A & 731 & 418 & 305 & 0 & 8 & 685 & 46 \\ 
Orion B L1622 & 11 & 4 & 7 & 0 & 0 & 11 & 0 \\ 
Orion B N2023 & 187 & 143 & 44 & 0 & 0 & 172 & 15 \\ 
Orion B N2068 & 169 & 113 & 55 & 0 & 1 & 146 & 23 \\ 
Perseus IC348 & 39 & 21 & 15 & 2 & 1 & 32 & 7 \\ 
Perseus West & 146 & 77 & 61 & 5 & 3 & 125 & 21 \\ 
Pipe B59 & 10 & 1 & 9 & 0 & 0 & 4 & 6 \\ 
Pipe E1 & 1 & 1 & 0 & 0 & 0 & 1 & 0 \\ 
Serpens East & 68 & 64 & 4 & 0 & 0 & 66 & 2 \\ 
Serpens Main & 73 & 31 & 38 & 4 & 0 & 63 & 10 \\ 
Serpens MWC297 & 19 & 9 & 9 & 1 & 0 & 13 & 6 \\ 
Serpens North & 23 & 18 & 5 & 0 & 0 & 22 & 1 \\ 
Taurus B18 East & 17 & 9 & 8 & 0 & 0 & 12 & 5 \\ 
Taurus B18 West & 1 & 0 & 1 & 0 & 0 & 0 & 1 \\ 
Taurus L1495 & 32 & 15 & 16 & 0 & 1 & 23 & 9 \\ 
Taurus TMC1 & 15 & 5 & 9 & 1 & 0 & 9 & 6 \\ 
\hline
Total & 2257 & 1321 & 847 & 70 & 19 & 2004 & 253 \\ 
\hline
\end{tabular}
\label{tab:regions}
\end{table*}

Of our 2257 sources, 1321 are classed as starless (`C'), 847 as containing an embedded protostar (`P'), 70 as heated (24\,$\upmu$m-bright; `H'), and 19 as potential extragalactic contaminants (`G').  We consider anything with a P or H classification as potentially protostellar in nature.  It is important to note that `protostellar' thus effectively means that the source has an infrared association.  Our aim is to create an uncontaminated catalogue of starless cores, and so some of the objects identified as protostellar may in fact be shocked knots, or other externally heated objects without embedded sources.  

\subsection{Derived properties}

The derived properties of our sources are listed in Table~\ref{tab:derived_sample}, and in the full catalogue which is supplied as an online resource.

\subsubsection{Source Temperature}
\label{sec:temperature}

To calculate core masses, a dust temperature estimate is required.  Dust temperatures vary between cores and within them, depending on the strength and penetration of the interstellar radiation field in the absence of internal heating \citep{evans2001}.  Isolated prestellar cores have been measured and modelled to have central dust temperatures as low as 7--9~K \citep[e.g.][]{leung1975, evans2001, Nielbock:2012}.  The {\em Herschel} surveys of Serpens and Aquila find core average temperatures of 10--11~K for robust prestellar cores and 15~K for unbound cores, from SED fitting with an opacity-modified blackbody \citep{fiorellino2021, konyves2015}.  From Bayesian modelling, core-containing filaments in Ophiuchus and Taurus have mass-weighted temperatures of 10--15~K in Taurus and 14--20~K in Ophiuchus \citep{Howard:2019, Howard:2021}.  Higher temperatures are produced by an enhanced radiation field, for example due to proximity to OB stars.  In particular, dust temperatures in Orion~A reach more than 40~K due to heating by the Orion Nebula Cluster and multiple other OB stars in the region \citep{schuller2021}.

The GBS has previously mapped dust colour temperature from the ratio of the SCUBA-2 450 \micron\ to 850 \micron\ intensities \citep{rumble2021}, assuming a fixed dust opacity index of $\beta=1.8$.  These maps do not cover the whole area mapped by the JCMT GBS, as this method requires a high-SNR 450 \micron\ detection, achievable only in bright regions observed in good weather, and so are insufficient to extract temperatures for the individual cores considered here.  However, where there is coverage, the average dust colour temperature of an unheated dust clump is 15~K, rising to 20~K or more in the vicinity of OB~stars (within 1~pc for early B type, 2.4~pc of O-type).  This effect can be modelled based on plane-of-the-sky proximity to the dominant OB~star and its stellar spectral classification.  Temperatures of 15~K are consistent with the {\em Herschel}-based estimates for unbound starless cores.  The lower, 10--11~K, temperatures for bound prestellar cores are not seen because the clumps in \citet{rumble2021} are typically larger (median flux-weighted clump size 0.08~pc) and estimates include the warmer dust in the surrounding filament.

For regions without OB stars, we assume $T=15$\,K for our cores, which is consistent with the mean temperature of unheated clumps in the GBS temperature maps and with the temperature of unbound cores in the {\em Herschel} studies of Serpens/Aquila \citep{fiorellino2021, konyves2015}.

For regions with OB stars, we estimate temperatures based on proximity to the irradiating source using the \citet{rumble2021} formula, which assumes a base temperature of 15~K and temperature increases based on proximity to the main OB star in the neighbourhood.

Orion A is complicated by the presence of many OB stars, so we assume temperatures of 40\,K in the vicinity of the Trapezium cluster, of 20\,K at declinations $> -5.277^{\circ}$, and of 15\,K at declinations $< -5.525^{\circ}$, following the NH$_{3}$ gas temperatures measured by \citet{friesen2017} and the dust temperatures measured by \citet{schuller2021}.

\subsubsection{Source Masses}
\label{sec:mass}

We calculate source masses from their 850-$\upmu$m flux densities using the \citet{hildebrand1983} relationship,
\begin{equation}
M = \dfrac{F_{\nu}D^{2}}{\kappa_{\nu}B_{\nu}(T)},
\label{eq:mass}
\end{equation}
where $F_{\nu}$ is integrated flux density at 850$\upmu$m, $D$ is distance to the source, $\kappa_{\nu}$ is the dust opacity and $B_{\nu}(T)$ is the Planck function, 
where $T$ is determined as discussed above. 
We determine dust opacity at 850\,$\upmu$m using the \citet{beckwith1990} relationship, 
\begin{equation}
    \kappa_{\nu} = 0.1\left(\frac{\nu}{10^{12}\,{\rm Hz}}\right)^{\beta}\,{\rm cm}^{2}\,{\rm g}^{-1},
    \label{eq:opacity}
\end{equation}
again taking $\beta = 1.8$ \citep[cf.][]{sadavoy2013,juvela2015,chen2016}, consistent with Planck observations \citep{juvela2015} and joint \textit{Herschel}/SCUBA-2 fits to GBS data \citep{chen2016,sadavoy2013}.  While some localised variation in $\beta$ around this value is seen in some GBS regions \citep{sadavoy2013,chen2016,pattle2025}, $\beta = 1.8$ is a good representative value for determining the masses of the cores in our catalogue.  In our choice of $\kappa_{\nu}(10^{12}\,{\rm Hz}) = 0.1\,{\rm cm}^{2}\,{\rm g}^{-1}$, we have implictly assumed a dust-to-gas mass ratio of 1:100 \citep{beckwith1990}.

Our assumed distances to the various clouds surveyed are taken from recent \textit{Gaia} measurements, and are listed in Table~\ref{tab:regions}.  Cloud distances vary from 134\,pc (Oph/Sco N2) to 751\,pc (IC 5146).  With the exception of IC 5146, all clouds are located within 500\,pc of the Solar System.

\subsubsection{Deconvolved source sizes}

We calculate deconvolved source sizes for sources extended relative to the beam at 850\,\um.  The deconvolved radius, $R$, is taken to be the geometric mean of the major and minor FWHMs measured at 850\,\um, with the JCMT 850\,\um\ effective beam FWHM subtracted in quadrature, i.e.
\begin{equation}
R = D\tan\left({\rm FWHM_{A}FWHM_{B}} - \theta_\mathrm{eff}  ^{2}\right)^{\frac{1}{2}},
\label{eq:deconv}
\end{equation}
where $\theta_\mathrm{eff}  =14.4$\arcsec\ is the effective FWHM beam size of the JCMT at 850 \um\ \citep{dempsey2013} and $D$ is the distance to the source.

Unresolved sources (i.e. those which are not extended along either axis relative to the 850\,\um\ beam) are classified as such in Table~\ref{tab:derived_sample}.  Their `deconvolved' size is taken to be the spatial extent of the FWHM 850 \um\ beam at the distance of the source, although we largely excluded these sources from our discussions of source size (see Section~\ref{sec:violins}, below).  We find that 266 sources are unresolved, 11.8\% of the catalogue.  Of these, 208 are protostellar sources (24.7\% of the protostellar sample), and 58 are starless cores (4.4\% of the starless core sample). 

\subsubsection{Column and volume density}

We determine mean H$_{2}$ column densities for our sources using the equation
\begin{equation}
N({\rm H}_{2}) = \dfrac{M}{\uppi \mu_{mol} R^{2}},
\label{eq:col}
\end{equation}
and mean H$_{2}$ volume densities using the equations
\begin{equation}
n({\rm H}_{2}) = \dfrac{M}{\frac{4}{3}\uppi \mu_{mol} R^{3}},
\label{eq:vol}
\end{equation}
taking a mean molecular weight of $\mu_{mol} = 2.86$ (assuming the gas is 70\% H$_{2}$ by mass; \citealt{kirk2013}).  Assuming a mean particle mass of 2.3 amu, our H$_{2}$ number densities can be converted to total gas particle number densities by multiplication by a factor of 1.24.

\subsubsection{Bonnor--Ebert mass}
\label{sec:be}

We assess the stability of the starless cores in our catalogue using the Bonnor--Ebert (BE) stability criterion.  The BE model (\citealt{ebert1955}; \citealt{bonnor1956}) treats a core as an isothermal, self-gravitating, polytropic sphere bounded by external pressure.  The mass at which an isothermal BE sphere of temperature $T$ is critically stable against gravitational collapse is a widely-used proxy for the virial mass (e.g. \citealt{konyves2015}, and refs. therein), and is given by
\begin{equation}
M_{\textsc{bec}} = 2.4\dfrac{c_{s}^{2}}{G}R_{\textsc{bec}} = 2.4\dfrac{k_{\textsc{b}}T}{\mu_{mol} m_{{\sc H}}G}R_{\textsc{bec}},
\label{eq:mbe}
\end{equation}
where $c_{s}=\sqrt{k_{\textsc{b}}T/\mu_{mol} m_{\textsc{h}}}$ is the sound speed and $R_{\textsc{bec}}$ is the radius at which the critically-stable Bonnor--Ebert (BEC) sphere is bounded by the gas pressure of its surroundings.

The ratio of a core's BEC mass to its measured mass,
\begin{equation}
\alpha_{\textsc{bec}} = \dfrac{M_{\textsc{bec}}}{M},
\label{eq:abe}
\end{equation}
is thus analogous to the virial stability ratio.  A value of $\alpha_{\textsc{bec}}<1$ implies that a core is gravitationally unstable, while $\alpha_{\textsc{bec}}>1$ implies that a core can be supported against collapse by its internal thermal pressure.  In keeping with standard practice, we consider those cores with $\alpha_{\textsc{bec}}<2$ as being potentially gravitationally unstable \citep[e.g.][]{konyves2015}.  This choice follows from a similar assumption, that cores with a virial ratio $<2$ can be considered likely to be gravitationally bound \citep[cf.][]{bertoldi1992}.  We emphasise that we are not suggesting that the cores in our catalogue have Bonnor--Ebert density profiles; we are simply comparing their masses to those of BEC spheres of the same size, and from this widely-used comparison identifying the cores most likely to be gravitationally bound.

\begin{figure}
    \centering
    \includegraphics[width=0.47\textwidth]{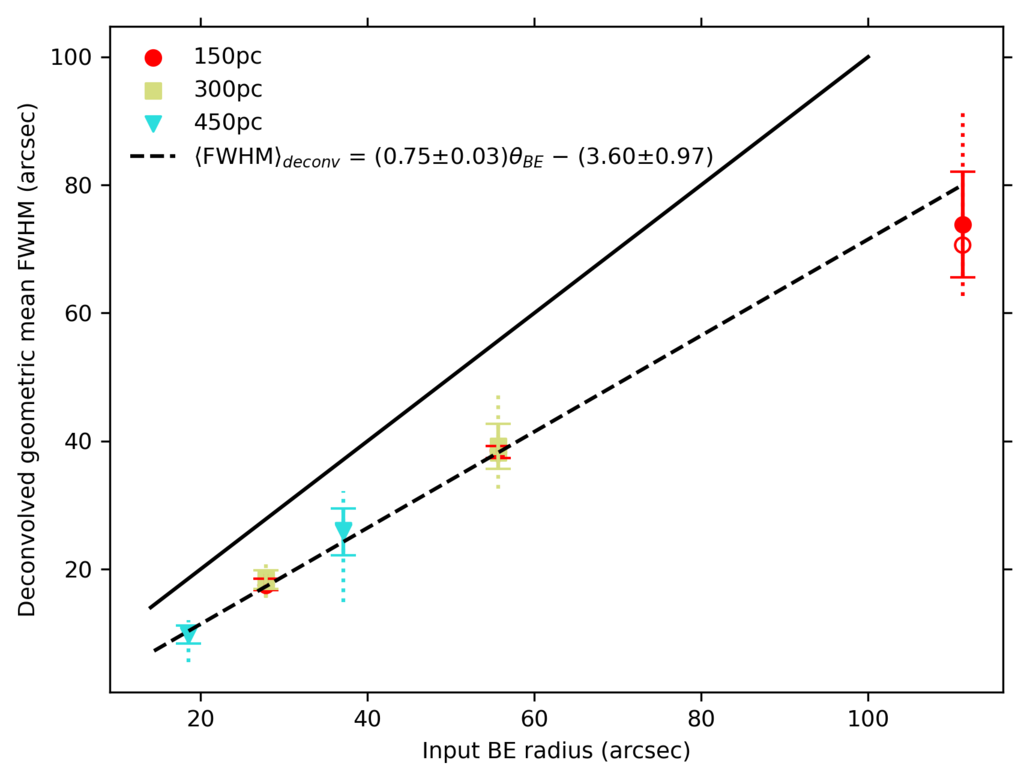}
    \caption{Measured deconvolved geometric mean FWHM vs input BEC radius, from completeness testing.  Solid symbols show means, open symbols show medians, error bars show 1-$\sigma$ uncertainties, dotted line shows full range. The straight lines show the 1:1 relationship (solid) and best fit (dashed).}
    \label{fig:rbe_fwhm}
\end{figure}

To infer the BEC masses of our cores, we first needed to determine the relationship between \textit{getsources} FWHM and $R_{\textsc{bec}}$.  For this purpose, we compared the geometric mean of the source sizes returned in our completeness fields, deconvolved with the JCMT beam, with their input BEC radii.  The results are shown in Figure~\ref{fig:rbe_fwhm}.  We found 
\begin{equation}
    \langle{\rm FWHM}\rangle_{\rm deconv} \,\,(^{\prime\prime}) = (0.75\pm 0.03)\theta_{\textsc{bec}}\,\,(^{\prime\prime})-(3.60\pm 0.97),
\end{equation}
and so,
\begin{equation}
    \theta_{\textsc{bec}}\,\,(^{\prime\prime}) = (1.33\pm 0.01)\langle{\rm FWHM}\rangle_{\rm deconv}\,\,(^{\prime\prime}) + (4.80\pm 0.97).
\end{equation}
Note that this equation gives an effective BEC radius for an unresolved source ($\langle{\rm FWHM}\rangle_{\rm deconv} = 0$) of 4.8$^{\prime\prime}$, approximately 1/3 of the 850\,$\upmu$m primary beam size.  We adopt this angular radius for unresolved sources in the following stability analysis, noting that only 4.4\% of our starless core sample is unresolved.

We further note that our completeness testing indicates that typically, $73\pm6$\% of the mass of a BEC sphere will be recovered by SCUBA-2.  We therefore empirically define the effective BEC masses of the sources we detect to be
\begin{equation}
    M_{\textsc{bec},\mathrm{eff}} = 0.73\times 2.4\dfrac{c_{s}^{2}}{G}D\tan\theta_{\textsc{bec}}.
\end{equation}

Although 15\,K is a typical temperature for unheated cores, a gravitationally-bound prestellar core may have temperatures as low as $\sim 10\,$K, as discussed in Section~\ref{sec:temperature}.  In order to identify starless cores which are good candidates for being gravitationally bound (`prestellar'), we therefore calculate $\alpha_\textsc{bec}$ if the core were at a temperature of 10\,K,
\begin{equation}
    \alpha_{\textsc{bec}, 10\,{\rm K}} = \dfrac{M_{\textsc{bec},\mathrm{eff}}(T=10\,{\rm K})}{M(T=10\,{\rm K})}.
\end{equation}
We consider cores with $\alpha_{\textsc{bec}, 10\,K} < 2$ (i.e. bound at 10\,K) to be `candidate' prestellar cores.

We further consider a more stringent criterion for core boundedness by calculating $\alpha_\textsc{bec}$ at the temperature assigned to the core in our catalogue (15\,K, or greater if the core is heated), such that
\begin{equation}
    \alpha_{\textsc{bec}, \geq15\,{\rm K}} = \dfrac{M_{\textsc{bec},\mathrm{eff}}(T)}{M(T)},
\end{equation}
where $T$ is the temperature listed in Table~\ref{tab:derived_sample}.  We consider those cores with $\alpha_{\textsc{bec}, \geq15\,{\rm K}} < 2$ to be `robust' prestellar cores.

\begin{figure}
    \centering
    \includegraphics[width=\columnwidth]{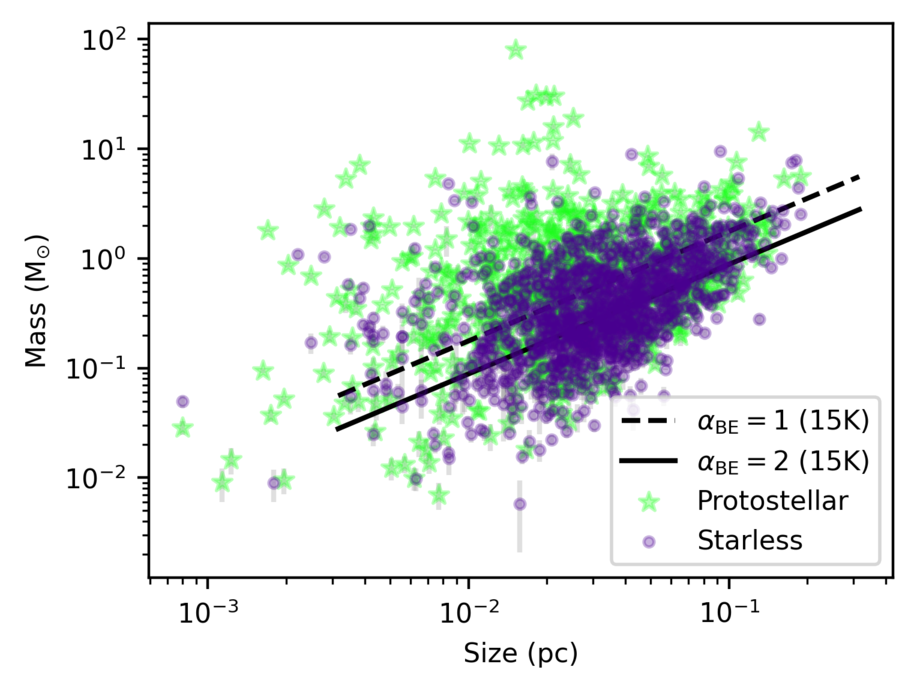}
    \caption{Source mass as a function of source size for the resolved sources in our catalogue.  Blue circles mark starless cores; green stars mark protostellar sources.  The solid black line marks the locus of a 15\,K core with a BEC mass ratio of 2, while the dashed black line marks that of a 15\,K core with a BEC mass ratio of 1.}
    \label{fig:mass-size}
\end{figure}

\begin{figure}
    \centering
    \includegraphics[width=0.47\textwidth]{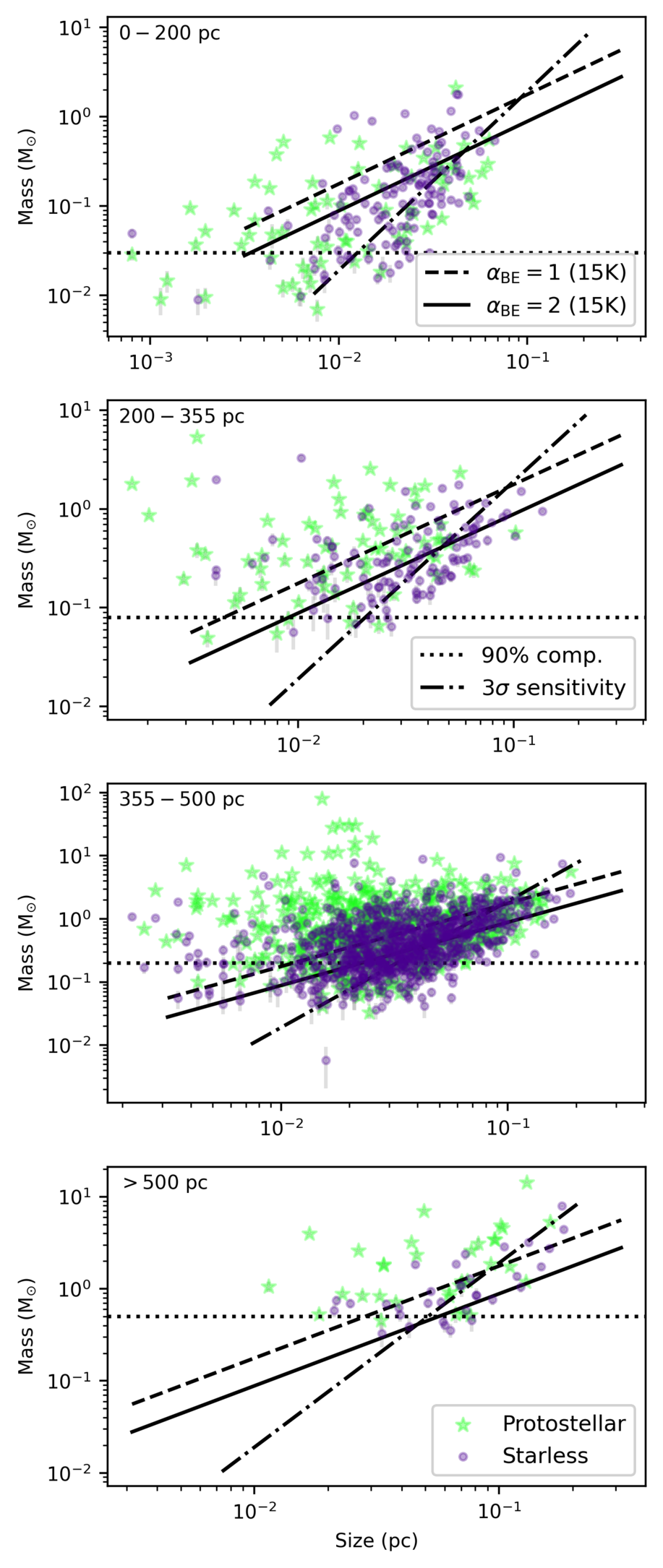}
    \caption{Source mass as a function of source size for resolved sources, separated by source distance.  \textit{Top panel:} near ($<200$\,pc), \textit{second panel:} mid-distance (200--355\,pc), \textit{third panel:} far (355--500\,pc), \textit{bottom panel:} very far ($>500$\,pc).  Data points are as in Figure~\ref{fig:mass-size}.  Solid black lines mark the locus of a 15\,K core with a BEC mass ratio of 2, dashed black lines mark that of a 15\,K core with a BEC mass ratio of 1, and the dotted black lines mark 90\% mass completeness limit in each distance range.}
    \label{fig:mass-size_by_dist}
\end{figure}

\section{Discussion of core properties}
\label{sec:properties}

To increase the sample sizes of starless cores to statistically meaningful levels, we considered our set of observed fields to represent 12 `cloud complexes': Aquila, Auriga, Cepheus (Cepheus L1228, L1251 and South), IC 5146, Lupus, Ophiuchus (Ophiuchus L1688, L1689/1709/1712, and Oph/Sco N2), Orion A, Orion B (L1622, N2023 and N2068), Perseus (IC348 and Perseus West), Pipe, Serpens (Serpens East, Main and MWC297), and Taurus (Taurus B18, L1495 and TMC-1).
We estimated the mass of potentially star-forming gas in each cloud complex using column density maps created from \textit{Herschel} SPIRE and PACS measurements by the Herschel Gould Belt Survey \citep{andre2010}\footnote{Column density maps are available at \url{http://www.herschel.fr/cea/gouldbelt/en/}}, except for the Auriga molecular cloud, for which we used column density maps published by \citet{harvey2013}.  We summed the mass at column densities $N({\rm H}_{2})> 7\times 10^{21}$\,cm$^{-2}$, to encompass the densest gas that is likely involved with star formation \citep[e.g.][]{konyves2015,difrancesco2020,konyves2020,Pezzuto2021}.  These `star-forming' gas masses are listed in Table~\ref{tab:cloud_masses} for each cloud complex.

\begin{table}
    \caption{Potentially star-forming gas mass in each cloud complex, as measured from \textit{Herschel} column density maps.} 
    \centering
    \begin{tabular}{ccccc}
    \hline
    Cloud & Average & \textit{Herschel} & \textit{Herschel}  \\
    complex & distance & Mass &  Reference \\
     & (pc) & (M$_{\odot}$) &  \\
    \hline
    Aquila & 484 & 14312 &\citet{konyves2015} \\
    Auriga & 470 & 6661 &\citet{harvey2013} \\
    Cepheus & 347 & 431 &\citet{difrancesco2020} \\
    CrA & 151 & 102 &\citet{bresnahan2018} \\
    IC 5146 & 751 & 902 &\citet{arzoumanian2011} \\
    Lupus & 151 & 26 &\citet{rygl2013} \\
    Ophiuchus & 140 & 598 &\citet{ladjelate2020} \\
    Orion A & 432 & 12918 &\citet{roy2013} \\
    & & & \citet{polychroni2013} \\
    Orion B & 423 & 3919 & \citet{schneider2013} \\
    Perseus & 308 & 1543 & \citet{sadavoy2014} \\
     & & & \citet{Pezzuto2021}\\
    Pipe & 180 & 34 &\citet{peretto2012} \\
    Serpens & 447 & 5443 &\citet{fiorellino2021} \\
    Taurus & 141 & 406 & \citet{kirk2013} \\
     & & & \citet{marsh2016} \\
     & & & \citet{kirk2024} \\
     & & & J. Kirk (priv. comm.) \\
    \hline
    \end{tabular}
    \label{tab:cloud_masses}
\end{table}

\begin{table*}
\caption{Core counts and classifications in each cloud complex.}
\centering
\begin{tabular}{c cc@{\extracolsep{4pt}}ccc}
\hline
 & \multicolumn{2}{c}{Core Counts} & \multicolumn{3}{c}{Starless Core Classes} \\ \cline{2-3} \cline{4-6}
 & & & Robust & Candidate & \\ 
Complex & Protostellar & Starless & Prestellar & Prestellar & Unbound \\ 
\hline
Aquila & 119 & 191 & 47 & 141 & 3 \\ 
Auriga & 49 & 44 & 14 & 15 & 15 \\ 
Cepheus & 20 & 22 & 5 & 5 & 12 \\ 
CrA & 18 & 15 & 2 & 10 & 3 \\ 
IC5146 & 37 & 33 & 17 & 1 & 15 \\ 
Lupus & 4 & 5 & 1 & 0 & 4 \\ 
Ophiuchus & 71 & 82 & 14 & 29 & 39 \\ 
Orion A & 305 & 418 & 157 & 149 & 112 \\ 
Orion B & 106 & 260 & 95 & 103 & 62 \\ 
Perseus & 83 & 98 & 32 & 22 & 44 \\ 
Pipe & 9 & 2 & 0 & 0 & 2 \\ 
Serpens & 61 & 122 & 66 & 4 & 52 \\ 
Taurus & 35 & 29 & 4 & 5 & 20 \\ 
\hline
Total & 917 & 1321 & 454 & 484 & 383 \\ 
\hline
\end{tabular}
\label{tab:core_stats}
\end{table*}

\begin{figure}
    \centering
    \includegraphics[width=\columnwidth]{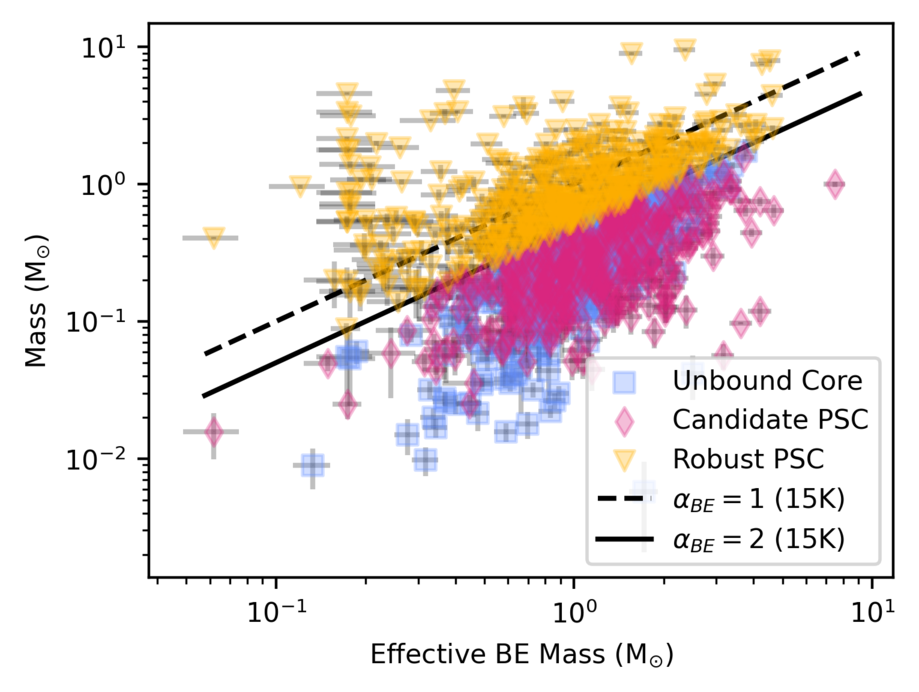}
    \caption{Source mass as a function of effective BEC mass, for the resolved starless cores in our sample.  Blue squares mark unbound cores, red diamonds mark candidate prestellar cores ($\alpha_{\textsc{bec}, 10\,K} < 2$), and yellow triangles mark robust starless cores ($\alpha_{\textsc{bec}, \geq15\,{\rm K}} < 2$).  The solid black line marks the locus of a 15\,K core with a BEC mass ratio of 2, while the dashed black line marks that of a 15\,K core with a BEC mass ratio of 1.}
    \label{fig:BE_starless}
\end{figure}

\begin{figure}
    \centering
    \includegraphics[width=0.47\textwidth]{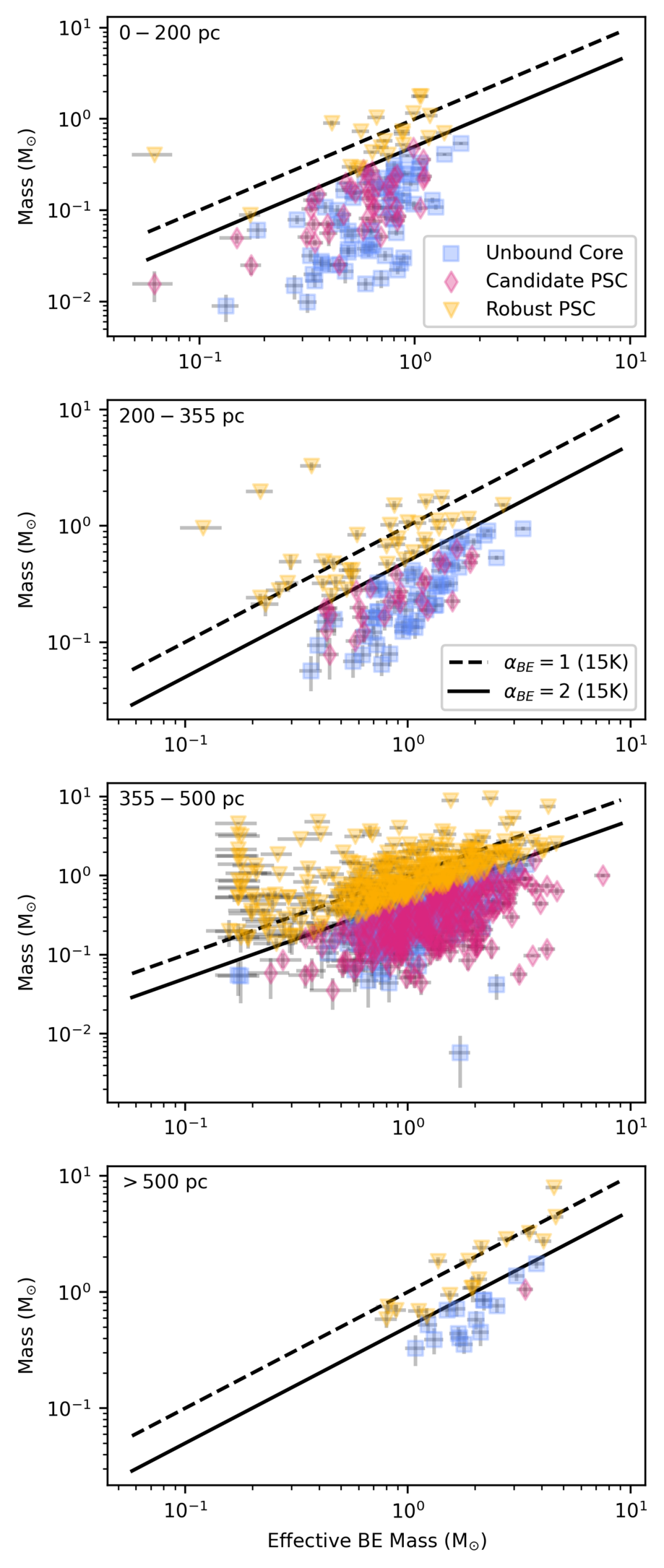}
    \caption{Source mass as a function of effective BEC mass, for the resolved starless cores in our sample, separated by source distance.  \textit{Top panel:} Near ($<200$\,pc). \textit{Second panel:} Mid-distance (200--355\,pc). \textit{Third panel:} Far (355--500\,pc). \textit{Bottom panel:} Very far ($>500$\,pc).  Data points are as in Figure~\ref{fig:BE_starless}.  The solid black lines mark the locus of a 15\,K core with a BEC mass ratio of 2, while the dashed black lines mark that of a 15\,K core with a BEC mass ratio of 1.}
    \label{fig:BE_starless_by_dist}
\end{figure}

\subsection{Mass vs. size}

For our resolved sources, we plotted mass as a function of deconvolved geometric mean size.  The mass/size diagram for our full sample of sources is shown in Figure~\ref{fig:mass-size}, while the mass/size diagrams for each of our distance bins are shown in Figure~\ref{fig:mass-size_by_dist}.  For each panel in Figure~\ref{fig:mass-size_by_dist}, our estimated 90\% completeness limit is shown as a dotted line, while our estimated 3-$\sigma$ sensitivity is shown as a dashed line.

We calculated our 3-$\sigma$ mass sensitivity using an assumed per-pixel 1-$\sigma$ mass sensitivity of 0.047 mJy\,arcsec$^{-2}$, which we measured using aperture photometry on the Oph/Sco N6 field (the field into which fake sources were inserted for the completeness testing, as described in Section~\ref{sec:completeness} and Appendix~\ref{app_compl}).  The dashed lines shown in Figure~\ref{fig:mass-size_by_dist} show the mass of a 15\,K source at the nominal distance, in which each pixel has a 3-$\sigma$ flux density.

We note that some sources appear below our 3-$\sigma$ mass sensitivity limit.  This apparent incongruity is likely to be due to some combination of non-uniform flux densities across real sources, 
many sources sitting on bright backgrounds, and so being easier to detect than sources sitting on the noisy background of the map, and slight differences in sensitivity between maps, due to combinations of mosaicking strategy, weather conditions in which the observations were made, and small differences in exposure time.

Gravitationally bound objects are expected to occupy the upper left-hand portion of the mass-size diagram, being massive and relatively compact.  This characteristic is demonstrated in Figure~\ref{fig:mass-size_by_dist} by the diagonal lines marking the effective BEC mass (and half of the effective BEC mass) for sources at 15\,K.  As distance increases, our completeness and mass sensitivity limits progressively exclude a larger area of the gravitationally unbound region of the mass-size plane, and so at greater distances we preferentially detect gravitationally bound and collapsing cores.
         
\subsection{Core stability}
 
We calculated effective BEC masses for all of the starless cores in our sample, as described in Section~\ref{sec:be}.  Core mass is plotted as a function of effective BEC mass in Figure~\ref{fig:BE_starless}, with both measured mass and BEC mass calculated for $T\geq 15\,$K.  Both robust ($\alpha_{\textsc{bec}, \geq15\,{\rm K}} < 2$) and candidate ($\alpha_{\textsc{bec}, 10\,K} < 2$) prestellar cores are marked on the figure.  Core masses as a function of BEC mass are also plotted for each of our distance ranges in Figure~\ref{fig:BE_starless_by_dist}.  As expected, fewer unbound cores are detected in the more distant clouds, likely as a matter of sensitivity.  Although more distant cloud complexes have a higher bound core fraction, there is no clear correlation with cloud mass within any given distance range (cf. Figure~\ref{fig:be_fracs}).

Note also that a large fraction of starless cores found in SCUBA-2 GBS maps have previously been found to be stable according to the BE criterion, or, in a virial analysis, confined by pressure rather than gravity \citep{pattle2015,pattle2017}.  This finding has been replicated in molecular line studies of dense cores \citep[e.g.][]{kirk2017,kerr2019}.  Magnetic fields may also play a significant role in supporting these cores against gravitational collapse (e.g. \citealt{myers2021}; \citealt{pattle2021}).  Thus, additional information is required to determine the exact virial state of the starless cores in our sample.

\begin{figure}
    \centering
    \includegraphics[width=0.47\textwidth]{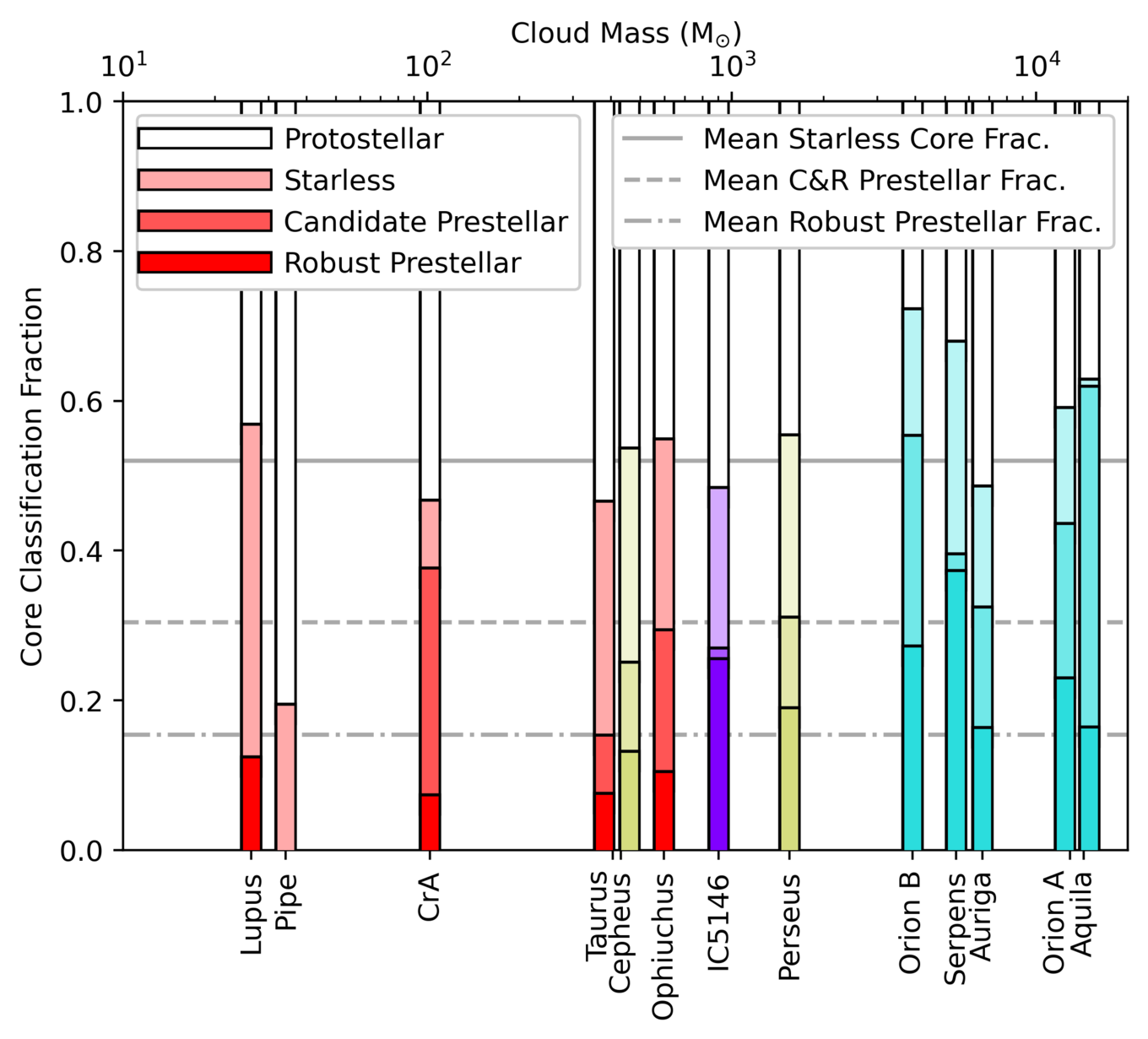}
    \caption{The fraction of cores within a cloud complex which are starless cores, candidate prestellar cores, robust prestellar cores, and protostellar cores as a function of cloud complex mass.  Grey lines show the average fractions of starless (solid line), prestellar (candidate and robust; dashed line) and robust prestellar (dot-dashed line) cores averaged over all regions, as discussed in Section~\ref{sec:core_fracs}.  Data points are colour-coded by cloud distance.  Note that the bars for Taurus, Cepheus, Orion A and Aquila are slightly offset from their true positions, to avoid overlap.  The tick marks associated with these bars show the exact cloud masses.}     \label{fig:starless_frac}
\end{figure}

\subsection{Relative numbers of protostellar and starless cores}
\label{sec:core_fracs}

The fraction of starless cores is plotted as a function of cloud mass for each cloud complex in Figure~\ref{fig:starless_frac}, based on numbers in Table~\ref{tab:core_stats}.  On average, 41\% of the detected cores are protostellar and 59\% are starless.  Of the starless cores, 34\% are robust prestellar cores, 37\% are candidate prestellar cores, and 29\% are unbound.  Thus, 41\% of our cores are protostellar and 42\% are prestellar (either candidate or robust).  Assuming all prestellar cores (candidate and robust) are the precursors of protostars, then similar source counts indicate similar lifetimes to Class 0/I embedded YSOs (0.5 Myr; \citealt{evans2009}), which are consistently detected by the JCMT GBS as protostellar cores.  If every core passes through the `robust' prestellar core phase, then it lasts half as long on average (0.25~Myr).

The starless and prestellar core fractions vary between complexes, and the differences are statistically significant.  We tested for consistency using a binomial distribution, checking core counts against the 95\% confidence interval and the two-tailed binomial test with 5\% significance, and with two hypotheses, setting the starless or prestellar core probability equal to the mean starless or prestellar core fraction, (1) averaged over all cores ($p=0.59, 0.42, 0.20$ for starless, prestellar and robust prestellar cores respectively; note the core counts are dominated by Orion~A);  (2) averaged over all regions, with equal weight for each region ($p=0.52, 0.30, 0.15$ for starless, prestellar and robust prestellar cores respectively).   For each hypothesis and type of core selected, at least four regions and up to eight regions had source counts that were not consistent with the average probability.  Under both hypotheses, Serpens and Orion B have an excess of starless and prestellar cores, whereas Pipe and Taurus show a deficit.

Under Hypothesis (1), where average core counts are dominated by Orion~A, many more of the lower and intermediate mass regions had a deficit of starless or prestellar cores (Auriga for starless cores; Auriga, Cepheus, Ophiuchus and Perseus for candidate prestellar cores; CrA and Ophiuchus for robust prestellar cores) while IC 5146 showed an excess but only for robust prestellar cores.  Under Hypothesis (2), Orion A also had an excess of starless and prestellar cores (IC 5146 showed an excess and Ophiuchus a deficit for robust prestellar cores only).

The trend is for higher-mass regions to have a higher ratio of starless (or prestellar) cores compared to protostars (Fig.~\ref{fig:starless_frac}; see also Figure~\ref{fig:starless_proto}).  Assuming candidate prestellar cores are the precursors of protostars, then a higher ratio indicates a longer lifetime for prestellar cores in higher mass regions, on average,  compared to those in lower mass regions.  This could be due to longer average freefall times in larger, lower average density clouds \citep[see][and references therein]{pokhrel2021}.  If this is the case, then prestellar cores should show a wider spatial distribution than protostars, with less concentration in high (column) density regions.  There is already some evidence that this is the case: in Orion~B, the surface density of prestellar cores follows a linear relationship with column density whereas the surface density of protostellar cores scales nonlinearly as the square of column density \citep{konyves2020, lombardi2014, pokhrel2020, retter2021}.

One might expect clouds with a large fraction of Class 0/I sources compared to more evolved YSOs also to have a large number of starless cores, as that could indicate a ramp-up in recent star formation.  From the \textit{Spitzer} Gould Belt survey, the highest fraction of Class 0/I YSOs occur in Auriga, IC 5146, and Perseus \citep{dunham2015}.  From the data presented here, there is nothing special about the starless core fractions in these clouds -- only a slight indication that it might be low in Auriga -- suggesting that star formation will continue at the same rate in the future.   

\begin{figure}
	\includegraphics[width=\columnwidth]{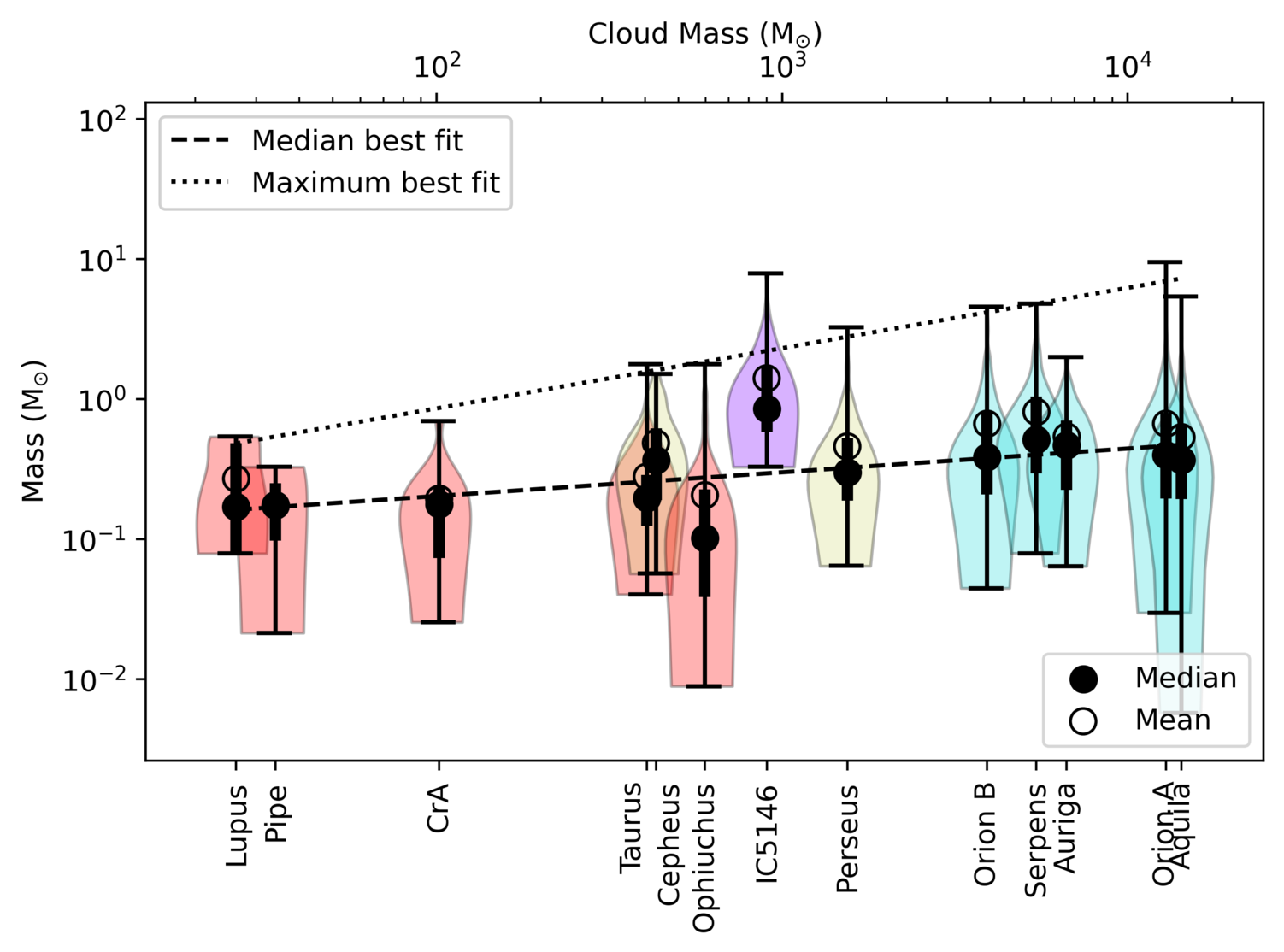}
	\includegraphics[width=\columnwidth]{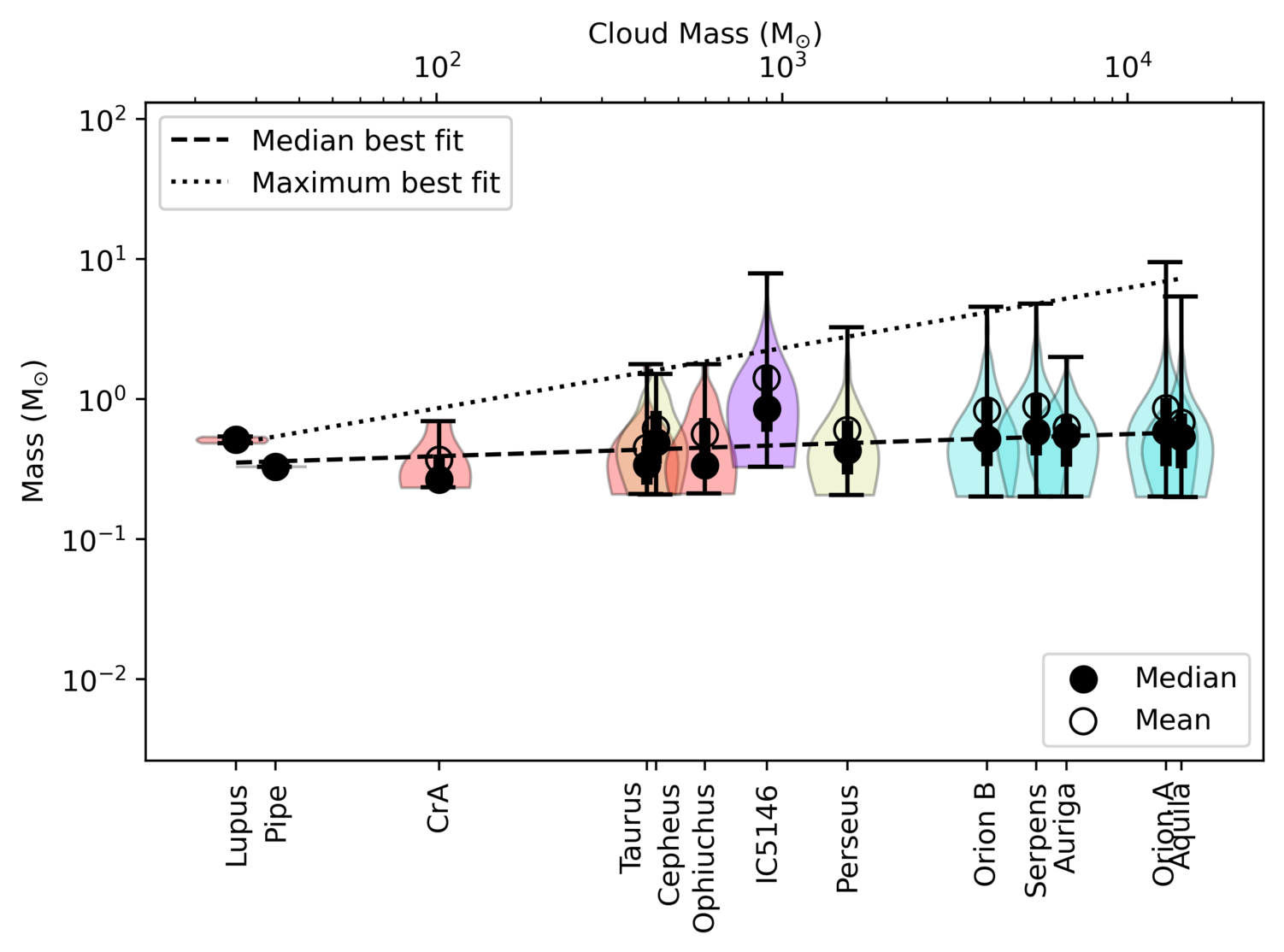}
    \caption{Mass distribution for each cloud complex, as a function of star-forming gas mass. Top: for all starless cores.  Bottom: for starless cores with masses above the 90\% completeness limit at 450\,pc of 0.2\,M$_{\odot}$.  Solid circles show median values; open circles show means.  Thick black lines show the interquartile ranges.  Dotted line shows the line of best fit to the maximum values in each cloud complex; dashed line shows the line of best fit to the median values.}
    \label{fig:mass_violin}
\end{figure}

\begin{figure}
	\includegraphics[width=\columnwidth]{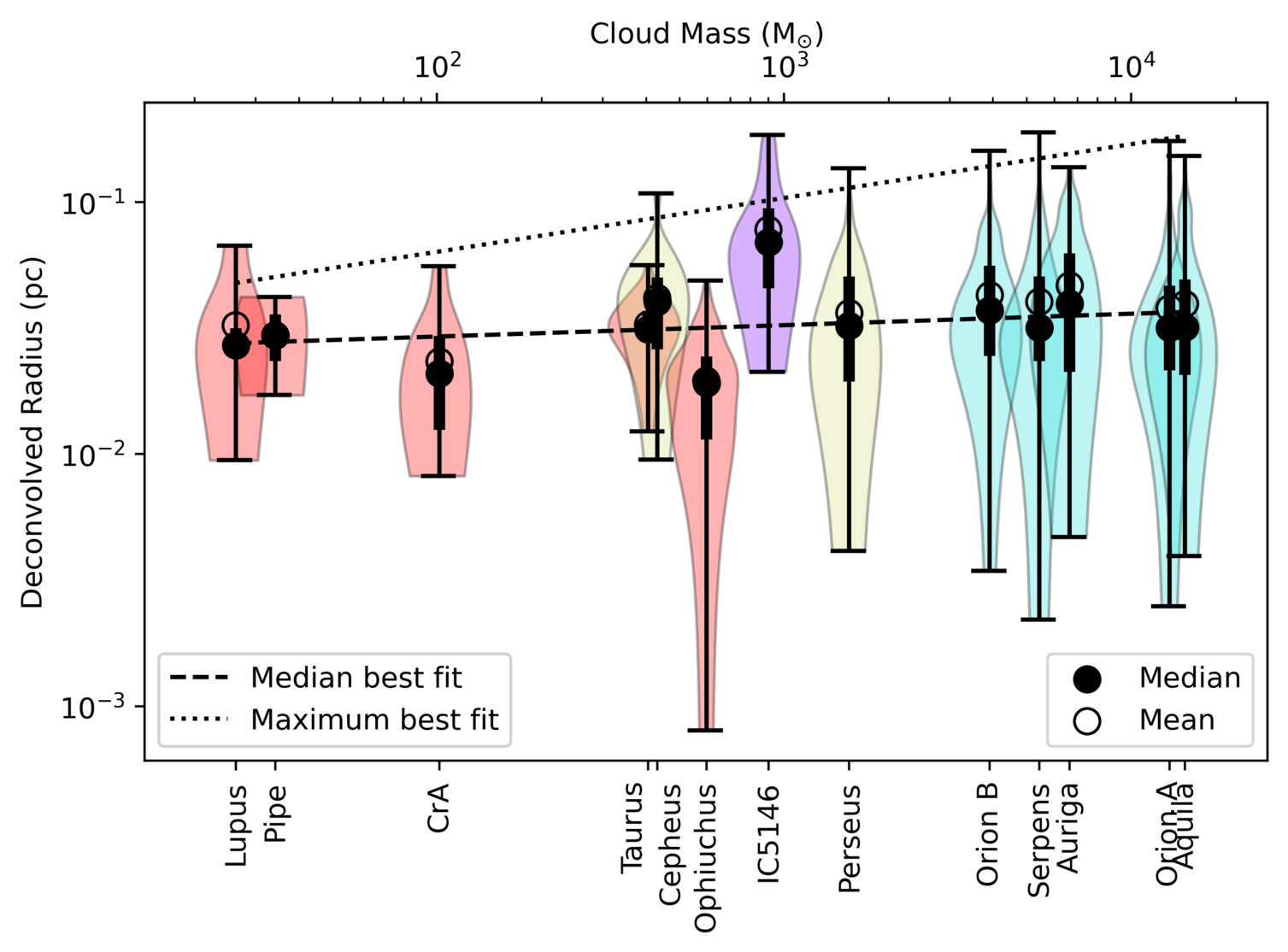}
	\includegraphics[width=\columnwidth]{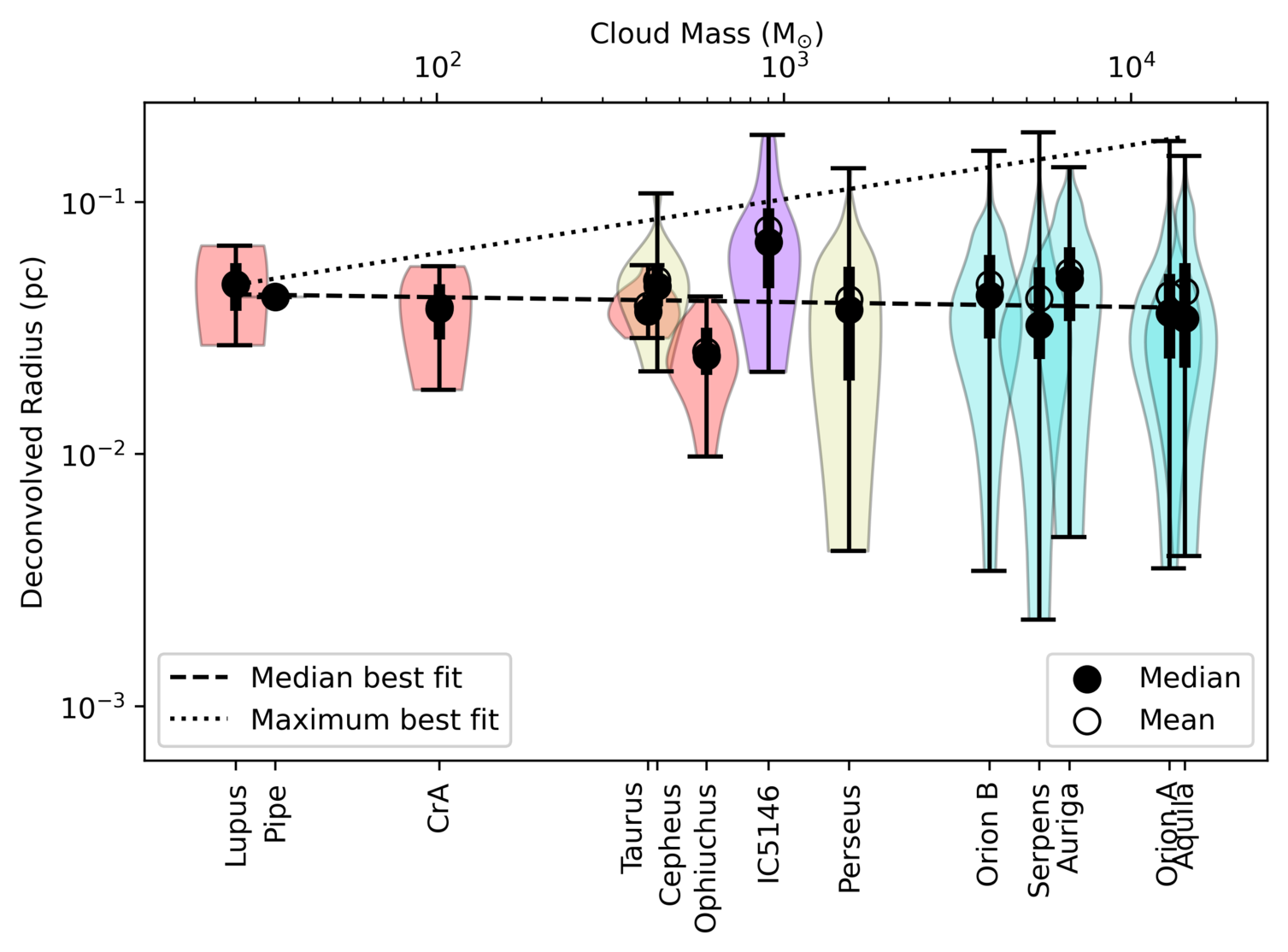}
    \caption{Deconvolved radius distribution for starless cores for each cloud complex, as a function of star-forming gas mass.  Top: for all starless cores.  Bottom: for starless cores with masses above the 90\% completeness limit at 450\,pc of 0.2\,M$_{\odot}$.  Solid circles show median values; open circles show means.  Thick black lines show the interquartile ranges.  Dotted line shows the line of best fit to the maximum values in each cloud complex; dashed line shows the line of best fit to the median values.}
    \label{fig:radius_violin}
\end{figure}

\subsection{Mass and radius distributions}
\label{sec:violins}

We plot the distribution of starless core masses as a function of potentially star-forming gas mass for each complex in Figure~\ref{fig:mass_violin}.  We find a moderately strong correlation between median starless core mass and cloud complex mass, with $M_{median}\propto M_{cloud}^{0.17\pm 0.06}$ ($r=0.62$, $p = 0.02$), determined using least-squares fitting.  However, this relationship may be influenced by the differing mass completeness limits between the different cloud complexes, with more massive clouds typically being more distant, and so we plotted the starless core mass distributions for each complex above the mass completeness limit at 450\,pc of 0.2\,M$_{\odot}$.  The best-fit relationship between cloud complex mass and median mass becomes shallower and less robust, with $M_{median}(>0.2\,{\rm M}_{\odot})\propto M_{cloud}^{0.08\pm 0.04}$ ($r=0.53$, $p=0.06$).  

At a later evolutionary stage, the maximum stellar mass in a cluster increases with embedded cluster mass \citep{WeidnerKroupaBonnell:2010,Elmegreen:2006}.  The relationship between stellar and cluster mass was examined by \citet{BonnellVineBate:2004}, who found the mass of the most massive star in a cluster scaled with cluster mass to a power of 2/3.  We investigated whether we see a similar relationship between cloud mass and maximum core mass in the cloud complexes that we consider.  We find a strong correlation between maximum starless core mass ($M_{max}$) and cloud complex mass ($M_{cloud}$) with $M_{max}\propto M_{cloud}^{0.58\pm 0.13}$ ($r=0.80$, $p < 0.005$), consistent with the \citet{BonnellVineBate:2004} index. The mass of the most massive core in a cloud thus typically increases with cloud mass (Figure \ref{fig:mass_violin}), although the extent to which this can be directly linked to the relationship between stellar and cluster mass is unclear, since the larger cloud complexes in our sample typically contain larger numbers of cores, and so are more likely to contain high-mass outliers.

We also see significant cloud-to-cloud variation in mass distributions in clouds at comparable distances.  Notably, the Ophiuchus molecular cloud has significantly more low-mass sources than are seen in any of the other nearby clouds, and a significantly lower minimum core mass.

We note that there is also a correlation between maximum and median mass and cloud distance, as might be expected from eq.~\ref{eq:mass}.  These relationships are shown in Figure~\ref{fig:mass_violin_dist} in Appendix~\ref{sec:appendix_mass_radius}.  However, the relationships between mass and distance are shallower than $D^{2}$, indicating that the correlation that we see between core and cloud masses is not solely due to more massive clouds on average being at a greater distance from us.

We further plotted the distribution of starless core deconvolved radii as a function of potentially star-forming gas mass for each complex, both for all starless cores and for cores above the 450\,pc mass completeness limit, as shown in Figure~\ref{fig:radius_violin}. 
We see no correlation between the median deconvolved core size and the mass of the cloud in either case.  However, in both cases we find a correlation between maximum deconvolved core size and cloud mass: for all cores, $R_{max}\propto M_{cloud}^{0.21\pm 0.05}$ ($r=0.81$, $p < 0.005$).  For cores above the 450\,pc mass completeness limit, $R_{max}(> 0.2\,{\rm M}_{\odot})\propto M_{cloud}^{0.21\pm 0.05}$ ($r=0.79$, $p < 0.005$).

We also investigated the relationship between maximum and median deconvolved core radius, as shown in Figure~\ref{fig:radius_violin_dist} in Appendix~\ref{sec:appendix_mass_radius}, and find that both median and maximum core radii are correlated with distance when all cores are considered, while when only cores above the 450\,pc mass completeness limit are considered, there is no correlation between median core radius and cloud distance.  However, in both cases, we find a relationship consistent within error with $R_{max}\propto D$.  Despite this, we find considerable variation between core radius distributions for clouds at similar distances to one another: Ophiuchus has a significantly smaller minimum source size and a considerable excess of small cores compared to other nearby clouds, while Cepheus has a significantly smaller spread in core radii than does Perseus, despite the two clouds being at comparable distances.

\section{Core Mass Functions}
\label{sec:cmfs}

The mass distribution of starless, or prestellar, cores is typically characterised using the Core Mass Function (CMF).  The form of the CMF is similar to that of the Initial Mass Function (IMF; \citealt{salpeter1955}; \citealt{kroupa2001}; \citealt{chabrier2003}), with a log-normal behaviour at low masses and a Salpeter-like power-law behaviour at high masses (e.g. \citealt{konyves2015}) leading to the suggestion of a causal link between the CMF and the IMF \citep{motte1998}.  \textit{Herschel} studies of dense cores in the Aquila molecular cloud have found that the characteristic mass of the prestellar CMF is $3\times$ that of the system IMF, suggesting a $\sim 33$\% core-to-star mass conversion efficiency \citep{konyves2010,konyves2015}.  However, this relies on there being a 1:1 relationship between the CMF and the IMF, and so does not account for further core fragmentation (other than into bound multiple systems), or for the potential for further accretion of mass onto cores \citep{offner2014}.

In keeping with standard practice, we visualised our measured CMFs by plotting histograms of the mass distributions of the starless cores in our catalogue, as shown in Figure~\ref{fig:cmf_sless}, and Figures~\ref{fig:cmf_aquila}--\ref{fig:cmf_taurus} in Appendix~\ref{sec:appendix_cmfs}. In each case we used logarithmically-spaced bins, with the number of non-empty bins determined using Sturges's Law, $N_{\rm bins} = 1 + \log_{2}N$, where $N$ is the number of cores in the sample.

We modelled our measured starless core mass distributions, and their candidate and robust prestellar subsets, using a log-normal mass distribution,
\begin{equation}
    \dfrac{\Delta N}{\Delta\log_{10} M} \propto \exp\left(-\dfrac{(\log_{10}M - \log_{10}\mu)^{2}}{2\sigma^{2}} \right),
\end{equation} 
where $\mu$ is the mean core mass in units of M$_{\odot}$ and $\sigma$ is the width of the mass distribution in units of $\log_{10}{\rm M}_{\odot}$.  This distribution is comparable to the log-normal part of the \citet{chabrier2003} and \citet{chabrier2005} IMF.

\subsection{Core Mass Functions by distance}
\label{sec:cmfs_distance}

CMFs for our four distance ranges, $<200\,$pc, $200 - 355$\,pc, $355-500$\,pc, $>500\,$pc, are shown in Figure~\ref{fig:cmf_sless}.  For the 355--500\,pc distance range, we also plotted the CMF with the heated cores of Orion A excluded, as shown in Figure~\ref{fig:cmf_sless_far_oflag} in Appendix~\ref{sec:appendix_cmfs}.

\subsubsection{Least-squares fitting of CMFs by distance}
\label{sec:ls_distance}

\begin{figure*}
\includegraphics[width=0.8\textwidth]{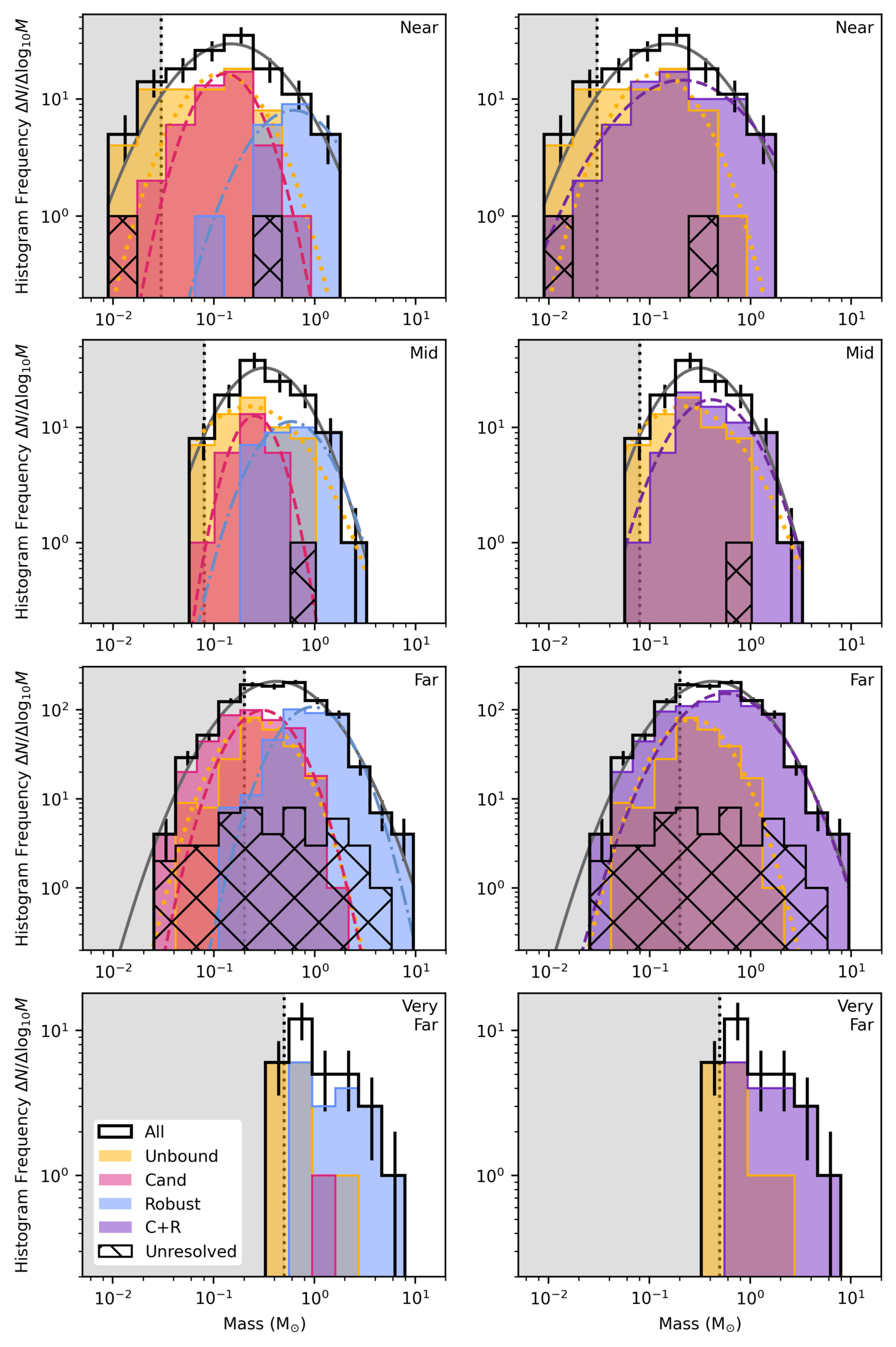}
\caption{Starless CMFs for the distance ranges which we consider.  Top to bottom: near, mid-distance, far and very far.  Left column: CMFs with fits to full, unbound, candidate prestellar and robust prestellar samples shown.  Right column: CMFs with fits to full, unbound, and combined candidate and robust (C \& R) prestellar samples.  In both columns, the unresolved sources are shown as a hatched histogram.}
\label{fig:cmf_sless}
\end{figure*}

We fitted a log-normal distribution to each of the CMFs for our four distance ranges, using the \textit{scipy} least-squares fitting routine \textit{curve\_fit}.  We fitted (i) the full distribution of starless cores, (ii) the distribution of prestellar cores (the combined candidate and robust samples; referred to as `C \& R'), (iii) the distribution of unbound cores, (iv) the distribution of candidate prestellar cores, and (v) the distribution of robust prestellar cores.  Note that samples (ii)--(v) are subsets of sample (i).  The best-fit model CMFs for each distance range are plotted on Figures~\ref{fig:cmf_sless} and \ref{fig:cmf_sless_far_oflag}.  The fitting results are shown in Table~\ref{tab:starless_cmfs_1} for all starless cores and for prestellar cores (C \& R).  Fitting results for the unbound, candidate prestellar and robust prestellar samples are given in \ref{tab:starless_cmfs_2}.  In each case we fitted only those bins whose centres are above the mass completeness limit in that distance range.  Where it is possible to fit the 355--500\,pc range with heated cores in Orion A excluded, in every case the results agree within error with those of the full far sample, and so we conclude that our choice of temperatures in Orion A makes little difference to the statistical properties of our sample.  Henceforth, we use all cores in Orion A in our CMF fitting.

\begin{table*} 
\caption{Least-squares best-fit log-normal CMFs for each of our distance ranges, for all starless cores, and for prestellar cores (combined candidate and robust samples).  The fitted parameters are $A$, maximum value of the log-normal, $\mu$, mass at which the log-normal distribution peaks, and $\sigma$, log-normal width.  The final two rows of the table show least-squares best fits to the near- and mid-distance samples, for cores above the far-distance mass completeness limit (0.2\,M$_{\odot}$) only.} 
\centering 
\begin{tabular}{@{\extracolsep{4pt}}c ccc ccc@{}} 
\hline & \multicolumn{3}{c}{Starless} & \multicolumn{3}{c}{Prestellar (C \& R)} \\ \cline{2-4} \cline{5-7} 
 & $A$ & $\mu$ & $\sigma$ & $A$ & $\mu$ & $\sigma$  \\ 
Range & & (M$_{\odot}$) & ($\log_{10}{\rm M}_{\odot}$) & & (M$_{\odot}$) & ($\log_{10}{\rm M}_{\odot}$) \\ 
\hline 
0 -- 200\,pc & 30$\pm$3 & 0.15$\pm$0.02 & 0.48$\pm$0.05 & 15$\pm$2 & 0.21$\pm$0.04 & 0.53$\pm$0.09 \\ 
200 -- 355\,pc & 33$\pm$4 & 0.32$\pm$0.04 & 0.37$\pm$0.04 & 17$\pm$3 & 0.41$\pm$0.06 & 0.35$\pm$0.05 \\ 
355 -- 500\,pc & 208$\pm$13 & 0.42$\pm$0.05 & 0.42$\pm$0.03 & 150$\pm$10 & 0.56$\pm$0.05 & 0.39$\pm$0.03 \\ 
no heated Orion A & 205$\pm$4 & 0.48$\pm$0.02 & 0.41$\pm$0.01 & 151$\pm$3 & 0.64$\pm$0.02 & 0.38$\pm$0.01 \\ 
$>500$\,pc & -- & -- & -- & $6\pm 2$ & $0.7\pm0.7$ & $0.6\pm0.3$  \\ 
\hline 
\multicolumn{7}{c}{For cores with masses $>0.2\,$M$_{\odot}$ only} \\
\hline 
0 -- 200\,pc & -- & -- & -- & -- & -- & --  \\ 
200 -- 355\,pc & 36$\pm$7 & 0.2$\pm$0.1 & 0.4$\pm$0.1 & 19$\pm$4 & 0.3$\pm$0.2 & 0.4$\pm$0.1 \\ 
\hline 
\end{tabular} 
\label{tab:starless_cmfs_1} 
\end{table*} 

The near-, mid- and far-distance core mass distributions are well-modelled with log-normal distributions.  No good log-normal fit could be found for the very far-distance starless CMF (i.e. IC 5146), likely because the peak of the starless CMF is below the completeness limit of the region.  A log-normal fit can be found for the prestellar very far-distance CMF, albeit with very large uncertainties.  Due to the difficulty in constraining the CMF of the very-far distance cores, and due to the high mass completeness limit at this distance, we do not consider this distance range further.

Our best-fit models show that the peak of the log-normal CMF model (the most probable core mass) increases with distance.  The peak of the prestellar CMF is consistently higher than that of the starless core CMF, as expected, as more massive cores are more likely to be gravitationally bound.  The starless and prestellar CMFs have similar widths in each case.  We find that the near-distance CMFs are $\simeq 0.1$ dex -- 0.15 dex wider than those of the mid- and far-distance CMFs, whose widths are consistent with one another.

The peak of the CMF is well above the mass completeness limit for both the near- and the mid-distance samples.  We can see a clear downturn in the distribution of core masses before the completeness limit is reached, suggesting that we are accurately characterising the low-mass ends of these CMFs.  However, the peak mass of the far-distance sample is near the completeness limit, suggesting that the low-mass end of the CMF is less well-characterised here.  The best-fit starless CMF for the far-distance sample follows the data below the completeness limit well, perhaps suggesting that our completeness limit is conservative. However, the best-fit prestellar CMF is narrower, and does not encompass the cores detected below the 90\% completeness limit.

\subsubsection{Similarity of CMFs at different distances}

We wish to determine whether or not the near-, mid- and far-distance core samples could be drawn from the same underlying CMF, i.e. whether differences in the best-fit CMFs result from their differing completeness limits.  To test this hypothesis, we first attempted to fit log-normal distributions to the cores in the near- and mid-distance samples only for masses above the 0.2\,M$_{\odot}$ completeness limit of the far-distance sample.  In the near-distance case, there were too few cores with masses $> 0.2\,$M$_{\odot}$ to produce a good fit, while in the mid-distance case the fits produced were poorly constrained, with lower peak masses and broader widths.  The parameters of these fits are listed in Table~\ref{tab:starless_cmfs_1}.

\subsubsection{Monte Carlo modelling of CMFs by distance}
\label{sec:mc_distance}

\begin{table*}
\centering
\caption{The most probable starless and prestellar CMFs for each distance range that we consider, as determined from Monte Carlo estimation and two-sided KS tests, \textbf{using matched-mass sampling}.   $p$ values show the probability that this model and our sample are drawn from the same underlying distribution.}
\begin{tabular}{@{\extracolsep{4pt}}c cccccc@{}}
\hline
 & \multicolumn{3}{c}{Starless} & \multicolumn{3}{c}{Prestellar (C \& R)} \\ \cline{2-4} \cline{5-7}
Range & $\mu$ & $\sigma$ & $p$ & $\mu$ & $\sigma$ & $p$ \\ 
 & $({\rm M}_{\odot})$ & $(\log_{10}{\rm M}_{\odot})$ &  & $({\rm M}_{\odot})$ & $(\log_{10}{\rm M}_{\odot})$ & \\ 
\hline
0--200 pc & $0.15$ & $0.41$ & $0.78$ & $0.19$ & $0.49$ & $0.75$\\ 
200--355 pc & $0.32$ & $0.36$ & $0.70$ & $0.41$ & $0.36$ & $0.68$\\ 
355--500 pc & $0.42$ & $0.42$ & $0.65$ &  $0.56$ & $0.39$ & $0.67$\\ 
\hline
\end{tabular}
\label{tab:mc_region_cmfs_dist}
\end{table*}

To further test the consistency of the starless and prestellar CMFs as a function of distance, we constructed a grid of log-normal CMFs, with parameters in the range $-1.5 \leq \log_{10}\mu \leq 0.0$ and $0.05\leq\sigma\leq 1.2$, in steps of 0.0125. From each of these CMFs we randomly drew a sample of `cores' 100 times larger than the size of the far-distance starless or prestellar sample (this number was chosen arbitrarily to ensure that the sample had a total mass significantly greater than that of the far-distance sample).  We then calculated the cumulative sum of the sample, and selected the cores whose cumulative mass was closest to the total mass of the near-distance starless or prestellar sample \citep{bonnell2011}.  We further selected the cores whose cumulative mass was closest to the total mass of the mid- and far-distance starless or prestellar samples.  

\begin{figure*}
    \centering
    \includegraphics[width=0.5\textwidth]{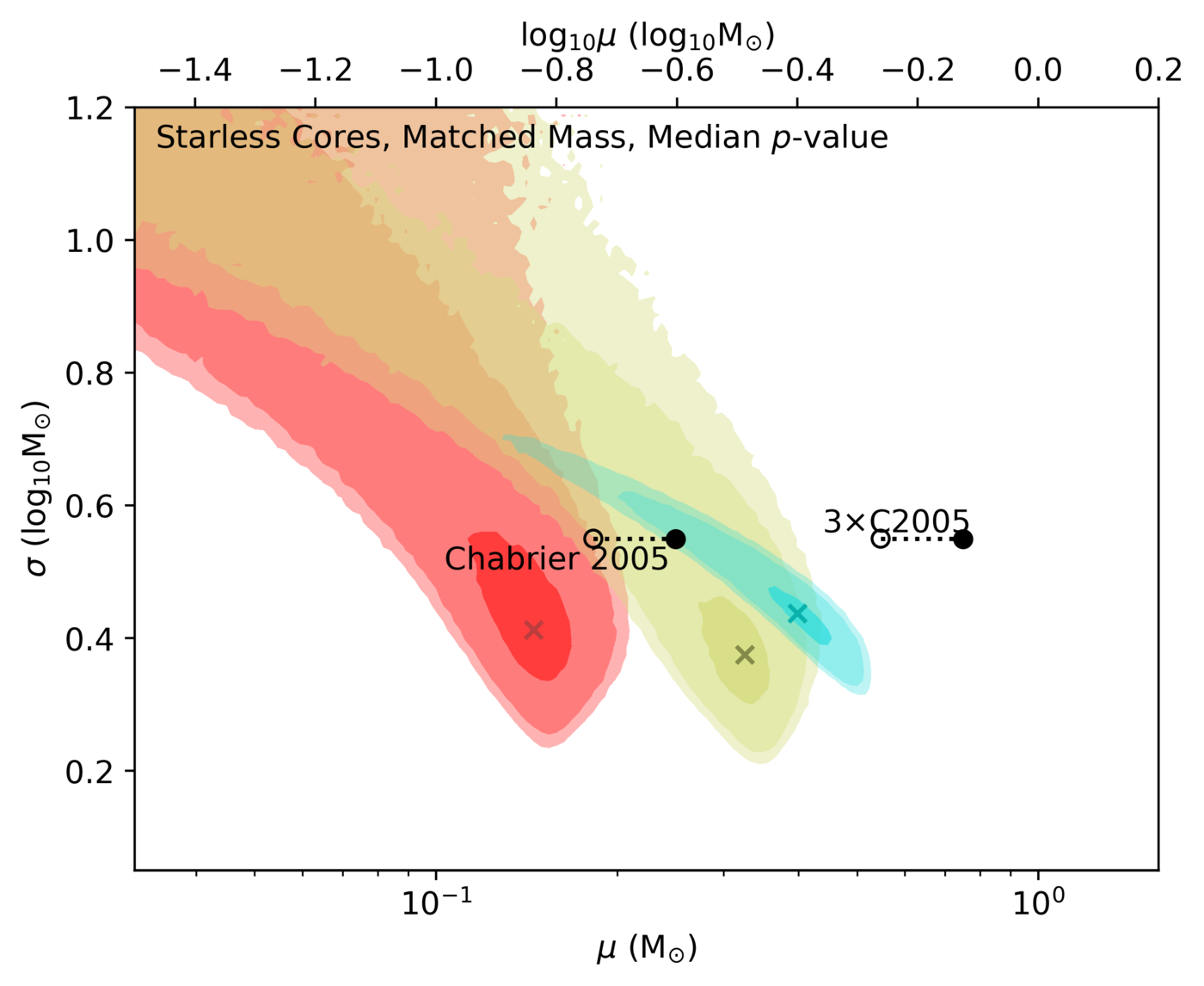}\includegraphics[width=0.5\textwidth]{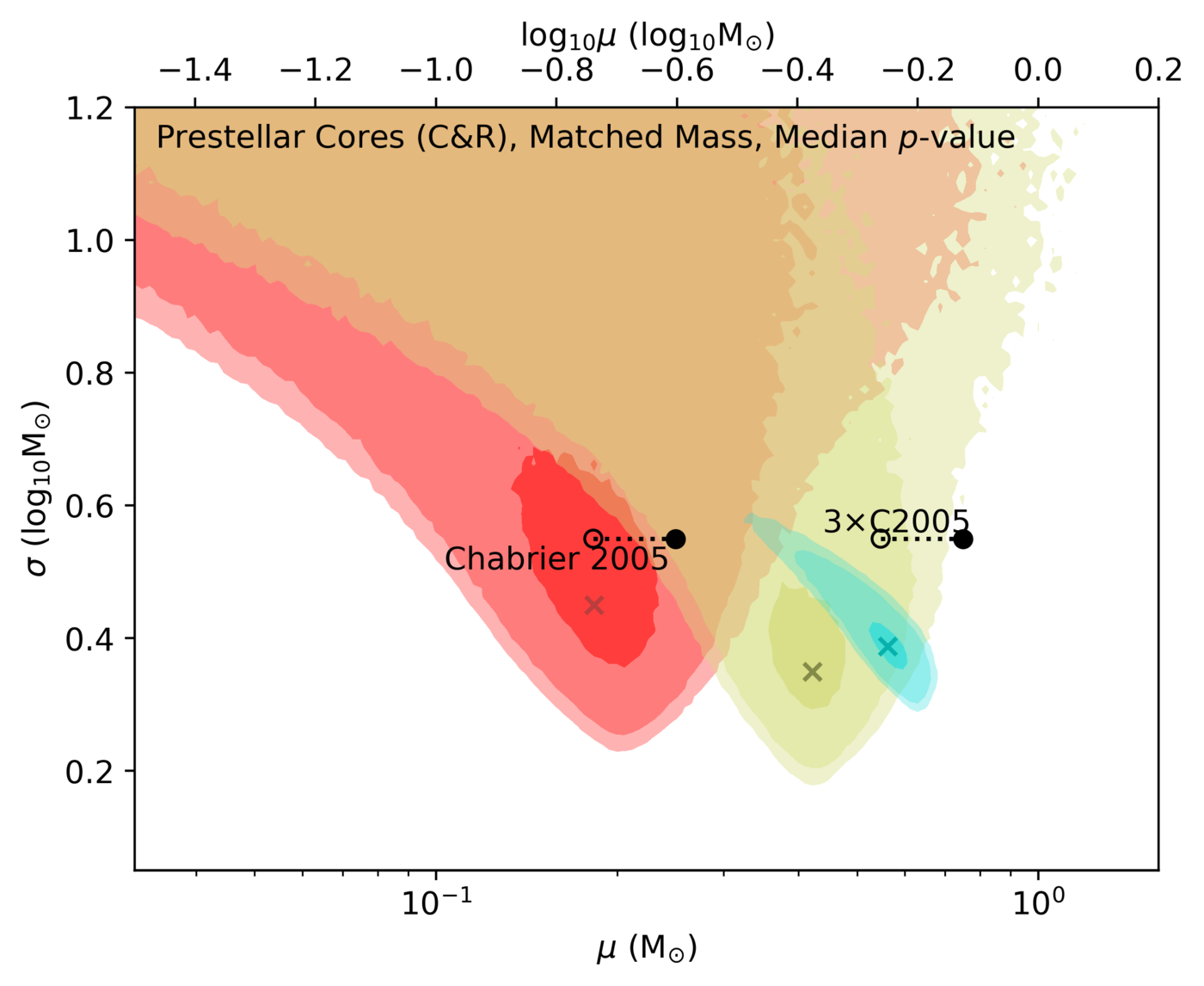}
    \caption{Median $p$-values for two-sided KS test between model CMFs and starless (left) and prestellar (right) CMFs, for matched-mass sampling.  Contours show $p$ values of 0.05, 0.1 and 0.5.  Red marks near-, yellow mid- and cyan far-distance CMFs.  Filled circles mark 1 and 3$\times$ the \citet{chabrier2005} peak system mass (0.25\,M$_{\odot}$).  Open circles mark 0.73$\times$ these values, accounting for the typical flux loss in our SCUBA-2 observations.}
    \label{fig:mc_3way}
\end{figure*}

For each of these mass distributions, we selected masses above the near-, mid-, or far-distance completeness limits as appropriate.  We then performed two-sided Kolmogorov-Smirnov (KS) tests between the sub-sample and the the above-completeness-limit core masses in the observed near-, mid- and far-distance starless or prestellar sample, respectively.  We repeated this exercise 1000 times for each CMF in the grid, and recorded the median $p$ values for the three distance bins. These median values are shown in Figure~\ref{fig:mc_3way}, and the most probable (highest $p$-value) combinations of $\mu$ and $\sigma$ are listed in Table~\ref{tab:mc_region_cmfs_dist}.  In each case, there is a well-defined most probable log-normal CMF, with a long tail of marginally consistent CMFs with lower peak masses and larger widths.  We note that the most probable values that we find for each distance range are consistent with the best-fit values from our least-squares fitting, suggesting that both methods are robustly characterising the CMFs that we measure.

The three prestellar CMFs are only marginally consistent at the $p=0.05$ level with being drawn from the same underlying log-normal distribution. 
In the starless case, the three CMFs are slightly more consistent at the $p=0.05$ level, although the area of parameter space over which the probability distributions overlap remains small.  For both the starless and the prestellar samples, the mid-distance CMF is consistent at $p>0.1$ with being drawn from the same underlying distribution as either the near- or the far-distance CMF, but the three cannot be simultaneously reconciled with each other.  The $p>0.5$ values of $\mu$ and $\sigma$ for the three distributions do not overlap in any case.  It therefore seems unlikely that the CMFs of the three distance bins are drawn from the same underlying log-normal distribution, although we cannot rule this out entirely.  

\subsection{Core Mass Functions by cloud complex}
\label{sec:cmfs_complex}

\subsubsection{Least-squares CMFs by cloud complex}
\label{sec:cmfs_complex_ls}

\begin{table*} 
\caption{Least-squares best-fit core mass functions for each of the cloud complexes that we consider.}
\centering 
\begin{tabular}{@{\extracolsep{4pt}}c ccc ccc@{}} 
\hline & \multicolumn{3}{c}{Starless} & \multicolumn{3}{c}{C\&R} \\ \cline{2-4} \cline{5-7} 
 & $A$ & $\mu$ & $\sigma$ & $A$ & $\mu$ & $\sigma$ \\ 
Complex & & (M$_{\odot}$) & ($\log_{10}{\rm M}_{\odot}$) & & (M$_{\odot}$) & ($\log_{10}{\rm M}_{\odot}$) \\ 
\hline 
\multicolumn{7}{c}{Near} \\
\hline
CrA & 3$\pm$3 & 0.2$\pm$0.2 & 0.7$\pm$1.2 & 4.2$\pm$1.1 & 0.19$\pm$0.03 & 0.33$\pm$0.10 \\ 
Lupus & -- & -- & -- & -- & -- & -- \\ 
Pipe & -- & -- & -- & -- & -- & -- \\ 
Ophiuchus & 19$\pm$3 & 0.08$\pm$0.06 & 0.6$\pm$0.2 & 10$\pm$2 & 0.13$\pm$0.07 & 0.7$\pm$0.3 \\ 
Taurus & 10$\pm$1 & 0.20$\pm$0.02 & 0.39$\pm$0.04 & -- & -- & -- \\ 

\hline 
\multicolumn{7}{c}{Mid} \\
\hline
Cepheus & 6$\pm$1 & 0.4$\pm$0.1 & 0.4$\pm$0.1 & -- & -- & -- \\ 
Perseus & 25$\pm$3 & 0.30$\pm$0.03 & 0.38$\pm$0.03 & 15$\pm$3 & 0.42$\pm$0.05 & 0.36$\pm$0.05 \\ 
\hline 
\multicolumn{7}{c}{Far} \\
\hline
Aquila & 48$\pm$5 & 0.52$\pm$0.06 & 0.32$\pm$0.04 & 52$\pm$3 & 0.45$\pm$0.05 & 0.35$\pm$0.03 \\ 
Auriga & 12$\pm$4 & 0.5$\pm$0.1 & 0.3$\pm$0.1 & 8$\pm$5 & 0.6$\pm$0.2 & 0.2$\pm$0.1 \\ 
Orion A & 100$\pm$10 & 0.46$\pm$0.08 & 0.41$\pm$0.05 & 70$\pm$10 & 0.6$\pm$0.1 & 0.41$\pm$0.07 \\ 
Orion B & 55$\pm$4 & 0.25$\pm$0.07 & 0.59$\pm$0.07 & 43$\pm$1 & 0.55$\pm$0.04 & 0.45$\pm$0.02 \\
Serpens & 31$\pm$2 & 0.54$\pm$0.08 & 0.43$\pm$0.05 & 21$\pm$5 & 1.0$\pm$0.1 & 0.29$\pm$0.04 \\ 
\hline 
\multicolumn{7}{c}{Very Far} \\
\hline
IC5146 & -- & -- & -- & 6$\pm$2 & 0.7$\pm$0.7 & 0.6$\pm$0.3 \\ 
\hline
    \end{tabular}
    \label{tab:cloud_cmfs}
\end{table*}

We further fitted log-normal distributions to the CMFs of each of the cloud complexes which we observed.  These distributions are shown in Figures~\ref{fig:cmf_aquila}--\ref{fig:cmf_taurus} in Appendix~\ref{sec:appendix_cmfs}, and their best-fit log-normal distributions are listed in Table~\ref{tab:cloud_cmfs}.  The Lupus and Pipe regions contain too few starless cores ($<10$) to fit a CMF, and no prestellar cores.  The Taurus region contains too few prestellar cores for a fit to be found.  As discussed above, good log-normal fits cannot be found for IC 5146 (the very far-distance cores).  We thus exclude IC 5146 from further consideration.

\subsubsection{Monte Carlo CMFs by cloud complex}

\begin{figure*}
    \centering
    \includegraphics[width=0.33\textwidth]{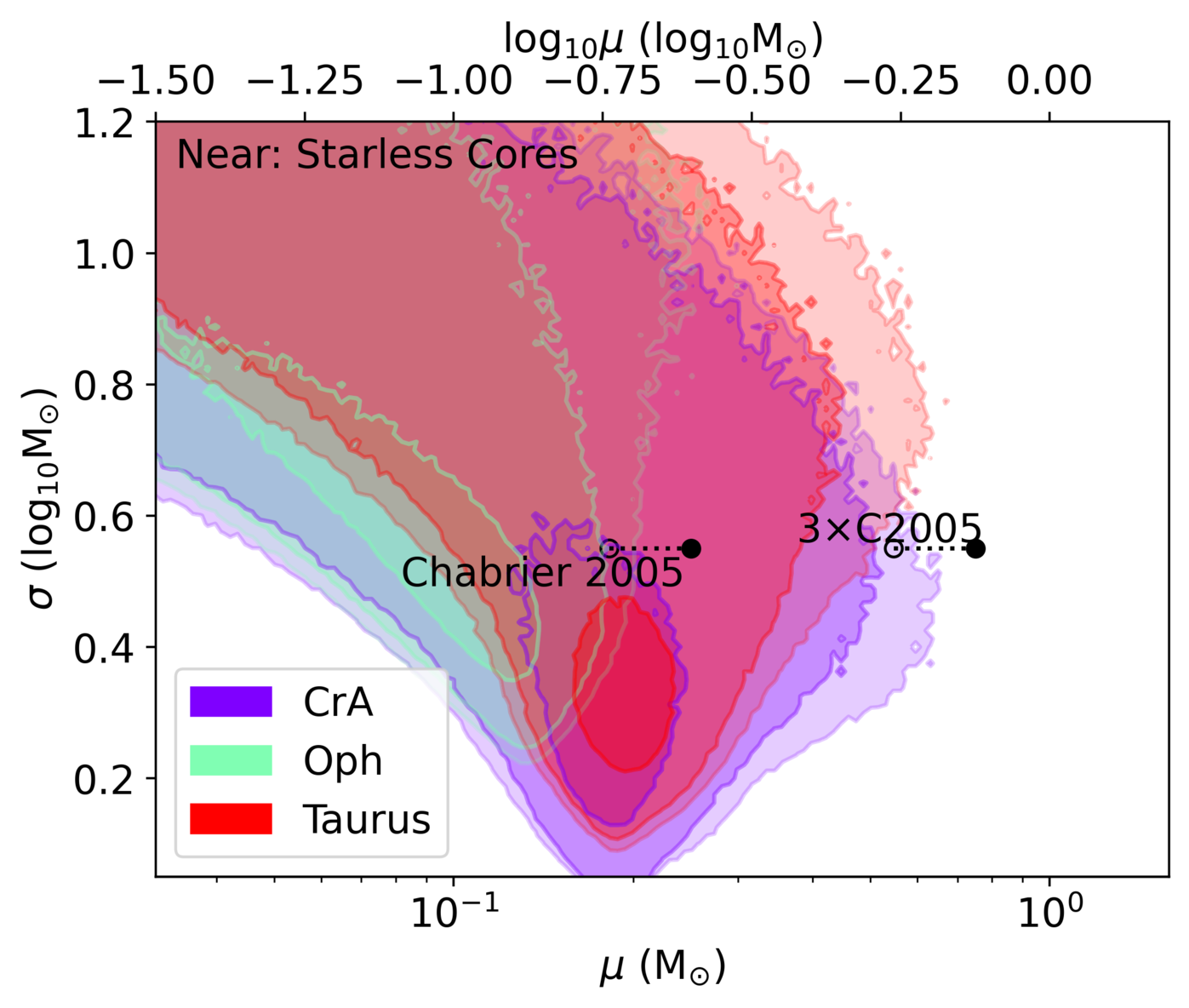}\includegraphics[width=0.33\textwidth]{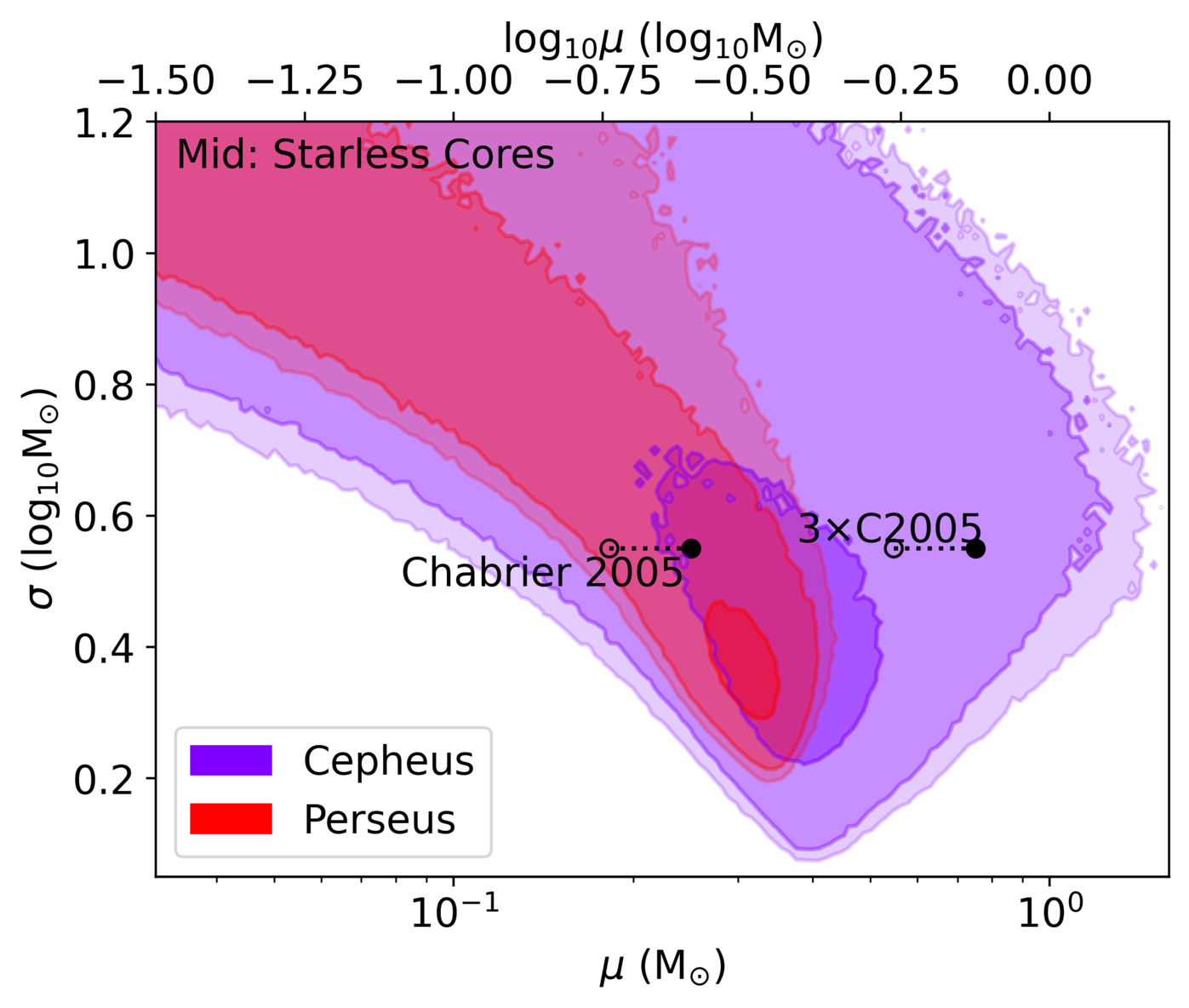}\includegraphics[width=0.33\textwidth]{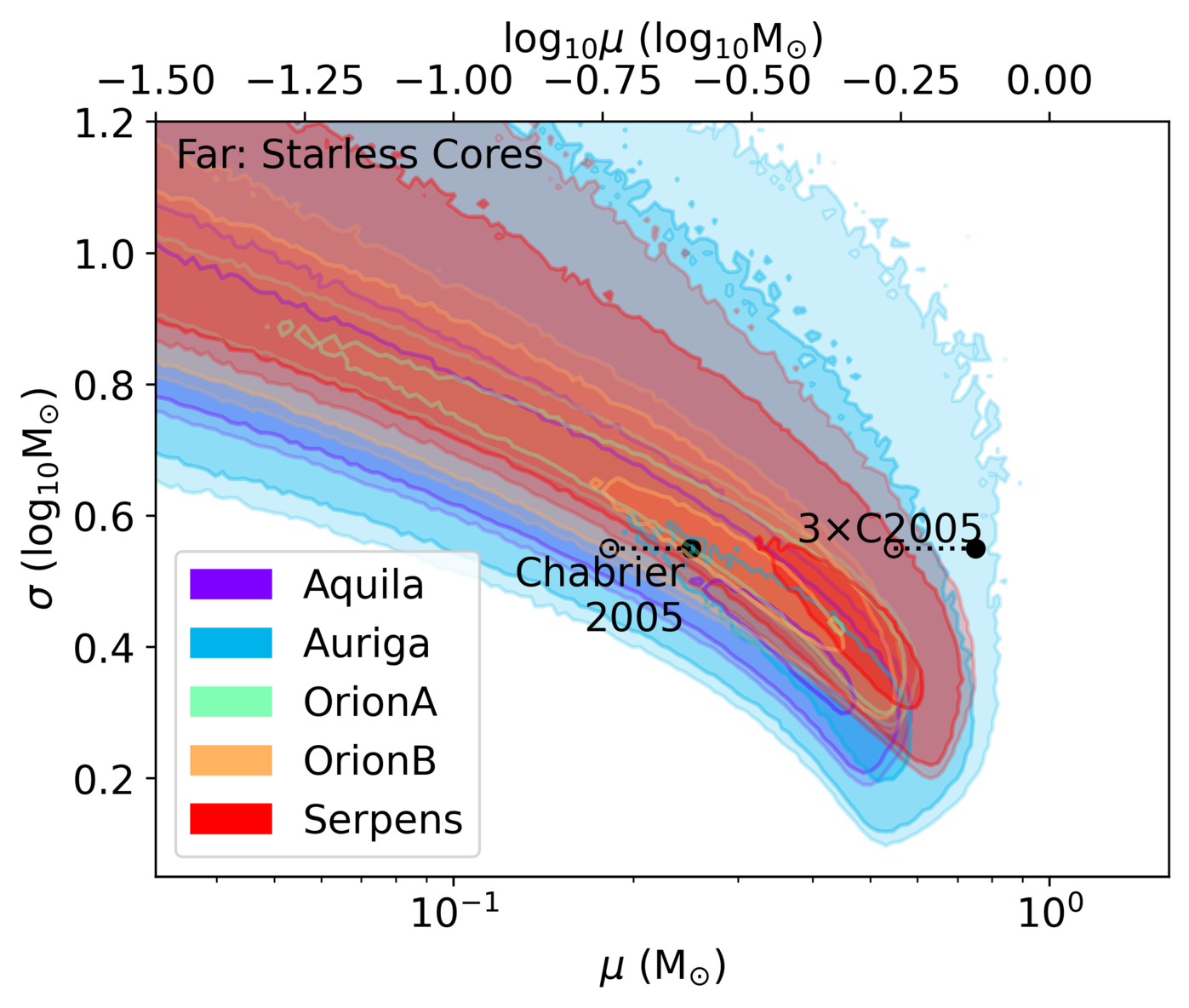}
    \caption{Median $p$-values for two-sided KS tests between model CMFs and starless CMFs, for matched-mass sampling.  Left: near clouds.  Centre: mid-distance clouds.  Right: Far clouds.  Contours show $p$ values of 0.05, 0.1 and 0.5.  Filled circles mark 1$\times$ and 3$\times$ the \citet{chabrier2005} peak system mass (0.25\,M$_{\odot}$).  Open circles mark 0.73$\times$ these values, accounting for the typical flux loss in our SCUBA-2 observations.}
    \label{fig:mc_cloud_starless}
    \includegraphics[width=0.33\textwidth]{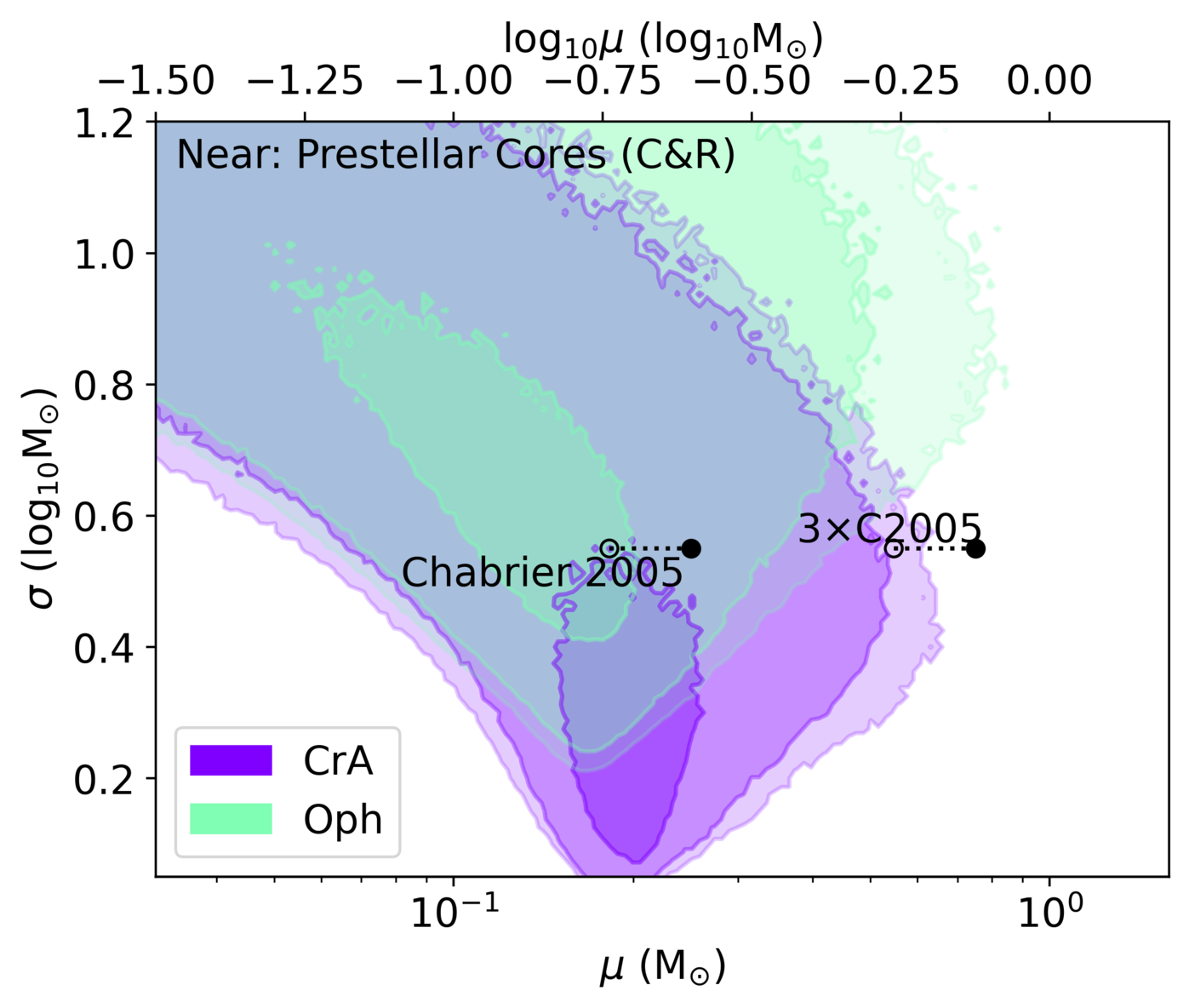}\includegraphics[width=0.33\textwidth]{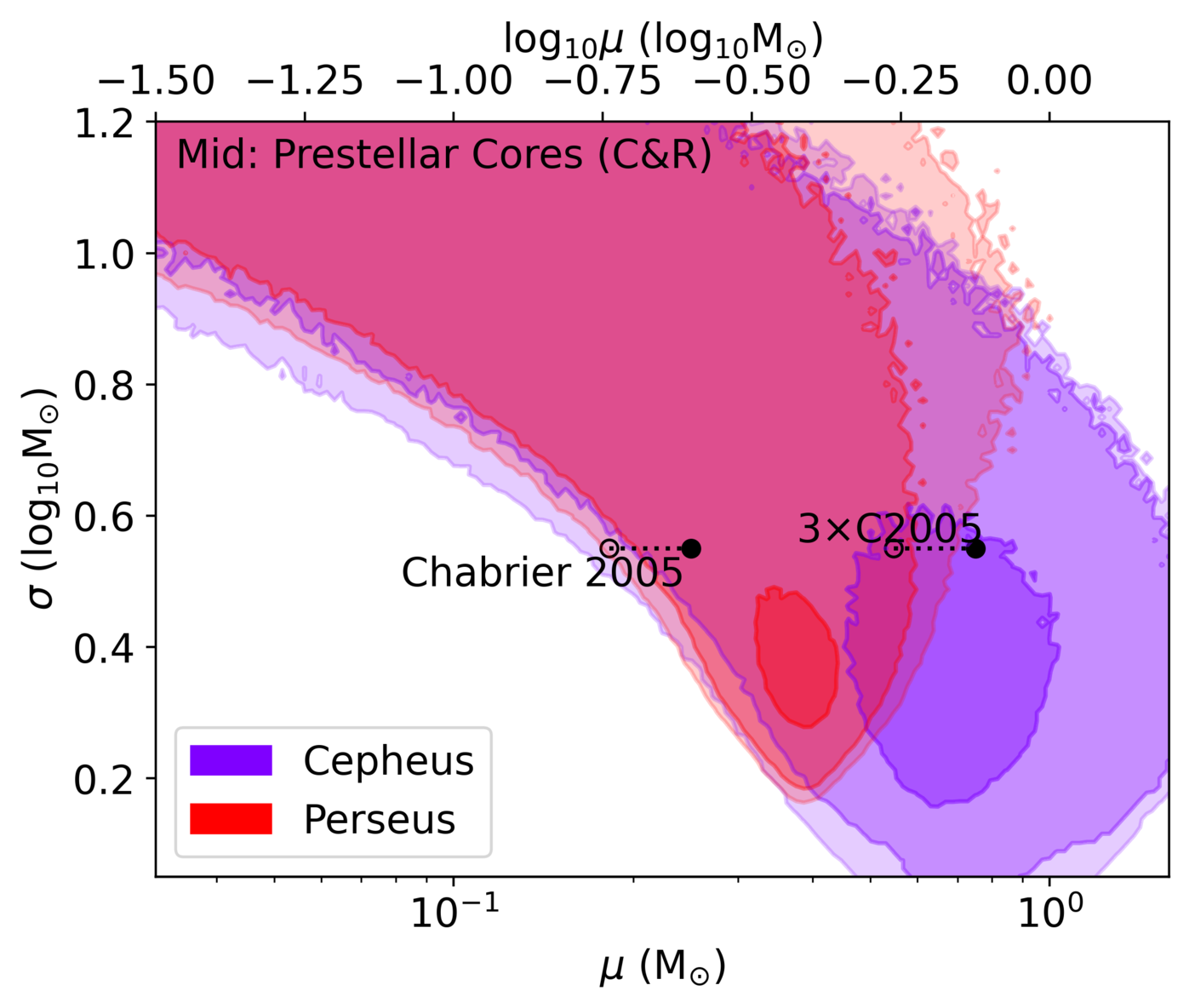}\includegraphics[width=0.33\textwidth]{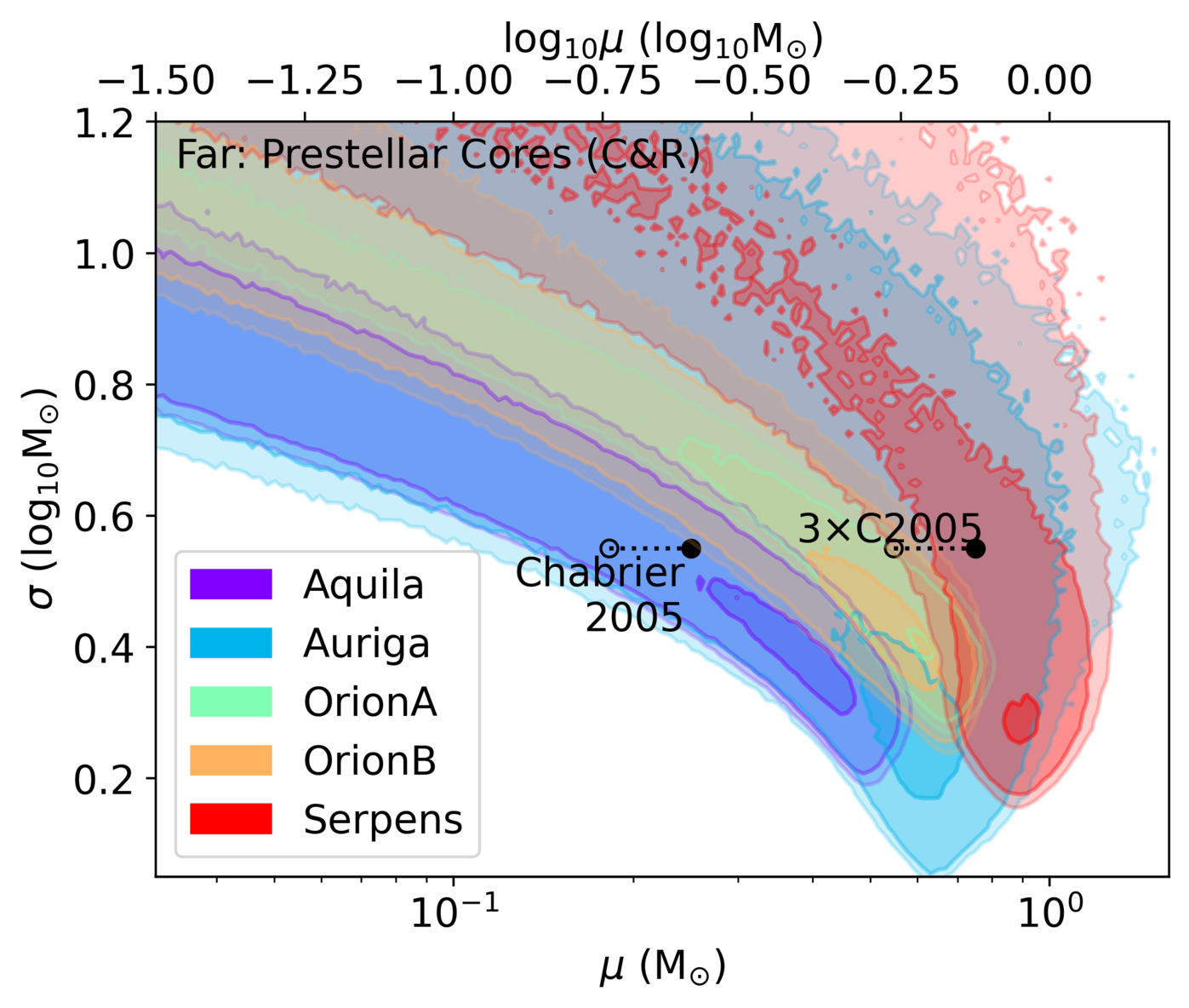}
    \caption{Median $p$-values for two-sided KS tests between model CMFs and prestellar CMFs, for matched-mass sampling.  Left: near clouds.  Centre: mid-distance clouds.  Right: Far clouds.  Filled circles mark 1$\times$ and 3$\times$ the \citet{chabrier2005} peak system mass (0.25\,M$_{\odot}$).  Open circles mark 0.73$\times$ these values, accounting for the typical flux loss in our SCUBA-2 observations.}
    \label{fig:mc_cloud_candr}
\end{figure*}

\begin{table*}
\caption{The most probable starless and prestellar CMFs for each cloud complex that we consider, as determined from Monte Carlo estimation and two-sided KS tests, \textbf{using matched-mass sampling}.  Values of $\mu$ and $\sigma$ are given to the number of decimal places specified in our grid of input models.  Median $p$ values, showing the probability that this model and our sample are drawn from the same underlying distribution, are given.}
\centering
\begin{tabular}{@{\extracolsep{4pt}}c cccccc@{}}
\hline
 & \multicolumn{3}{c}{Starless} & \multicolumn{3}{c}{Prestellar (C \& R)} \\ \cline{2-4} \cline{5-7}
Region & $\mu$ & $\sigma$ & $p$ & $\mu$ & $\sigma$ & $p$ \\ 
 & $({\rm M}_{\odot})$ & $(\log_{10}{\rm M}_{\odot})$ &  & $({\rm M}_{\odot})$ & $(\log_{10}{\rm M}_{\odot})$ & \\ 
\hline
\multicolumn{7}{c}{Near} \\ 
\hline
CrA& $0.17$ & $0.26$ & $0.77$ & $0.19$ & $0.20$ & $0.80$ \\ 
Ophiuchus& $0.11$ & $0.49$ & $0.78$ & 0.15 & $0.54$ & $0.69$ \\ 
Taurus& $0.19$ & $0.29$ & $0.76$ &  -- & -- & -- \\ 
\hline
\multicolumn{7}{c}{Mid} \\ 
\hline
Cepheus& $0.40$ & $0.35$ & $0.75$ & $0.63$ & $0.29$ & $0.74$ \\ 
Perseus&$0.31$ & $0.36$ & $0.70$ & $0.39$ & $0.35$ & $0.71$ \\ 
\hline
\multicolumn{7}{c}{Far} \\ 
\hline 
Aquila& $0.37$ & $0.40$ & $0.66$ & $0.37$ & $0.40$ & $0.64$ \\ 
Auriga& $0.47$ & $0.32$ & $0.71$ & $0.60$ & $0.25$ & $0.69$ \\ 
Orion A& $0.45$ & $0.42$ & $0.51$ & $0.61$ & $0.40$ & $0.53$ \\ 
Orion B& $0.38$ & $0.46$ & $0.67$ & $0.58$ & $0.40$ & $0.72$ \\ 
Serpens& $0.52$ & $0.40$ & $0.72$ & $0.89$ & $0.29$ & $0.58$ \\ 
\hline
\end{tabular}
\label{tab:mc_region_cmfs}
\end{table*}

We performed the Monte Carlo modelling described in Section~\ref{sec:mc_distance} for each of the cloud complexes which we observed.  We did not attempt this modelling for complexes with sample sizes of less than 10.  The results of this analysis are given in Table~\ref{tab:mc_region_cmfs}.  The matched-mass sample contour plots are shown in Figures~\ref{fig:mc_cloud_starless} (starless cores) and \ref{fig:mc_cloud_candr} (prestellar cores).  

\begin{figure}
    \centering
    \includegraphics[width=0.47\textwidth]{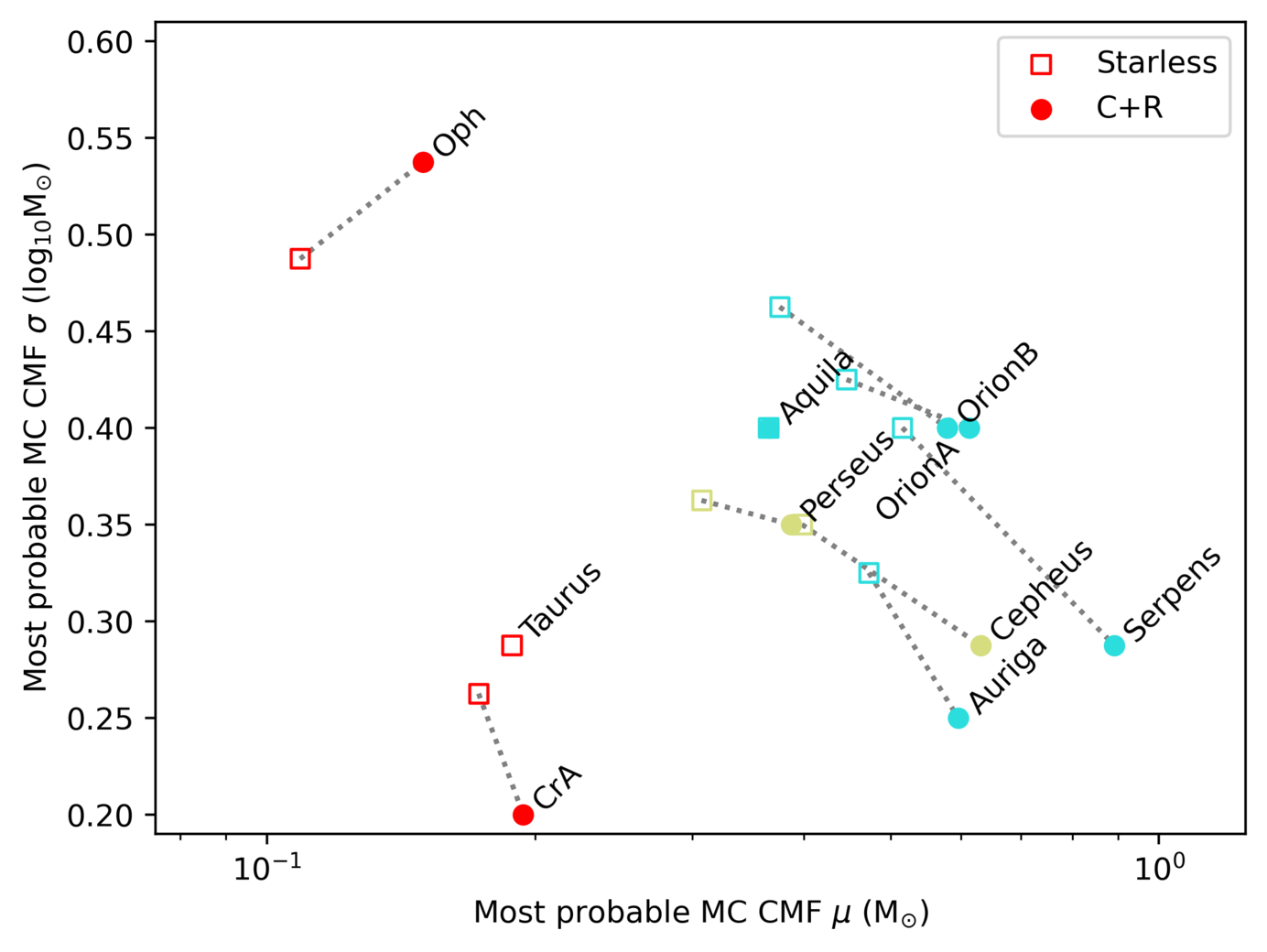}
    \caption{The most probable starless and prestellar CMF properties (as determined from median $p$-value) measured from matched-mass Monte Carlo modelling for each cloud complex.  Open squares show starless core CMFs, closed circles show prestellar CMFs.  Cloud complexes are colour-coded by their distance range.  The parameter space explored in Figures~\ref{fig:mc_3way}--\ref{fig:mc_cloud_candr} is here restricted to the region of interest.}
    \label{fig:mc_starless_candr_compare}
\end{figure}

The most probable starless and prestellar CMFs are compared in Figure~\ref{fig:mc_starless_candr_compare}.  The most probable prestellar CMF has a higher peak mass than the starless CMF in every cloud complex, as is expected as higher-mass cores are more likely to be gravitationally bound.  In most complexes, the most probable prestellar CMF has a narrower width than the starless CMF, with the exceptions of Ophiuchus and Perseus, both of which show a slight increase in width.

The best-fit least squares (LS) and most-probable Monte Carlo (MC)-derived CMFs are similar to one another; we compare them in more detail in Appendix~\ref{sec:appendix_cmfs_lsmc}.

\subsubsection{CMF properties as a function of cloud mass}

The peak starless and prestellar CMF masses, as determined from LS fitting, are plotted as a function of cloud complex mass ($M_{cloud}$) in Figure~\ref{fig:peak_cmf_ls}.  The equivalent plots for the MC case are shown in Appendix~\ref{sec:appendix_cmfs_lsmc}.  We see that in both cases there is a weak trend for peak mass to increase with cloud mass.  Ophiuchus again has a considerably lower peak core mass than would be expected for a cloud of its mass. 

\begin{table}
    \caption{Results of linear regressions of the function $\mu \propto M_{cloud}^{\gamma}$ to our best-fit LS and most-probable MC CMFs.  $p_{LR}$ indicates the probability for a null hypothesis test that $\gamma = 0$ (i.e. no correlation between cloud mass and peak mass of the CMF).  Unlike the $p$ value used in previous discussions of the two-sided KS test, a lower $p_{LR}$ indicates better agreement between the data and the model; we consider values $p_{LR} < 0.05$ to be statistically robust.}
    \centering
    \begin{tabular}{c cc @{\extracolsep{4pt}} cc}
        \hline
         & \multicolumn{2}{c}{With Oph} & \multicolumn{2}{c}{No Oph} \\ \cline{2-3} \cline{4-5}
        Case & $\gamma$ & $p_{LR}$ & $\gamma$ & $p_{LR}$ \\
        \hline
         Starless, LS & $0.25\pm0.10$ & 0.03 & $0.20\pm0.06$ & 0.02 \\
         Starless, MC & $0.22\pm0.08$ & 0.02 & $0.18\pm0.05$ & 0.01 \\
         Prestellar, LS & $0.30\pm0.10$ & 0.03 & $0.23\pm0.08$ & 0.04 \\
         Prestellar, MC & $0.21\pm0.11$ & 0.09 & $0.15\pm0.09$ & 0.14 \\
         \hline 
         \end{tabular}
    \label{tab:gamma}
\end{table}

We performed linear regressions on the data, fitting a power-law model, $\mu \propto M_{cloud}^{\gamma}$, to the data in logarithmic space.  The best-fit values of $\gamma$ are listed in Table~\ref{tab:gamma}, and the fits are plotted on Figures~\ref{fig:peak_cmf_ls} and \ref{fig:peak_cmf_mc}.  We perform fits both with and without Ophiuchus; excluding Ophiuchus slightly reduces the best-fit value of $\gamma$ in every case.

The values of $\gamma$ listed in Table~\ref{tab:gamma} are consistent with the relationship between cloud mass and median core mass, $M_{median}\propto M_{cloud}^{0.17\pm 0.06}$, as shown in Figure~\ref{fig:mass_violin}.  However, as discussed in Section~\ref{sec:violins}, this latter trend is not robust when only cores above the 450\,pc mass completeness limit are considered.  The trend which we see in $\mu$ with $M_{cloud}$ may therefore be in part a selection effect resulting from poor completeness of low-mass cores in the more distant clouds in our sample.  Nonetheless, there is a lack of high-mass cores in the most nearby (typically lower-mass) clouds, suggesting that this trend is to some extent physical in origin.

Starless and prestellar CMF widths from LS fitting are plotted as a function of cloud complex mass in Figure~\ref{fig:width_cmf_ls}, with the equivalent plots for the MC case again shown in Appendix~\ref{sec:appendix_cmfs_lsmc}.  There is no clear correlation with cloud complex mass or with distance in any case.

\subsubsection{Comparisons between clouds}
\label{sec:cmf_clouds}

We find that the nearby clouds have different most-probable CMFs to those of the mid- and far-distance clouds.  The mid-distance CMFs, while not particularly well-constrained, are broadly consistent with the far-distance CMFs, all of which fall in the same area of $\mu-\sigma$ parameter space, as shown in Figures~\ref{fig:mc_starless_candr_compare} and \ref{fig:cmf_params_ls_mc}.

\textbf{Nearby clouds:}  The starless core distributions of all of the nearby clouds are consistent with one another at the $p=0.1$ level.  Ophiuchus and Taurus are inconsistent with one another at the $p=0.5$ level; however, CrA is consistent with both.  In the prestellar case, CrA and Ophiuchus are again consistent at $p=0.5$, while Taurus does not contain enough prestellar cores to be considered.  It should be noted that CrA is not well-characterised; its low number statistics (15 starless cores) give it a very broad distribution, and so it is consistent with both Taurus and Ophiuchus.  However, Taurus and Ophiuchus appear to have different low-mass behaviours.  Specifically, there is a lack of low-mass cores in Taurus, and a significant excess in Ophiuchus, despite the two clouds having a similar maximum core mass.  Taurus is generally noted as a region of relatively dispersed star formation \citep[e.g.][]{marsh2016}, while Ophiuchus is considered to be relatively clustered \citep[e.g.][]{friesen2009}.

\textbf{Mid-distance clouds:} Cepheus and Perseus have quite similar starless core distributions, despite their different morphologies.  There is a somewhat higher fraction of bound cores in Perseus than in Cepheus.  The starless core CMFs are consistent at the $p=0.5$ level, while the prestellar CMFs are consistent at the $p=0.1$ level.  However, the CMFs of Cepheus, which only contains 22 starless cores compared to 98 in Perseus, are not very well-constrained.

\textbf{Far clouds:} In the far clouds, the starless CMFs are quite consistent with one another, while the prestellar CMFs show more distinct variation.  Auriga (44 starless cores) is less well-constrained than the other far-distance clouds, and so is consistent with all of the other far-distance cloud complexes.  Orion A and Orion B agree well with each other in both the starless and prestellar cases (consistent at $p>0.5$), as might be expected as they are physically associated with one another.  However, Orion, Serpens and Aquila do not agree well with one another.  In the prestellar case, Orion and Aquila agree at $p=0.1$, as do Orion and Serpens.  However, Aquila and Serpens do not agree with one another at the $p = 0.05$ level in the prestellar case, and are only consistent at the $p = 0.1$ level in the starless case, despite the fact that Aquila is sometimes considered to be a subset of the Serpens Molecular Cloud \citep[e.g.,][]{pillai2020}.

\begin{figure*}
    \centering
    \includegraphics[width=0.5\textwidth]{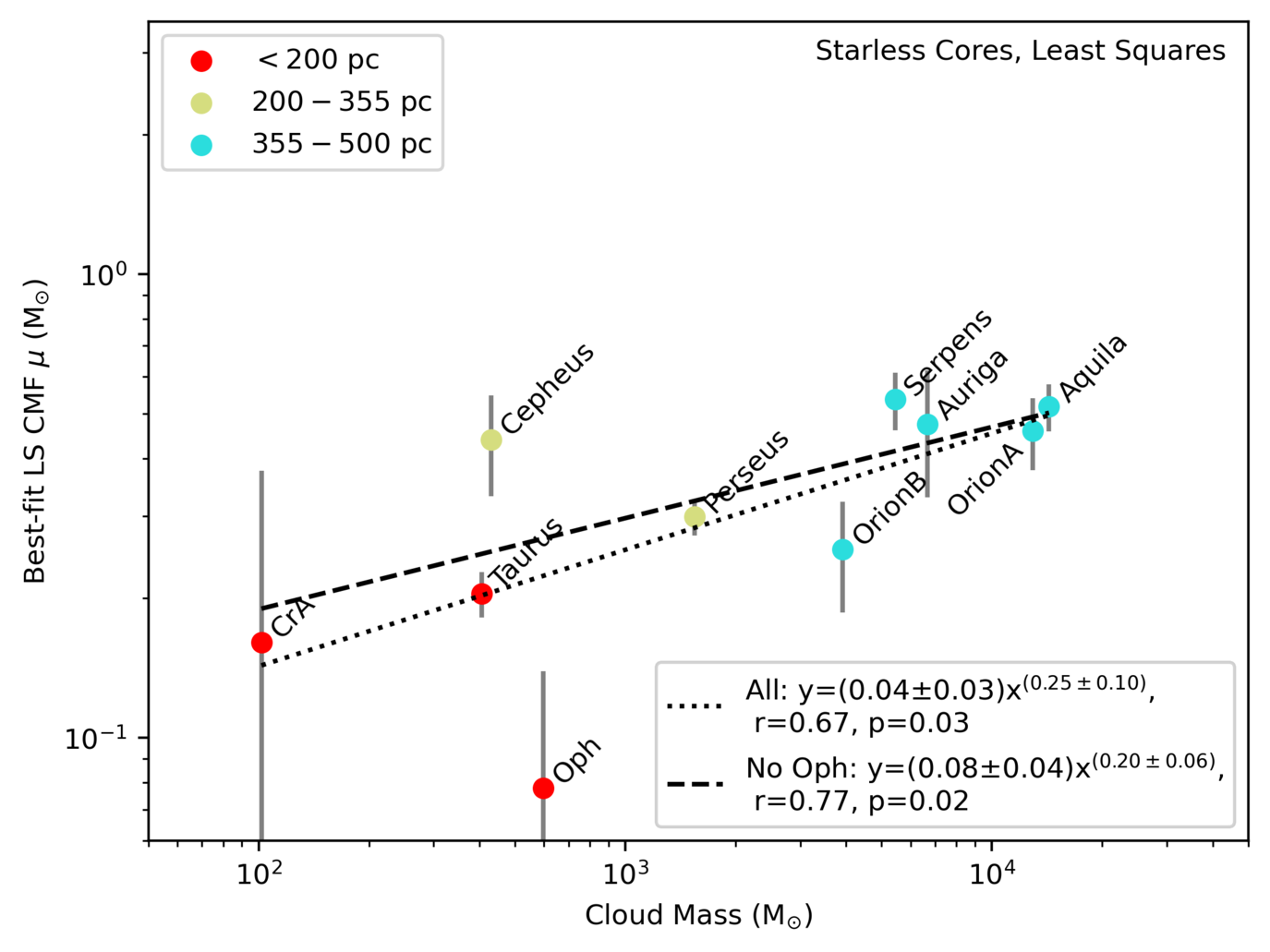}\includegraphics[width=0.5\textwidth]
    {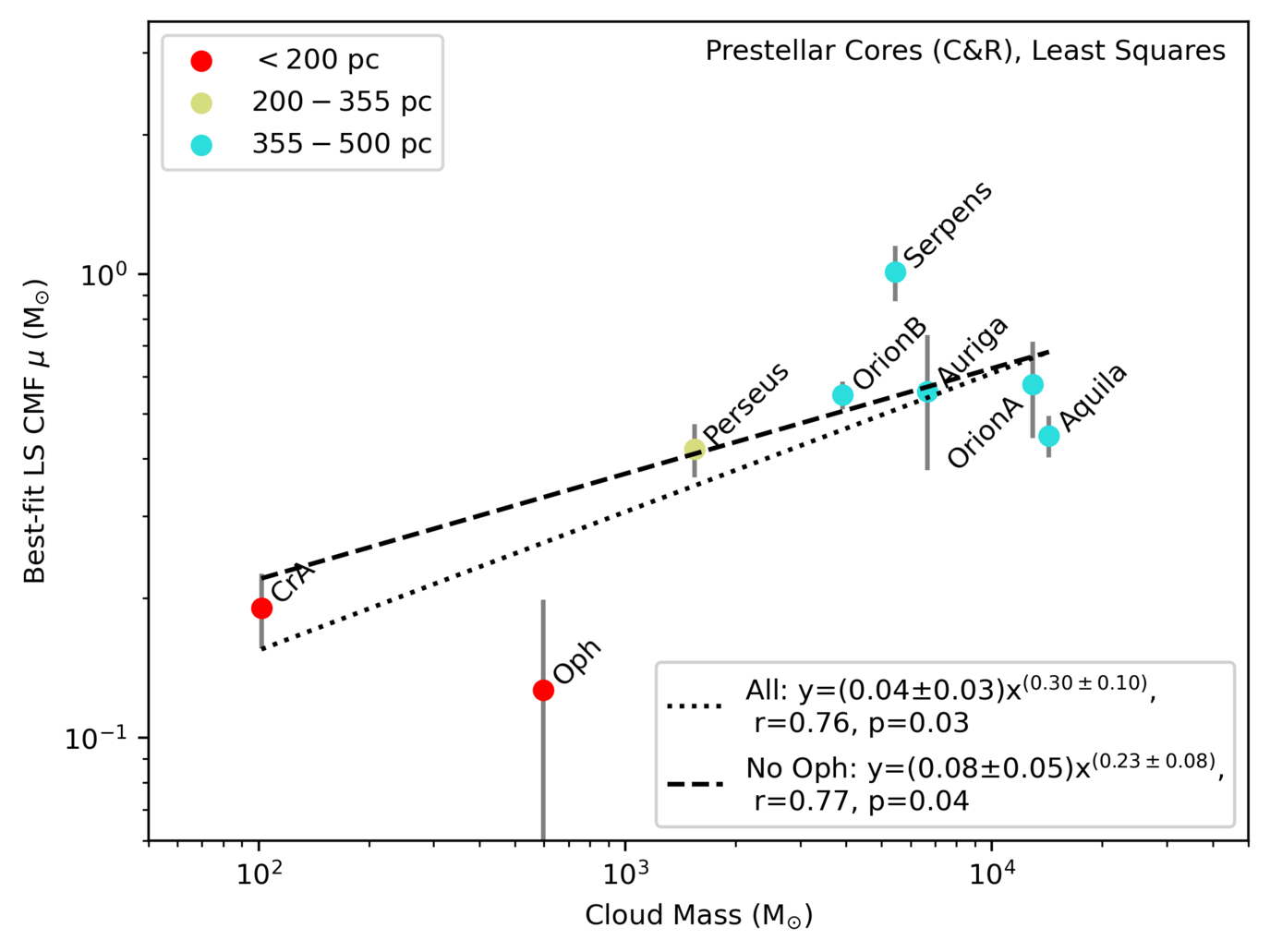}  
    \caption{The peak mass of the starless (left) and prestellar (right) CMFs as a function of cloud complex mass, determined using least-squares fitting.  Dotted line shows the power-law model producing the best fit to all data points; dashed line shows the power-law model producing the best fit to the data points with Ophiuchus, which has a notably low peak mass, excluded.  Data points are colour-coded by their distance range.} 
    \label{fig:peak_cmf_ls}
\end{figure*}

\begin{figure*}
    \centering
    \includegraphics[width=0.5\textwidth]{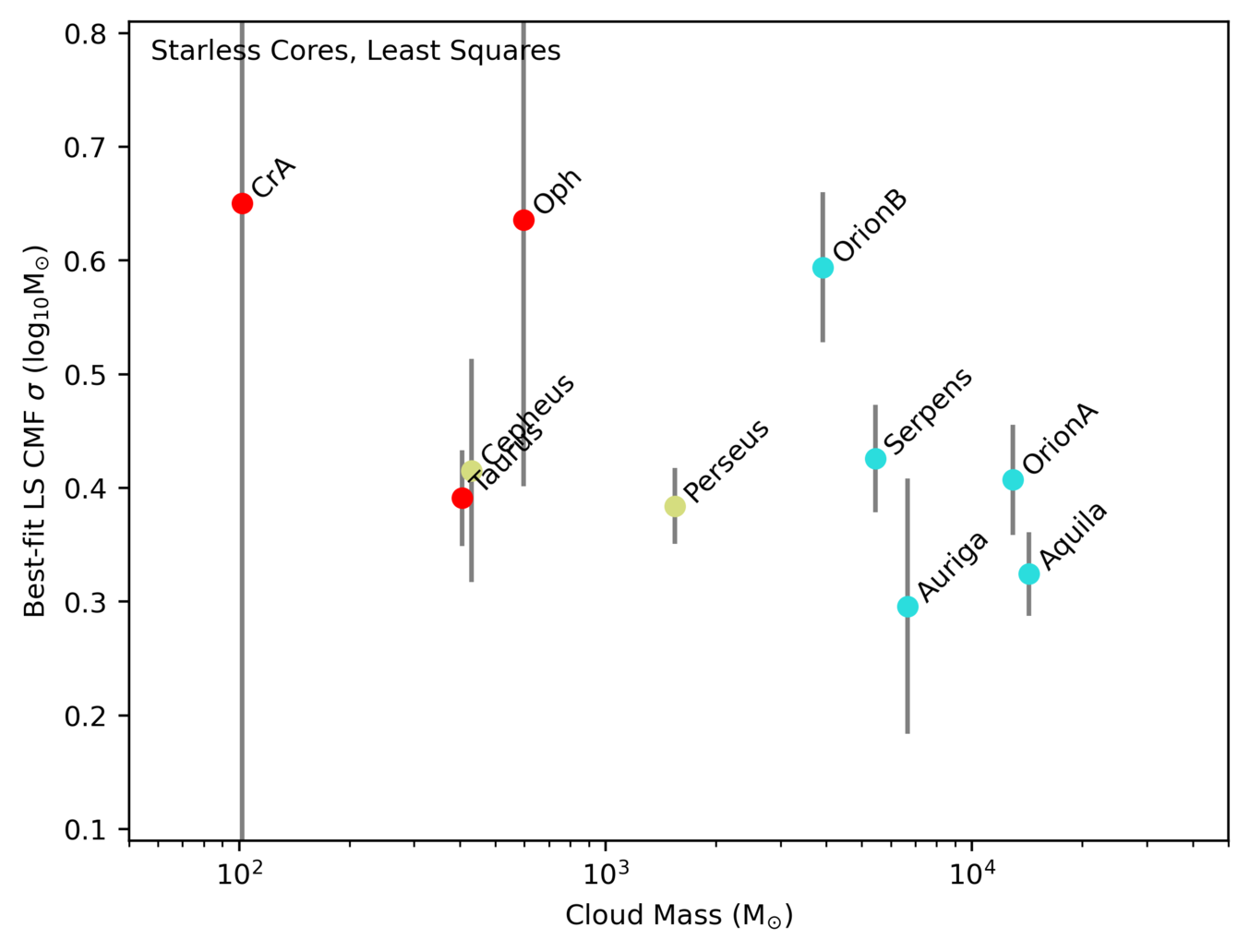}\includegraphics[width=0.5\textwidth]{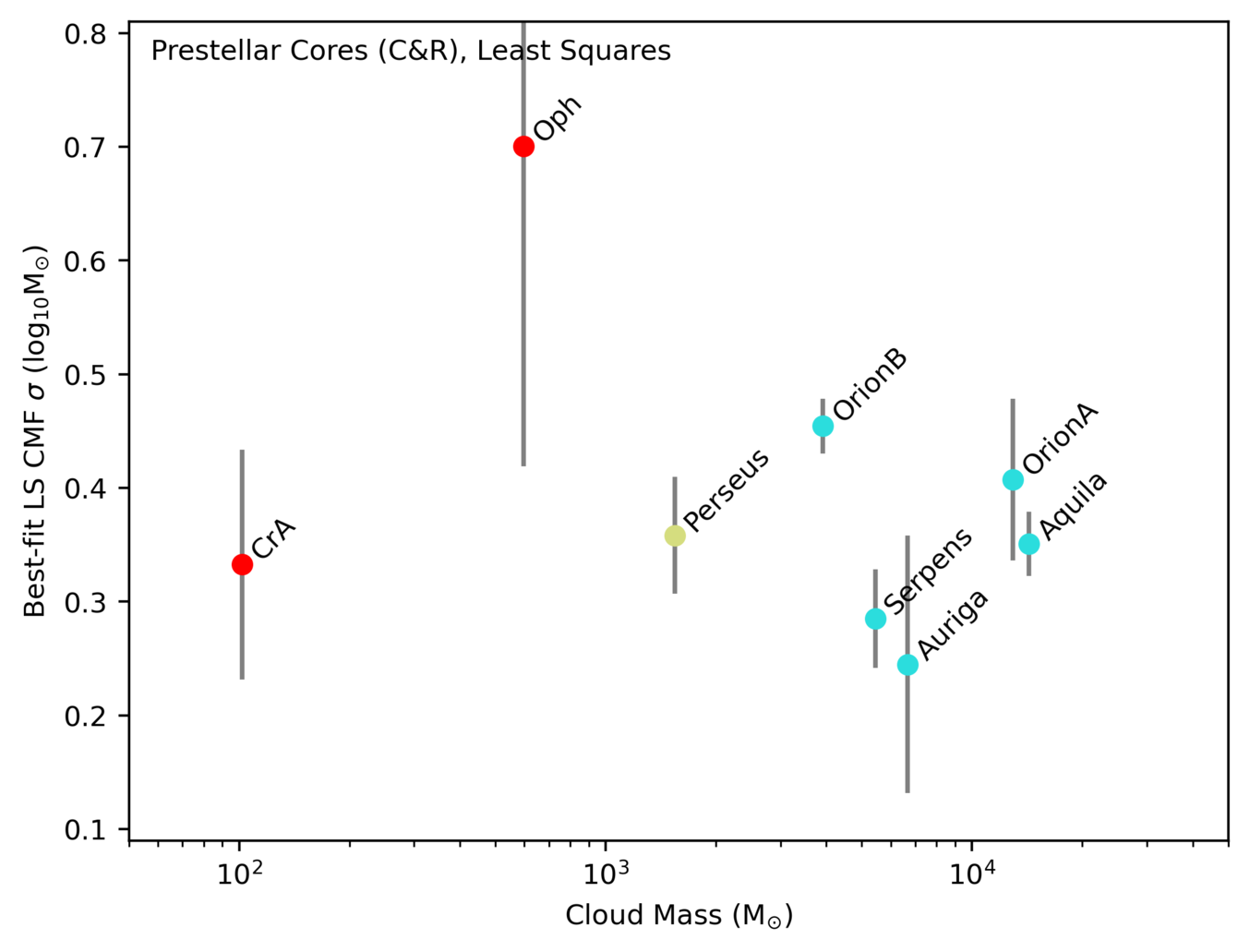}
    \caption{The width of the starless (left) and prestellar (right) CMFs as a function of cloud complex mass, determined using least-squares fitting.  Data points are colour-coded by their distance range.}
    \label{fig:width_cmf_ls}
\end{figure*}

\section{Discussion of Core Mass Functions} 
\label{sec:cmf_implications}

\subsection{Comparison with the stellar IMF}

\citet{chabrier2005} characterized the low-mass part of the system IMF with a log-normal distribution with peak mass $\mu_{Chabrier} = 0.25\,$M$_{\odot}$, and width $\sigma_{Chabrier} = 0.55\,\log_{10}$\,M$_{\odot}$.

The far-distance prestellar CMF which we measure, which is marginally consistent with the near- and mid- prestellar CMFs, has $\mu = 0.60\pm 0.05$ M$_{\odot}$ and $\sigma = 0.37\pm 0.02$ $\log_{10}$M$_{\odot}$ (best-fit LS), and $\mu=0.58$ M$_{\odot}$ and $\sigma = 0.39$ $\log_{10}$M$_{\odot}$ (most probable MC).  The LS and MC estimates are thus consistent with one another. 

We estimate a typical mass recovery fraction in our observations of 73\%, which would imply a corrected $\mu \approx 0.8\,$M$_{\odot}$.  This value suggests a prestellar CMF which peaks at $\sim 3\times$ the peak stellar system mass, and which is $\sim 0.17$ dex narrower.  If we have correctly characterised the prestellar CMF, this implies a core-to-star star formation efficiency of $\approx 33$\%.  This is very similar to that found by \citet{konyves2010} in \textit{Herschel} observations of the Aquila molecular cloud. 

\subsubsection{Sampling of the high-mass CMF}
\label{sec:cmf_sampling_highmass}

Our sample does not contain a sufficient number of high-mass cores to capture the power-law behaviour associated with the high-mass end of the IMF \citep{salpeter1955}.  We can model the CMF adequately using only a log-normal, with the exception of two very massive sources in Orion A.  The \citet{chabrier2005} IMF breaks to a power-law behaviour at 1 M$_{\odot}$, while we can fit our CMFs with log-normal distributions up to masses $> 10$\,M$_{\odot}$.

\subsubsection{Sampling of the low-mass CMF}
\label{sec:cmf_sampling_lowmass}

It is important to note that we may not be sampling the low-mass end of the CMF well enough to characterise it fully, particularly in the more distant clouds.  It is possible to both broaden a CMF and lower its characteristic mass by adding more low-mass cores.  Moreover, the MC plots above show a degeneracy between $\mu$ and $\sigma$; many of the CMFs that we find could plausibly be drawn from distributions with a lower $\mu$ and broader $\sigma$ than those of the highest-probability distribution.  However, in both the near- and the mid-distance cases, the peak of the CMF is significantly above the mass completeness limit, and so should be well-characterised (see Figure~\ref{fig:cmf_sless}).  Additionally, the values of $\mu$ and $\sigma$ determined for the CMFs of the mid-distance clouds are quite consistent with those determined for the far-distance clouds (see Figure~\ref{fig:cmf_params_ls_mc}).  These results suggest that incomplete sampling of the low-mass end of the CMF cannot fully explain the differing widths of the IMF and the CMF which we measure.

\subsubsection{Effect of unresolved cores}

We note that we have a number of unresolved cores in our sample, for which we have assumed a uniform BEC angular radius of 4.8$^{\prime\prime}$ (see Section~\ref{sec:be}).  These cores could potentially create a bias in our results.  However, we consider this effect to be minimal, because we see a fairly even spread in masses of unresolved cores (see Figure~\ref{fig:cmf_sless}).  As a result, these cores are unlikely to be creating any significant biases in the CMFs.  Moreover, only 4\% of the starless cores in our sample are unresolved, so their impact on the statistics of our sample is likely small.  The large majority of the unresolved cores are in the more distant clouds: there is only one unresolved core in the middle distance range, and the mid-distance Perseus and Cepheus prestellar CMFs are broadly consistent with those for the more distant clouds.  These points again suggest that the effect of unresolved cores on our CMFs is minimal. 

\subsection{CMF variation between clouds}

As discussed in Section~\ref{sec:cmf_clouds} above, the CMFs of the nearby clouds appear to be different both from one another other and from the mid- and far-distance clouds.  The mid- and far-distance clouds all appear to be consistent with having similar underlying starless core CMFs.  The mid- and far-distance clouds all have statistically similar prestellar core CMFs, but different most probable prestellar core CMFs.

The fact that the CMFs of the associated Orion A and Orion B clouds are extremely consistent with each another in both the starless and prestellar cases suggests that the variation between cloud CMFs which we see may not be only statistical scatter.  There may be genuine physical differences in how the clouds fragment into cores, or how the cores in these clouds acquire further mass. 

One such physical difference could be the timescale on which cores evolve.  If higher-mass prestellar cores last longer than their low-mass counterparts \citep{offner2014}, then the observed CMF would be weighted towards the longer-lived, more massive cores.  The median mass of the CMF would also be biased upwards in more massive regions (and downwards in lower-mass regions).  As the far-distance clouds contain more massive sources than the near- or mid-distance regions in Figure~\ref{fig:mc_3way}, evolutionary timescales might go some way towards reconciling the differences in CMF and median mass between clouds.

\subsubsection{Nearby clouds}

The nearby clouds appear to occupy a different part of the CMF parameter space than do the mid-distance and far clouds (cf. Figure~\ref{fig:mc_starless_candr_compare}), likely with a lower peak mass.  The nearby clouds also lack high-mass cores (cf. Figure~\ref{fig:mass_violin}).  These differences suggest that, depending on the relative number of low-mass and high-mass clouds, low-mass clouds may be able to contribute a significant number of lower-mass stars to the IMF, potentially both broadening the IMF and lowering its peak mass.  However, it should be noted that we are less sensitive to lower-mass sources in the more distant clouds, and so these more distant clouds are likely also contributing significant numbers of lower-mass stars to the IMF.

The nearby clouds, as well as lacking high-mass cores, also appear to have more significant intrinsic difference from one another than do the more distant clouds, although this finding is largely based on the well-known significant differences between the Ophiuchus and Taurus molecular clouds.

The Ophiuchus molecular cloud in particular appears to occupy a different part of the log-normal CMF parameter space than the rest of the clouds that we consider here, with a broader CMF and a lower peak mass (Figure~\ref{fig:mc_starless_candr_compare}).  The cloud has a significant excess of low-mass cores (Figure~\ref{fig:mass_violin}) and small core radii (Figure~\ref{fig:radius_violin}).  However, its maximum core mass and size are consistent with those of other clouds.

If the excess of low-mass cores in Ophiuchus is physical, fragmentation is proceeding differently in Ophiuchus than in the other clouds, and particularly the other nearby clouds.  Alternatively, or additionally, the combination of being both nearby and undergoing clustered star formation may make the detection of low-mass cores easier in Ophiuchus than elsewhere, as small, low-mass cores are more likely to be positioned against a background of cloud material, and so be easier to detect than comparable cores in a region of more dispersed star formation.

Previous work has shown that Class I and II disc masses in Ophiuchus are low compared to those in other clouds.  \citet{williams2019} found that disc masses in Class II sources in Ophiuchus are on average lower-mass than those in Lupus, despite Lupus being older and disc masses being expected to decrease with age. \citet{tobin2020} found that Class I and II disc masses in Ophiuchus are significantly lower than those in Orion, Taurus and Perseus.  They consider whether this difference is due to mis-classification of disc types in Ophiuchus due to heavy foreground extinction, or to the Class I disc population in Ophiuchus being systematically older than that in Orion.  However, they also consider the possibility that disc masses in Ophiuchus are genuinely physically lower than those in Orion (and other regions) due to differences in their initial formation conditions.  Our finding that core masses are also systematically lower in Ophiuchus than elsewhere supports this latter interpretation, suggesting that fragmentation into cores may be proceeding differently in Ophiuchus than in other nearby clouds.

We note that \citet{cazzoletti2019} similarly found low disc masses in CrA compared to the Chamaeleon and Lupus clouds.  Although the CMF of CrA is not very well-constrained, we note that it has the second lowest peak mass of the clouds in our sample, after Ophiuchus.  If CrA and Ophiuchus both have both a low average core mass and a low disc mass compared to other clouds at similar distance, we speculate that the lower core and disc masses in Ophiuchus may be due to differences in fragmentation (or in subsequent mass acquisition) between cores in regions of clustered and dispersed star formation.

We note that while Taurus has been thought to have an unusual IMF \citep[e.g.][]{luhman2003,goodwin2004,luhman2009,kraus2017}, recent reanalysis with \textit{Gaia} data has shown that its IMF is consistent with those of other nearby star-forming regions \citep{luhman:2018}.  Despite this, we find that like the other nearby clouds, the CMF of Taurus occupies a different part of parameter space than do the more distant clouds.

The differences both between the nearby low-mass clouds and the rest of our sample, and between the nearby clouds themselves, suggests that none can confidently be treated as a `typical' star-forming region.  The variation in behaviour that we see between the CMFs in these clouds may provide insights into the effects of environment and clustering on the properties of cores.  The proximity of these clouds allows the effect of environment on core properties to be investigated with good mass sensitivity and physical resolution.

\subsection{Effects of sample size}

As discussed in Section~\ref{sec:intro}, if a subsample is drawn from the stellar IMF, then the maximum stellar mass increases with the size of the sample \citep{weidnerkroupa:2005, WeidnerKroupaBonnell:2010, Elmegreen:2006}.  Only the largest samples (above around $10^4$\,M$_{\odot}$) sample the full mass distribution and include the highest-mass stars.  The sum of the masses of the starless cores in our sample is 917\,M$_{\odot}$, and so for a core-to-star mass efficiency of $\sim 1/3$, we would expect them to form a total stellar mass of $\sim 300$\,M$_{\odot}$, two orders of magnitude too low to sample the full IMF.

As discussed in Section~\ref{sec:violins}, we find that the maximum starless core mass in a cloud complex scales with cloud complex mass as $M_{max}\propto M_{cloud}^{0.58\pm 0.13}$.  This relationship is consistent with the 2/3-power scaling between the mass of the most massive star in stellar cluster and the cluster mass found by \citet{BonnellVineBate:2004}.  Hence, there may be a relatively constant core-to-star mass conversion efficiency in the clouds in our sample.  Furthermore, we need a significantly larger sample size in order to capture the behaviour of the CMF fully -- if indeed there is a single CMF representative of all star-forming regions.

To characterise the high-mass CMF accurately, we require observations of higher-mass (and perforce more distant) star-forming regions.  Orion A and Aquila, the most massive cloud complexes that we consider, while undergoing some high-mass star formation, are not high-mass star-forming regions in the sense that more distant massive hub-filament systems are \citep[e.g.,][]{motte2018}.  The form of the high-mass CMF in high-mass star-forming regions is being investigated by the ALMA-IMF Survey \citep{motte2022}.  The ALMA-IMF Survey has characterised the high-mass power-law slope of the CMF in some massive star-forming clouds \citep{pouteau2022}, but their mass completeness limits (e.g. 0.8\,M$_{\odot}$ in W43) are too high to capture the behaviour of the low-mass CMF in these more distant regions.

In our sample, we see that the distributions of masses in nearby clouds appear to be more disparate than those in more distant, typically more massive, star-forming regions.  This difference begs the question of whether larger and higher-mass clouds are genuinely more similar to one another than are smaller clouds, perhaps due to being less influenced by their local environment, or whether by virtue of their size they encompass the range of behaviours seen in low-mass clouds at different locations within them.

To answer these questions, and understand fully the form of the CMF, and how it varies between star-forming regions, we require the ability to detect, and ideally resolve, low-mass cores in more distant high-mass star-forming regions.  To do so in sufficient numbers to sample the full CMF adequately will require the sensitivities and mapping speeds which are planned for next-generation submillimetre instrumentation such as 50-m-class single-dish telescopes or focal plane array interferometers.

\section{Conclusions}
\label{sec:conclusions}

In this paper we have presented a catalogue of dense cores identified in SCUBA-2 observations of nearby molecular clouds made as part of the JCMT Gould Belt Survey.  We identified 2257 dense cores using the \textit{getsources} algorithm, along with further selection criteria.  Of these, 2004 were resolved.  We identified 1323 sources as starless cores, 845 sources as protostellar cores, 70 sources as being heated (24-\um-bright) starless cores, and 19 sources as being potential extragalactic contaminants.  Of the starless cores, 456 were identified as robust prestellar cores, 484 as candidate prestellar cores, and 383 as unbound cores using the critical Bonnor--Ebert criterion.

Our key conclusions are as follows:
\begin{enumerate}
    \item On average, 59\% of the detected cores are starless, and 41\% are protostellar.  71\% of the starless cores are prestellar cores (candidate or robust).   This breakdown suggests that the lifetime for prestellar cores is $\sim 0.5$\,Myr, similar to that of Class 0/I embedded YSOs \citep{evans2009}.
    \item We see statistically significant differences in starless and prestellar core fractions between cloud complexes.  We found that both Serpens and Orion B have an excess of starless and prestellar cores, while Taurus and the Pipe Nebula have a deficit.  The trend is for higher-mass regions to have a higher fraction of starless (or prestellar) cores compared to protostars, suggesting a longer average lifetime for prestellar cores in higher-mass clouds.
    \item  There is a weak correlation between median starless core mass and cloud complex mass.  We find that maximum starless core mass scales with cloud complex mass, such that $M_{max}\propto M_{cloud}^{0.58\pm 0.13}$.  This relationship is consistent with the $2/3$ scaling between maximum stellar mass in a cluster and cluster mass \citep{BonnellVineBate:2004}.
    \item We found that the CMFs of clouds in our survey can be characterised using log-normal distributions.  However, we do not sample a sufficiently large number of sources to recover the expected high-mass Salpeter-like power-law slope.
    \item We found that the CMFs of cores in our sample are not consistent with all being drawn from the same underlying distribution, both when considered as a function of distance and when considered by cloud complex.  The mid-distance (200--355\,pc) and far-distance (355--500\,pc) CMFs are somewhat consistent with one another.  However, the near-distance ($<200\,$pc) starless CMFs are only marginally consistent with the mid- and far-distance starless CMFs, while the near-distance prestellar CMF is marginally consistent with the mid-distance prestellar CMF, and almost entirely inconsistent with the far-distance prestellar CMF.  Starless core CMFs for individual cloud complexes are typically consistent with one another in the mid- and far-distance ranges, but the prestellar core CMFs show greater variation.  The CMFs of the near-distance clouds are less consistent both with one another and with the mid- and far-distance clouds.
    \item The prestellar CMF of the far-distance clouds has a peak mass ($\mu$) of approximately $3\times$ the \citet{chabrier2005} log-normal peak for the system IMF, consistent with the value seen in Aquila by the \textit{Herschel} Gould Belt Survey \citep{konyves2015}.  This implies a prestellar core-to-star efficiency of $\sim 1/3$.  The prestellar and starless CMFs of both the mid- and far-distance clouds have a width systematically $\simeq 0.15$ dex lower than that of the \citet{chabrier2005} log-normal IMF.
    \item We found that the CMF of the nearby Ophiuchus molecular cloud is noticeably different from those of the other clouds in our survey, being wide, and with a significantly lower peak mass.  This difference appears to be due to an excess of low-mass cores in this cloud, but may be the result of a selection effect due to the proximity of the cloud and its clustered star formation.  However, Ophiuchus also has unusually low disc masses, compared to region of non-clustered star formation at similar distances.  This difference suggests that the excess of low-mass sources in Ophiuchus may be indicative of differences in fragmentation between regions of clustered and dispersed star formation.
\end{enumerate}

We present this catalogue as a resource to the community, noting that many further analyses of the cores are possible, such as full energetic balance analyses for those cores for which appropriate spectroscopic and polarimetric data is available, and comparison with \textit{Herschel} Gould Belt Survey catalogues, and other catalogues of Gould Belt sources. 

Our analysis shows that the CMFs of the Gould Belt clouds are not consistent with being drawn from a single underlying CMF.  The Gould Belt clouds do not contain a number -- or mass -- of cores sufficient to sample the full range of core masses needed to create the stellar IMF.  However, they do give us insight into the variation of properties of low-mass cores within and between molecular clouds.  Thus, the ability to detect cores with masses down to the brown dwarf mass limit in more distant, higher-mass, clouds, is required to understand the form, and the variability, of the Core Mass Function.

\section*{Acknowledgements}

We thank Alexander Men'shchikov for helpful discussions on the \textit{getsources} algorithm, and Jason Kirk for providing a complete \textit{Herschel} column density map of the Taurus Molecular Cloud. 

K.P. is a Royal Society University Research Fellow, supported by grant number URF\textbackslash R1\textbackslash 211322, and at various points in the long history of this project has been supported by the Ministry of Science and Technology of Taiwan under grant number 106-2119-M-007-021-MY3, and by the Science and Technology Facilities Council (STFC) of the United Kingdom under grant numbers ST/K002023/1 and ST/M000877/1.  J.D.F., H.K. and D.J. acknowledge support from separate NSERC Discovery Grants.  H.K., J.D.F., S.N., and R.K. acknowledge support from the National Research Council of Canada co-op program at the Herzberg Astronomy and Astrophysics Research Centre which enabled S.N. and R.K. to participate in this research.  D.W.T.  acknowledges support from the UK STFC on grant number ST/R000786/1.  The Dunlap Institute is funded through an endowment established by the David Dunlap family and the University of Toronto.

The James Clerk Maxwell Telescope (JCMT) is operated by the East Asian Observatory on behalf of the National Astronomical Observatory of Japan; the Academia Sinica Institute of Astronomy and Astrophysics; the Korea Astronomy and Space Science Institute; the National Astronomical Research Institute of Thailand; the Center for Astronomical Mega-Science (as well as the National Key R\&D Program of China with grant no. 2017YFA0402700). Additional funding support is provided by the STFC of the UK and participating universities and organizations in the UK, Canada and Ireland.  Additional funds for the construction of SCUBA-2 were provided by the Canada Foundation for Innovation. The JCMT has historically been operated by the Joint Astronomy Centre on behalf of the STFC of the UK, the NRC of Canada and the Netherlands Organisation for Scientific Research.

This research has made use of:\\
\textbullet\ The Canadian Advanced Network For Astronomy Research \citep[CANFAR;][]{Gaudet2010}, operated in partnership by the Canadian Astronomy Data Centre and The Digital Research Alliance of Canada with support from the National Research Council (NRC) of Canada, the Canadian Space Agency, CANARIE and the Canadian Foundation for Innovation.\\
\textbullet\ Starlink software \citep{currie2014}, currently supported by the East Asian Observatory.\\
\textbullet\ Astropy (\url{http://www.astropy.org}), a community-developed core Python package and an ecosystem of tools and resources for astronomy \citep{astropy:2013, astropy:2018, astropy:2022}.\\
\textbullet\ The NASA/IPAC Extragalactic Database (NED), which is funded by the National Aeronautics and Space Administration and operated by the California Institute of Technology.\\
\textbullet\ The SIMBAD database, operated at CDS, Strasbourg, France.\\
\textbullet\ Data products from the Wide-field Infrared Survey Explorer, which is a joint project of the University of California, Los Angeles, and the Jet Propulsion Laboratory/California Institute of Technology, funded by the National Aeronautics and Space Administration (NASA).\\
\textbullet\ Data from the Herschel Gould Belt Survey (HGBS) project (\url{http://gouldbelt-herschel.cea.fr}). The HGBS is a Herschel Key Programme jointly carried out by SPIRE Specialist Astronomy Group 3 (SAG 3), scientists of several institutes in the PACS Consortium (CEA Saclay, INAF-IFSI Rome and INAF-Arcetri, KU Leuven, MPIA Heidelberg), and scientists of the Herschel Science Center (HSC).\\
\textbullet\ NASA's Astrophysics Data System.

The authors wish to recognize and acknowledge the very significant cultural role and reverence that the summit of Maunakea has always had within the indigenous Hawaiian community.  We are most fortunate to have the opportunity to conduct observations from this mountain.

\section*{Data availability}

The raw data used to create the maps used in this paper are available in the JCMT data archive hosted by the Canadian Astronomy Data Centre (CADC) \url{https://www.cadc-ccda.hia-iha.nrc-cnrc.gc.ca/en/}, under project codes MJLSG31 (Orion A), MJLSG32 (Ophiuchus), MJLSG33 (Aquila and Serpens), MJLSG34 (Lupus), MJLSG35 (Corona Australis), MJLSG36 (IC 5146), MJLSG37 (Auriga and Taurus), MJLSG38 (Perseus), MJLSG39 (Pipe Nebula), MJLSG40 (Cepheus) and MJLSG41 (Orion B).  The maps of all regions used in this paper are available for public download through \citet{kirk2018}, or directly at the DOI {\tt \url{https://doi.org/10.11570/18.0005}}.  The JCMT Gould Belt Survey Core Catalogue is available at [DOI TBC],
along with the output of the \textit{getsources} algorithm without our selection criteria applied.  The code used to create model Bonnor--Ebert spheres for completeness testing is available at \url{https://github.com/KatePattle/bonnor-ebert-sphere}.



\bibliographystyle{mnras}



\appendix

\section{JCMT GBS Data}
\label{sec:appendix_images}

\begin{figure*}
    \centering
    \includegraphics[width=\textwidth]{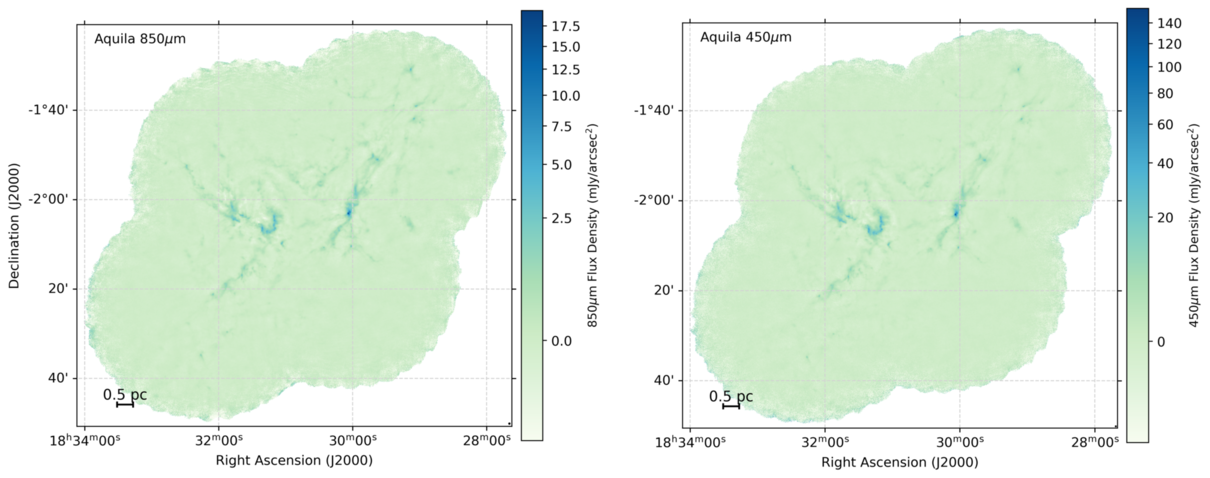}
    \caption{SCUBA-2 850\,$\upmu$m (left) and 450\,$\upmu$m (right) IR3 images of Aquila.  Colour scale is cube-root stretched.}
    \label{fig:ir3_aquila}
\end{figure*}

\begin{figure*}
    \centering
    \includegraphics[width=\textwidth]{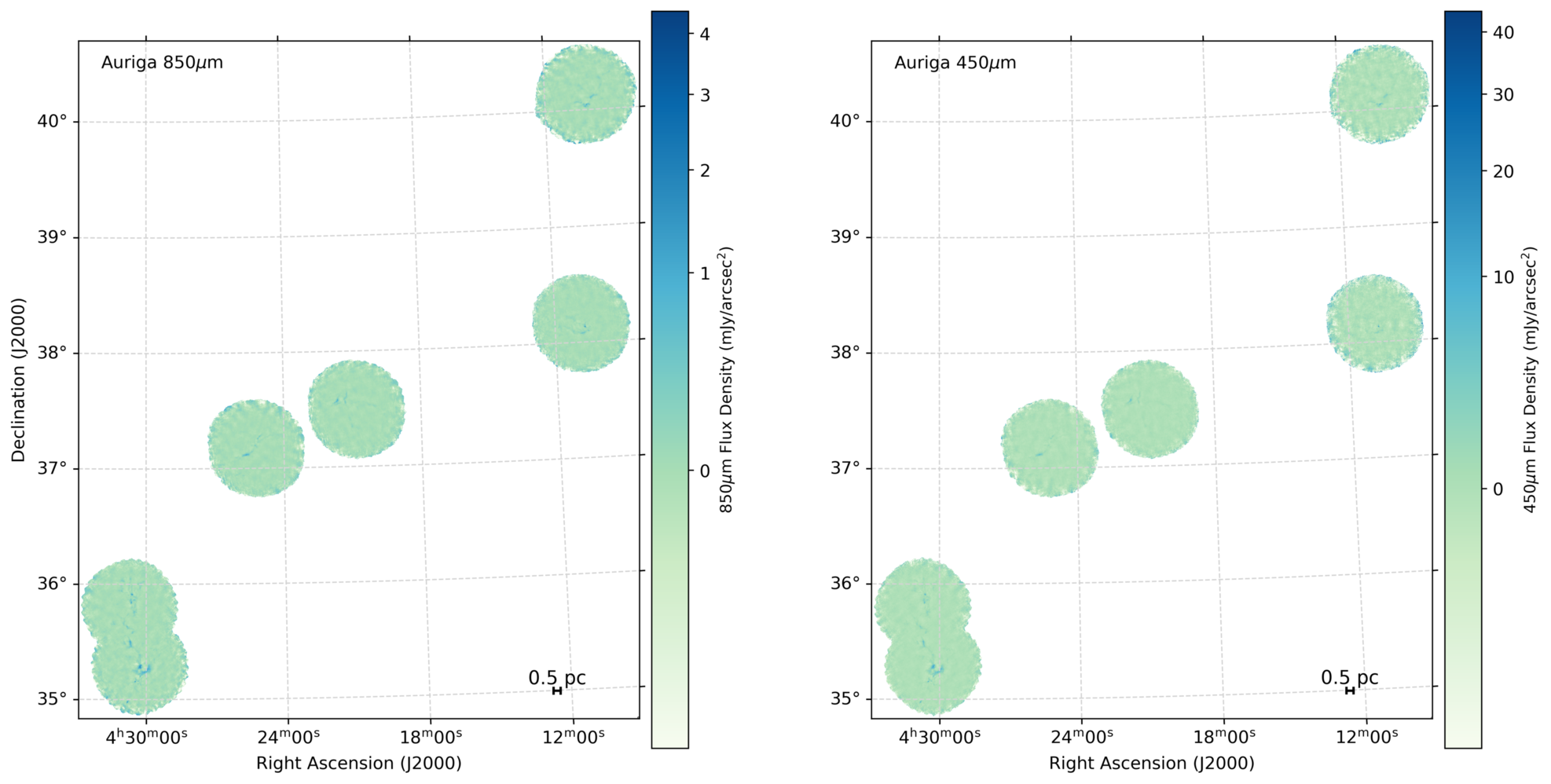}
    \caption{SCUBA-2 850\,$\upmu$m and 450\,$\upmu$m IR3 images of Auriga.}
    \label{fig:ir3_auriga}
\end{figure*}

\begin{figure*}
    \centering
    \includegraphics[width=\textwidth]{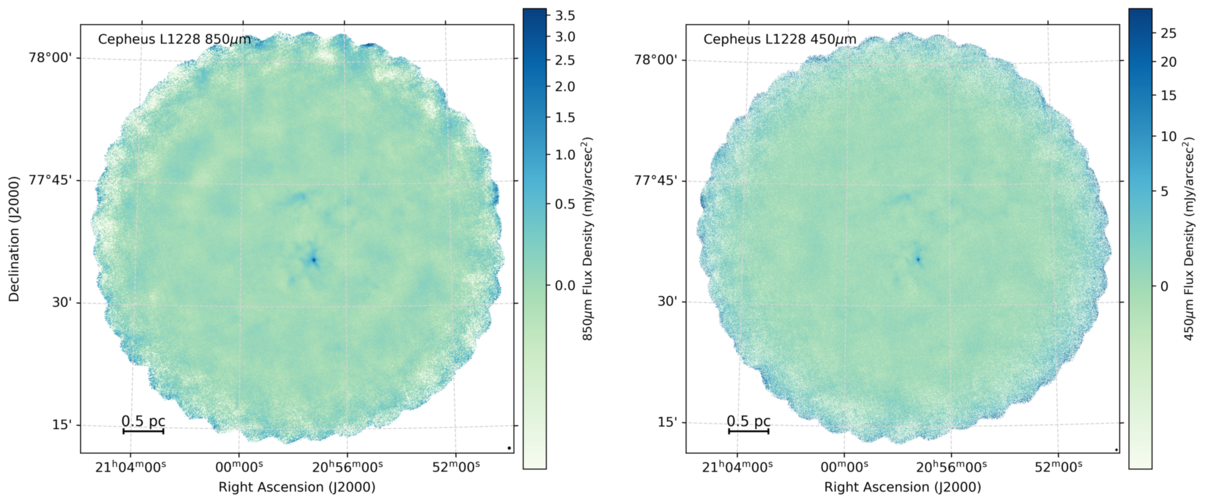}
    \caption{SCUBA-2 850\,$\upmu$m and 450\,$\upmu$m IR3 images of Cepheus L1228.}
    \label{fig:ir3_cephL1228}
\end{figure*}

\begin{figure*}
    \centering
    \includegraphics[width=0.8\textwidth]{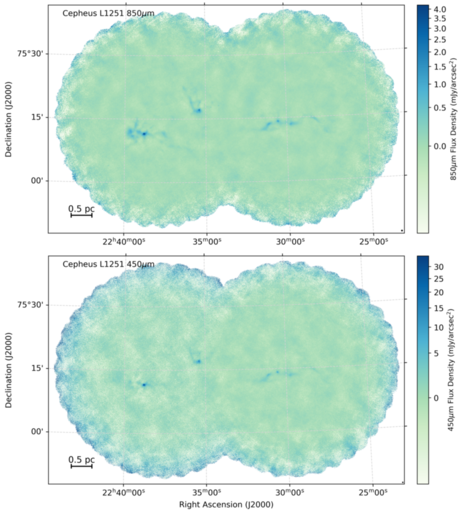}
    \caption{SCUBA-2 850\,$\upmu$m and 450\,$\upmu$m IR3 images of Cepheus L1251.}
    \label{fig:ir3_cephL1251}
\end{figure*}

\begin{figure*}
    \centering
    \includegraphics[width=\textwidth]{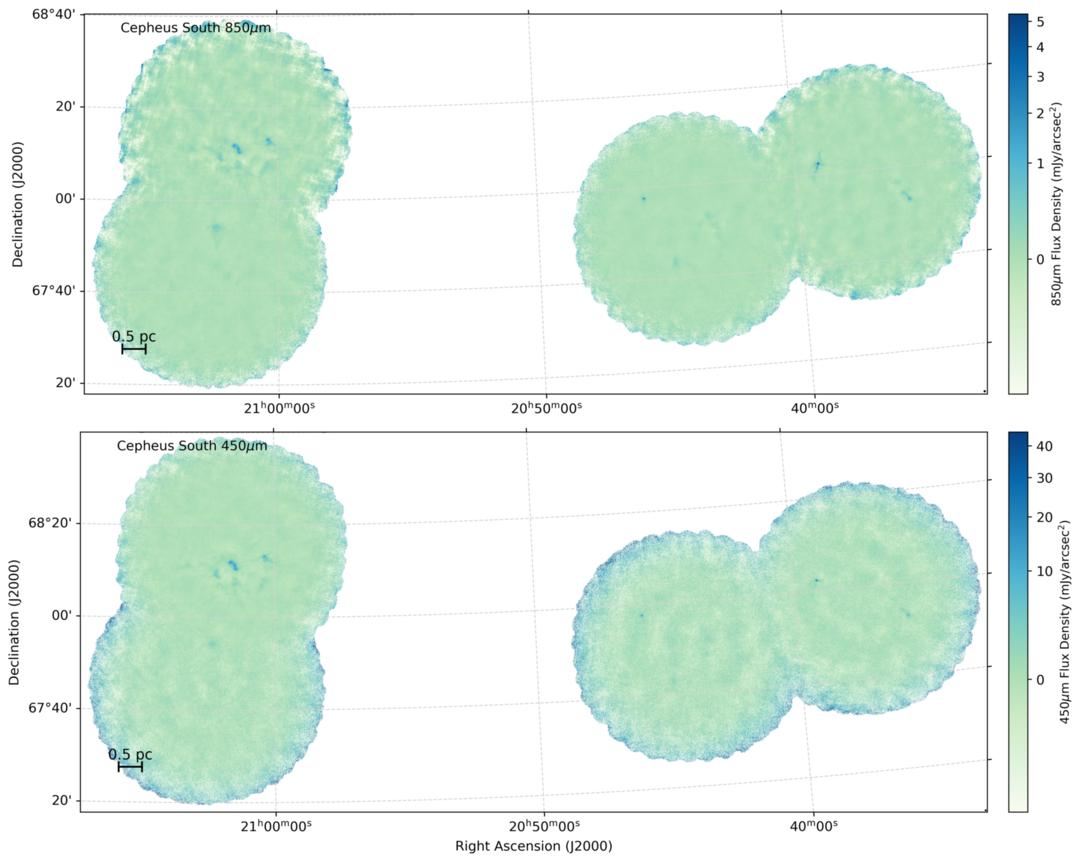}
    \caption{SCUBA-2 850\,$\upmu$m and 450\,$\upmu$m IR3 images of Cepheus South.}
    \label{fig:ir3_cephsouth}
\end{figure*}

\begin{figure*}
    \centering
    \includegraphics[width=\textwidth]{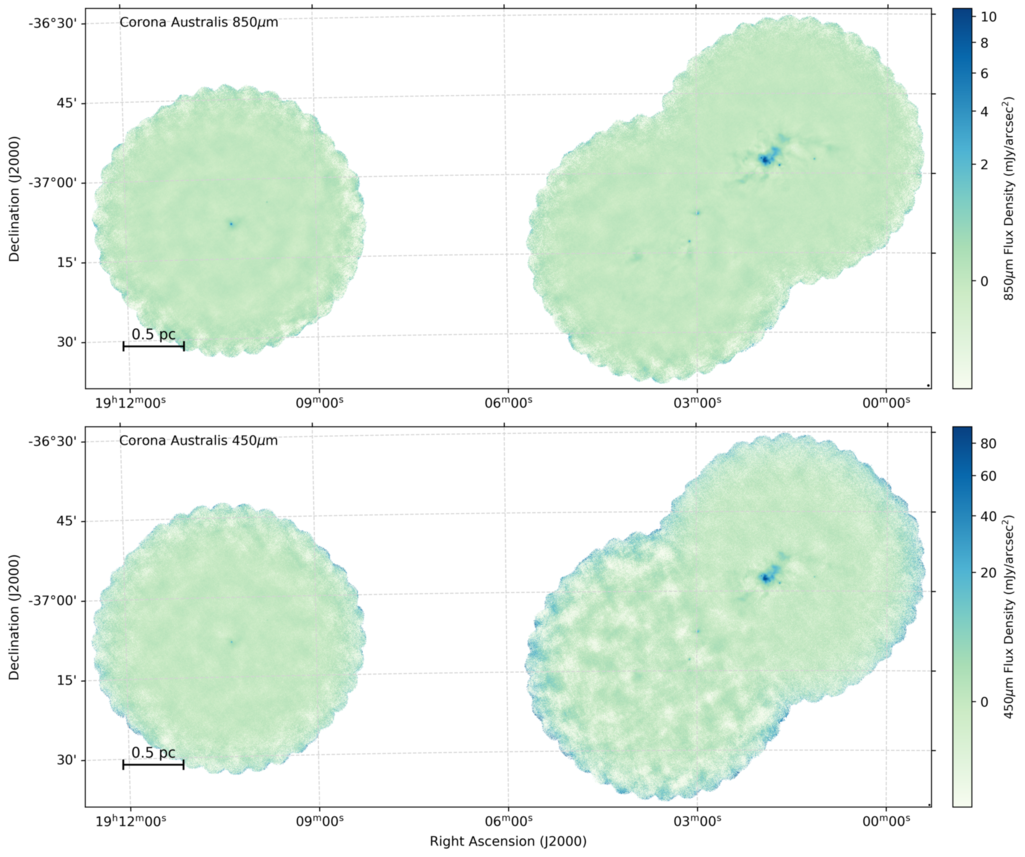}
    \caption{SCUBA-2 850\,$\upmu$m and 450\,$\upmu$m IR3 images of Corona Australis.}
    \label{fig:ir3_CrA}
\end{figure*}

\begin{figure*}
    \centering
    \includegraphics[width=\textwidth]{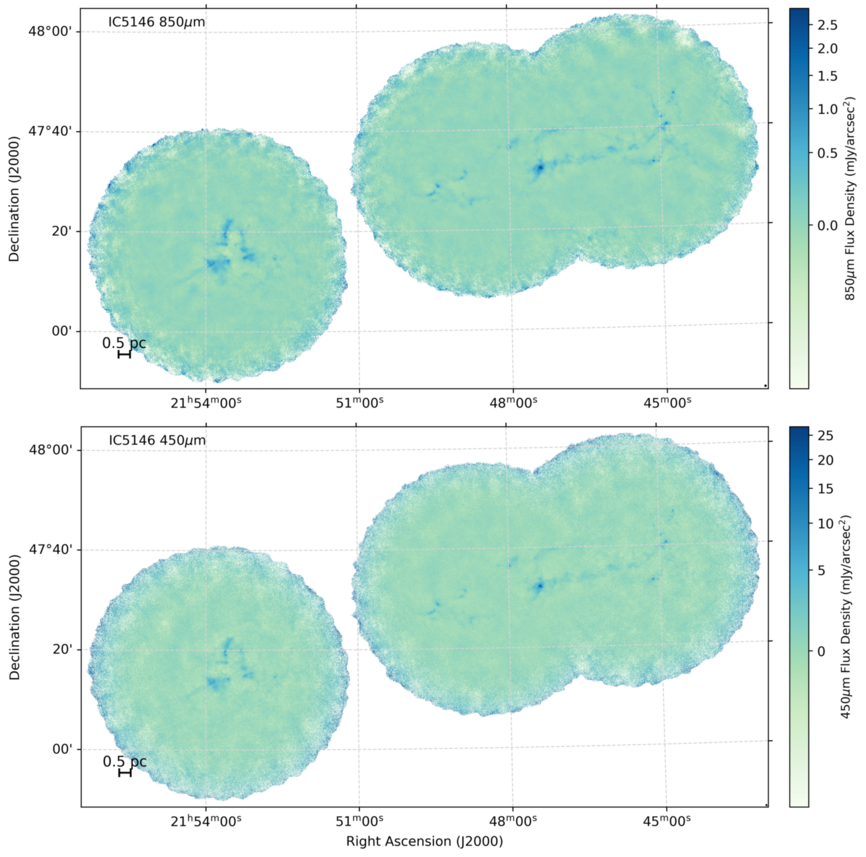}
    \caption{SCUBA-2 850\,$\upmu$m and 450\,$\upmu$m IR3 images of IC 5146.}
    \label{fig:ir3_IC5146}
\end{figure*}

\begin{figure*}
    \centering
    \includegraphics[width=\textwidth]{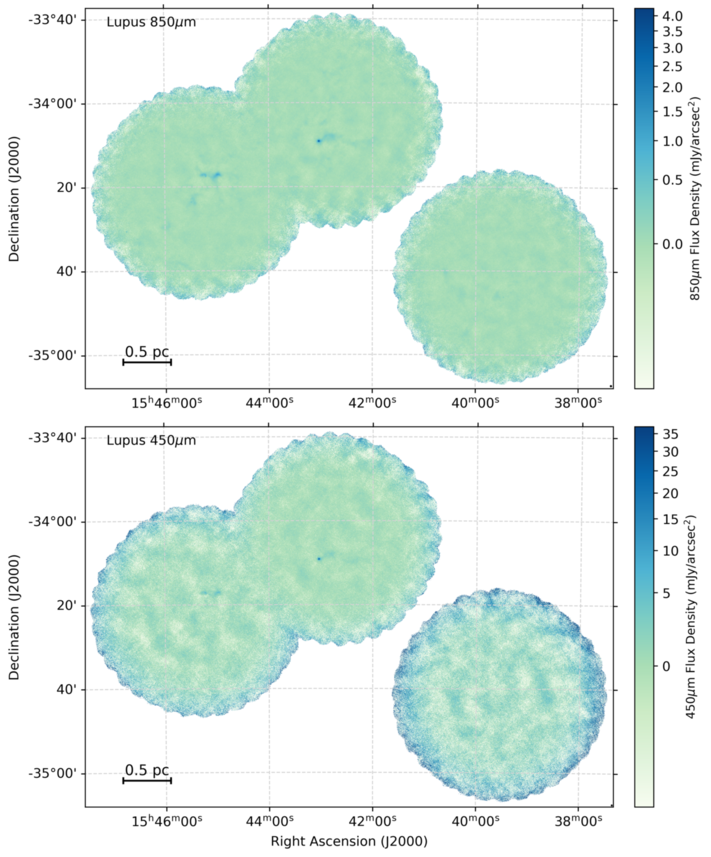}
    \caption{SCUBA-2 850\,$\upmu$m and 450\,$\upmu$m IR3 images of Lupus.}
    \label{fig:ir3_lupus}
\end{figure*}

\begin{figure*}
    \centering
    \includegraphics[width=\textwidth]{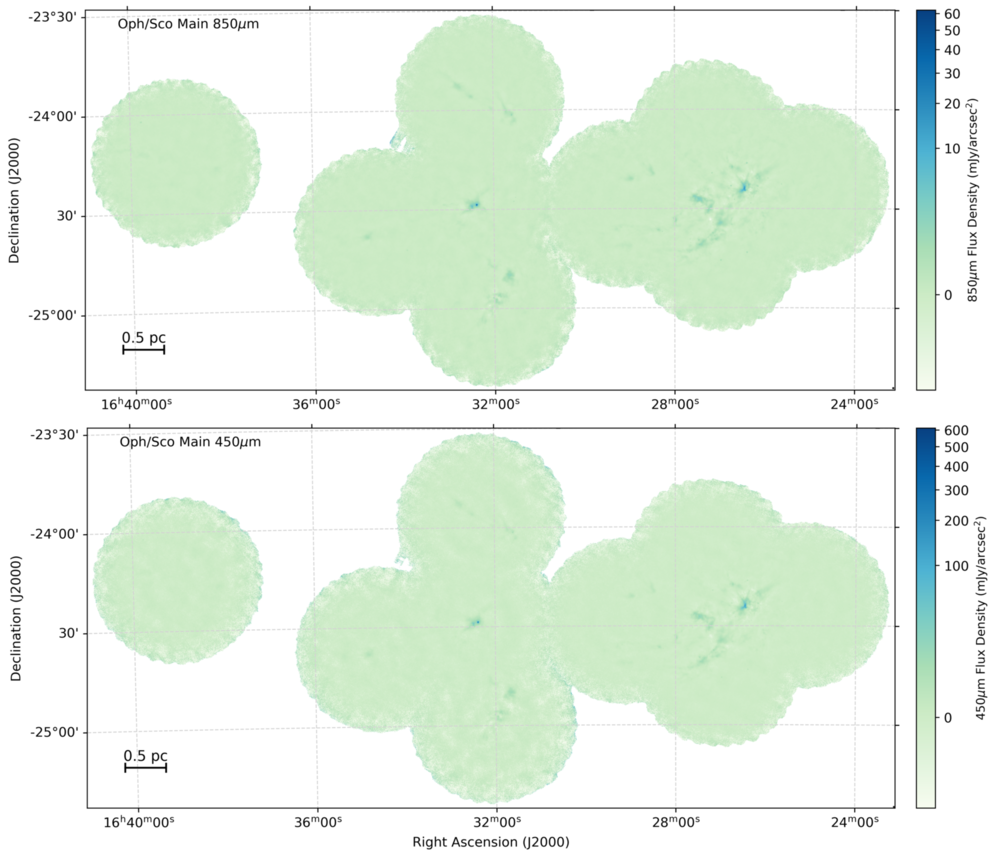}
    \caption{SCUBA-2 850\,$\upmu$m and 450\,$\upmu$m IR3 images of Ophiuchus L1688 (right) and L1689, L1709 and L1712 (left).  Colour map is fourth-root scaled.}
    \label{fig:ir3_ophscomain}
\end{figure*}

\begin{figure*}
    \centering
    \includegraphics[width=\textwidth]{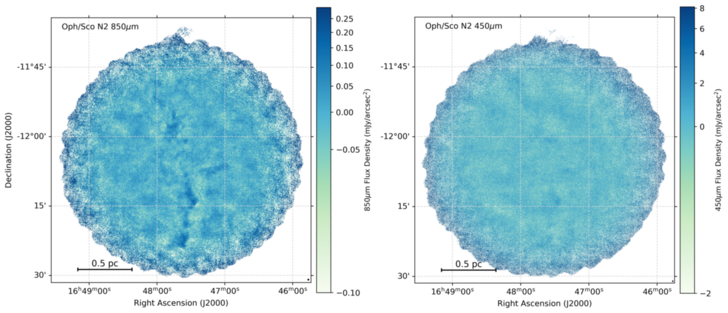}
    \caption{SCUBA-2 850\,$\upmu$m and 450\,$\upmu$m IR3 images of Oph/Sco N2.}
    \label{fig:ir3_ophscoN2}
\end{figure*}

\begin{figure*}
    \centering
    \includegraphics[width=\textwidth]{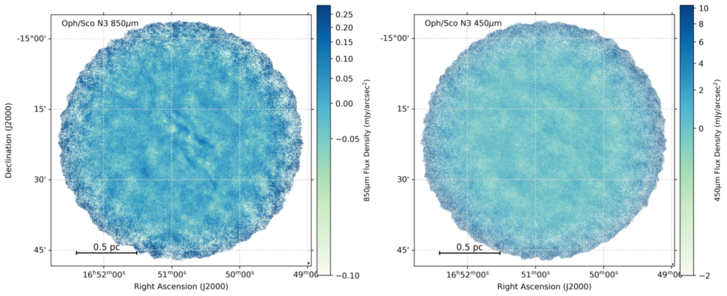}
    \caption{SCUBA-2 850\,$\upmu$m and 450\,$\upmu$m IR3 images of Oph/Sco N3.}
    \label{fig:ir3_ophscoN3}
\end{figure*}

\begin{figure*}
    \centering
    \includegraphics[width=\textwidth]{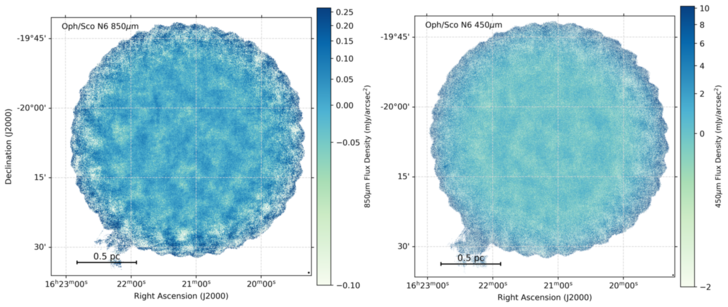}
    \caption{SCUBA-2 850\,$\upmu$m and 450\,$\upmu$m IR3 images of Oph/Sco N6.}
    \label{fig:ir3_ophscoN6}
\end{figure*}

\begin{figure*}
    \centering
    \includegraphics[width=\textwidth]{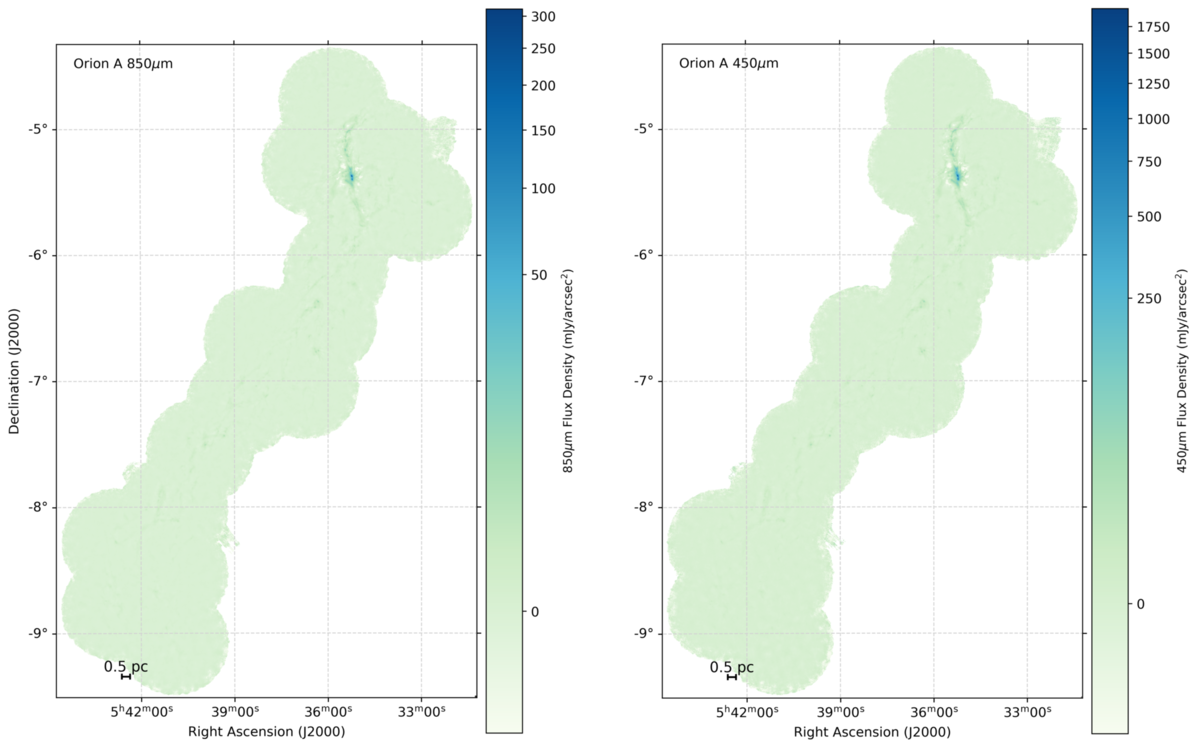}
    \caption{SCUBA-2 850\,$\upmu$m and 450\,$\upmu$m IR3 images of Orion A.  Colour map is fourth-root scaled.}
    \label{fig:ir3_orionA}
\end{figure*}

\begin{figure*}
    \centering
    \includegraphics[width=\textwidth]{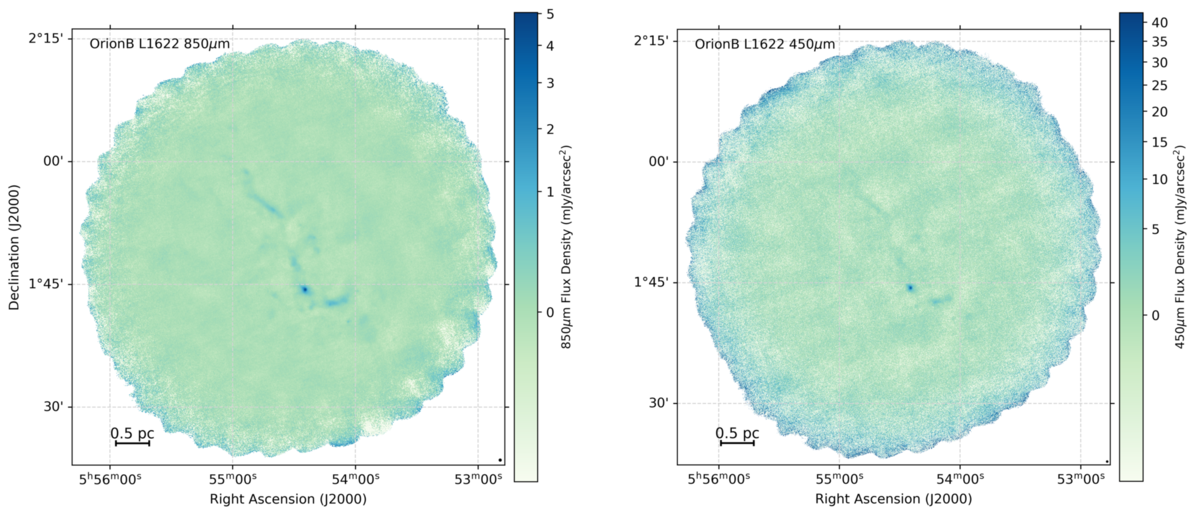}
    \caption{SCUBA-2 850\,$\upmu$m and 450\,$\upmu$m IR3 images of Orion B L1622.}
    \label{fig:ir3_OrionB_L1622}
\end{figure*}

\begin{figure*}
    \centering
    \includegraphics[width=\textwidth]{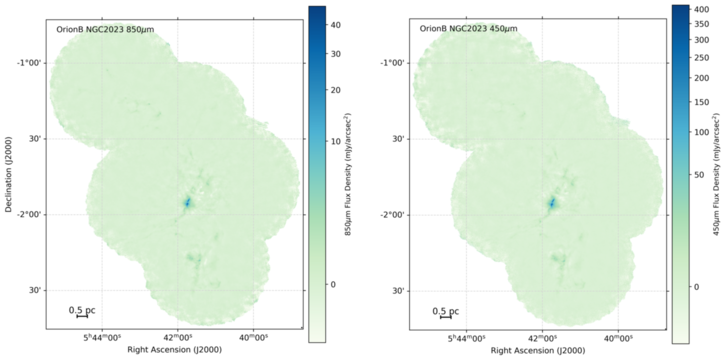}
    \caption{SCUBA-2 850\,$\upmu$m and 450\,$\upmu$m IR3 images of Orion B NGC2023.}
    \label{fig:ir3_OrionB_N2023}
\end{figure*}

\begin{figure*}
    \centering
    \includegraphics[width=\textwidth]{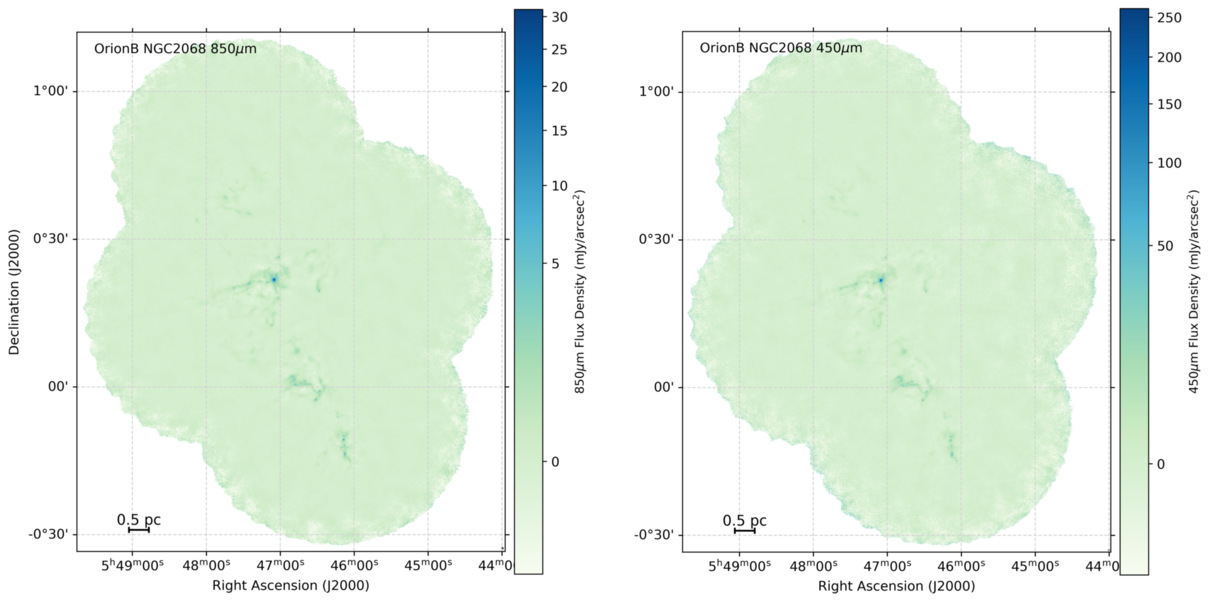}
    \caption{SCUBA-2 850\,$\upmu$m and 450\,$\upmu$m IR3 images of Orion B NGC2068.}
    \label{fig:ir3_OrionB_N2068}
\end{figure*}

\begin{figure*}
    \centering
    \includegraphics[width=\textwidth]{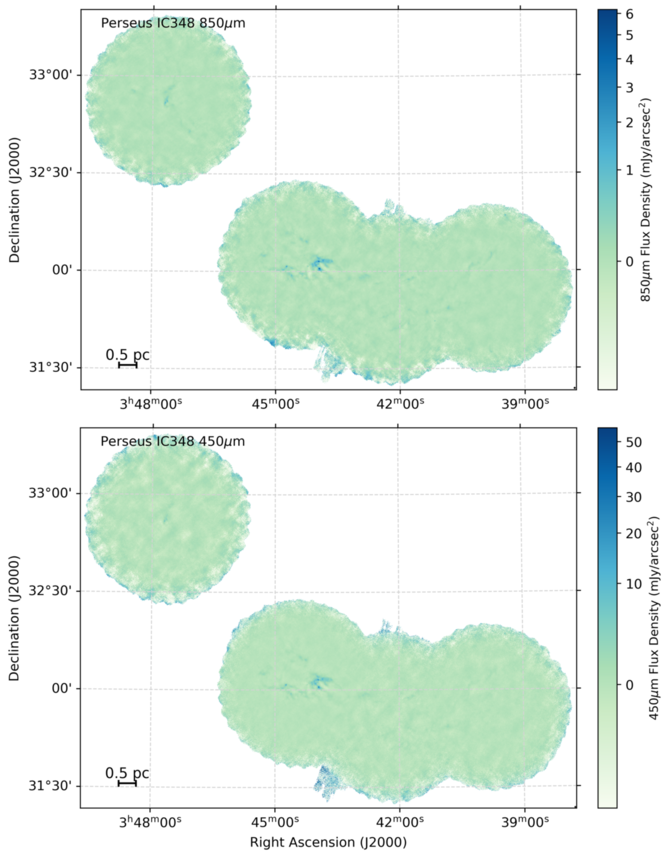}
    \caption{SCUBA-2 850\,$\upmu$m and 450\,$\upmu$m IR3 images of Perseus IC348.}
    \label{fig:ir3_perseusIC348}
\end{figure*}

\begin{figure*}
    \centering
    \includegraphics[width=\textwidth]{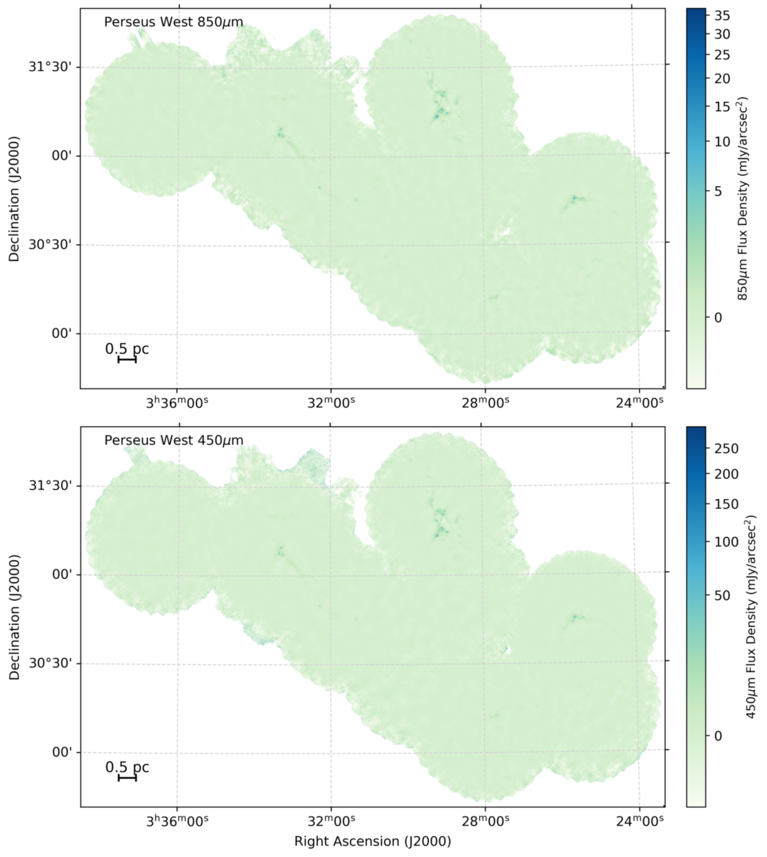}
    \caption{SCUBA-2 850\,$\upmu$m IR4 image of Perseus West.}
    \label{fig:ir4_perseusWest}
\end{figure*}

\begin{figure*}
    \centering
    \includegraphics[width=\textwidth]{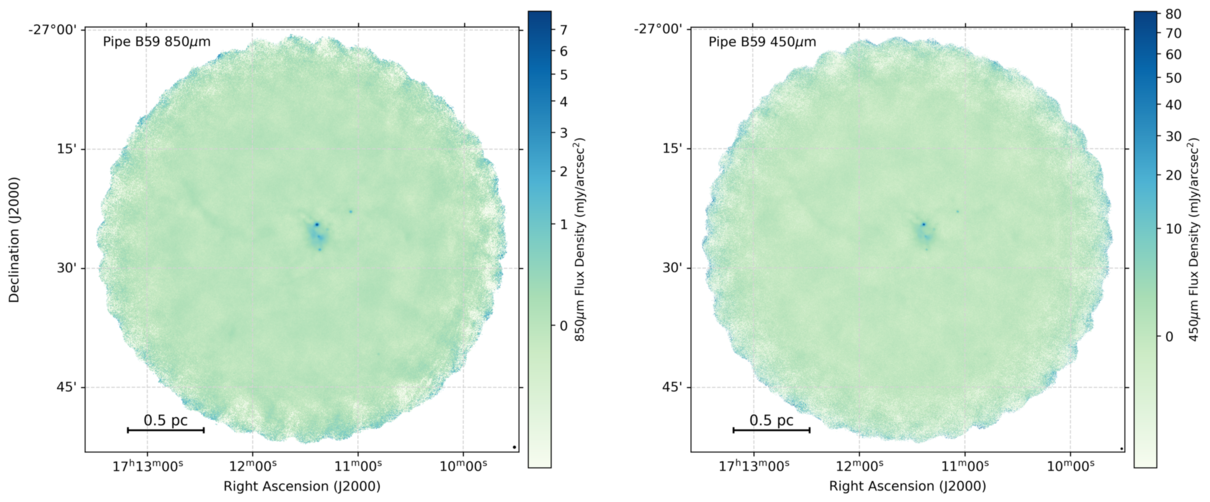}
    \caption{SCUBA-2 850\,$\upmu$m and 450\,$\upmu$m IR3 images of Pipe B59.}
    \label{fig:ir3_pipeB59}
\end{figure*}

\begin{figure*}
    \centering
    \includegraphics[width=\textwidth]{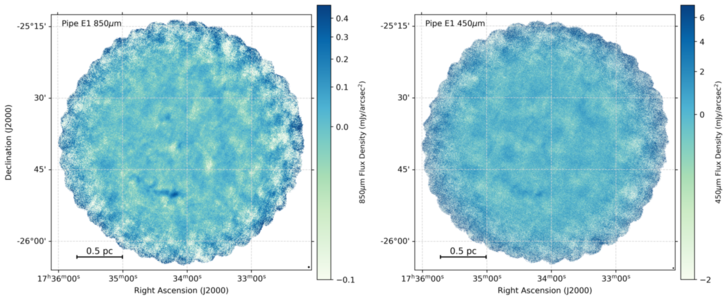}
    \caption{SCUBA-2 850\,$\upmu$m and 450\,$\upmu$m IR3 images of Pipe E1.}
    \label{fig:ir3_pipeE1}
\end{figure*}

\begin{figure*}
    \centering
    \includegraphics[width=\textwidth]{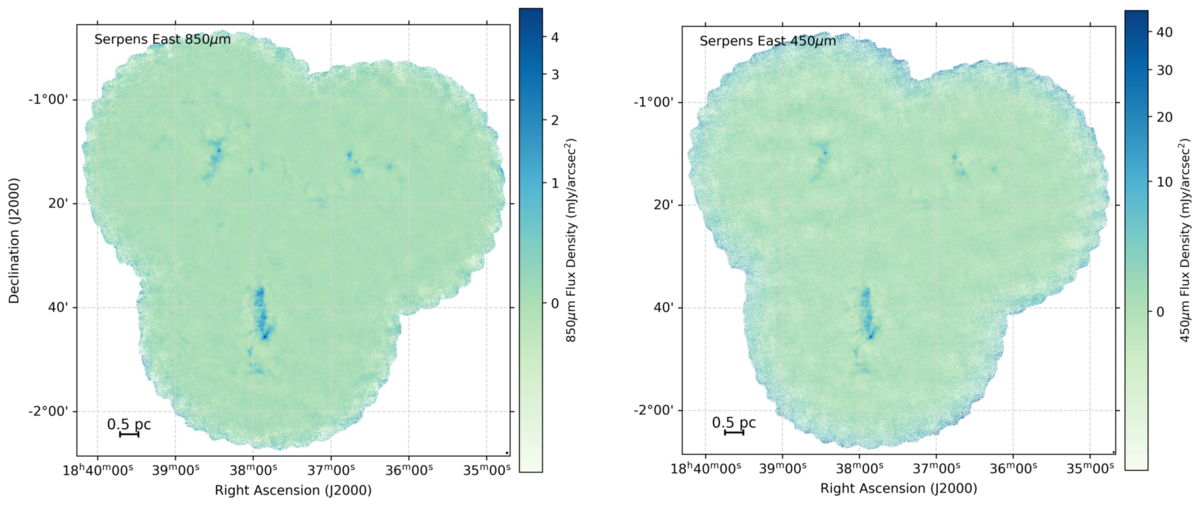}
    \caption{SCUBA-2 850\,$\upmu$m and 450\,$\upmu$m IR3 images of Serpens East.}
    \label{fig:ir3_serpensEast}
\end{figure*}

\begin{figure*}
    \centering
    \includegraphics[width=\textwidth]{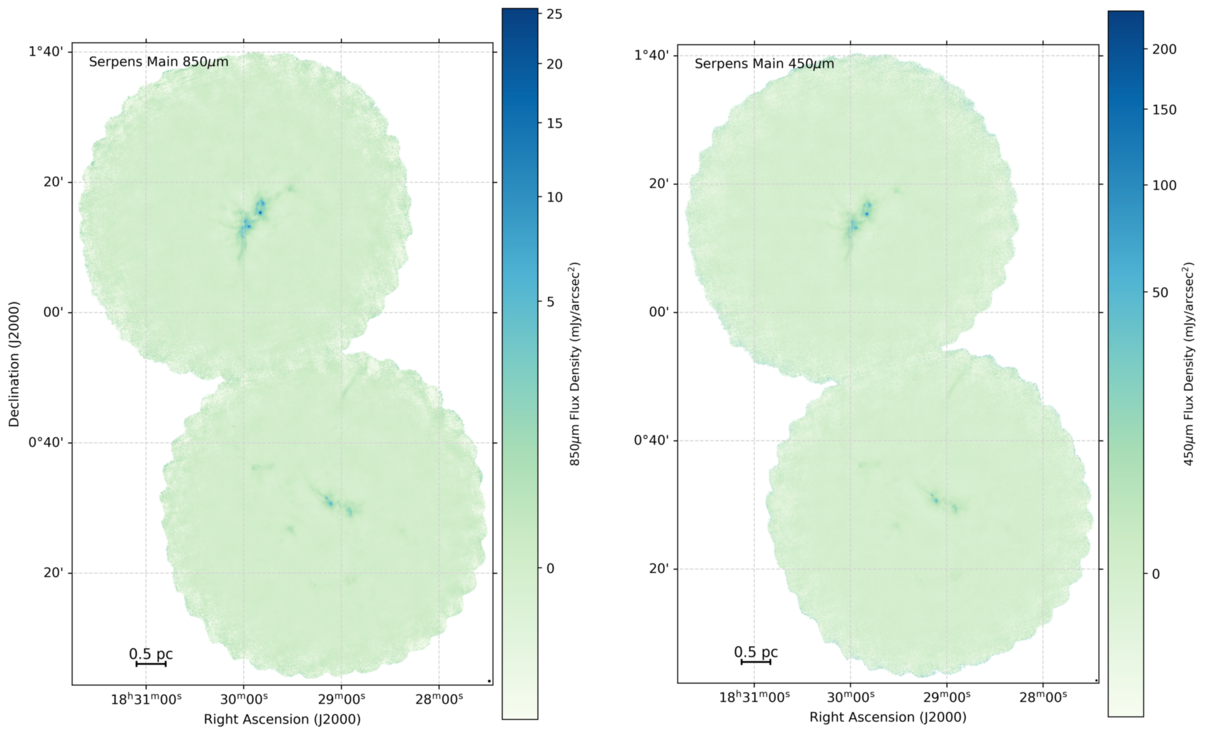}
    \caption{SCUBA-2 850\,$\upmu$m and 450\,$\upmu$m IR3 images of Serpens Main.}
    \label{fig:ir3_serpensMain}
\end{figure*}

\begin{figure*}
    \centering
    \includegraphics[width=\textwidth]{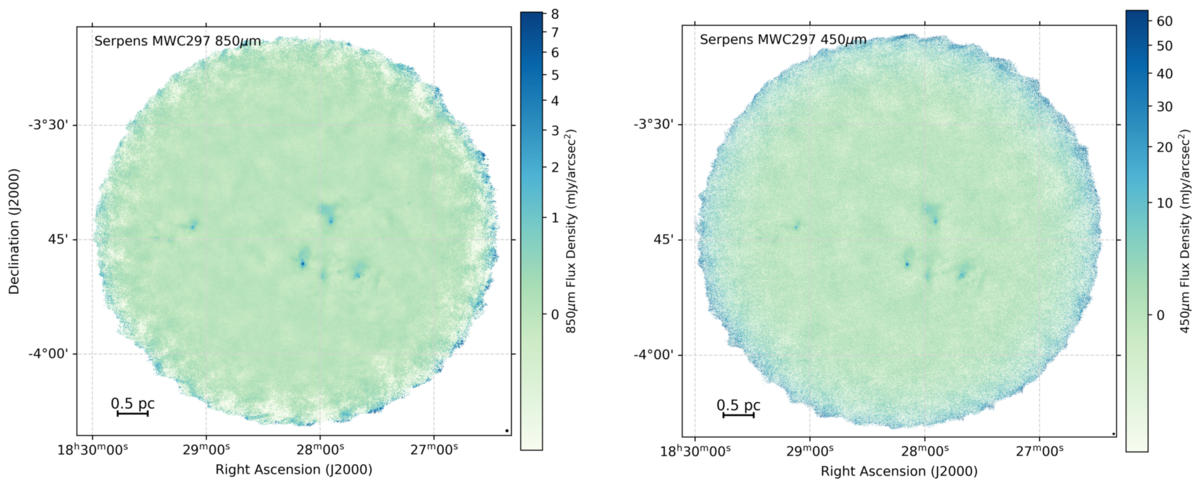}
    \caption{SCUBA-2 850\,$\upmu$m and 450\,$\upmu$m IR3 images of Serpens MWC297.}
    \label{fig:ir3_serpensMWC297}
\end{figure*}

\begin{figure*}
    \centering
    \includegraphics[width=\textwidth]{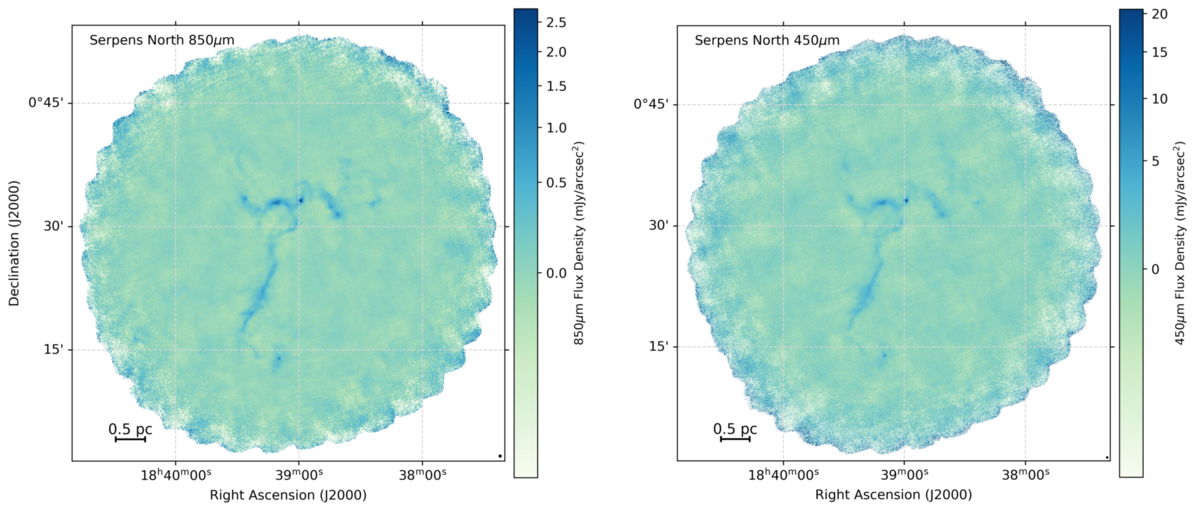}
    \caption{SCUBA-2 850\,$\upmu$m and 450\,$\upmu$m IR3 images of Serpens North.}
    \label{fig:ir3_serpensNorth}
\end{figure*}

\begin{figure*}
    \centering
    \includegraphics[width=\textwidth]{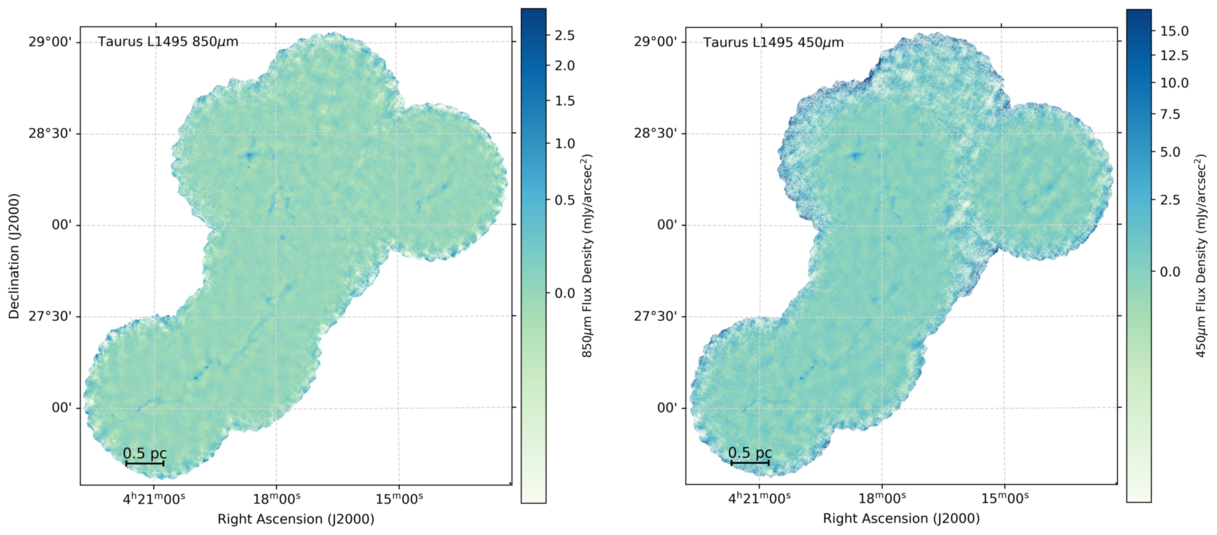}
    \caption{SCUBA-2 850\,$\upmu$m and 450\,$\upmu$m IR3 images of Taurus L1495.}
    \label{fig:ir3_taurusL1495}
\end{figure*}

\begin{figure*}
    \centering
    \includegraphics[width=\textwidth]{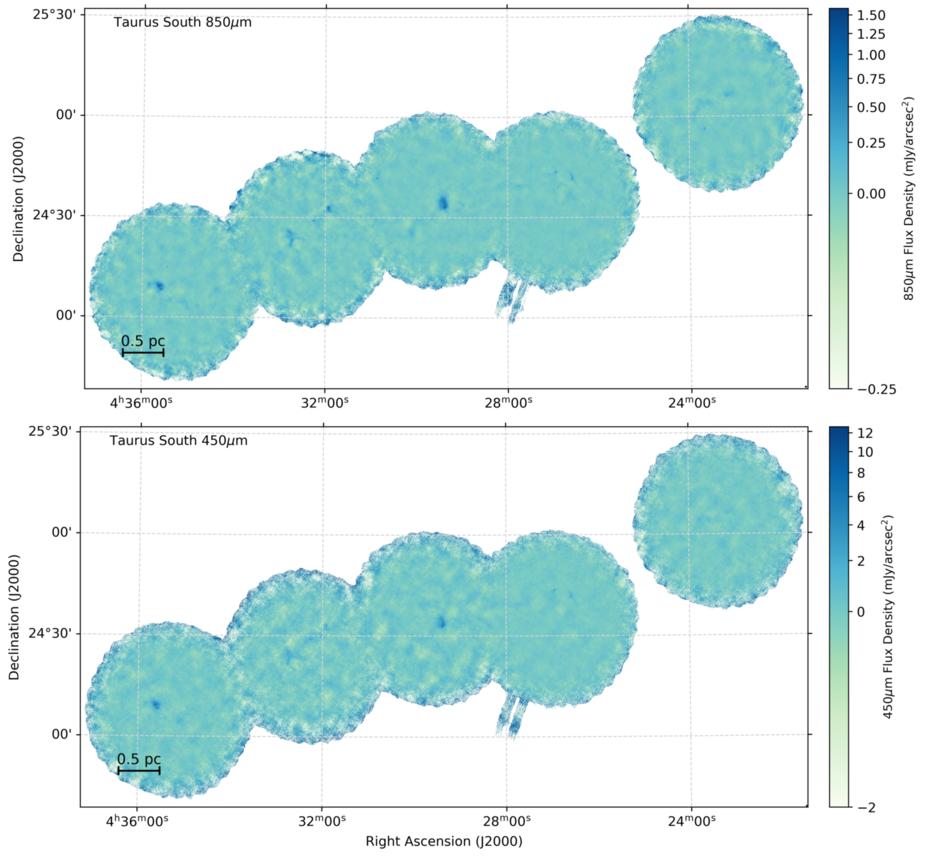}
    \caption{SCUBA-2 850\,$\upmu$m and 450\,$\upmu$m IR3 images of Taurus South.}
    \label{fig:ir3_taurusSouth}
\end{figure*}

\begin{figure*}
    \centering
    \includegraphics[width=\textwidth]{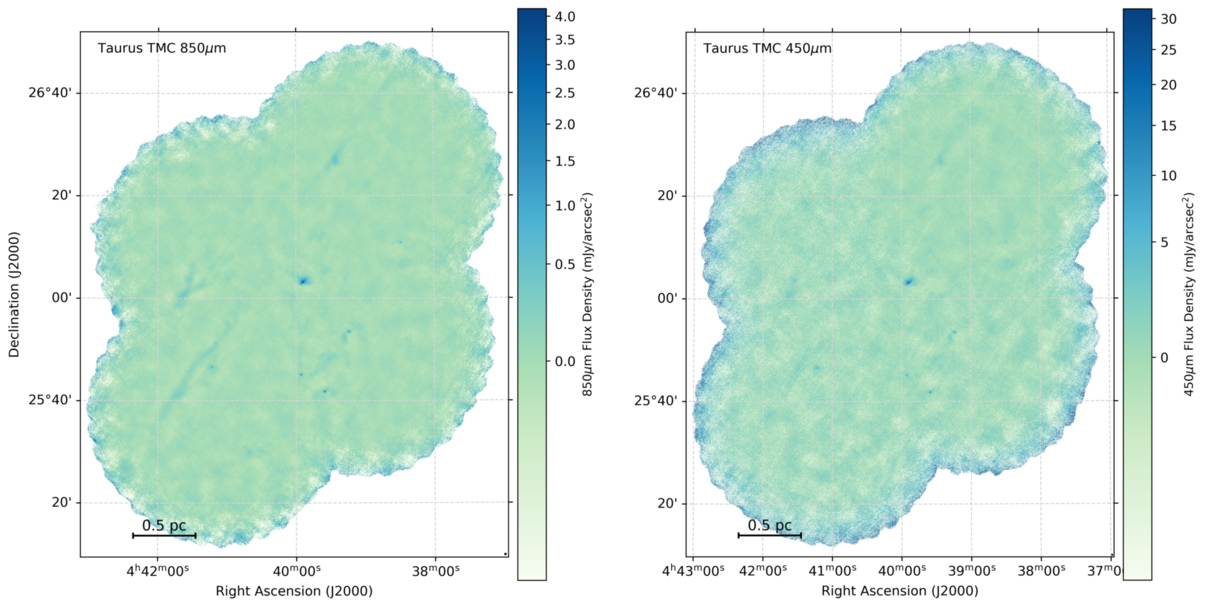}
    \caption{SCUBA-2 850\,$\upmu$m and 450\,$\upmu$m IR3 images of Taurus TMC1.}
    \label{fig:ir3_taurusTMC1}
\end{figure*}

In this appendix we present the JCMT GBS data that forms the basis of the catalogue in this paper.  Table~\ref{tab:map_completeness} presents the mapping completeness of the survey as a function of Herschel GBS column density.  Figures~\ref{fig:ir3_aquila}--\ref{fig:ir3_taurusTMC1} present the 850\,\um\ and 450\,\um\ JCMT GBS data used in this work, available for public download through \citet{kirk2018}, or directly at {\tt \url{https://doi.org/10.11570/18.0005}}.

\begin{table*}
\centering
\caption{JCMT GBS mapping completeness as a function of Herschel-derived column density.  The mapping completeness is calculated as the fraction of pixels/area
in the Herschel column density map at or above the listed value which is covered
by the JCMT GBS map.}
\label{tab:map_completeness}
\begin{tabular}{clllllllllllllll}
\hline
Cloud$^{a}$ & \multicolumn{15}{c}{$N(H_{2})^b$ $(\times 10^{21}$cm$^{-3}$)} \\
\hline
 & 5 & 6 & 7 & 8 & 9 & 10 & 12 & 
14 & 16 & 18 & 20 & 25 & 30 & 40 & 50 \\
\hline
    Aquila  & 0.42  & 0.67  & 0.84  & 0.92  & 0.95  & 0.97  & 0.98  & 0.99  & 0.99  & 1.00  & 1.00  & 1.00  & 1.00  & 1.00  & 1.00 \\
    Auriga  & 0.41  & 0.48  & 0.60  & 0.69  & 0.74  & 0.77  & 0.82  & 0.85  & 0.86  & 0.87  & 0.89  & 0.92  & 0.92  & 0.93  & 0.94 \\
   Cepheus  & 0.90  & 0.93  & 0.96  & 0.99  & 1.00  & 1.00  & 1.00  & 1.00  & 1.00  & 1.00  & 1.00  & 1.00  & 1.00  & 1.00  & 1.00 \\
       CrA  & 0.98  & 0.99  & 1.00  & 1.00  & 1.00  & 1.00  & 1.00  & 1.00  & 1.00  & 1.00  & 1.00  & 1.00  & 1.00  & 1.00  & 1.00 \\
    IC 5146  & 0.99  & 0.99  & 0.99  & 0.99  & 0.99  & 1.00  & 1.00  & 1.00  & 1.00  & 1.00  & 1.00  & 1.00  & 1.00    & -1    & -1 \\
     Lupus  & 0.99  & 1.00  & 1.00  & 1.00  & 1.00  & 1.00  & 1.00  & 1.00  & 1.00  & 1.00  & 1.00  & 1.00  & 1.00  & 1.00    & -1 \\
 Ophiuchus  & 0.99  & 1.00  & 1.00  & 1.00  & 1.00  & 1.00  & 1.00  & 1.00  & 1.00  & 1.00  & 1.00  & 1.00  & 1.00  & 1.00  & 1.00 \\
    Orion A  & 0.93  & 0.94  & 0.95  & 0.95  & 0.96  & 0.96  & 0.97  & 0.98  & 0.99  & 0.99  & 1.00  & 1.00  & 1.00  & 1.00  & 1.00 \\
    Orion B  & 0.90  & 0.92  & 0.93  & 0.94  & 0.95  & 0.95  & 0.96  & 0.96  & 0.95  & 0.95  & 0.95  & 0.96  & 0.97  & 1.00  & 1.00 \\
   Perseus  & 1.00  & 1.00  & 1.00  & 1.00  & 1.00  & 1.00  & 1.00  & 1.00  & 1.00  & 1.00  & 1.00  & 1.00  & 1.00  & 1.00  & 1.00 \\
      Pipe  & 0.93  & 0.94  & 0.96  & 0.99  & 1.00  & 1.00  & 1.00  & 1.00  & 1.00  & 1.00  & 1.00  & 1.00  & 1.00    & -1    & -1 \\
   Serpens  & 0.77  & 0.83  & 0.87  & 0.89  & 0.90  & 0.92  & 0.95  & 0.98  & 0.99  & 0.99  & 1.00  & 1.00  & 1.00  & 1.00  & 1.00 \\
    Taurus  & 0.89  & 0.90  & 0.91  & 0.92  & 0.93  & 0.94  & 0.96  & 0.96  & 0.96  & 0.95  & 0.95  & 0.93  & 0.93  & 1.00    & -1 \\
\hline
\end{tabular}
\raggedright

a: Most clouds have a one-to-one correspondence between the area on the sky included in the Herschel GBS (HGBS) map and the JCMT GBS map, but there are some exceptions which are noted here, in alphabetical cloud order.  Aquila: the HGBS `Aquila' field covers both the JGBS `Aquila' and `SerpensMWC297' fields, so we include both in the calculation.  Auriga: we excluded some noisy artefacts from the edge of the HGBS map.  Cepheus: This includes the HGBS map `cepl1228' matching the JGBS `CepheusL1228' field, and similarly the HGBS map `cepl1251' matching the JGBS `CepheusL1251' field.  The JGBS 'CepheusSouth' map is covered partially in each of the HGBS maps `cep1172' and `cep1157'.  The HGBS map `cep1241' was not covered by JGBS.  Lupus: JGBS only covered part of the HGBS `LupI' map, so only that field is used for comparison.  We excluded some noisy artefacts from the edge of the HGBS LupI map.  Ophiuchus: The HGBS `ophiuchus' map does not include coverage for the L1712 area within the JGBS `OphScoMain' map \citep[see, e.g.,][]{Wilking08}, and the JGBS maps `OphScoN2', `OphScoN3' and `OphScoN6' are also not included in the HGBS.  Orion A: we excluded some noisy artefacts from the edge of the HGBS map.  Taurus: we used the Herschel-based column density map derived by J. Kirk (priv. comm.), as this covered all three of our JGBS fields, while the published HGBS map only matched the JGBS `TaurusL1495' map.\\
b: A value of -1 is used to denote instances when there are no pixels in the Herschel column density map at or above the specified value.
\end{table*}

\clearpage

\section{Completeness Testing}
\label{app_compl}


\begin{figure*}
    \centering
        \includegraphics[width=0.49\textwidth]{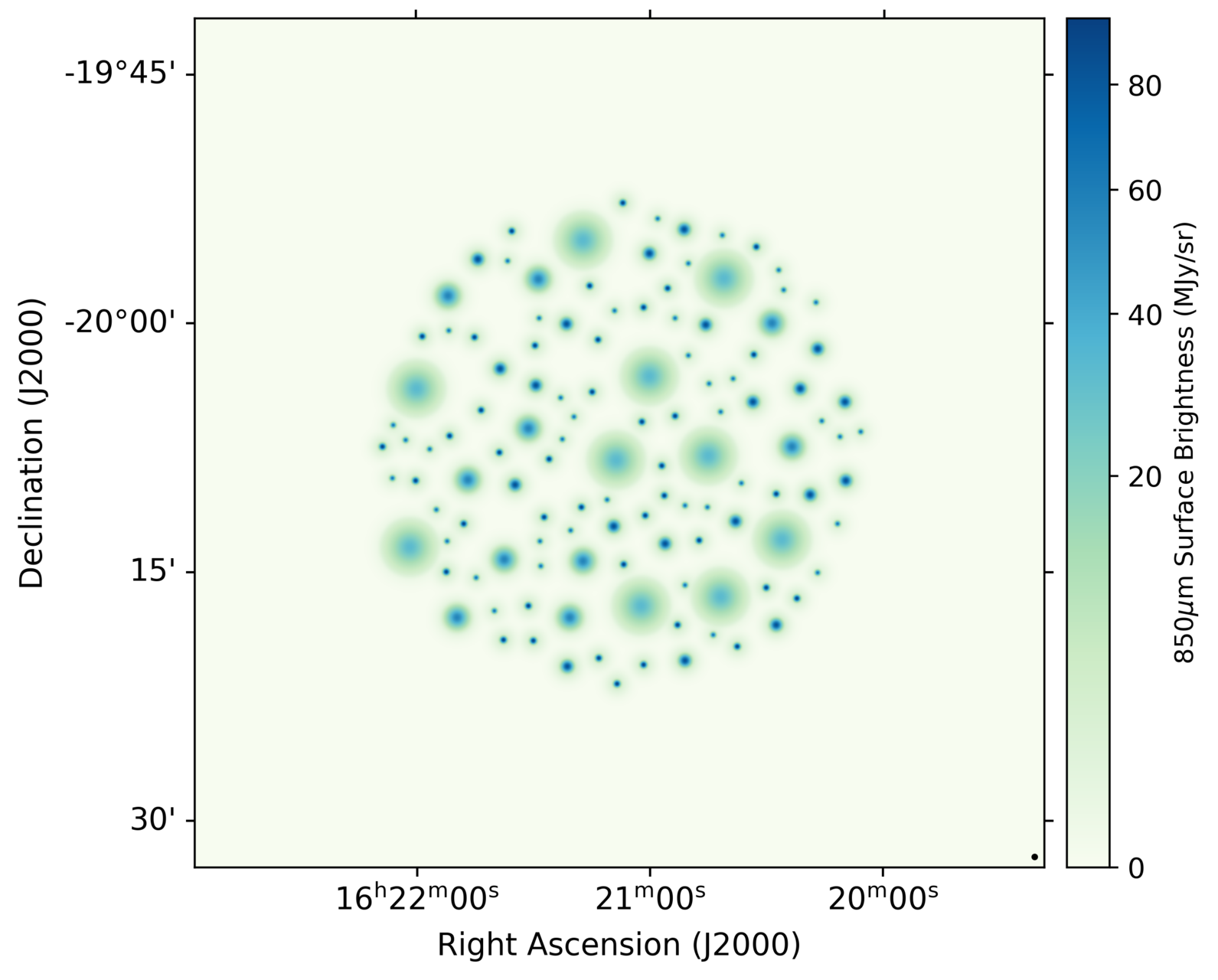}
        \hfill
        \includegraphics[width=0.49\textwidth]{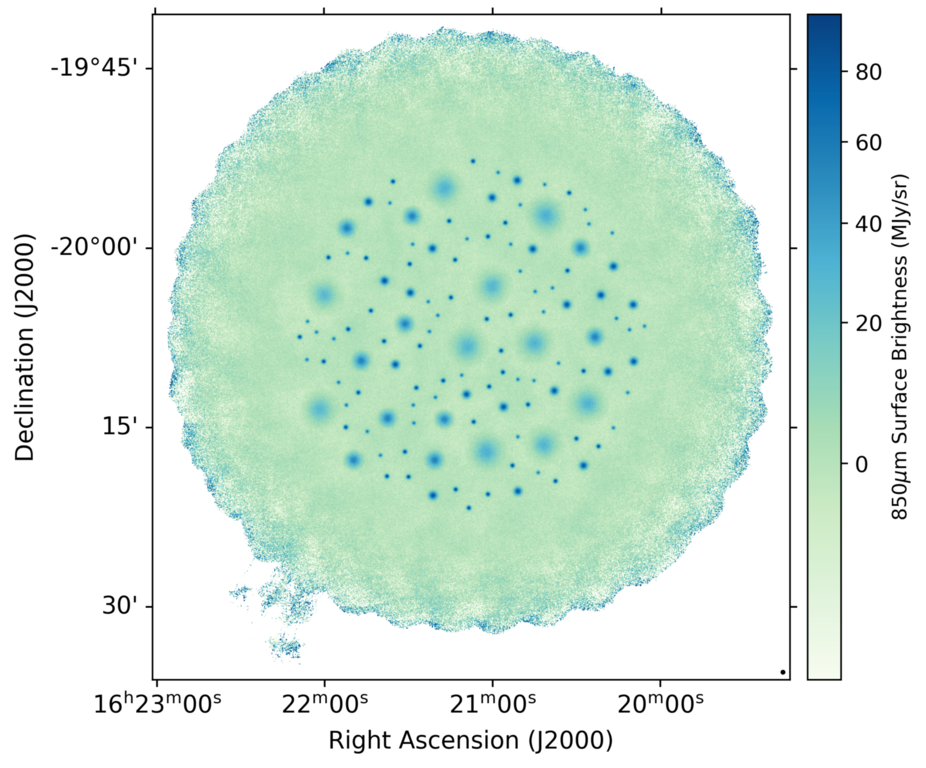}
        \includegraphics[width=0.49\textwidth]{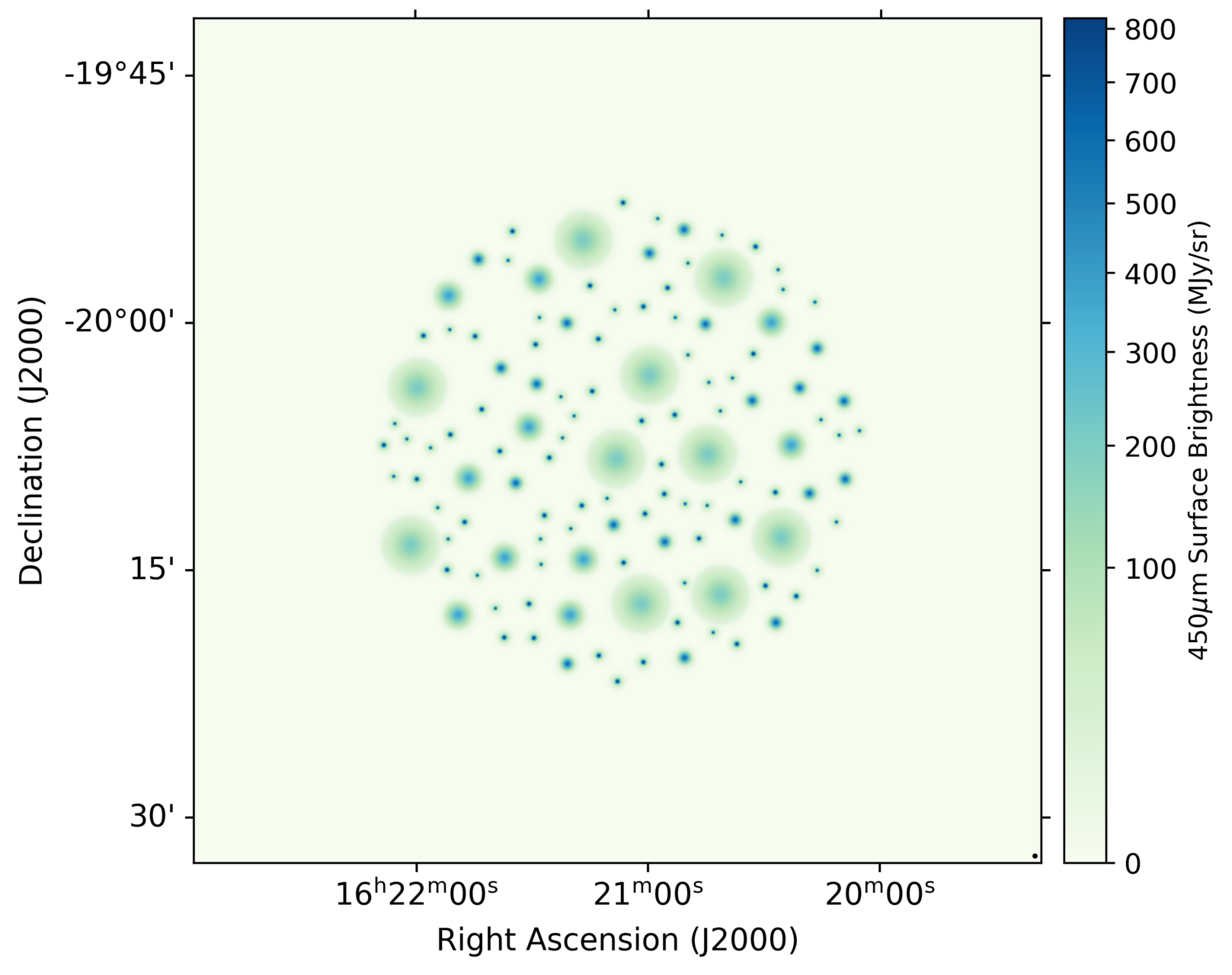}
	\includegraphics[width=0.49\textwidth]{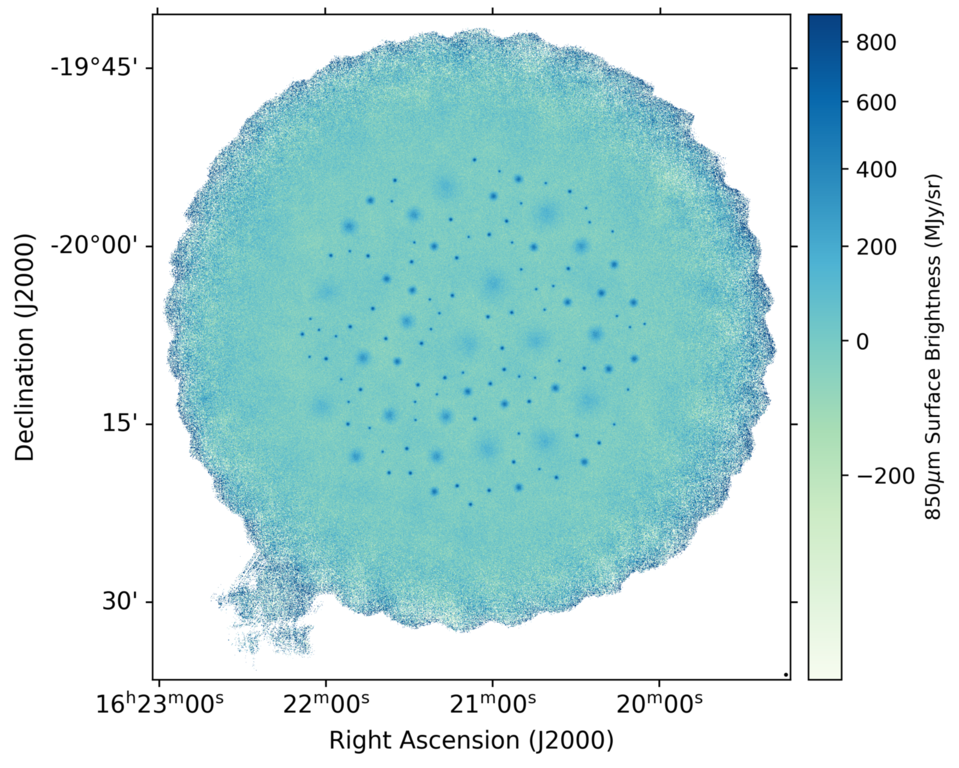}
        \caption{BEC spheres in the mass range 0.1--2.0\,M$_{\odot}$, placed at a distance of 150\,pc.  \textit{Top left:} 850\,\um\ model sources.  \textit{Top right:} 850\,\um\ model sources, after having been inserted into the Oph/Sco N6 field and processed through the SCUBA-2 pipeline.  \textit{Bottom left:} 450\,\um\ model sources.  \textit{Bottom right:} 450\,\um\ model sources, after processing through the SCUBA-2 pipeline.}
        \label{fig:complete_150pc_hi}
\end{figure*}


\begin{figure*}
	\centering
	\includegraphics[width=0.49\textwidth]{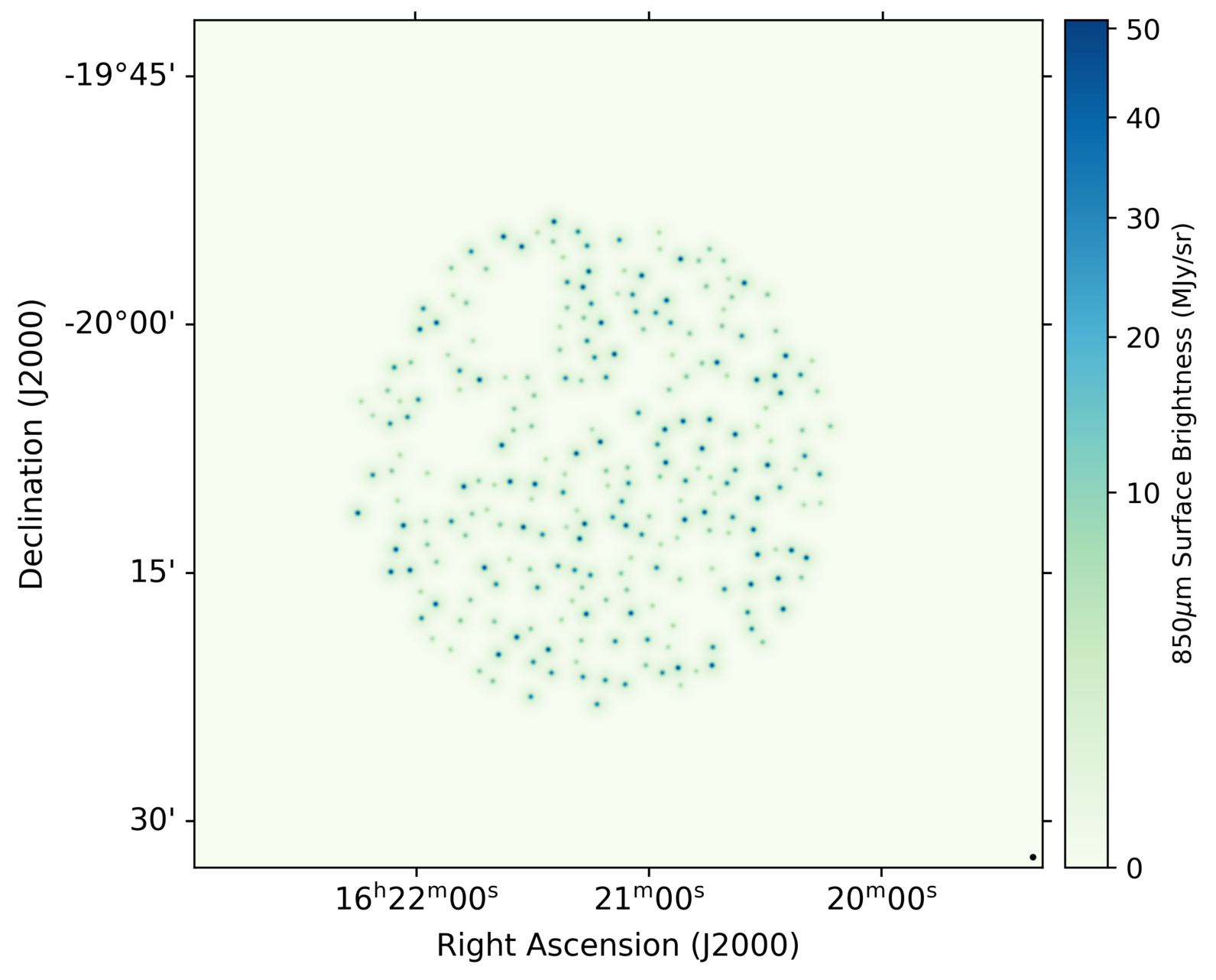}
	\hfill
         \includegraphics[width=0.49\textwidth]{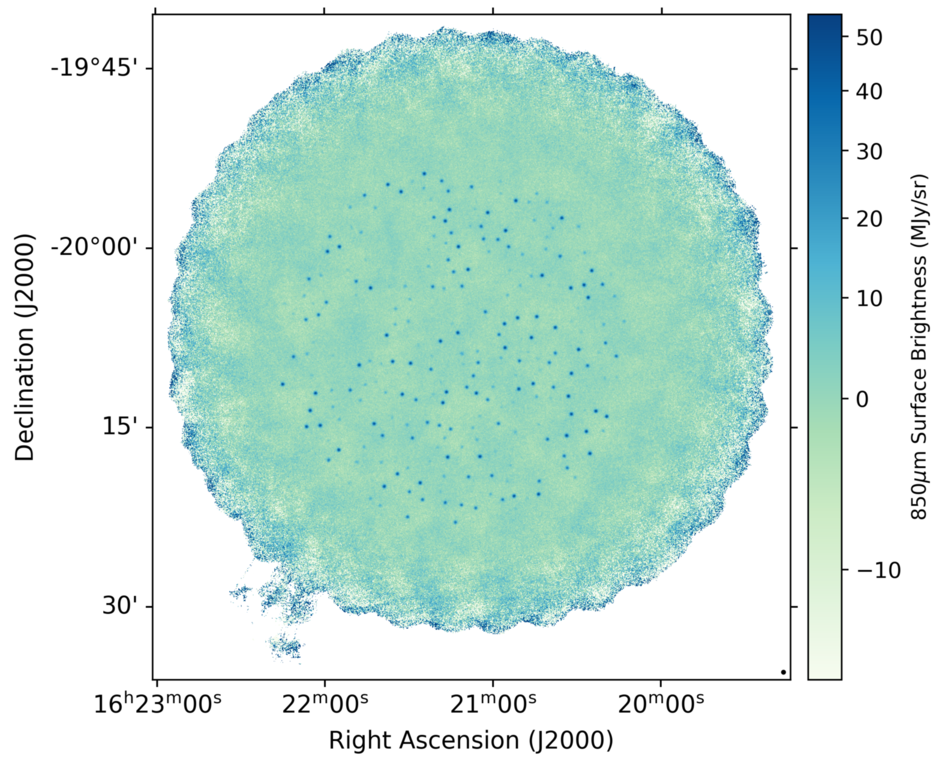} 
         \includegraphics[width=0.49\textwidth]{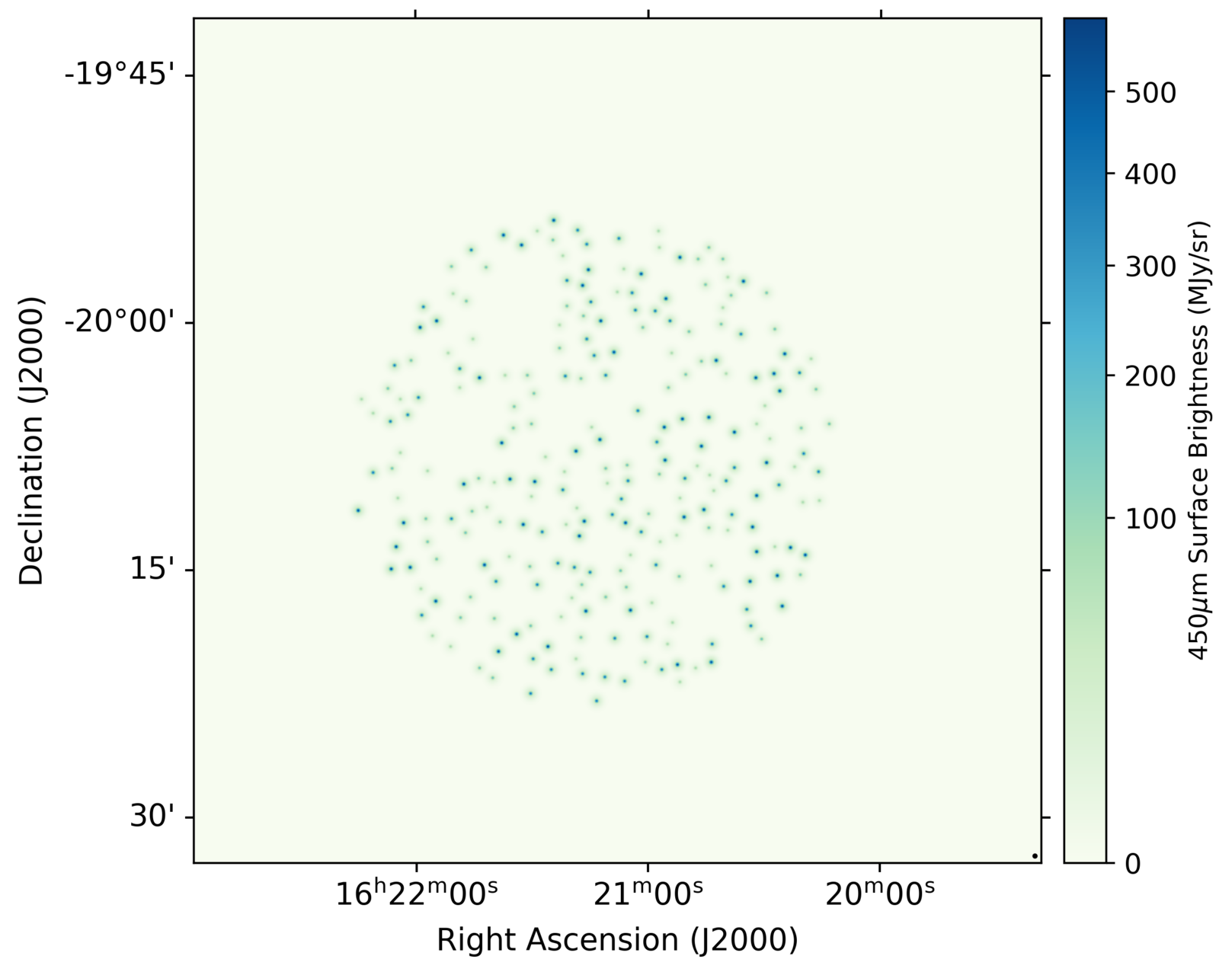}
	\hfill
      	\includegraphics[width=0.49\textwidth]{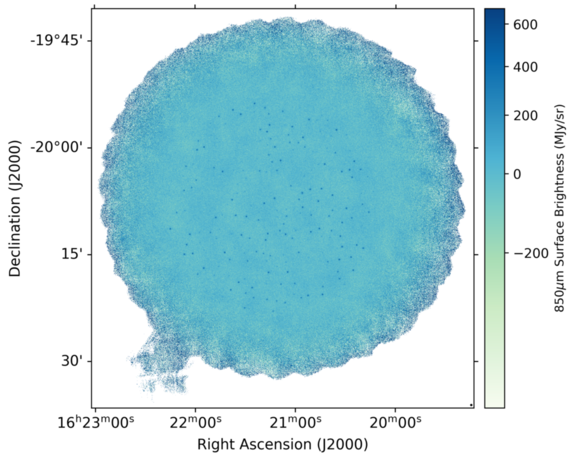}
        \caption{BEC spheres in the mass range 0.01--0.075\,M$_{\odot}$, placed at a distance of 150\,pc.  Panels as in Figure~\ref{fig:complete_150pc_hi}.}
        \label{fig:complete_150pc_lo}
\end{figure*}


\begin{figure*}
    \centering
    \includegraphics[width=0.49\textwidth]{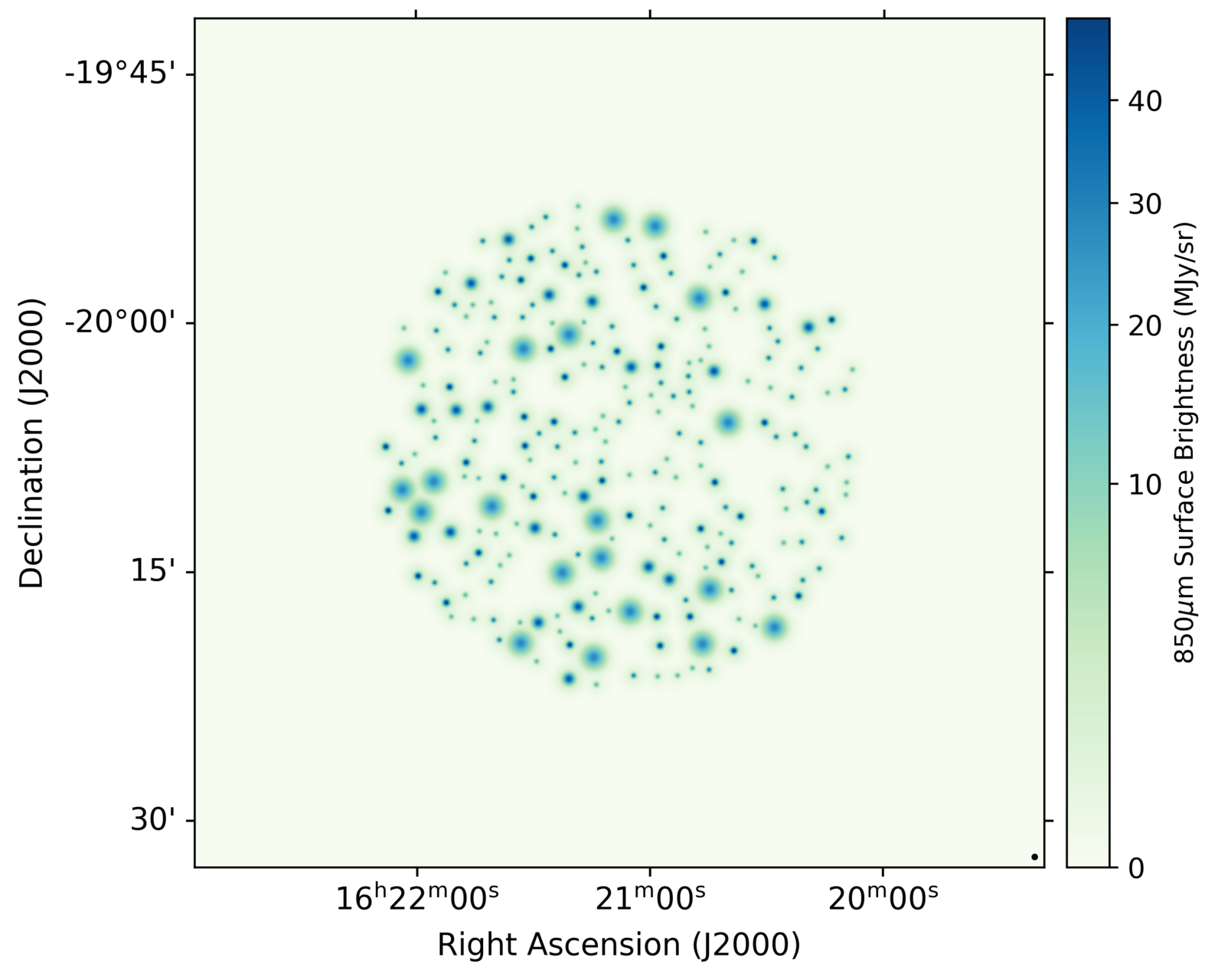}
	\hfill
	\includegraphics[width=0.49\textwidth]{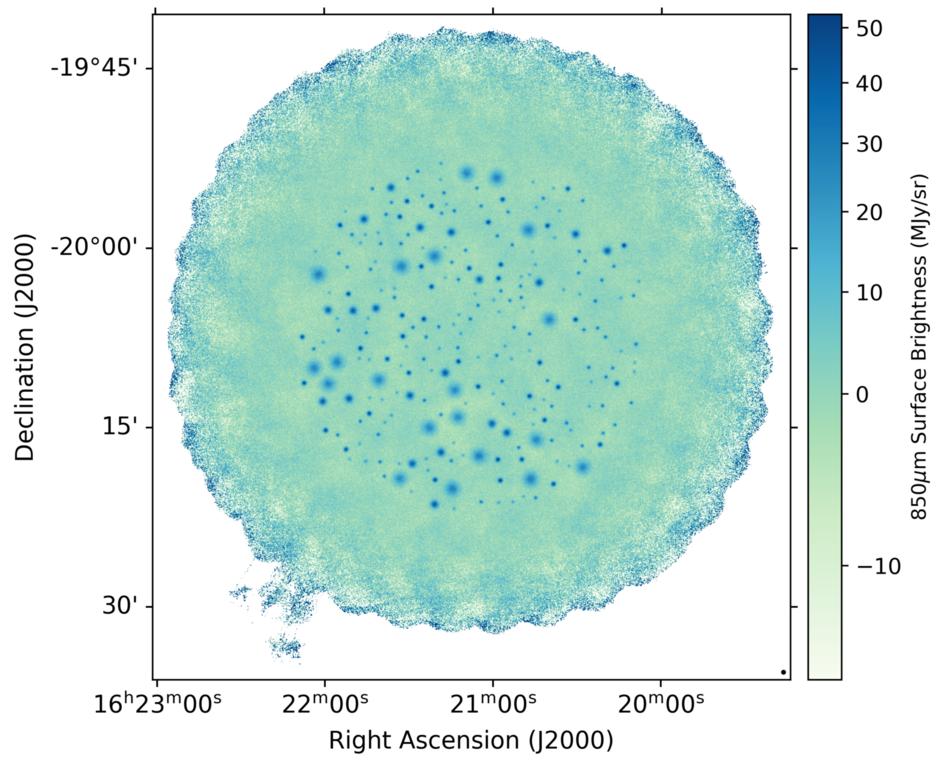}
         \includegraphics[width=0.49\textwidth]{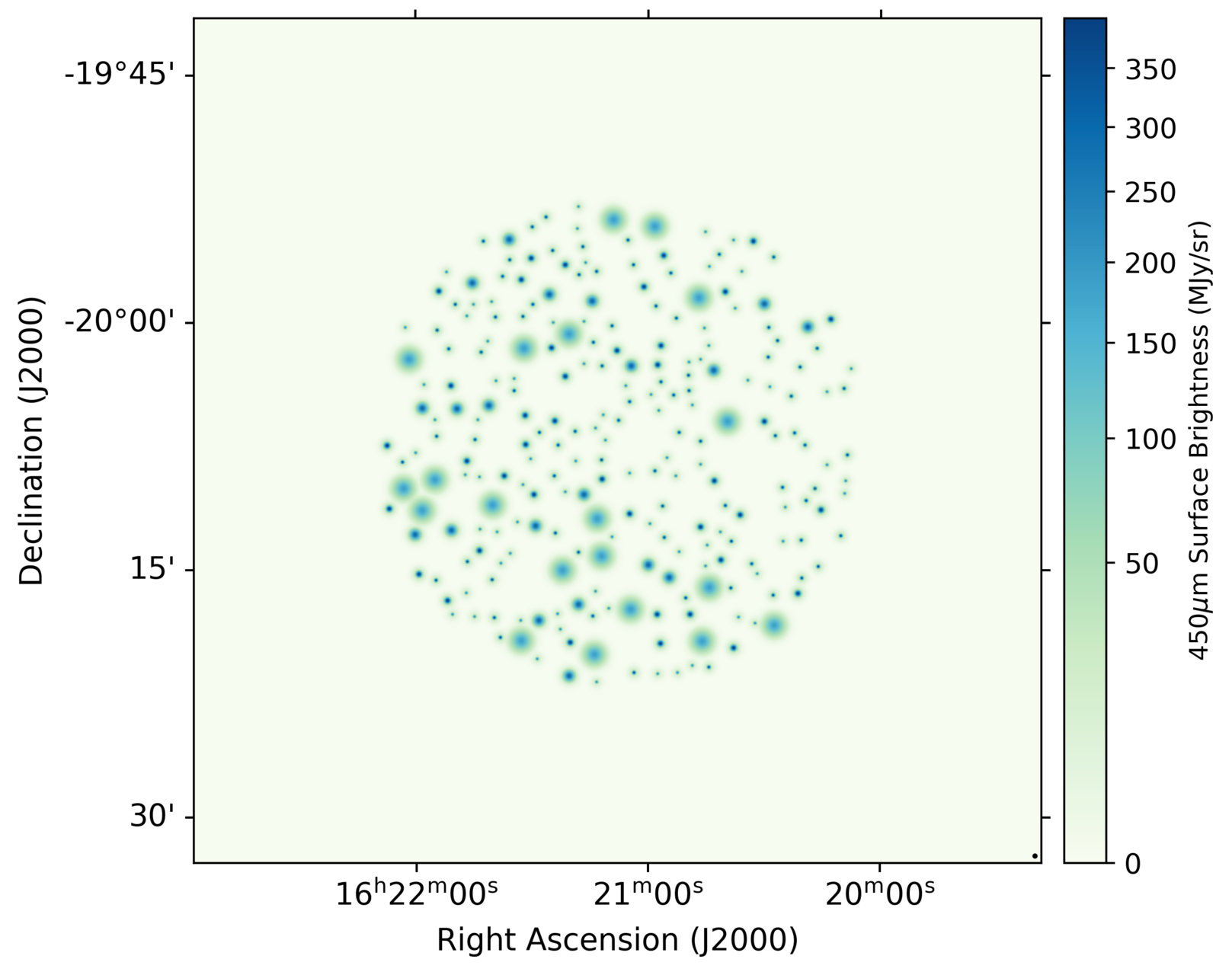}
        	\hfill
        \includegraphics[width=0.49\textwidth]{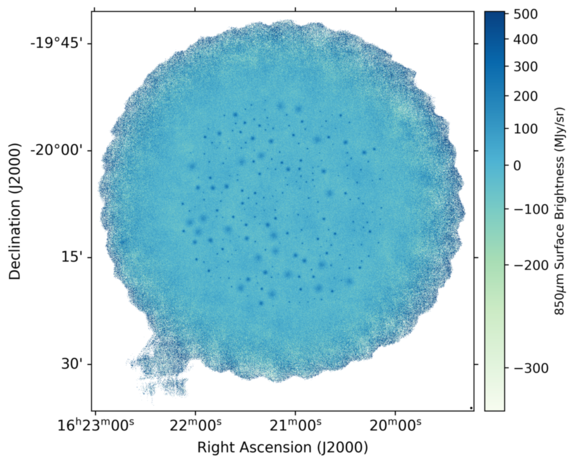}
        \caption{BEC spheres in the mass range 0.1--2.0\,M$_{\odot}$, placed at a distance of 300\,pc.  Panels as in Figure~\ref{fig:complete_150pc_hi}.}
        \label{fig:complete_300pc_hi}
\end{figure*}


\begin{figure*}
    \centering
        \centering 
        \includegraphics[width=0.49\textwidth]{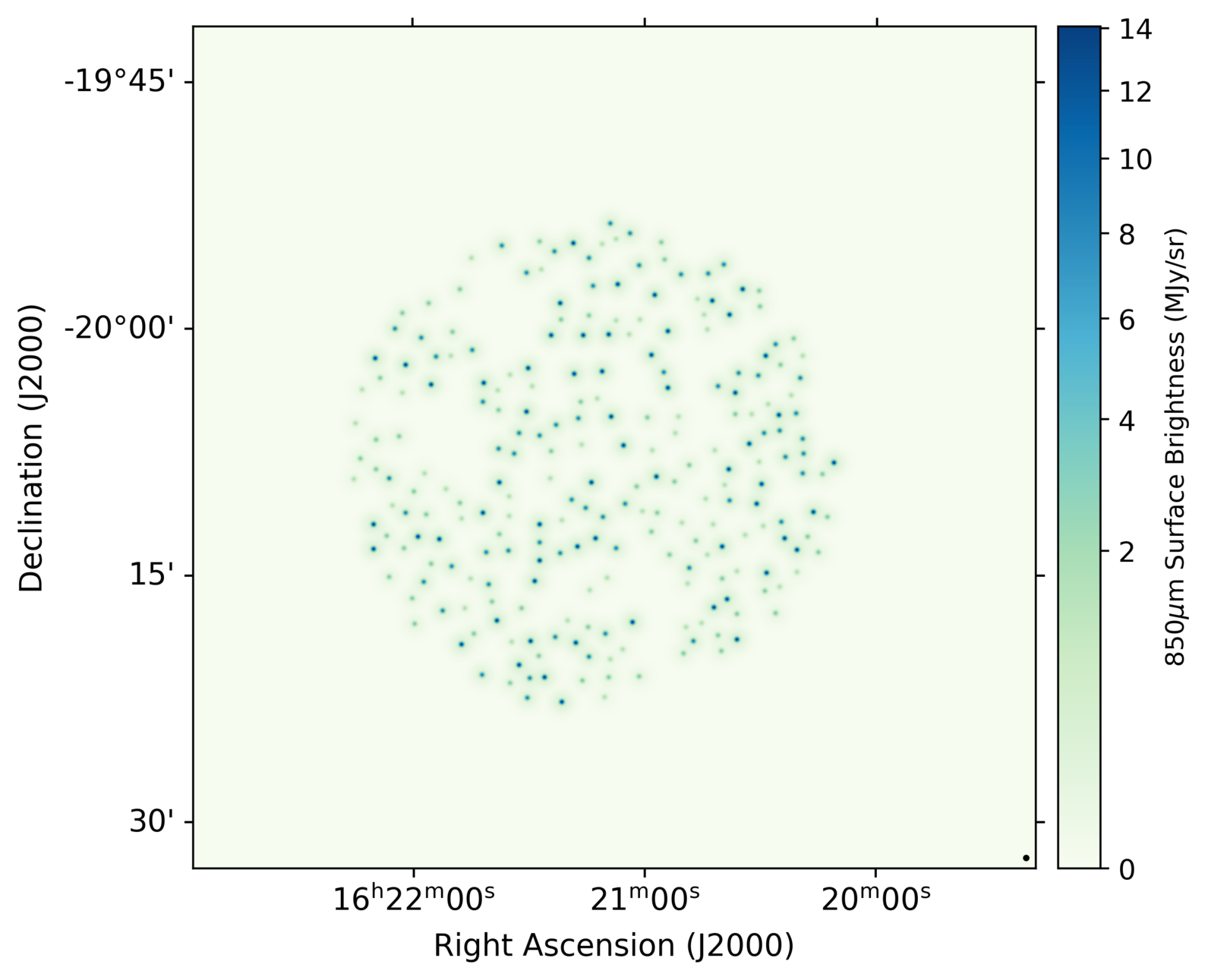}
        \hfill
            \includegraphics[width=0.49\textwidth]{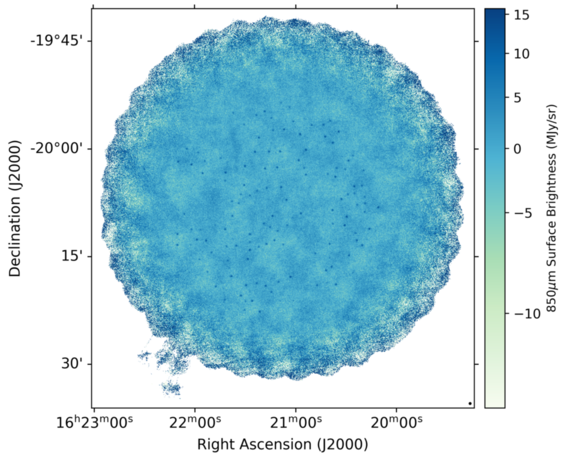}
            \includegraphics[width=0.49\textwidth]{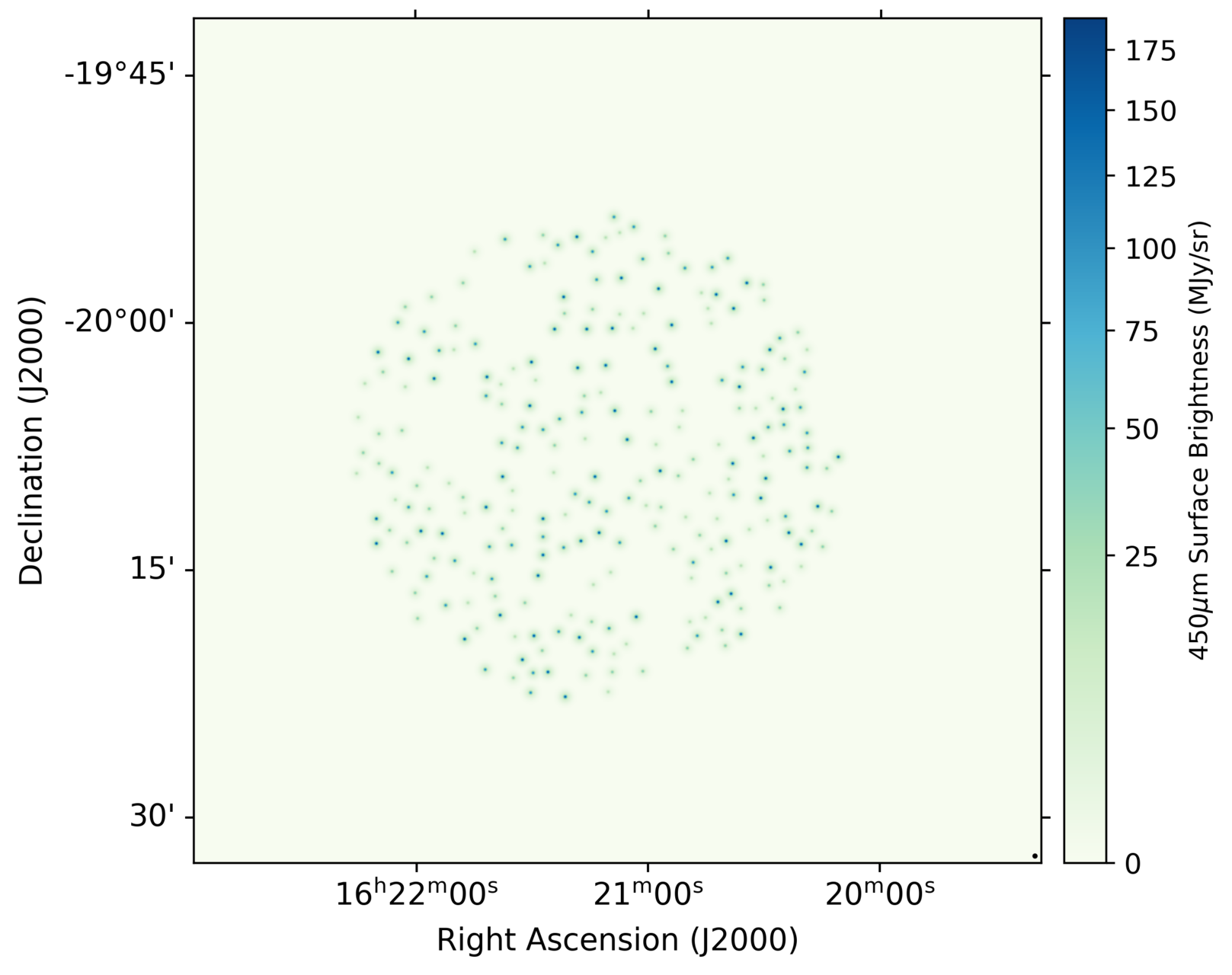}
            \includegraphics[width=0.49\textwidth]{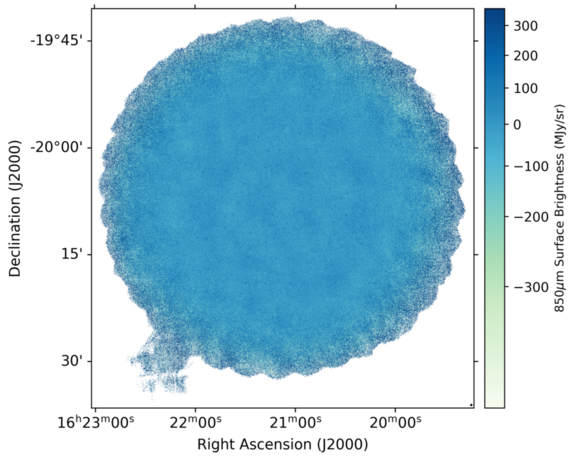}
        \caption{BEC spheres in the mass range 0.01--0.075\,M$_{\odot}$, placed at a distance of 300\,pc.  Panels as in Figure~\ref{fig:complete_150pc_hi}.}
        \label{fig:complete_300pc_lo}
\end{figure*}


\begin{figure*}
    \centering
        \includegraphics[width=0.49\textwidth]{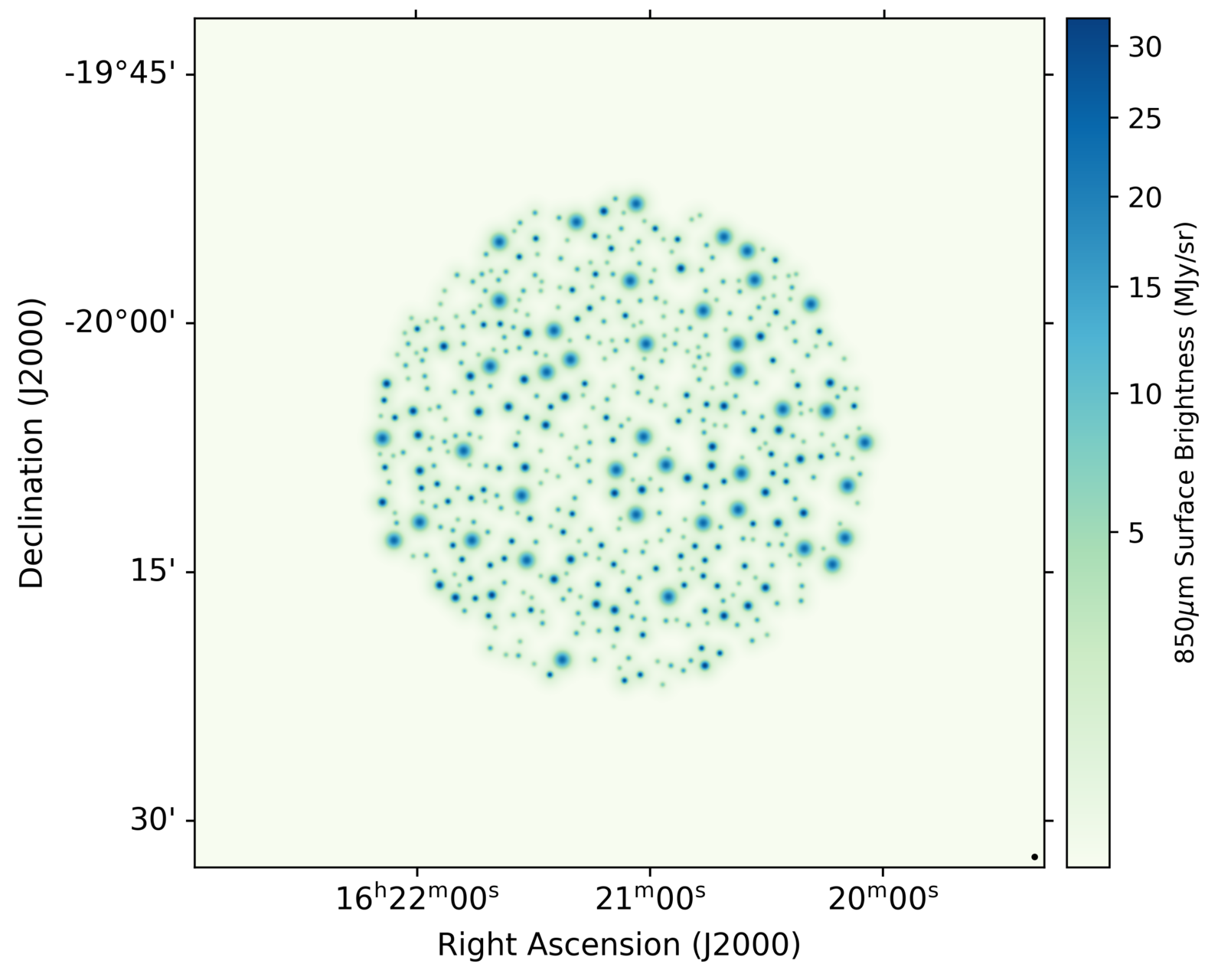}
        \hfill
            \includegraphics[width=0.49\textwidth]{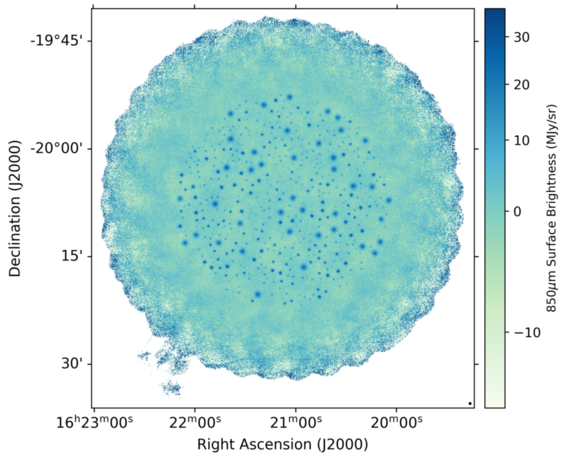}
            \includegraphics[width=0.49\textwidth]{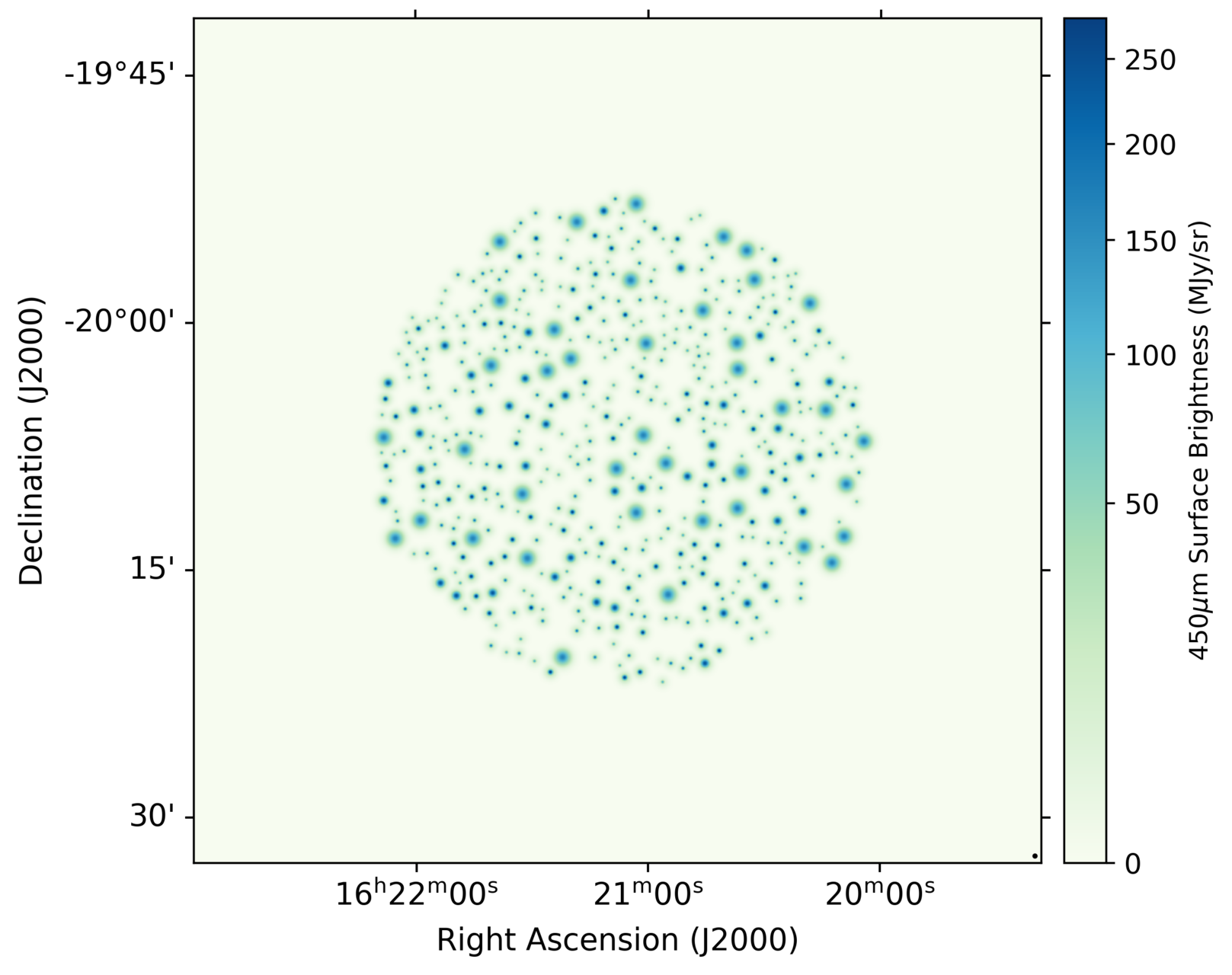}
        \hfill
            \includegraphics[width=0.49\textwidth]{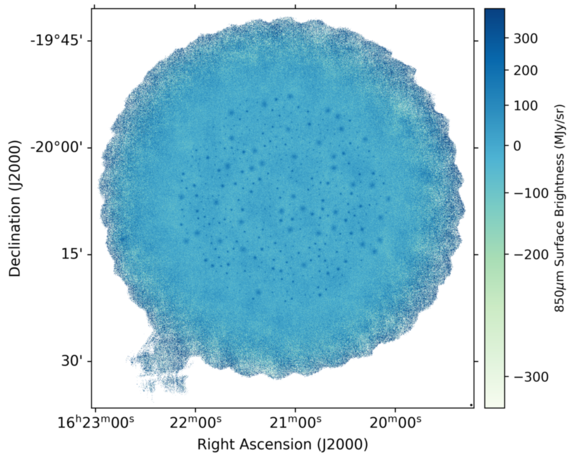}
        \caption{BEC spheres in the mass range 0.1--2.0\,M$_{\odot}$, placed at a distance of 450\,pc.  Panels as in Figure~\ref{fig:complete_150pc_hi}.}
        \label{fig:complete_450pc_hi}
\end{figure*}

One question that must be addressed with all ground-based submillimetre observations is the impact of the insensitivity to large-scale emission structures on the accuracy of measurements of smaller-scale emission structures.  A first attempt to quantify this for the JCMT GBS was made in \citet{kirk2018}, but this was done under artificial and idealized conditions: artificial Gaussian emission sources of varying peak brightness and size were added to a noise-only field, and were then examined using knowledge of their expected properties.  The overall conclusion was that compact sources with Gaussian width $\le$ 30\arcsec\ and peak brightness at least five times the local noise level were reliably detected, with measured properties lying within 15\% of their true input values.

Here, we add several important layers of realism to our completeness testing.  First, we model sources as critical Bonnor--Ebert (BEC) spheres, a model which is often found to describe dense cores reasonably well \citep[e.g.][]{alves2001}.  Second, instead of guiding a source-finding algorithm to the input cores, we employ one of the most commonly-used source finders, {\it getsources} \citep{menshchikov2012}, the same algorithm used to produce the GBS core catalogue itself.  We also create both 850 \um\ and 450 \um\ model emission, and provide both wavelengths as input to \textit{getsources}.  This work is the first time that JCMT GBS completeness testing has included model 450\,$\upmu$m data.

These additional layers of testing are essential to interpret our observed core catalogue correctly.  For example, interpreting the observed core mass function requires knowledge of the source completeness as a function of mass, as well as by the fraction of mass recovered relative to the true core mass.  \citet{kirk2018} provided results in the best-case scenario, while here we employ a more realistic approach which includes the effects of the source identification algorithm.  We note that these current tests still neglect the influence of source crowding and variable backgrounds which could be additional important factors in some of our observed regions.

For our completeness testing, we inserted model Bonnor--Ebert spheres into the Oph/Sco N6 field, an observed GBS field which appears to be devoid of real source emission.  This field provides a good worst-case scenario, in that it was one of many areas observed in JCMT Weather Band 2 ($0.05 < \tau_{225\,{\rm GHz}} < 0.08$; \citealt{dempsey2013}), and so has poorer 450 \um\ noise properties than do fields observed in JCMT Weather Band 1 ($\tau_{225\,{\rm GHz}} < 0.05$).  We might expect \textit{getsources} to perform slightly better in regions observed in Band 1, since the 450 \um\ data would have better SNR.  However, Band 1 weather was used to observe regions known to contain many bright sources, and so we do not have an equivalent `noise-only' Band 1 pointing to use for completeness testing.

The following subsections describe our artificial source testing procedure in more detail.

\subsection{General Setup}
Much of the basic set-up mimics what was used in \citet{kirk2018}, and the reader is referred there for further details.  We inserted model Bonnor--Ebert spheres into the Oph/Sco N6 field, as discussed above.  There are two areas in Oph/Sco N6 where potential ($<3\sigma)$ emission is present, which we excluded from the area where artificial sources could be placed.  We additionally excluded the outer 3\arcmin\ edge of the map for placement of artificial sources as the noise is significantly higher there.  These areas are shown in Figure~2 of \citet{kirk2018}.  Within the remaining map area, for each set of Bonnor--Ebert sphere model parameters, we placed model sources as described in Section~\ref{sec:be_placement}.  The artificial sources were added directly to the time stream of the raw observations, and the data then processed using the procedure used to reduce the maps used in this work --  i.e. that which the GBS has found best recovers extended emission (DR2).  Each of the six independent observations of the Oph/Sco N6 field were reduced independently using the DR2 automask procedure.  The reduced maps were then mosaicked and masks (identifying areas of likely real emission) created.  A second round of reduction was then run using the masks, again following the DR2 external mask procedure, and the resulting images were mosaicked together once more.  We did not introduce any telescope pointing errors between the six observations, hence the additional post-processing corrections employed for DR3 were not applicable.

\subsection{BEC Sphere Models}

\subsubsection{Making critical Bonnor--Ebert sphere models}

We generated the flux density distribution of a BEC sphere at 850\,$\upmu$m and 450\,$\upmu$m for each of the range of masses $M$ and distances $D$ which we wished to test.  In each case we assumed $T=15$ K, $\beta = 1.8$, $\kappa_0 = 0.01$\,m$^2$kg$^{-1}$ at 1 THz.

The density distribution of a BEC sphere of central density $\rho_c$ is given by \citep{ebert1955,bonnor1956}
\begin{equation}
    \rho = \rho_c\exp(-\psi(\xi)),
\end{equation}
where $\xi$ is a function of core radius $r$, such that
\begin{equation}
\xi = r \left(\frac{4\pi G\rho_c}{c_s^2}\right)^{\frac{1}{2}},
\end{equation}
and $\psi(\xi)$ is implicitly defined via the Lane-Emden equation,
\begin{equation}
    \xi^{-2}\frac{\rm d}{{\rm d}\xi}\left(\xi^2\,\frac{{\rm d}\psi}{{\rm d}\xi}\right) = \exp(-\psi)                                           
\end{equation}
for boundary conditions $\psi(\xi=0)=0$ and ${\rm d}\psi/{\rm d}\xi|_{\xi=0} = 0$.
The mass of the sphere within a dimensionless radius $\xi_0$ is given by
\begin{equation}
    M = \frac{c_s^3}{(4\pi G^3 \rho_c)^\frac{1}{2}}\int_{0}^{\xi_0}\exp(-\psi(\xi))\xi^2{\rm d}\xi
\end{equation}

For a source of given mass $M$, we numerically solved these equations for central radius $\rho_c$ and a dimensionless edge radius $\xi_0$, assuming a centre-to-edge density contrast of $\rho/\rho_c = 14.1$ -- that of a BEC sphere.  We then converted $\xi_0$ into an edge radius $r_0 = \xi_0(4\pi G\rho_c/c_s^2)^{-0.5}$, and made a lookup table for $\rho$ as a function of radius $r$.

We next determined the flux density distribution of the BEC sphere on the 850um (3$^{\prime\prime}$) pixel grid.  For a given temperature and constant dust properties, the flux density in a given pixel is proportional to the mass in that pixel, using $F_{\nu} = M\times(\kappa_\nu B_\nu(T)/D^2)$ \citep{hildebrand1983}.

We first calculated the angular size of $r_0$ at our chosen distance.  If this angular size was less than 3$^{\prime\prime}$, we simply placed all of the mass in the central pixel.

Otherwise, we found the mass in each pixel as follows.
We first calculated the distance of all four corners of the pixel from the centre of the map (the impact parameter $b$ of each corner).  If all corners of pixel had $b> r_0$, we set the mass in that pixel to zero.

Otherwise, we picked a random position inside the pixel, and calculated the impact parameter $b$ of that position.  If $b > r_0$, we set the mass at that position to zero.  Otherwise, we made an array of distances $x$ along the line of sight from 0 to the edge of the sphere, where $x$ is defined relative to the plane of the centre of the sphere, and calculated radii for each of these distances, such that $r = \sqrt{x^2 + b^2}$.  We then created an array of density values $\rho(x)$ by interpolating our $\rho(r)$ lookup table for our $r(x,b)$ values.  We integrated under this $\rho(x,b)$ curve using the trapezium rule, and then doubled this value to account for both hemispheres of the sphere, giving the mass along this line of sight.  We repeated this process 1000 time, and took the mean of these 1000 values to find the average mass in the pixel. Pixels with equal central impact parameters were filled by symmetry.

To check the accuracy of our calculation, we summed over all of the pixels in the array.  If the returned mass differed from the input BEC sphere mass by $> 1$\%, we repeated the calculation until our required tolerance was achieved.

To create the 850\,$\upmu$m flux density map, we multiplied the mass distribution by $\kappa_\nu B_\nu(T)/D^2$ where $\nu = \nu(850\upmu{\rm m})$.  We then smoothed the map with the two-component JCMT 850um beam model \citep{dempsey2013}.  This gave us the 850\,$\upmu$m flux density map of the BEC sphere at the resolution of the JCMT.

To create the 450um map, we reprojected the 850um mass distribution to the 450\,$\upmu$m pixel grid.  Since both the 850\,$\upmu$m (3$^{\prime\prime}$) and 450\,$\upmu$m (2$^{\prime\prime}$ pixel grids are significantly smaller than the 450\,$\upmu$m primary beam size, this approach does not cause us any loss of 450\,$\upmu$m resolution, while significantly shortening the time required for mass calculations and, through use of exact reprojection routines, the certainty that both wavelengths were centred on precisely the same position.  We then multiplied the mass distribution by $\kappa_\nu B_\nu(T)/D^2$ where $\nu = \nu(450\upmu{\rm m})$, and smoothed the map with the two-component JCMT 450um beam model \citep{dempsey2013}, giving us the 450\,$\upmu$m flux density map of the BEC sphere at the resolution of the JCMT.

\subsubsection{Placing BEC spheres on map}
\label{sec:be_placement}

\begin{table*}
    \setlength\tabcolsep{3pt}    
    \caption{BE sphere placement properties}
    \centering
    \begin{tabular}{cccccccccc}
    \hline
        \multicolumn{10}{c}{150\,pc} \\
    \hline
     & \multicolumn{9}{c}{Mass (M$_{\odot}$)} \\
     \hline
        &  0.01 & 0.02 & 0.05 & 0.075 & 0.1 & 0.2 & 0.5 & 1.0 & 2.0 \\
        \hline
        Target to place & 60 & 60 & 60 & 60 & 40 & 40 & 20 & 10 & 10 \\
        Maximum tries & 16000 & 8000 & 2000 & 2000 & 16000 & 16000 & 8000 & 2000 & 2000 \\ 
        Exclusion multiplier & 8 & 8 & 8 & 8 & 4 & 6 & 8 & 10 & 12 \\
        Sources placed  & 60 & 60 & 60 & 60 & 18 & 23 & 20 & 10 & 10 \\
        Sources recovered  & 0 & 51 & 60 & 60 & 18 & 23 & 20 & 10 & 10 \\
        Fraction recovered & 0.00 & 0.85 & 1.00 & 1.00 & 1.00 & 1.00 & 1.00 & 1.00 & 1.00 \\
        $\langle M_{\rm recovered}\rangle$ (M$_{\odot}$) & -- &  0.014$\pm$0.002 &  0.035$\pm$0.002 &  0.053$\pm$0.002 &  0.069$\pm$0.002 &  0.138$\pm$0.006 &  0.399$\pm$0.008 &  0.86$\pm$0.06 &  1.5$\pm$0.2  \\ 
        $\langle M_{\rm recovered}/M\rangle$ & -- &  0.7$\pm$0.1 &  0.70$\pm$0.04 &  0.71$\pm$0.03 &  0.69$\pm$0.02 &  0.69$\pm$0.03 &  0.80$\pm$0.02 &  0.86$\pm$0.06 &  0.74$\pm$0.09 \\ 
    \hline
        \multicolumn{10}{c}{300\,pc} \\
    \hline
     & \multicolumn{9}{c}{Mass (M$_{\odot}$)} \\
     \hline
        &  0.01 & 0.02 & 0.05 & 0.075 & 0.1 & 0.2 & 0.5 & 1.0 & 2.0 \\
        \hline
        Target to place & 60 & 60 & 60 & 60 & 80 & 80 & 40 & 20 & 20 \\
        Maximum tries & 16000 & 8000 & 2000 & 2000 & 16000 & 16000 & 8000 & 2000 & 2000 \\
        Exclusion multiplier & 8 & 8 & 8 & 8 & 3 & 9 & 12 & 15 & 18 \\
        Sources placed  & 60 & 60 & 60 & 60 & 60 & 75 & 40 & 20 & 20 \\
        Sources recovered  & 0 & 0 & 1 & 52 & 58 & 75 & 40 & 20 & 20 \\
        Fraction recovered & 0.00 & 0.00 & 0.02 & 0.87 & 0.97 & 1.00 & 1.00 & 1.00 & 1.00 \\
        $\langle M_{\rm recovered}\rangle$ (M$_{\odot}$) & -- &  -- &  0.042 &  0.06$\pm$0.02 &  0.069$\pm$0.008 &  0.14$\pm$0.01 &  0.34$\pm$0.02 &  0.79$\pm$0.04 &  1.7$\pm$0.1 \\ 
        $\langle M_{\rm recovered}/M\rangle$ & -- &  -- &  0.84 &  0.8$\pm$0.2 &  0.69$\pm$0.08 &  0.69$\pm$0.05 &  0.68$\pm$0.03 &  0.79$\pm$0.04 &  0.83$\pm$0.07 \\ 

    \hline
            \multicolumn{10}{c}{450\,pc} \\
    \hline
     & \multicolumn{9}{c}{Mass (M$_{\odot}$)} \\
     \hline
        &  0.01 & 0.02 & 0.05 & 0.075 & 0.1 & 0.2 & 0.5 & 1.0 & 2.0 \\
        \hline
        Target to place & -- & -- & -- & -- & 160 & 160 & 80 & 40 & 40 \\
        Maximum tries & -- & -- & -- & -- & 16000 & 16000 & 8000 & 2000 & 2000 \\ 
        Exclusion multiplier & -- & -- & -- & -- & 8 & 12 & 16 & 20 & 24 \\
        Sources placed  & -- &  -- & -- & -- & 62 & 27 & 22 & 29 & 40 \\
        Sources recovered  & -- &  -- & -- &  -- &  0 & 25 & 22 & 29 & 40 \\
        Fraction recovered & -- &  -- & -- &  -- & 0.00 & 0.93 & 1.00 & 1.00 & 1.00 \\
        $\langle M_{\rm recovered}\rangle$ (M$_{\odot}$) & -- &  -- &  -- &  -- & -- &  0.13$\pm$0.02 &  0.33$\pm$0.02 &  0.72$\pm$0.06 &  1.6$\pm$0.2 \\
        $\langle M_{\rm recovered}/M\rangle$ & -- &  -- &  -- &  -- & -- &  0.7$\pm$0.1 &  0.65$\pm$0.04 &  0.71$\pm$0.06 &  0.8$\pm$0.1\\
    \hline
    \end{tabular}
    \label{tab:be_placement_props}
\end{table*}

We began the process of placing cores by generating a suitable list of $x$ and $y$ coordinates.  For each field, we defined a target number of cores of each mass to place, and a maximum number of attempts to make to place sources of that mass.  We also wished to ensure that the sources placed did not overlap with one another significantly, and did not fall close to the edge of the map.  To this end, we defined an `exclusion radius', $r_{ex}$, for each mass of core, defined as the larger of $r_{0}$ and the 850\,$\upmu$m beam FWHM.  We further defined, for each mass and distance, an empirical factor $M_{ex}$ by which $r_{ex}$ should be multiplied, in order to prevent significant overlap of sources.

Until either the target number of sources or the maximum number of attempts was reached, the following process was performed:
\begin{enumerate}
    \item Randomly select a pair of $x$ and $y$ coordinates within the field, defined relative to the map centre.
    \item If $\sqrt{x^2 + y^2} < 900^{\prime\prime} - 3r_{ex}$, discard the position and retry.
    \item The $(x,y)$ coordinate pair was compared to the list of previously placed sources, of all masses.  If the distance between the source and any other placed source is less than the sum of the $M_{ex}\times r_{ex}$ values of the two sources, discard the position and retry.
    \item Otherwise, place a model source at these $(x,y)$ coordinates and record its mass.
\end{enumerate}
This process was then repeated for the next mass under consideration, progressing from highest to lowest mass.  The sources placed, along with the target number of sources, maximum number of tries, and $r_{ex}$ multipliers for each mass, are listed in Table~\ref{tab:be_placement_props}.  We then used the list of source positions and masses and create an equivalent 450\,$\upmu$m map.

This process was performed using the first of the six observations of the field.  This map was then exactly reprojected to the World Coordinate System (WCS) frames of the other five observations using the \textit{reproject} package in Python.  

Our fields of model BEC spheres are shown at 850\,\um\ and 450\,\um\ in the top and bottom left-hand panels of Figures~\ref{fig:complete_150pc_hi}--\ref{fig:complete_450pc_hi}, respectively.  

\subsubsection{Re-reducing SCUBA-2 field with fake sources added}

We produced SCUBA-2 ``observations" of our fields of model BEC spheres by adding the maps to our SCUBA-2 observations of the empty Oph/Sco N6 field.  To do so, we first `uncalibrated' the maps into pW using the SCUBA-2 flux conversion factors given by \citet{dempsey2013}.  The fields of fake sources were added to each of the observations of the Oph/Sco N6 field using the \textit{fakemap} parameter, which allows the user to provide an image of the sky that will produce corresponding additional astronomical signal in the SCUBA-2 bolometer time series, in \textit{makemap} \citep{chapin2013}, and the data reduction procedure for the field as described in Section~\ref{sec:dr} and by \citet{kirk2018} was repeated, including creation of a fixed mask.  This process is described in detail by \citet{sadavoy2013}.  We then calibrated and coadded the reduced maps to produce final maps of fake sources at 850\,$\upmu$m and 450\,$\upmu$m.  These final maps are shown at 850\,\um\ and 450\,\um\ in the top and bottom right-hand panels of Figures~\ref{fig:complete_150pc_hi}--\ref{fig:complete_450pc_hi}, respectively.  

\subsection{Fake Source Identification}

We next attempted to identify the model cores in our data.  We first ran \textit{getsources} on our fields of fake sources, with a set-up identical to that used on our real maps, as described in Section~\ref{sec:getsources}.  We then applied the selection criteria listed in Section~\ref{sec:source_selection} the source catalogue returned by \textit{getsources} in order to produce final catalogues of fake sources which have been treated identically to our real data throughout.

The number of sources recovered in each field and for each mass are listed in Table~\ref{tab:be_placement_props}.

\subsection{Results}

The results of our completeness testing are described in Section~\ref{sec:completeness} in the main body of the paper.

\clearpage


\section{Ancillary source classification plots}

\begin{figure}
    \centering
    \includegraphics[width=0.47\textwidth]{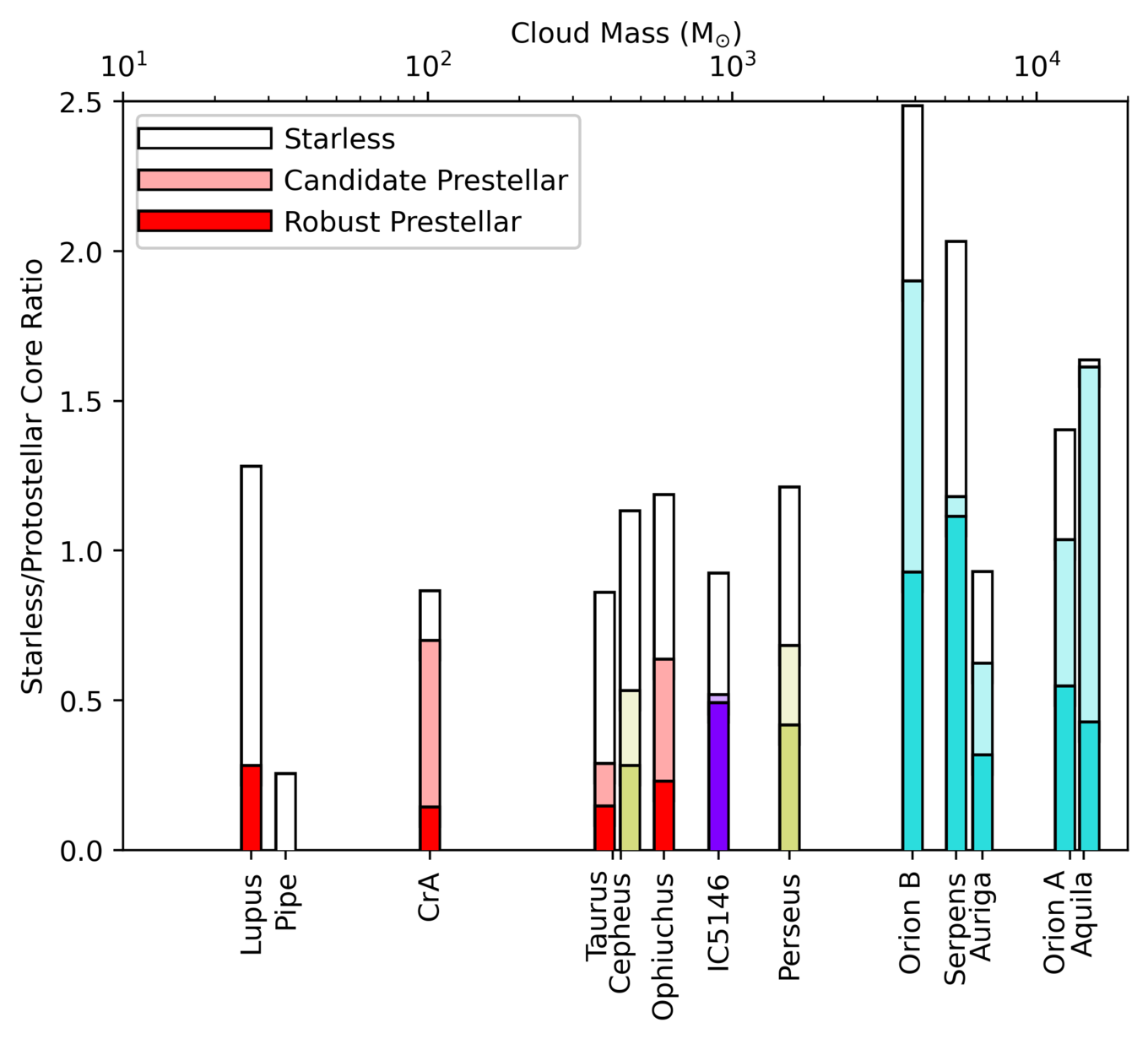}
    \caption{The ratio of starless cores, prestellar cores (candidate and robust prestellar cores), and robust prestellar cores to protostellar cores in each of the cloud complexes that we consider.  Note that the bars for Taurus, Cepheus, Orion A and Aquila are slightly offset from their true positions, to avoid overlap.  The tick marks associated with these bars show the exact cloud masses.}
    \label{fig:starless_proto}
\end{figure}

\begin{figure}
    \centering
    \includegraphics[width=0.47\textwidth]{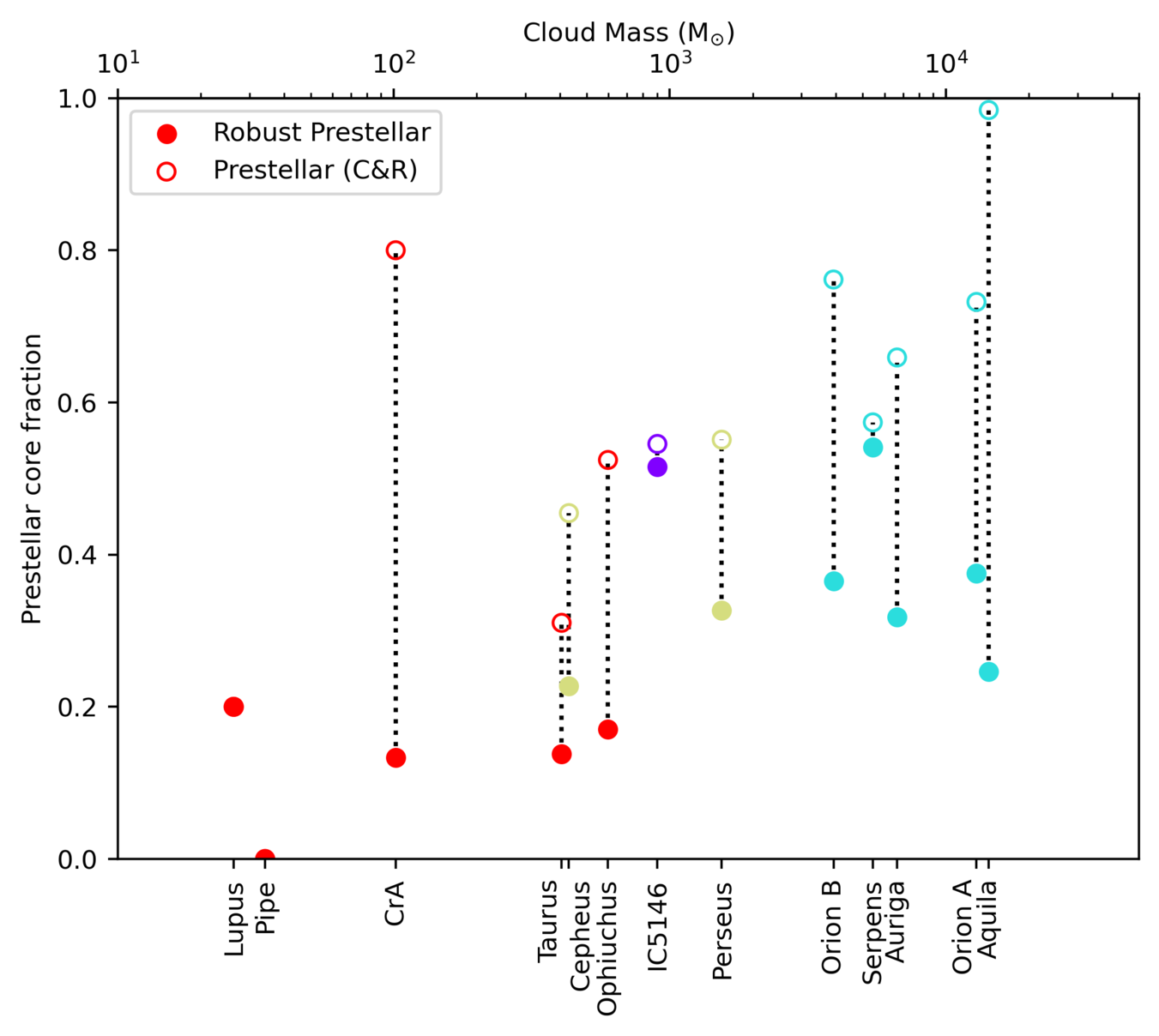}
    \caption{Fraction of starless cores which are prestellar cores (candidate and robust; open circles), or robust prestellar cores (filled circles) as a function of cloud complex mass.  Cloud complexes are colour-coded by their distance range.}
    \label{fig:be_fracs}
\end{figure}

The ratio of starless to protostellar cores is shown in Figure~\ref{fig:starless_proto}. The fraction of starless cores which are either robust or candidate prestellar cores is shown for each cloud complex in Figure~\ref{fig:be_fracs}.

\section{Ancillary mass and radius distributions}
\label{sec:appendix_mass_radius}

\begin{figure}
	\includegraphics[width=\columnwidth]{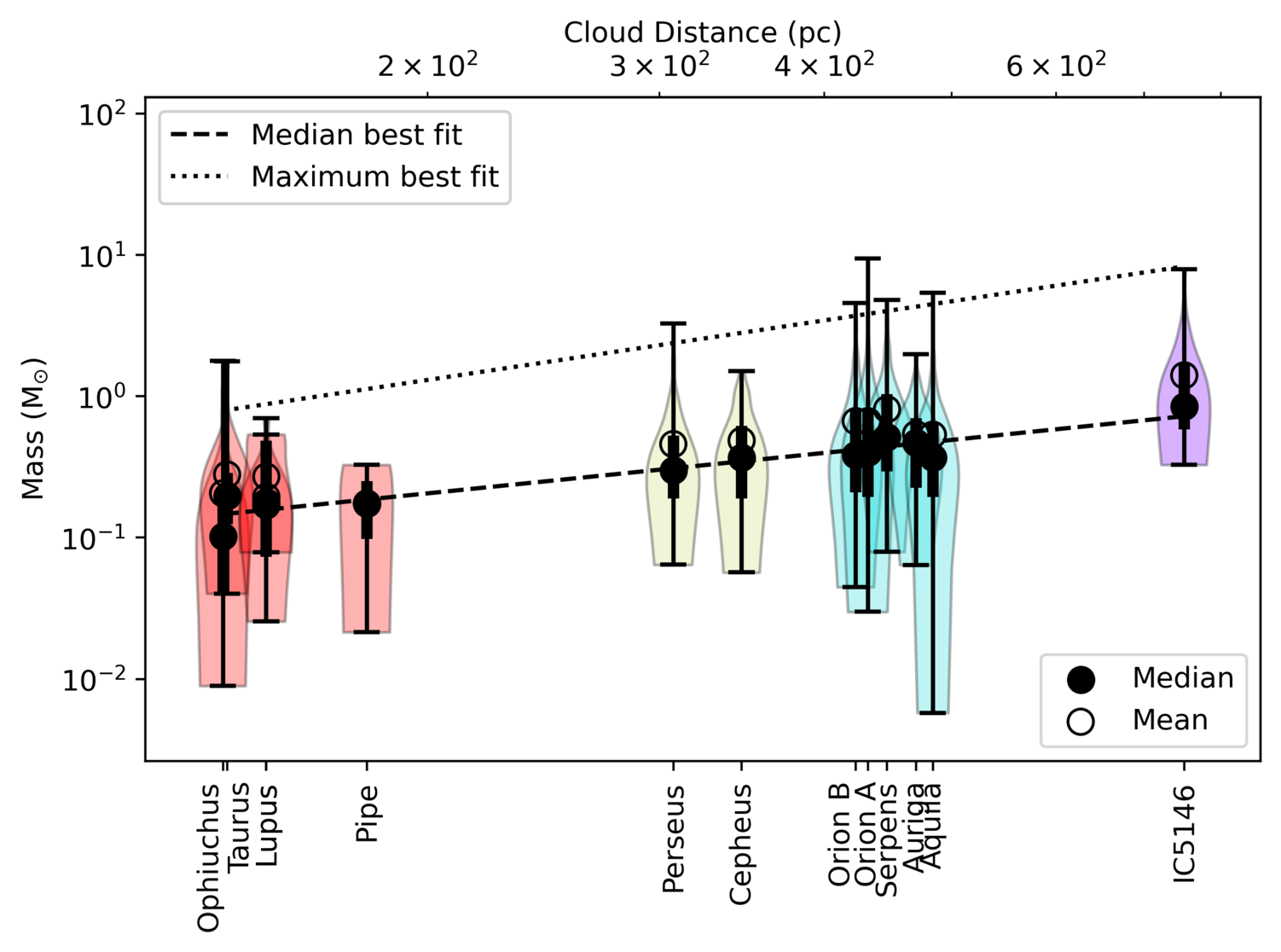}
	\includegraphics[width=\columnwidth]{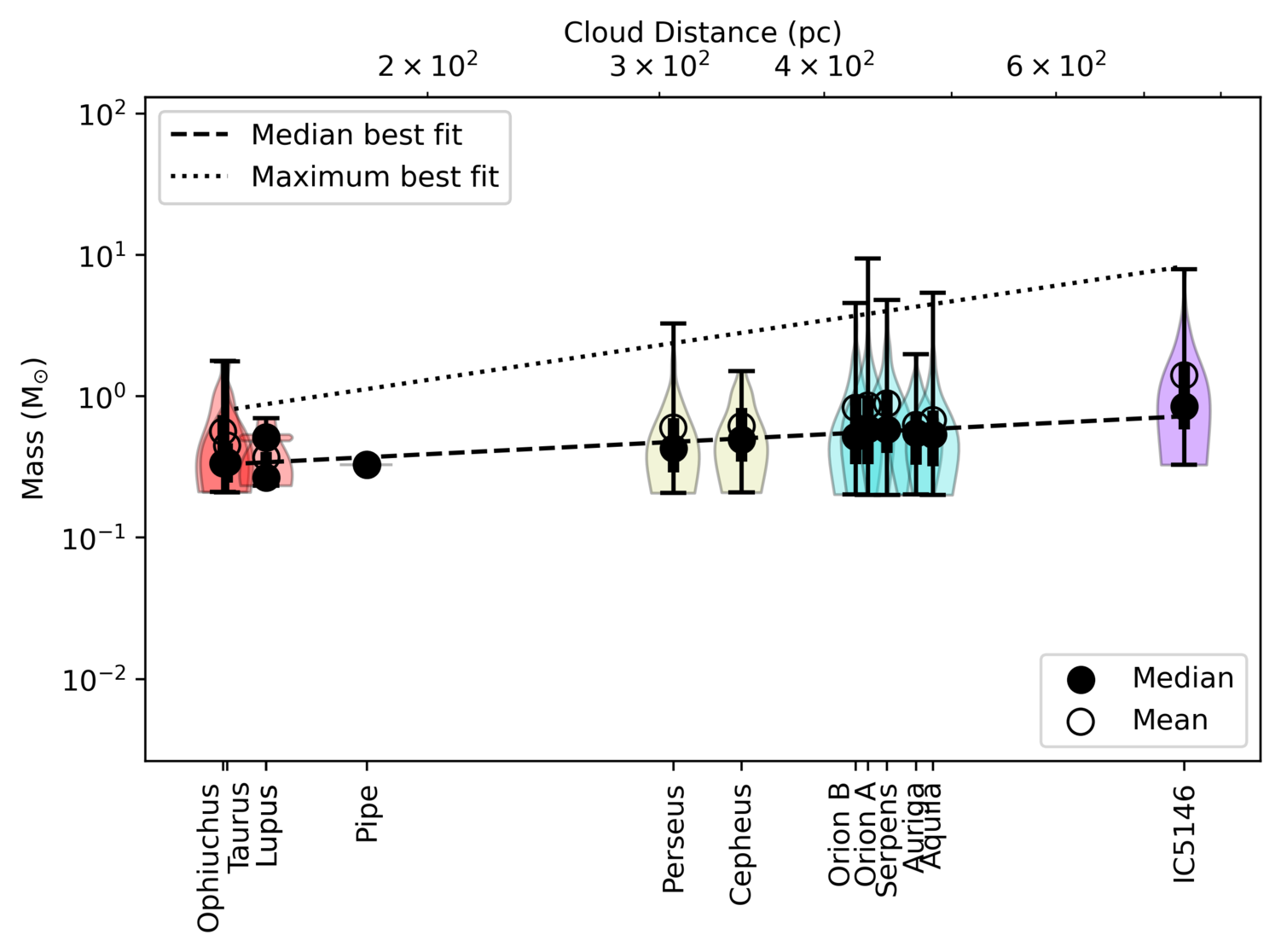}
    \caption{Mass distribution for each cloud complex, as a function of distance. Top: for all starless cores.  Bottom: for starless cores with masses above the 90\% completeness limit at 450\,pc of 0.2\,M$_{\odot}$.  Solid circles show median values; open circles show means.  Thick black lines show the interquartile ranges.  Dotted line shows the line of best fit to the maximum values in each cloud complex; dashed line shows the line of best fit to the median values.}
    \label{fig:mass_violin_dist}
\end{figure}

\begin{figure}
	\includegraphics[width=\columnwidth]{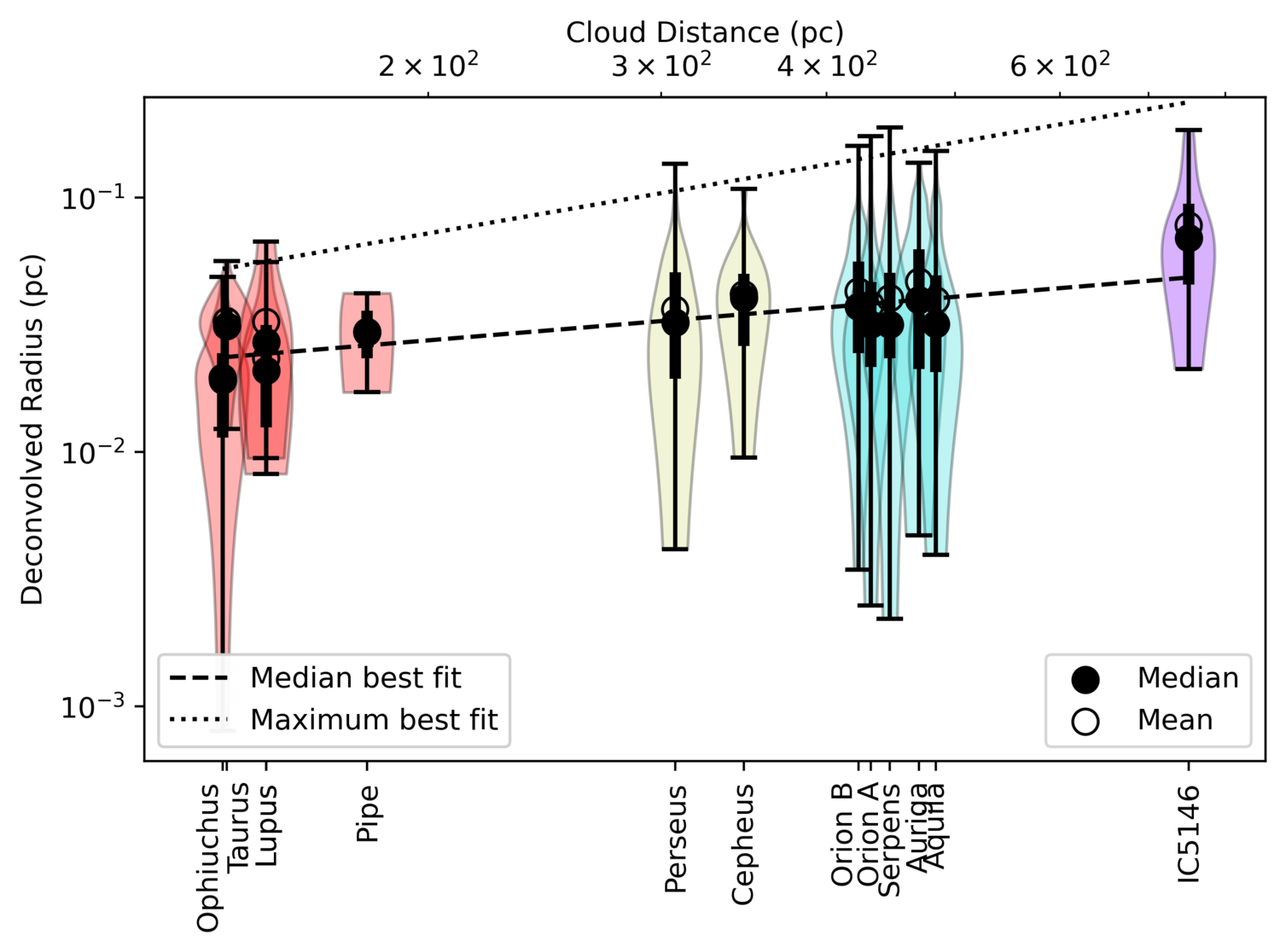}
	\includegraphics[width=\columnwidth]{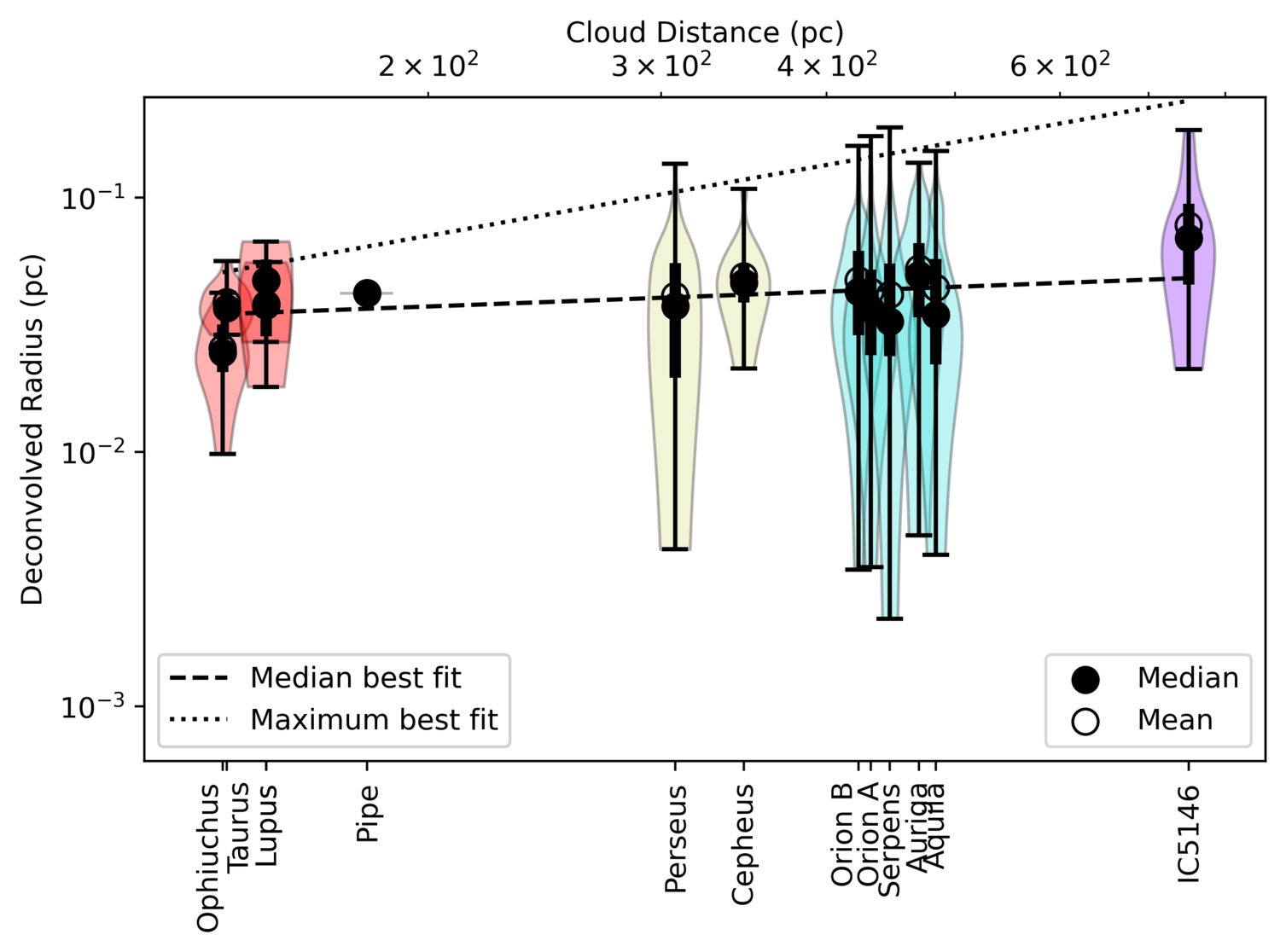}
    \caption{Deconvolved radius distribution for starless cores for each cloud complex, as a function of distance.  Top: for all starless cores.  Bottom: for starless cores with masses above the 90\% completeness limit at 450\,pc of 0.2\,M$_{\odot}$.  Solid circles show median values; open circles show means.  Thick black lines show the interquartile ranges.  Dotted line shows the line of best fit to the maximum values in each cloud complex; dashed line shows the line of best fit to the median values.}
    \label{fig:radius_violin_dist}
\end{figure}

The distributions of starless core masses for each cloud complex as a function of distance are shown in Figure~\ref{fig:mass_violin_dist}, for all cores and for cores with masses above the 90\% completeness limit at 450\,pc of 0.2\,M$_{\odot}$.  The equivalent distributions of deconvolved starless core radius are shown in Figure~\ref{fig:radius_violin_dist}.

\clearpage

\section{Ancillary Core Mass Functions and comparisons of least-squares and Monte-Carlo modelling}
\label{sec:appendix_cmfs}

\begin{figure*}
\includegraphics[width=0.8\textwidth]{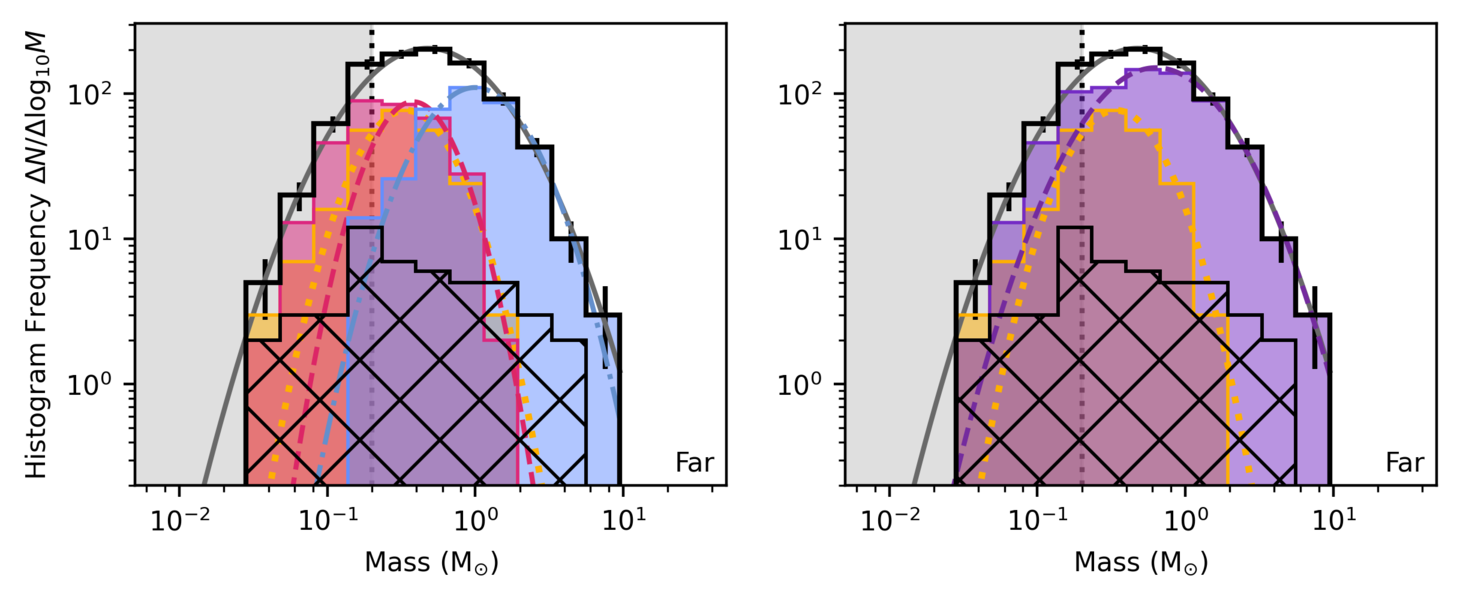}
\caption{Starless CMF for the far-distance clouds in our sample, with cores in Orion A with temperatures $> 15$\,K excluded.  Left: CMFs with fits to full, unbound, candidate prestellar and robust prestellar samples shown.  Right: CMFs with fits to full, unbound, and combined candidate and robust (C \& R) prestellar samples.  In both panels, the unresolved sources are shown as a hatched histogram.  Colours of histograms are as in Figure~\ref{fig:cmf_sless}.}
\label{fig:cmf_sless_far_oflag}
\end{figure*}

\begin{figure*}
	\includegraphics[width=\textwidth]{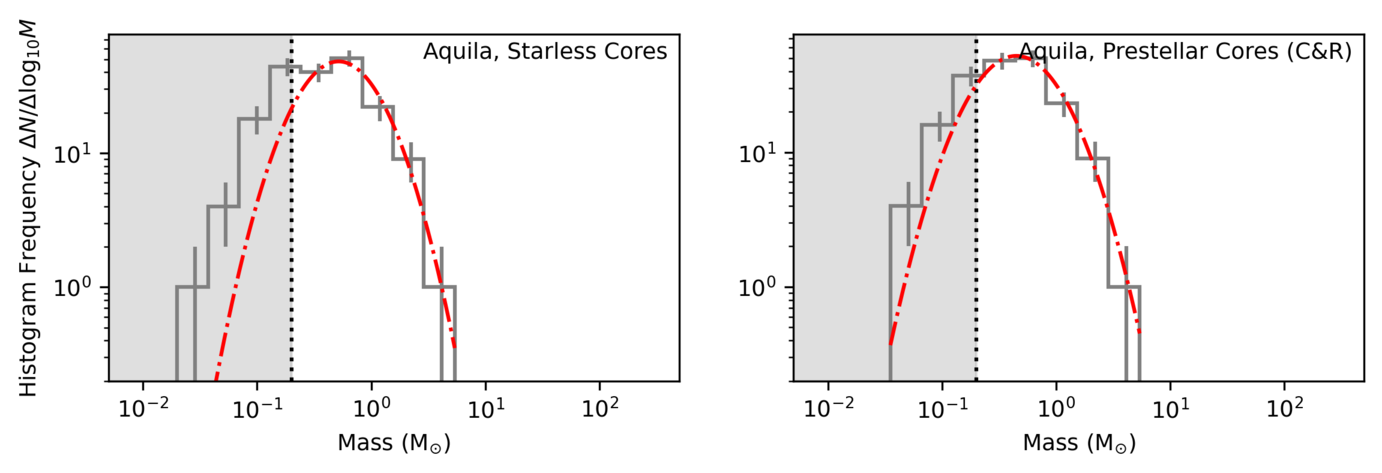}
    \caption{CMFs for Aquila.  Left: starless CMF.  Right: prestellar (C \& R) CMF.  Least-squares best-fit model CMFs are shown as red dot-dashed line.  Areas below the 90\% mass completeness limit are shaded in grey.} 
    \label{fig:cmf_aquila}
\end{figure*}

\begin{figure*}
	\includegraphics[width=\textwidth]{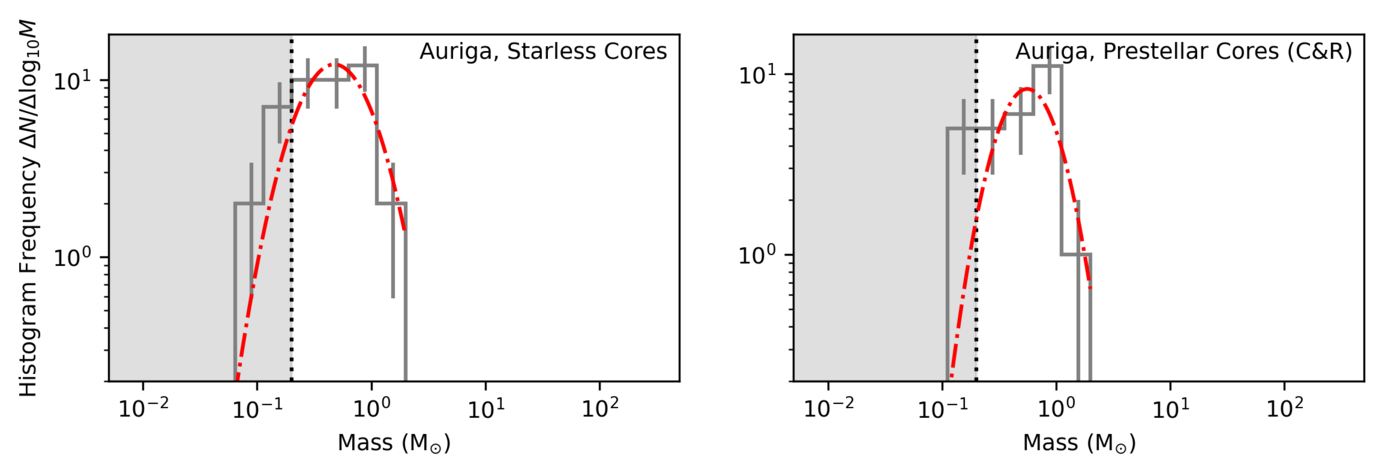}
    \caption{CMFs for Auriga.  Left: starless CMF.  Right: prestellar (C \& R) CMF.  Least-squares best-fit model CMFs are shown as red dot-dashed line.  Areas below the 90\% mass completeness limit are shaded in grey.}
    \label{fig:cmf_auriga}
\end{figure*}

\begin{figure*}
	\includegraphics[width=\textwidth]{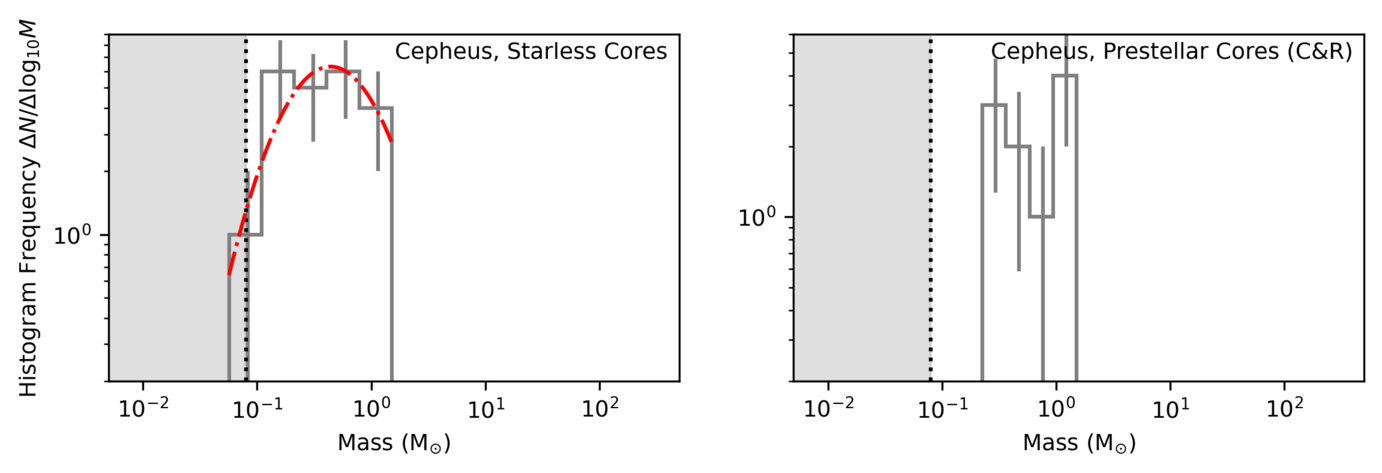}
    \caption{CMFs for Cepheus.  Left: starless CMF.  Right: prestellar (C \& R) CMF.  Least-squares best-fit model CMFs are shown as red dot-dashed line.  Areas below the 90\% mass completeness limit are shaded in grey.}
    \label{fig:cmf_cepheus}
\end{figure*}

\begin{figure*}
	\includegraphics[width=\textwidth]{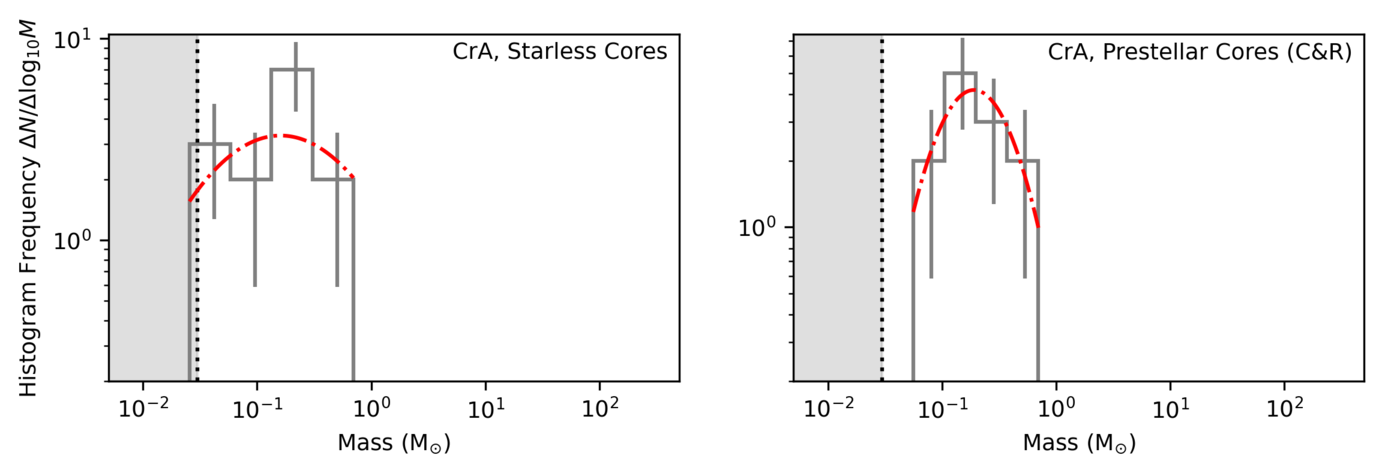}
    \caption{Core mass functions for Corona Australis.  Left: starless CMF.  Right: prestellar (C \& R) CMF.  Least-squares best-fit model CMFs are shown as red dot-dashed line.  Areas below the 90\% mass completeness limit are shaded in grey.}
    \label{fig:cmf_CrA}
\end{figure*}

\begin{figure*}
	\includegraphics[width=\textwidth]{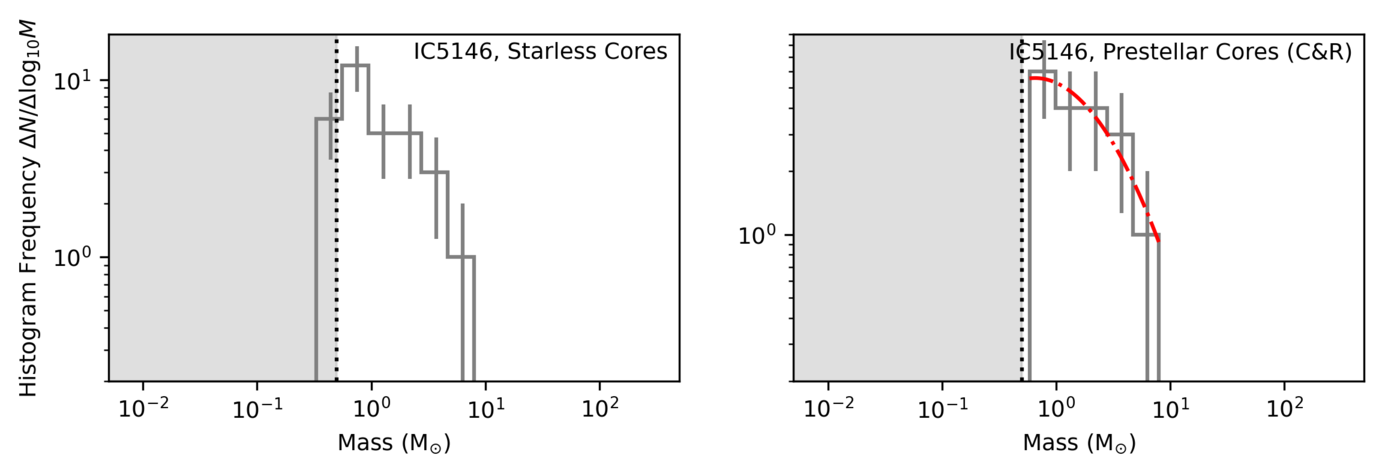}
    \caption{Core mass functions for IC 5146.  Left: starless CMF.  Right: prestellar (C \& R) CMF.  Least-squares best-fit model CMFs are shown as red dot-dashed line.  Areas below the 90\% mass completeness limit are shaded in grey.}
    \label{fig:cmf_IC5146}
\end{figure*}

\begin{figure*}
	\includegraphics[width=0.5\textwidth]{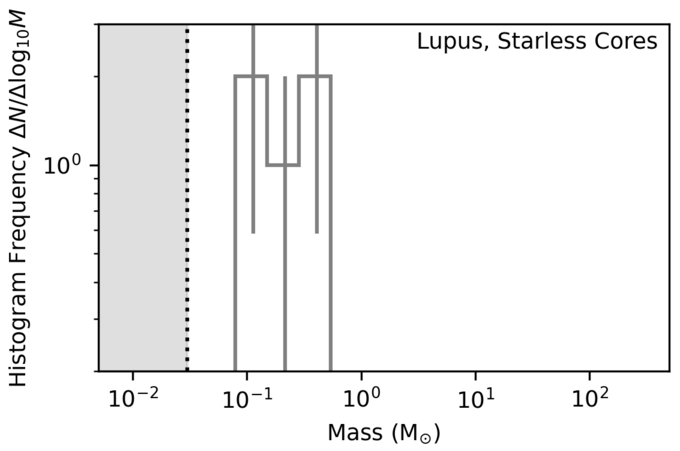}
    \caption{Starless core mass functions for Lupus.  Area below the 90\% mass completeness limit is shaded in grey.}
    \label{fig:cmf_lupus}
\end{figure*}

\begin{figure*}
	\includegraphics[width=\textwidth]{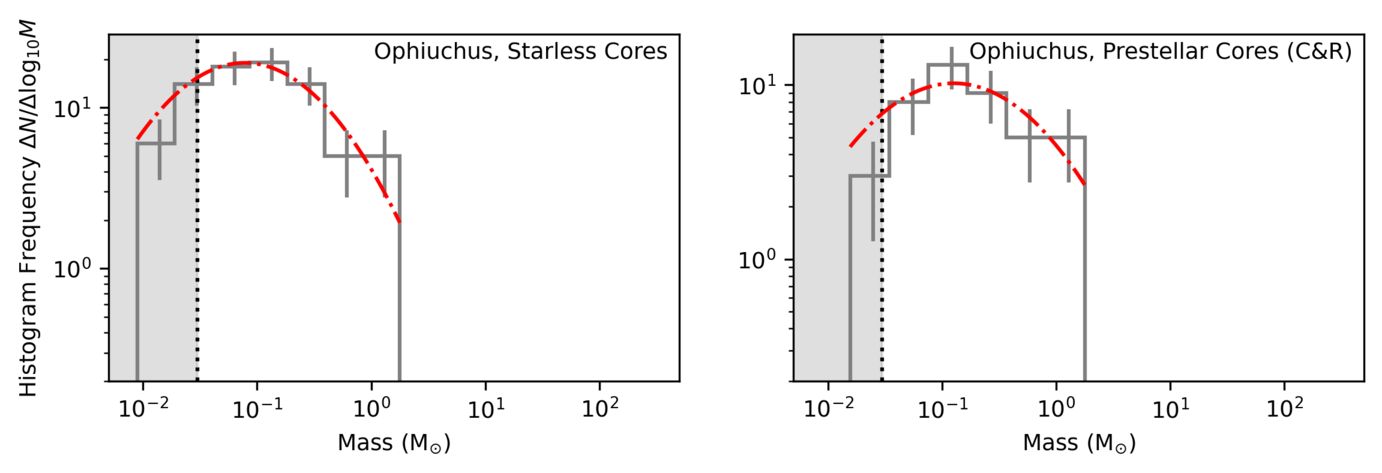}
    \caption{Core mass functions for Ophiuchus.  Left: starless CMF.  Right: prestellar (C \& R) CMF.  Least-squares best-fit model CMFs are shown as red dot-dashed line.  Areas below the 90\% mass completeness limit are shaded in grey.}
    \label{fig:cmf_oph}
\end{figure*}

\begin{figure*}
	\includegraphics[width=\textwidth]{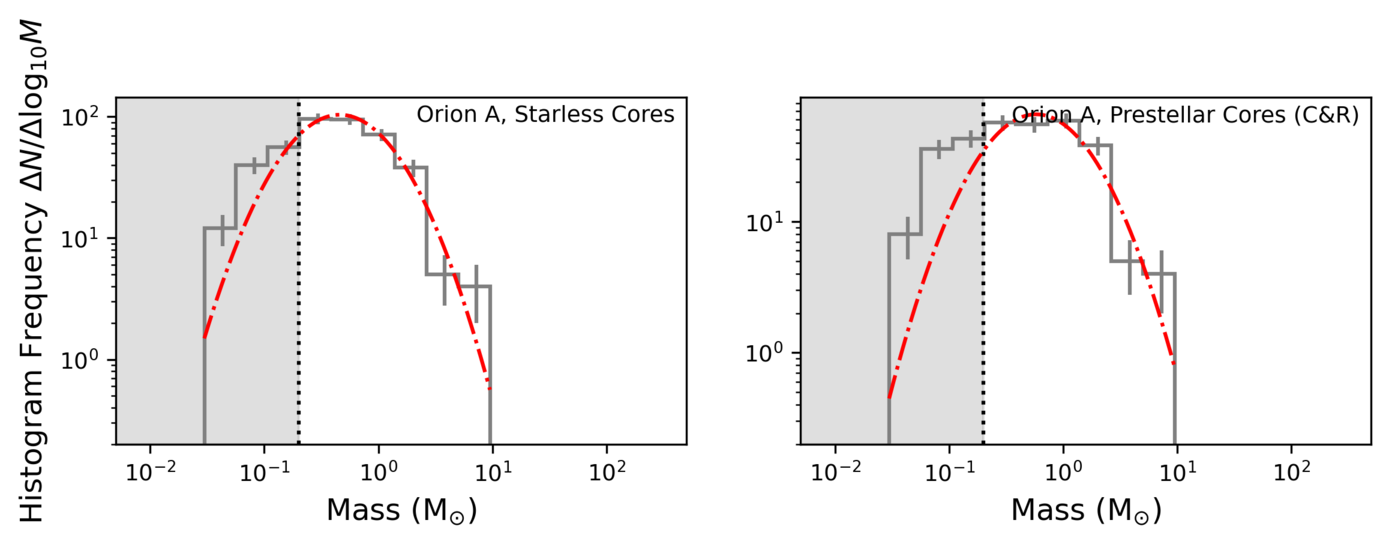}
    \caption{Core mass functions for Orion A.  Left: starless CMF.  Right: prestellar (C \& R) CMF.  Least-squares best-fit model CMFs are shown as red dot-dashed line.  Areas below the 90\% mass completeness limit are shaded in grey.}
    \label{fig:cmf_oriona}
\end{figure*}

\begin{figure*}
	\includegraphics[width=\textwidth]{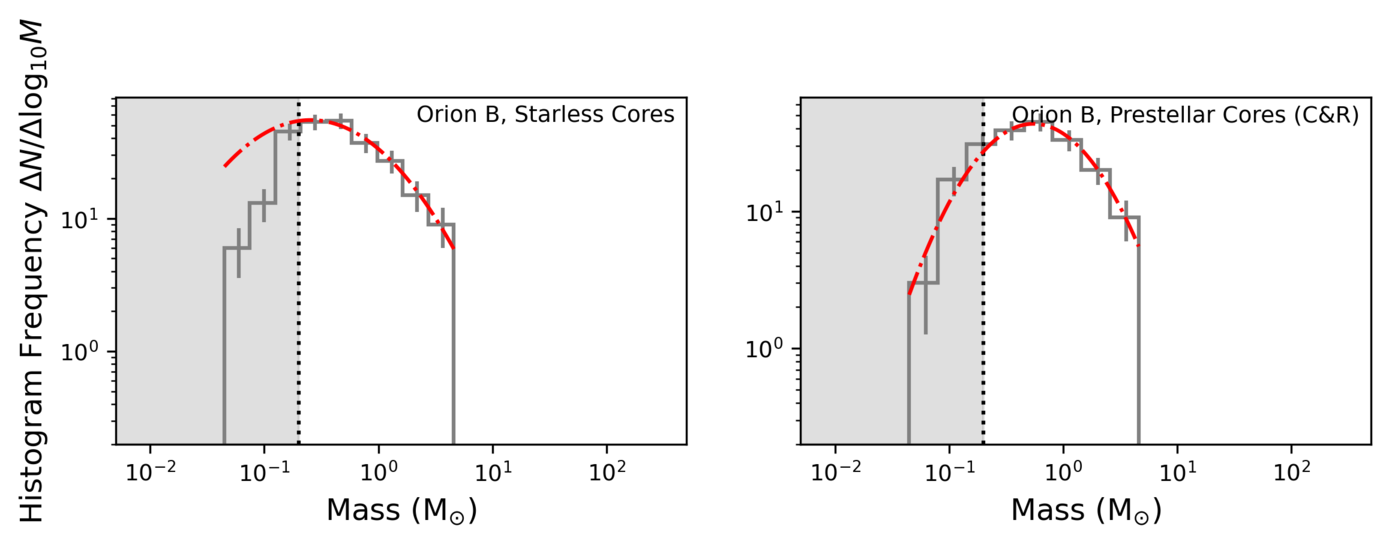}
    \caption{Core mass functions for Orion B.  Left: starless CMF.  Right: prestellar (C \& R) CMF.  Least-squares best-fit model CMFs are shown as red dot-dashed line.  Areas below the 90\% mass completeness limit are shaded in grey.}
    \label{fig:cmf_orionb}
\end{figure*}

\begin{figure*}
	\includegraphics[width=\textwidth]{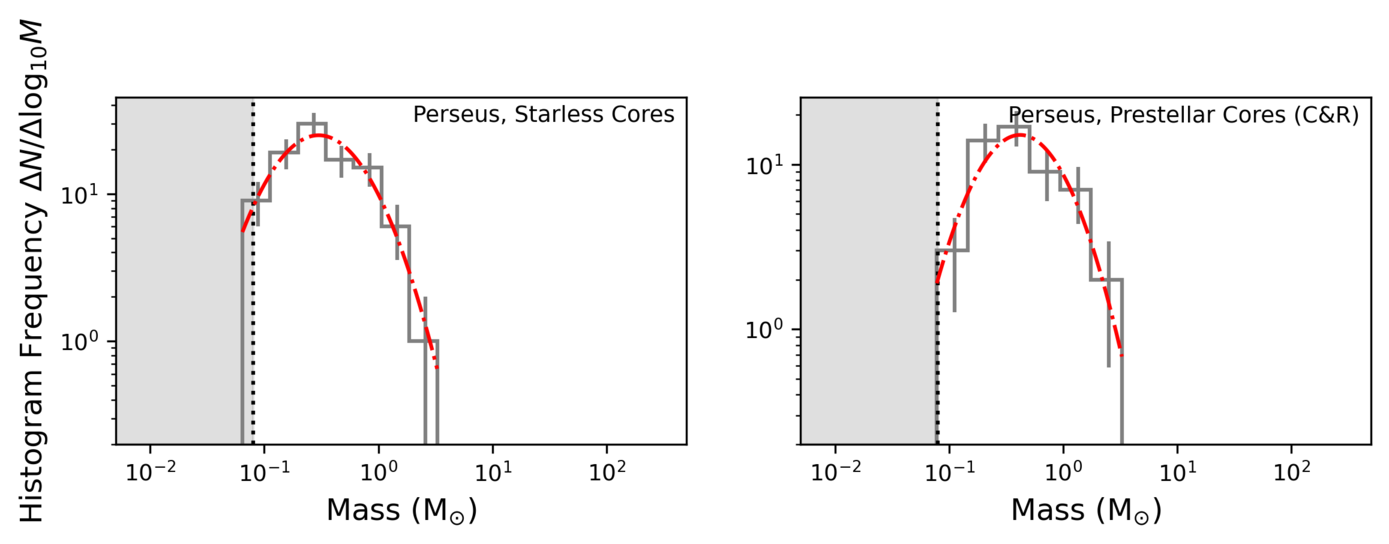}
    \caption{Core mass functions for Perseus.  Left: starless CMF.  Right: prestellar (C \& R) CMF.  Least-squares best-fit model CMFs are shown as red dot-dashed line.  Areas below the 90\% mass completeness limit are shaded in grey.}
    \label{fig:cmf_perseus}
\end{figure*}

\begin{figure*}
	\includegraphics[width=0.5\textwidth]{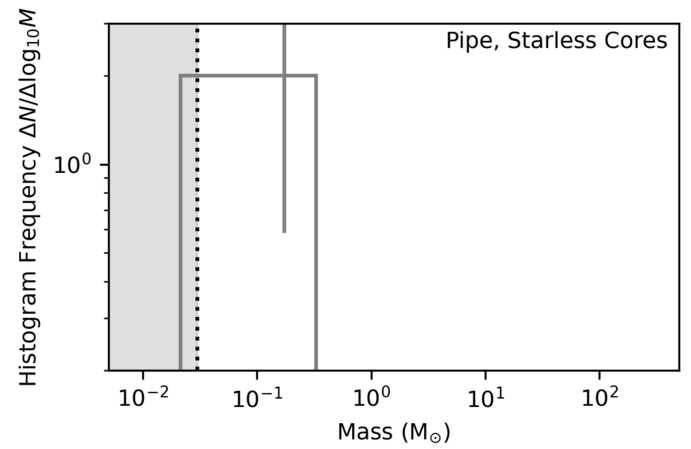}
    \caption{Starless core mass function for the Pipe Nebula.  Area below the 90\% mass completeness limit is shaded in grey.}
    \label{fig:cmf_pipe}
\end{figure*}

\begin{figure*}
	\includegraphics[width=\textwidth]{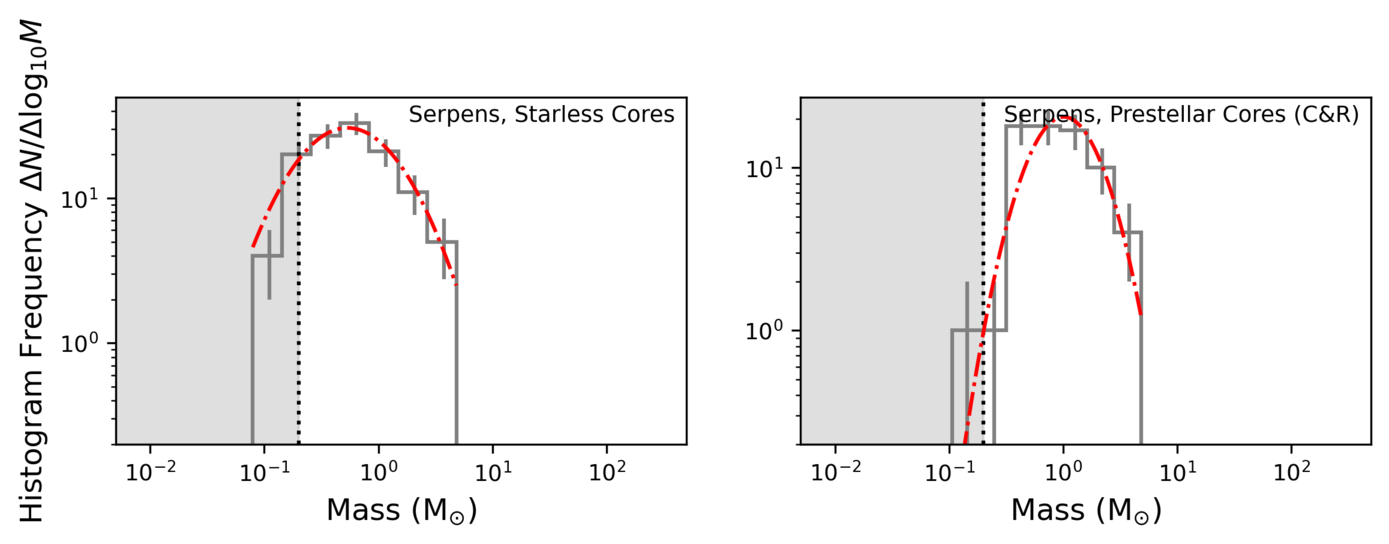}
    \caption{Core mass functions for Serpens.  Left: starless CMF.  Right: prestellar (C \& R) CMF.  Least-squares best-fit model CMFs are shown as red dot-dashed line.  Areas below the 90\% mass completeness limit are shaded in grey.}
    \label{fig:cmf_serpens}
\end{figure*}

\begin{figure*}
	\includegraphics[width=\textwidth]{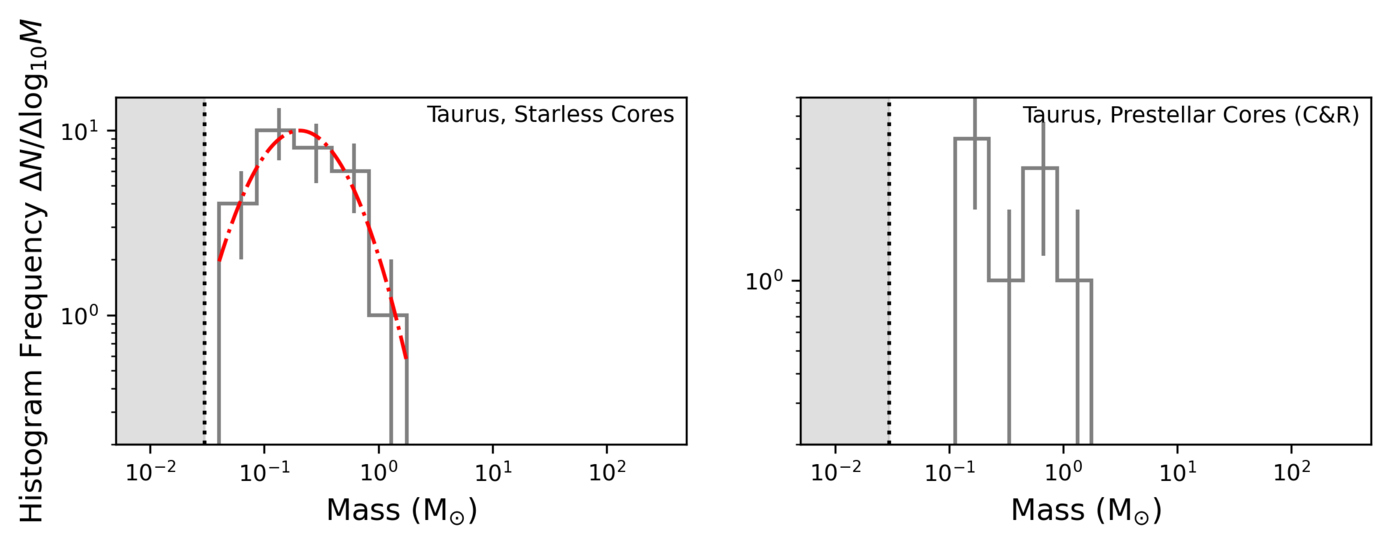}
    \caption{Core mass functions for Taurus.  Left: starless CMF.  Right: prestellar (C \& R) CMF.  Least-squares best-fit model CMFs are shown as red dot-dashed line.  Areas below the 90\% mass completeness limit are shaded in grey.}
    \label{fig:cmf_taurus}
\end{figure*}

In this section, we show ancillary Core Mass Functions for various subsets of our sample.  We first show ancillary results for CMFs plotted by distance (see Section~\ref{sec:cmfs_distance}).  Table~\ref{tab:starless_cmfs_2} presents the least-squares best fits to the unbound, candidate prestellar, and robust prestellar cores in our sample.  Figure~\ref{fig:cmf_sless_far_oflag} shows the CMF for the far-distance clouds, with cores with temperatures $> 15$\,K excluded.  We then show the individual CMFs for each molecular cloud complex that we consider, in Figures~\ref{fig:cmf_aquila}--\ref{fig:cmf_taurus}.  Where relevant, the best-fit histogram produced by least-squares (LS) fitting is overplotted.  These fits are listed in Table~\ref{tab:cloud_cmfs} in the text.

\begin{table*} 
\caption{Least-squares best-fit CMFs for each of our distance ranges, for unbound starless cores, candidate prestellar cores, and robust prestellar cores.  The final two rows of the table show least-squares best fits to the near- and mid-distance samples, for cores above the far-distance mass completeness limit (0.2\,M$_{\odot}$) only.} 
\centering 
\begin{tabular}{c ccc ccc ccc} 
\hline & \multicolumn{3}{c}{Unbound} & \multicolumn{3}{c}{Candidate} & \multicolumn{3}{c}{Robust} \\ \cline{2-4} \cline{5-7} \cline{8-10} 
 & $A$ & $\mu$ & $\sigma$ & $A$ & $\mu$ & $\sigma$ & $A$ & $\mu$ & $\sigma$ \\ 
Range & & (M$_{\odot}$) & ($\log_{10}{\rm M}_{\odot}$) & & (M$_{\odot}$) & ($\log_{10}{\rm M}_{\odot}$) &  & (M$_{\odot}$) & ($\log_{10}{\rm M}_{\odot}$) \\ 
\hline 
0 -- 200\,pc & 16$\pm$4 & 0.12$\pm$0.03 & 0.36$\pm$0.08 & 16$\pm$3 & 0.13$\pm$0.01 & 0.28$\pm$0.03 & 8$\pm$1 & 0.61$\pm$0.09 & 0.38$\pm$0.05 \\
200 -- 355\,pc & 15$\pm$3 & 0.2$\pm$0.1 & 0.5$\pm$0.2 & 12.5$\pm$0.9 & 0.25$\pm$0.01 & 0.21$\pm$0.01 & 11$\pm$3 & 0.6$\pm$0.1 & 0.32$\pm$0.07 \\
355 -- 500\,pc & 80$\pm$10 & 0.27$\pm$0.06 & 0.30$\pm$0.05 & 100$\pm$10 & 0.31$\pm$0.06 & 0.28$\pm$0.05 & 110$\pm$10 & 0.95$\pm$0.05 & 0.29$\pm$0.02 \\
no heated Orion A & 77$\pm$6 & 0.34$\pm$0.04 & 0.27$\pm$0.03 & 88$\pm$9 & 0.38$\pm$0.04 & 0.23$\pm$0.03 & 110$\pm$4 & 1.00$\pm$0.02 & 0.30$\pm$0.01 \\
$>500$\,pc & -- & -- & -- & -- & -- & --  & -- & -- & --  \\
\hline 
\multicolumn{10}{c}{For cores with masses $>0.2\,$M$_{\odot}$ only} \\
\hline
0 -- 200\,pc & -- & -- & -- & -- & -- & --  & -- & -- & -- \\
200 -- 355\,pc & -- & -- & -- & -- & -- & --  & 11$\pm$3 & 0.6$\pm$0.1 & 0.32$\pm$0.07 \\
\hline
\end{tabular} 
\label{tab:starless_cmfs_2} 
\end{table*} 

\subsection{Matched-count sampling}

As well as the Monte Carlo sampling of the CMFs described in Section~\ref{sec:mc_distance}, that matched the total mass of the cores in each region, we also investigated a `matched count' sample.  In this case, we randomly drew a sample of `cores' equal to the number of cores in the far-distance starless or prestellar sample.  From this sample, we then randomly drew a number of `cores' equal to the number of cores in the near-distance starless or prestellar sample, and repeated the exercise for the mid-distance starless or prestellar sample.

Broadly, the matched mass sample and the matched count sample produce very similar mean and median values, with the matched-mass method producing a larger area of the parameter space having at least some consistency with the observed sample.  

\begin{table*}
\caption{The most probable starless and prestellar CMFs for each distance range that we consider, as determined from Monte Carlo estimation and two-sided KS tests, \textbf{using matched-count sampling}.  $p$ values show the probability that this model and our sample are drawn from the same underlying distribution. }
\centering
\begin{tabular}{@{\extracolsep{4pt}}c cccccc@{}}
\hline
 & \multicolumn{3}{c}{Starless} & \multicolumn{3}{c}{Prestellar (C \& R)} \\ \cline{2-4} \cline{5-7}
Range & $\mu$ & $\sigma$ & $p$ & $\mu$ & $\sigma$ & $p$  \\ 
 & $({\rm M}_{\odot})$ & $(\log_{10}{\rm M}_{\odot})$ &  & $({\rm M}_{\odot})$ & $(\log_{10}{\rm M}_{\odot})$ &  \\ 
\hline
0--200 pc & $0.14$ & $0.44$ & $0.80$ & $0.19$ & $0.47$ & $0.77$ \\
200--355 pc & $0.33$ & $0.36$ & $0.72$ & $0.40$ & $0.35$ & $0.69$ \\
355--500 pc & $0.40$ & $0.44$ & $0.65$ & $0.56$ & $0.39$ & $0.68$ \\
\hline
\end{tabular}
\label{tab:mc_region_cmfs_dist_sampcount}
\end{table*}

\begin{table*}
\caption{The most probable starless and prestellar CMFs for each cloud complex that we consider, as determined from Monte Carlo estimation and two-sided KS tests, \textbf{using matched-count sampling}.  Median $p$ values, showing the probability that this model and our sample are drawn from the same underlying distribution, are given.}
\centering
\begin{tabular}{@{\extracolsep{4pt}}c cccccc@{}}
\hline
 & \multicolumn{3}{c}{Starless} & \multicolumn{3}{c}{Prestellar (C \& R)} \\ \cline{2-4} \cline{5-7}
Region & $\mu$ & $\sigma$ & $p$ & $\mu$ & $\sigma$ & $p$ \\ 
 & $({\rm M}_{\odot})$ & $(\log_{10}{\rm M}_{\odot})$ &  & $({\rm M}_{\odot})$ & $(\log_{10}{\rm M}_{\odot})$ &  \\ 
\hline
\multicolumn{7}{c}{Near} \\ 
\hline
CrA& $0.17$ & $0.27$ & $0.77$ & $0.16$ & $0.24$ & $0.87$  \\ 
Ophiuchus& $0.10$ & $0.49$ & $0.77$ & $0.15$ & $0.56$ & $0.71$  \\ 
Taurus& $0.17$ & $0.26$ & $0.79$ &  -- & -- & --  \\ 
\hline
\multicolumn{7}{c}{Mid} \\ 
\hline
Cepheus& $0.38$ & $0.37$ & $0.83$ & $0.45$ & $0.35$ & $0.79$  \\ 
Perseus& $0.31$ & $0.35$ & $0.70$ & $0.38$ & $0.35$ & $0.74$  \\ 
\hline
\multicolumn{7}{c}{Far} \\ 
\hline 
Aquila& $0.39$ & $0.37$ & $0.66$ & $0.41$ & $0.36$ & $0.66$ \\ 
Auriga& $0.46$ & $0.31$ & $0.72$ & $0.58$ & $0.24$ & $0.69$ \\ 
Orion A& $0.43$ & $0.42$ & $0.53$ & $0.60$ & $0.41$ & $0.53$ \\ 
Orion B& $0.35$ & $0.49$ & $0.67$ & $0.55$ & $0.42$ & $0.74$  \\ 
Serpens& $0.50$ & $0.40$ & $0.71$ & $0.89$ & $0.27$ & $0.56$ \\ 
\hline
\end{tabular}
\label{tab:mc_region_cmfs_sampcount}
\end{table*}

\begin{figure*}
    \centering
    \includegraphics[width=0.5\textwidth]{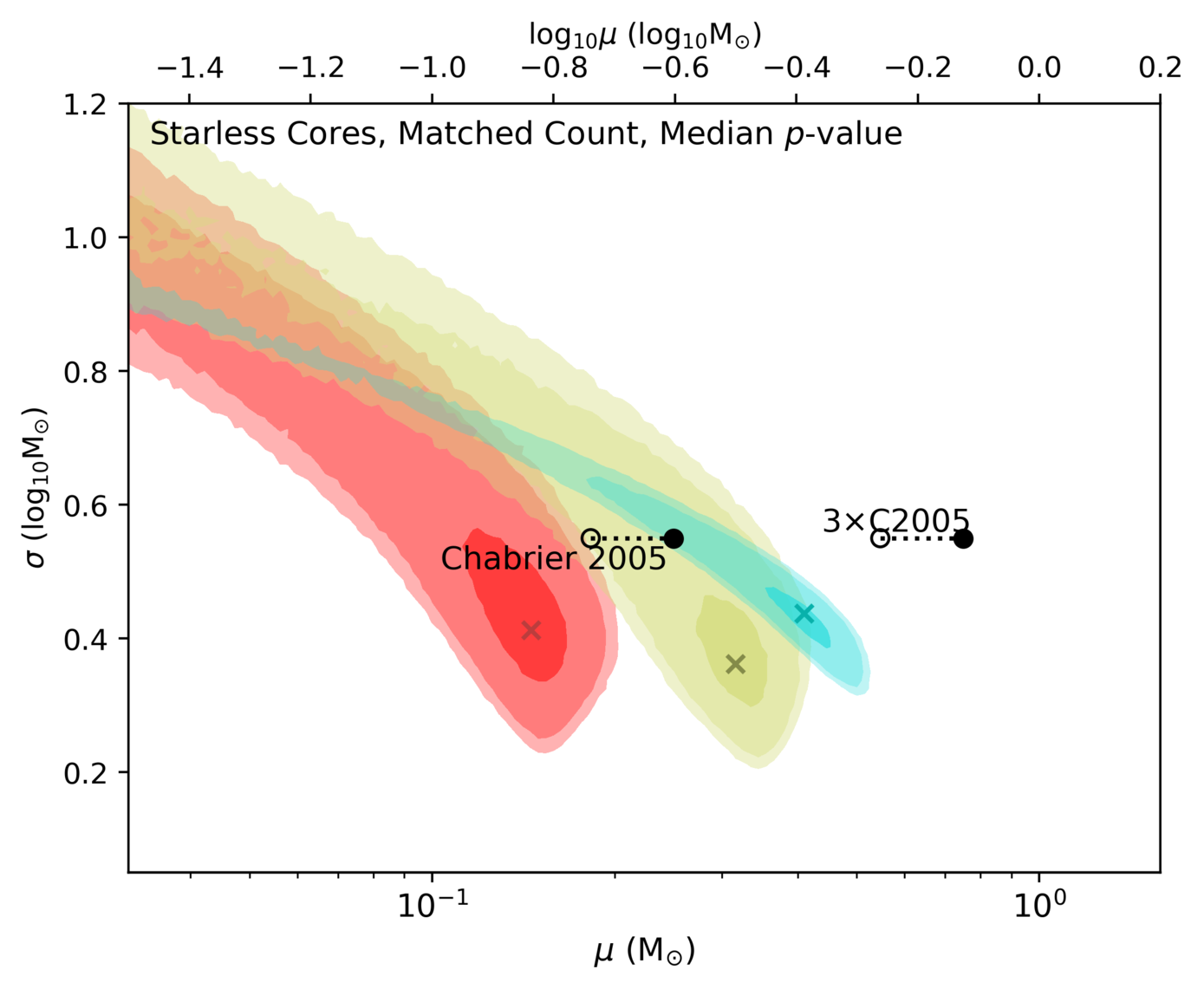}\includegraphics[width=0.5\textwidth]{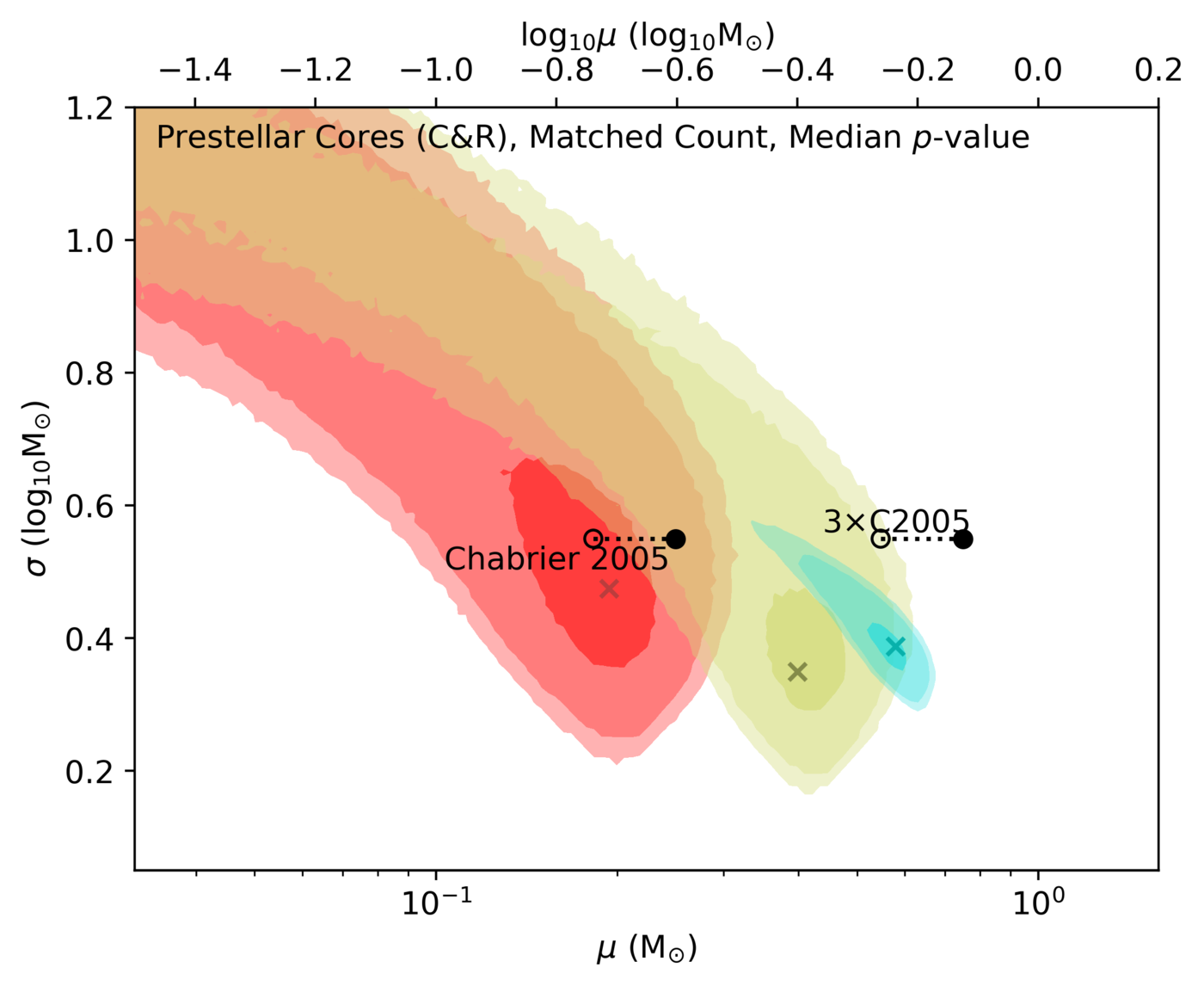}
    \caption{Median $p$-values for two-sided KS test between model CMFs and starless (left) and prestellar (right) CMFs, using the `matched count' sampling method.  Contours show $p$ values of 0.05, 0.1 and 0.5.  Red marks near-, green mid- and blue far-distance CMFs.  Filled circles mark 1 and 3$\times$ the \citet{chabrier2005} peak system mass (0.25\,M$_{\odot}$).  Open circles mark 0.73$\times$ these values, accounting for the typical flux loss in our SCUBA-2 observations.}
    \label{fig:mc_3way_sampcount}
\end{figure*}

\subsection{Comparing Least Squares fits and Monte-Carlo models of CMFs}
\label{sec:appendix_cmfs_lsmc}

\begin{figure*}
    \centering
    \includegraphics[width=0.5\textwidth]{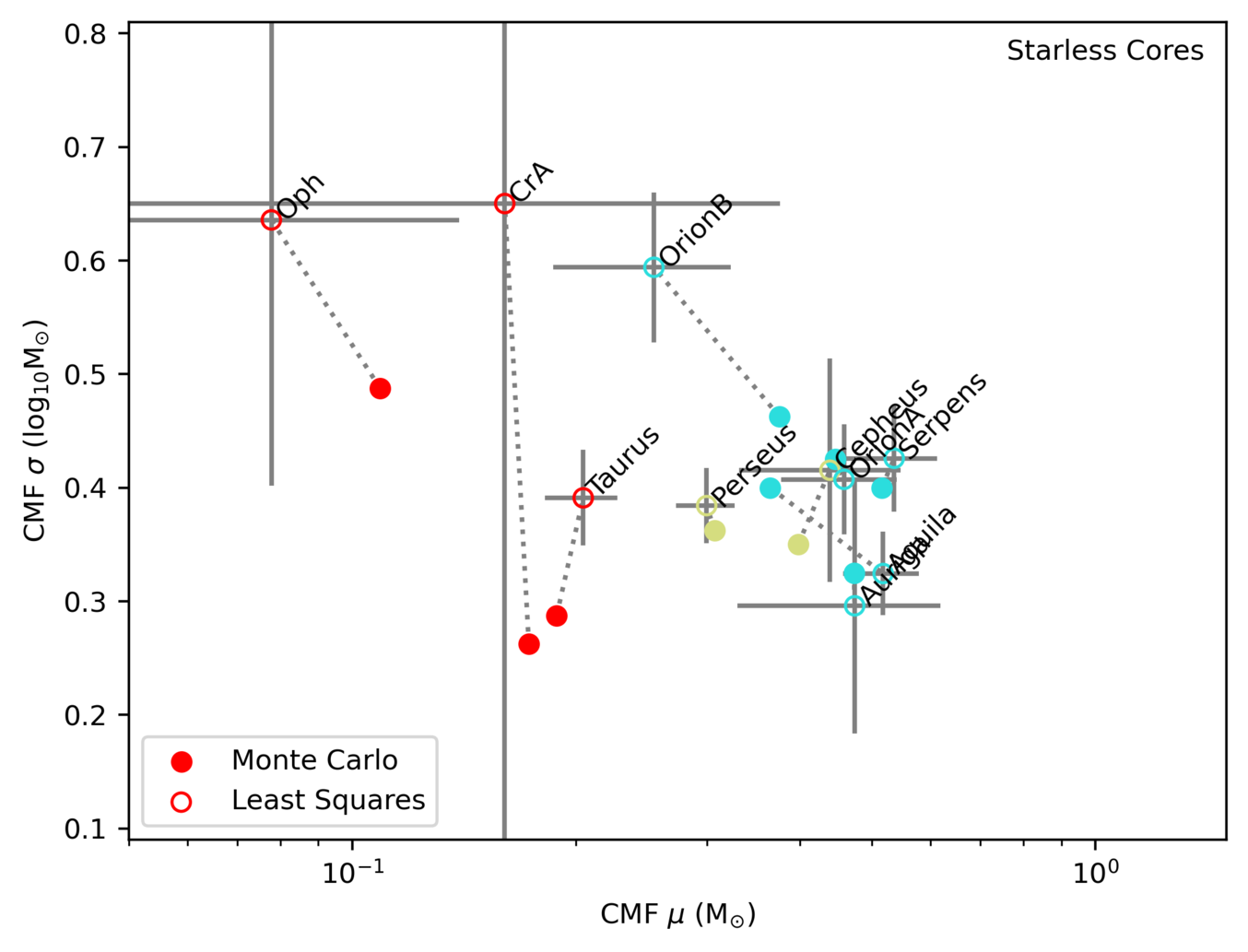}\includegraphics[width=0.5\textwidth]{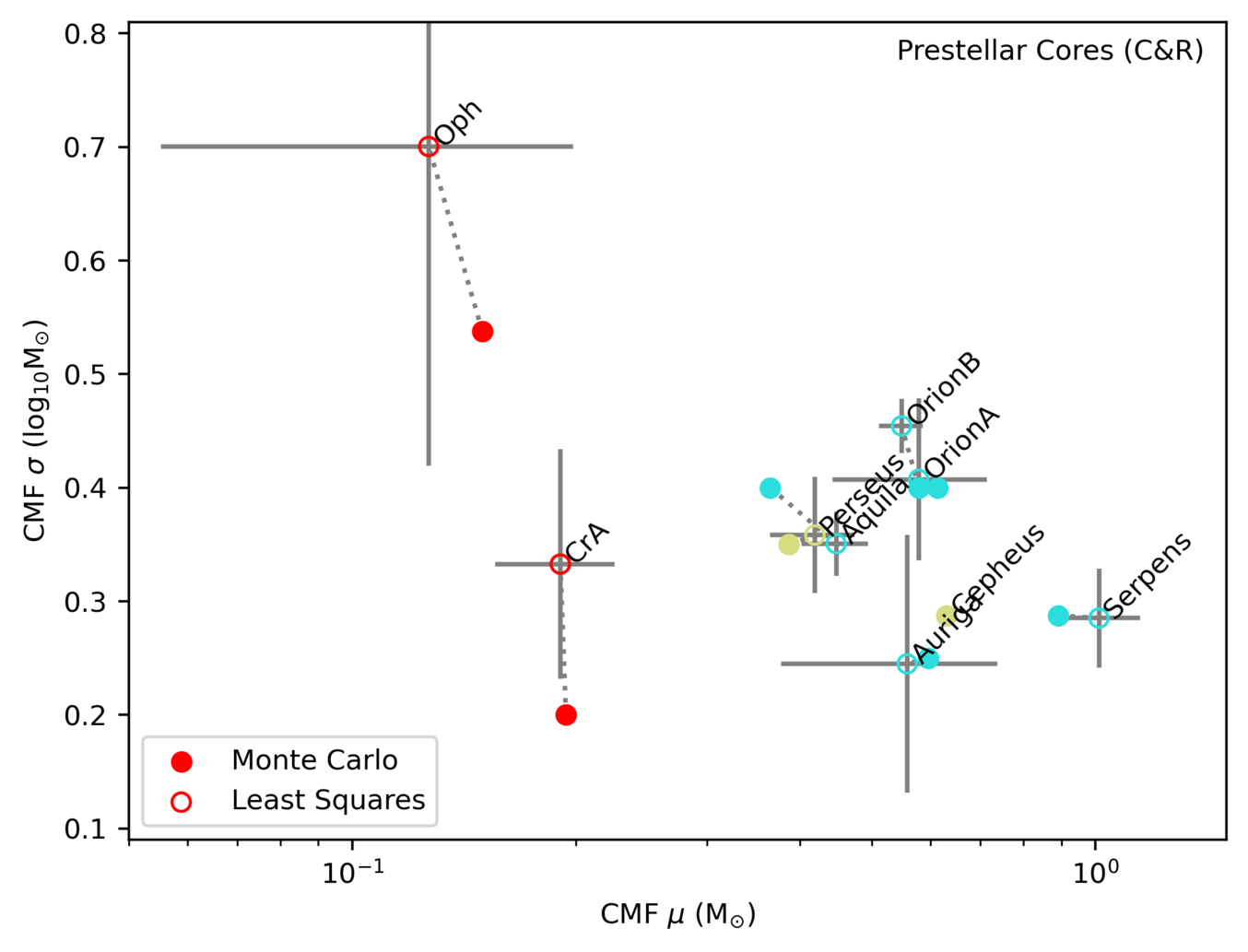}
    \caption{Comparison between the CMF properties (peak mass, $\mu$, and width, $\sigma$) determined from LS and MC methods for each cloud complex.  Left: starless cores.  Right: prestellar (C \& R) cores.  In both panels, open circles indicate the best-fit LS values, while closed circles indicate the most probable MC values.  Cloud complexes are colour-coded by their distance range.}
    \label{fig:cmf_params_ls_mc}
\end{figure*}

Figure~\ref{fig:cmf_params_ls_mc} compares the best-fit least squares (LS) and most-probable Monte Carlo (MC)-derived CMFs for each cloud complex.  The two methods produce similar results, although the most probable MC values do not always fall within the LS fitting errors.  The LS and MC prestellar CMFs are much more similar to one another than are the LS and MC starless CMFs, likely because the prestellar cores are typically higher-mass, and so their peak masses are better-constrained.

\begin{figure*}
    \centering
    \includegraphics[width=0.5\textwidth]{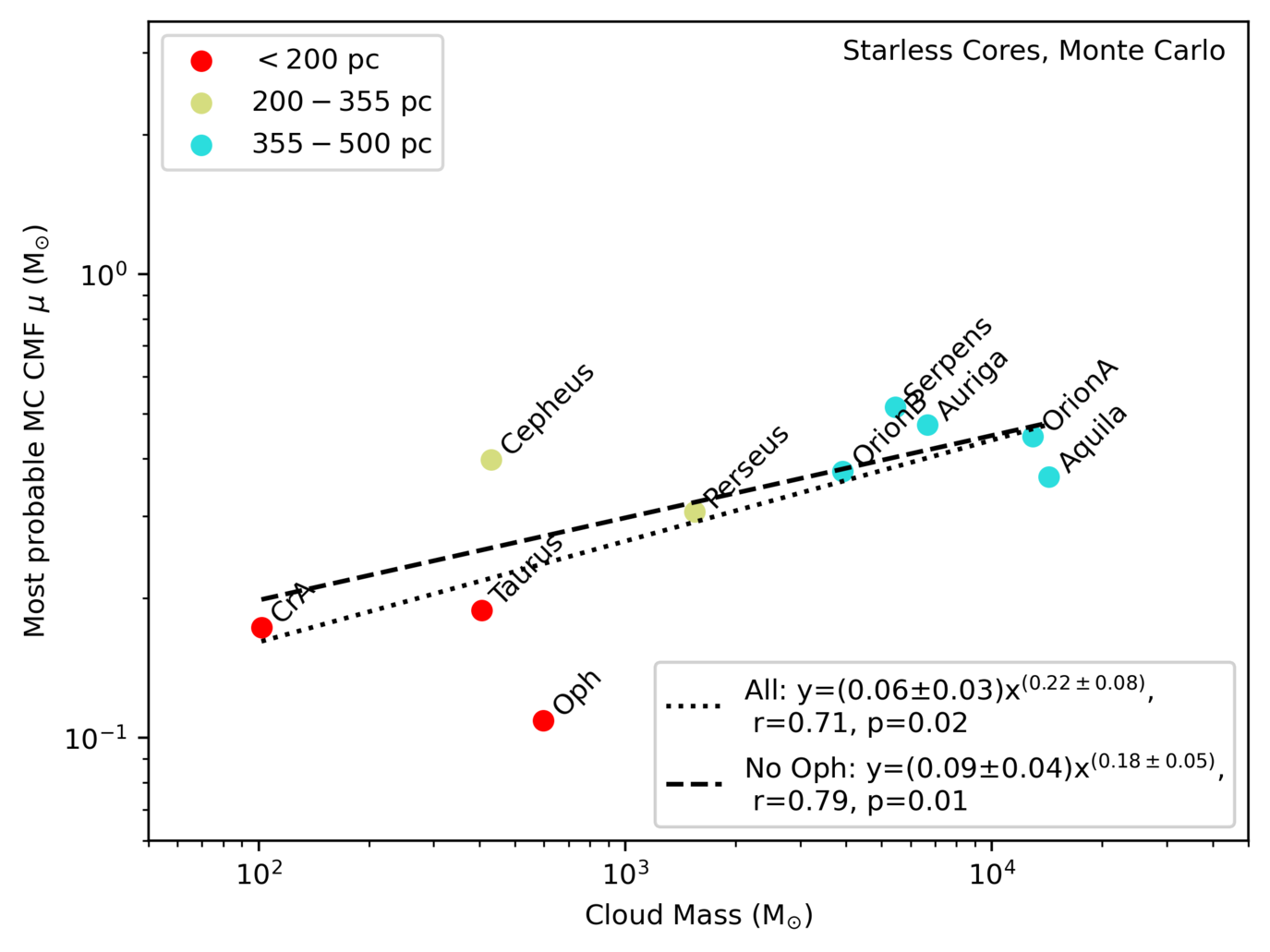}\includegraphics[width=0.5\textwidth]{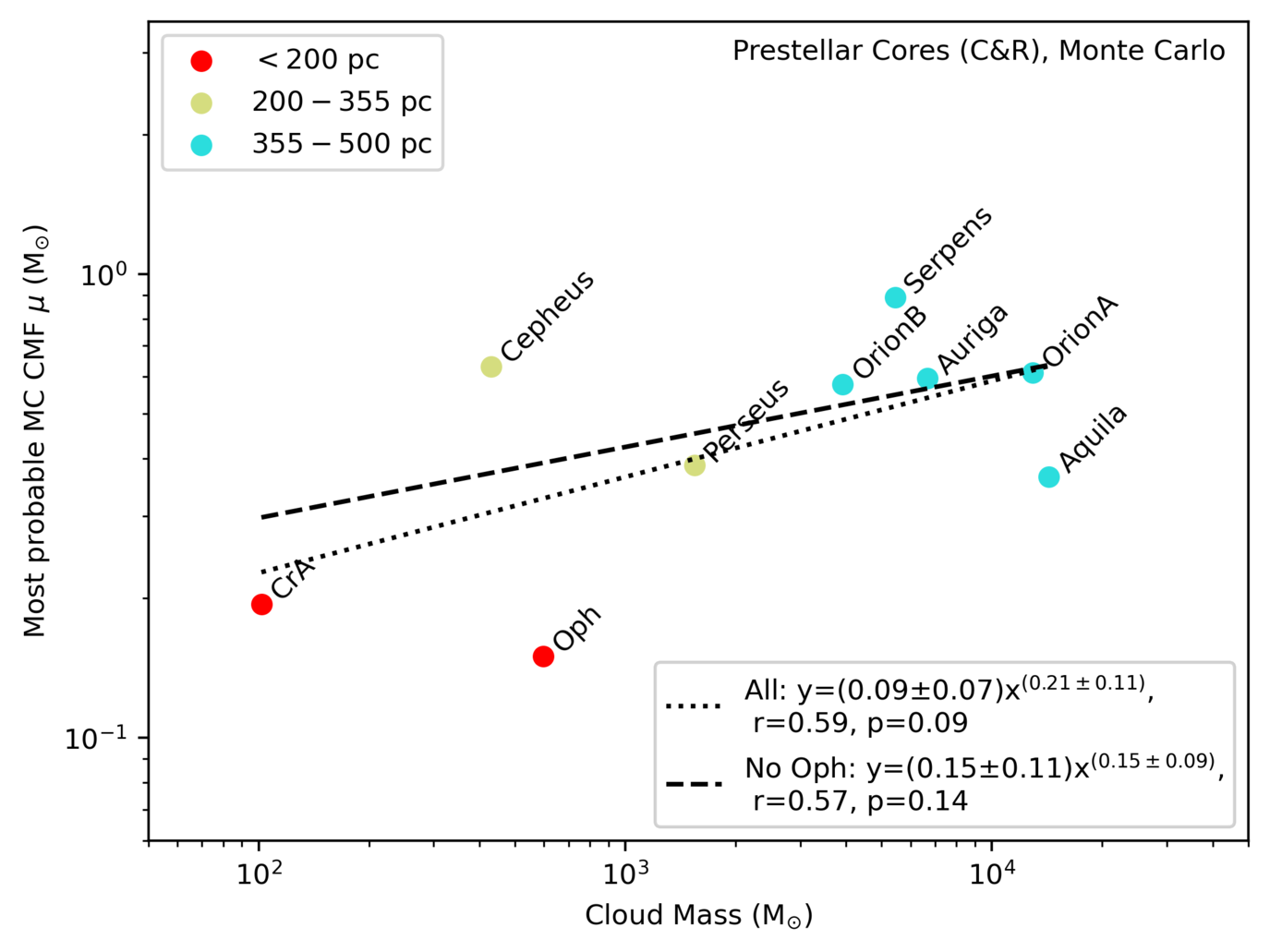}
    \caption{The peak masses of the starless and prestellar CMFs as a function of cloud complex mass, determined using Monte Carlo modelling.  Left: starless CMFs.  Right: prestellar CMFs.  Dotted line shows the power-law model producing the best fit to all data points; dashed line shows the power-law model producing the best fit to the data points with Ophiuchus, which has a notably low peak mass, excluded.  Data points are colour-coded by their distance range.}
    \label{fig:peak_cmf_mc}
\end{figure*}

\begin{figure*}
    \centering
    \includegraphics[width=0.5\textwidth]{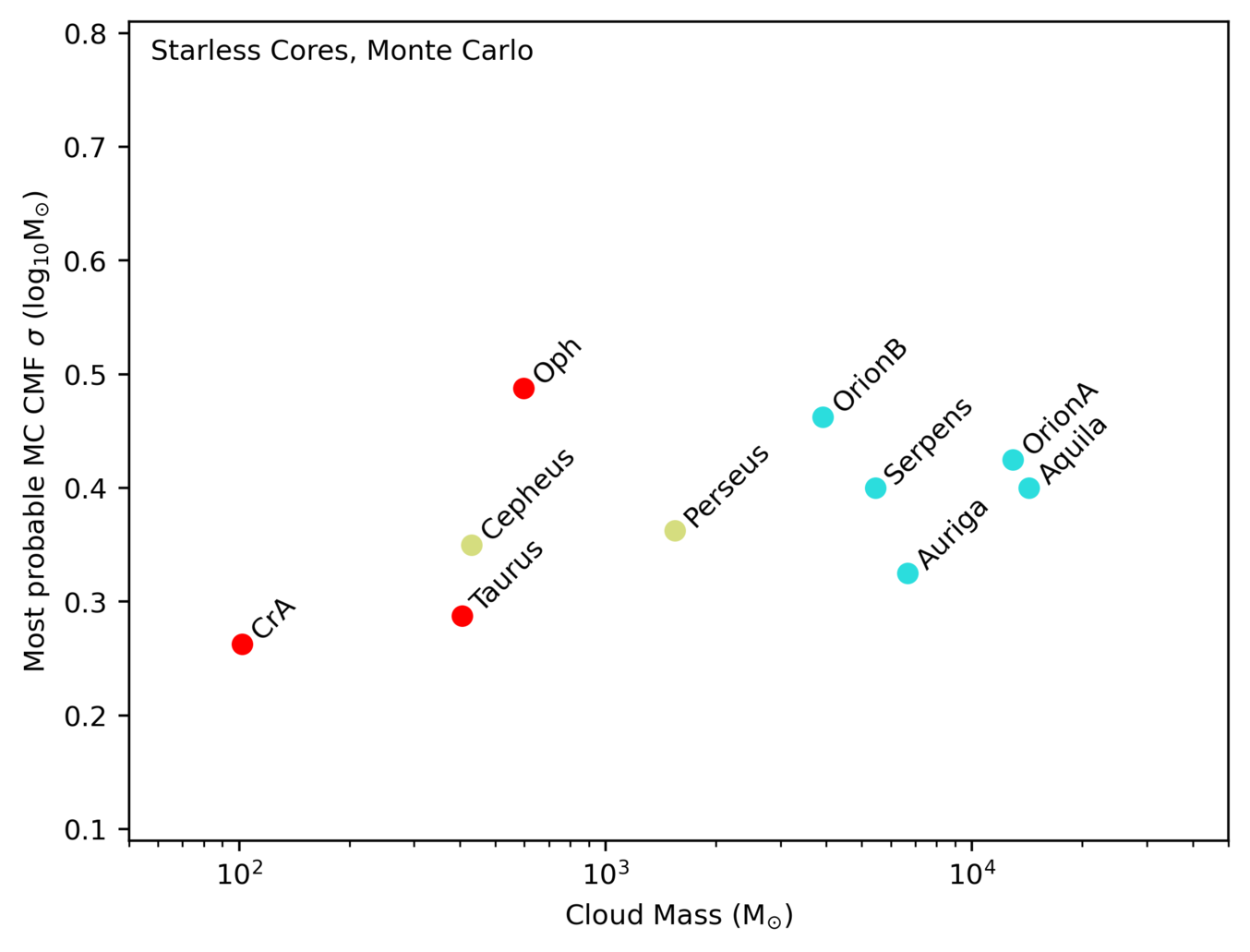}\includegraphics[width=0.5\textwidth]{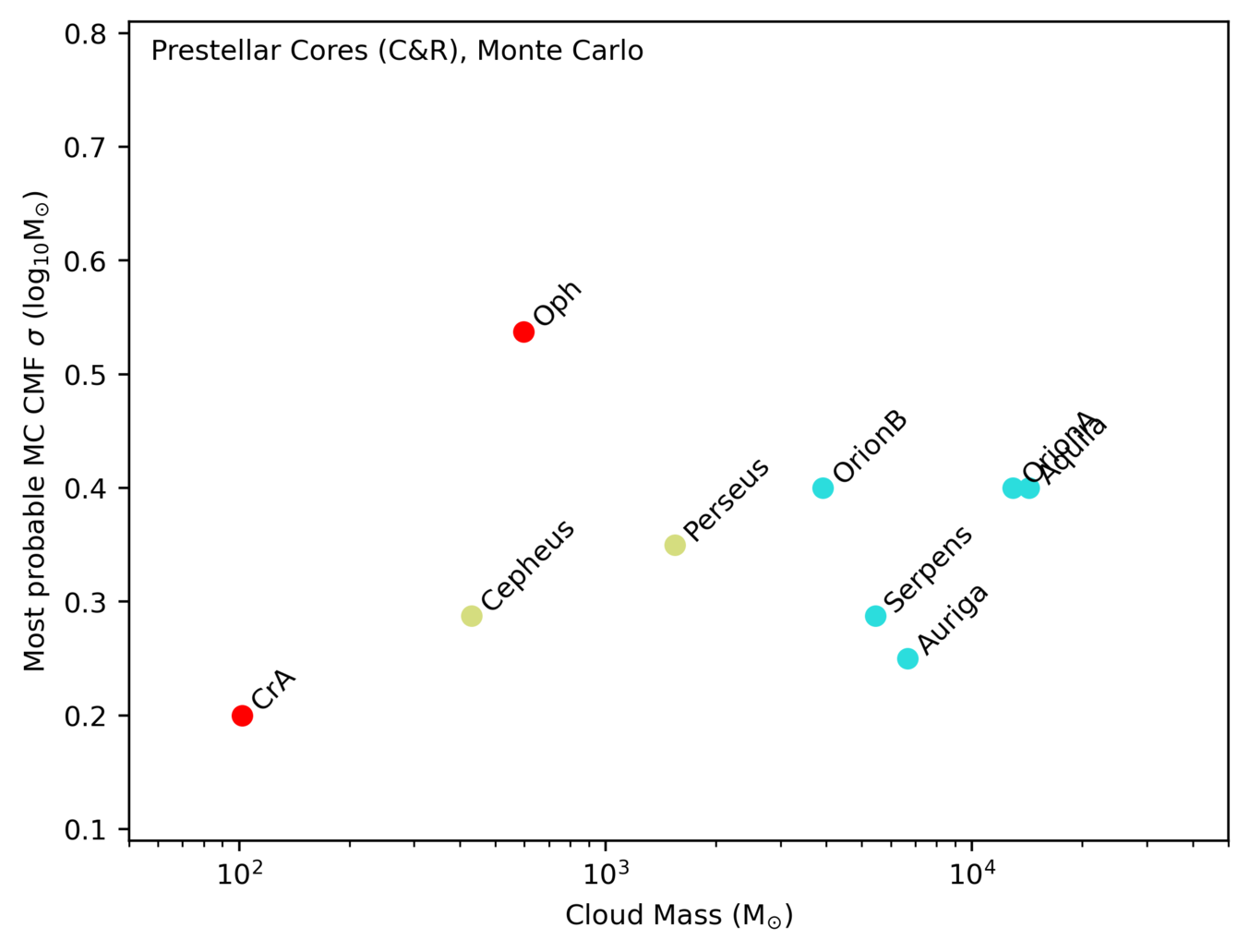}
    \caption{The width of the starless (left) and prestellar (right) CMFs as a function of cloud complex mass, determined using Monte Carlo modelling.   Data points are colour-coded by their distance range.}
    \label{fig:width_cmf_mc}
\end{figure*}

Figures~\ref{fig:peak_cmf_mc} and \ref{fig:width_cmf_mc} show the peak masses and widths of the starless and prestellar CMFs, determined using Monte Carlo modelling, as a function of cloud mass.

\bsp
\label{lastpage}
\end{document}